\documentclass[JHEP,12pt]{article}
\usepackage{amsmath,amssymb}
\usepackage{booktabs}
\usepackage{epsfig}
\usepackage{graphicx}
\usepackage{color}
\usepackage{shuffle}
\usepackage{wrapfig}
\usepackage{cases}
\usepackage[utf8]{inputenc}
\usepackage{mathtools,tabu}
\usepackage{slashed}
\usepackage[title]{appendix}
\usepackage[compat=1.1.0]{tikz-feynman}
\usepackage{tikz}
\usepackage{setspace}

\usepackage{jheppub}
\definecolor{NUpurple}{RGB}{078,042,132}

\newcommand\nothing[1]{}

\def\sect#1{Section~{\ref{#1}}}

\def\includegraph#1#2{\includegraphics[scale=1.3,trim=0 #1 0 0]{figs/#2.pdf}}

\author[1]{Tim Adamo,}
\author[2,3]{John Joseph M. Carrasco,}
\author[4]{Mariana Carrillo-González,}
\author[5]{Marco Chiodaroli,}
\author[6]{Henriette Elvang,}
\author[3,5,7]{Henrik Johansson,}
\author[8,3]{Donal O'Connell,}
\author[9]{Radu Roiban,}
\author[3,5]{Oliver Schlotterer}

\affiliation[1]{School of Mathematics and Maxwell Institute for Mathematical Sciences \\ University of Edinburgh, EH9 3FD, Scottland, UK}
\affiliation[2]{Department of Physics and Astronomy, Northwestern University, Evanston, Illinois 60208, USA}
\affiliation[3]{Kavli Institute for Theoretical Physics, University of California, Santa Barbara, CA 93106, USA}
\affiliation[4]{Theoretical Physics, Blackett Laboratory, Imperial College, London, SW7 2AZ, U.K.}
\affiliation[5]{Department of Physics and Astronomy, Uppsala University, Box 516, 75120 Uppsala, Sweden}
\affiliation[6]{
    Leinweber Center for Theoretical Physics,\\ 
    Randall Laboratory of Physics, Department of Physics, University of Michigan, \\
    450 Church St, Ann Arbor, MI 48109, USA}
\affiliation[7]{Nordita, Stockholm University and KTH Royal Institute of Technology, Hannes Alfvéns väg 12, 10691 Stockholm, Sweden}
\affiliation[8]{Higgs Centre for Theoretical Physics, School of Physics and Astronomy, The University of Edinburgh, EH9 3FD, Scotland, UK}
\affiliation[9]{Institute for Gravitation and the Cosmos, 
Pennsylvania State University, University Park, PA 16802, USA}
\begin{document}

\title{Snowmass White Paper: the Double Copy and its Applications}
\abstract{The double copy is, in essence, a map between scattering amplitudes in a broad variety of familiar field and string theories. In addition to the mathematically rich intrinsic structure, it underlies a multitude of active research directions and has a range of interesting applications in quantum, classical and effective field theories, including broad topics such as string theory, particle physics, astrophysics, and cosmology. This Snowmass white paper provides a brief introduction to the double copy, its applications, current research and future challenges.

\vspace{5mm}
\noindent {\bf Preprint Numbers:} LCTP-22-05, UUITP-21/22
}

\maketitle

\section{Introduction}
 
Scattering amplitudes provide a powerful arena for advancing our understanding of relativistic quantum field theory and its applications in particle physics and beyond. Traditional Feynman diagrammatics grow rapidly in complexity and computational difficulty with increasing number of external particles and loop order. In contrast, modern amplitudes methods have provided not only computational power to calculate otherwise intractable processes, but they have also shed new light on field theories, both classical and quantum. 

The {\em amplitudes program } takes the physical observables, the scattering amplitudes, as the central objects of interest and constructs them directly from knowledge about their analytical structure and symmetries, often without direct reference to Lagrangians. Crucially, the amplitudes program rests on two pillars: (1) its connection to and direct relevance for particle phenomenology and experiment and (2) its strong potential to further our fundamental understanding of quantum field theories. Moreover, the program  makes significant and lasting connections to adjacent fields: pure mathematics, string theory, and gravitational wave physics.\footnote{See e.g.~ref.~\cite{Travaglini:2022uwo} for a recent edited compilation of reviews on scattering amplitudes.}

In this Snowmass 2021 white paper, we describe recent progress, challenges, and opportunities in research on {\it the double copy}.
Conceptually, the double copy~\cite{Kawai:1985xq,Bern:2008qj,Bern:2010ue, Cachazo:2013gna, Cachazo:2013hca} is a way to calculate amplitudes in one theory using, as input,  amplitudes from one, or two different, simpler theories. It is a multiplicative and bi-linear operation, hence the name ``double copy" is appropriate. 

As an example, the double copy expresses graviton tree amplitudes in terms of sums of products of gluon tree amplitudes: this relation is obscured at the level of off-shell Lagrangians and Feynman rules, but takes a surprisingly simple form in terms of on-shell amplitudes. For example, the 4-graviton tree amplitude of Einstein gravity is simply
\begin{equation}
\label{M4A4sq}
 M_4 = -\frac{s_{12}\,s_{14}}{s_{13}} \,A_4[1234]^2\,,
\end{equation}
where $A_4[1234]$ is the color-stripped four-gluon\footnote{We use ``gluon" to denote the on-shell excitation of a massless non-abelian adjoint vector field, independent of whether the gauge group is the QCD $SU(3)$ or a more general gauge group.} amplitude of Yang-Mills (YM) theory and $s_{ij}=(p_i+p_j)^2$ are the Mandelstam variables. Explicit relations like (\ref{M4A4sq}) are known for any multiplicity of external states. This relationship is often summarized compactly as 
\begin{equation}
  \label{GRisYM2}
  \text{gravity} = \text{(Yang{\rm -}Mills)}^2\,.
\end{equation}

The double copy was first observed in string theory. In the mid-1980s it was found~\cite{Kawai:1985xq}  that closed string tree amplitudes could be written as sums of products of open string amplitudes. This is a non-trivial manifestation, at the level of physical observables, of the factorization of closed string states into products of left- and right-moving open string states. The low-energy limit  of the string-theory double copy is the field-theory double copy exemplified in~\eqref{M4A4sq}. 

The existence of a multiplicative structure on the space of certain classes of field theories raises many interesting questions of fundamental interest in field and string theory, and has sprouted several directions of research. At the same time, the double copy is also relevant for more practical applications  where it becomes a tool to make otherwise intractable calculations, such as the evaluation of higher-loop corrections, accessible. 

In the remainder of the Introduction we provide some background to motivate the interest in graviton scattering. Also, to give examples of the power of the double copy, we preview (i) its application to the binary inspiral problem relevant for gravitational wave physics, (ii) its ability to simplify complex multiloop calculation like determining the UV behavior of supergravity theories, (iii) its utility to organize and classify a web of theories.

\subsection{What is Graviton Scattering?}

Before getting too technical,
let us clarify what is meant by perturbative gravity and graviton amplitudes. Expanding the Einstein-Hilbert action around flat space
as $g_{\mu\nu} = \eta_{\mu\nu} + \kappa h_{\mu\nu}$ gives an action for a fluctuating spin-two field $h_{\mu\nu}$, with two-derivative interactions involving any number of fields controlled by the powers of the coupling $\kappa = \sqrt{8 \pi G}$, where $G = 6.67 \times 10^{-11}$\,N m$^2$/kg$^2$ is Newton's constant. Couplings to matter are introduced by expanding the covariant derivative and metric and are likewise controlled by powers of $\kappa$. The Feynman rules are exceedingly complicated in standard gauges, but nonetheless the sums of diagrams simplify to give relatively compact expressions for the on-shell amplitudes, such as the double-copy formula in (\ref{M4A4sq}).

Let us now address the physical relevance of perturbative graviton scattering.  
In natural units, $c= \hbar=1$, the scale of $G$ is set by the Planck mass  as
\begin{equation}
 \kappa \,\sim\, \sqrt{G} \,\sim\, 1/M_\text{Planck} \sim 10^{-19}\,\text{GeV}^{-1}\,.
\end{equation}
Since $\kappa$ is dimensionful, the gravitational interactions are controlled by the dimensionless effective coupling $E \kappa$, where $E$ is the typical energy of the scattering process. This means that even for a 100\,TeV scale process, the effective gravitational coupling $E \kappa \sim 10^{-14}$ is very small. This may make perturbative scattering look like a particularly exotic exercise, however, let us now show how perturbative graviton scattering connects directly to familiar gravitational physics. 

We know that gravity is an important force in our daily lives: it helps us keep our feet on the ground and makes our planet orbit the Sun. So suppose we consider massive (scalar) particles canonically coupled to gravity\footnote{At the level of the Lagrangian, the canonical coupling of massive scalars $\phi$ to gravity arises through the kinetic term $\sim  g^{\mu \nu}\partial_{\mu}\phi \partial_{\nu}\phi$, where the inverse metric $g^{\mu \nu}$ is an infinite series in the spin-two field.}.  The leading order diagram for two massive particles interacting via the exchange of a graviton is then
\begin{equation}
  \label{newton}
\begin{gathered}
\begin{tikzpicture}[scale=1.5]
\begin{feynman}
\vertex (a1) at (-1.2, -0.5) {\(m_2\)};
\vertex (a2) at (-1.2, 0.5) {\(m_1\)};
\vertex (mid1) at (0,-.5);
\vertex (mid2) at (0,.5);
\vertex (a3) at (1, 0.5);
\vertex (a4) at (1, -0.5);
\diagram{
(a1)--[very thick] (a4),
(a2) -- [very thick] (a3) ,
(mid1)--[thick,gluon,blue](mid2) 
};
\vertex [below left=.05em of mid2 ] {\(\kappa\)};
\vertex [above left=.05em of mid1 ] {\(\kappa\)};
\end{feynman}
\end{tikzpicture}
\end{gathered} \implies  V = -\frac{G m_1 m_2}{r}
\end{equation} 
The equivalence principle tells us that gravity couples universally, so the coupling of each vertex is $\kappa$ and hence the diagram is proportional to $G$.
In the non-relativistic limit, $v \to 0$, a Fourier transformation of this scattering amplitude results in the familiar Newton potential as indicated in (\ref{newton}). Thus we see how a familiar result from classical physics can arise from a perturbative scattering process. Let us now describe how gravity self-interactions are also relevant to study.

\subsection{Preview: Application to Gravitational Wave Physics}

Newton's potential (\ref{newton}) describes the leading order gravitational interaction between two static masses and to obtain it from the scattering amplitudes, we set the relative speed $v$ between the two particles to zero.
Corrections to this potential can depend on the dimensionless quantities $v^2/c^2$ and on $GM/(rc^2)$ and their relative importance depends on the nature of the motion (while time-reversal-non-invariant contributions to observables may depend on $v/c$).

For motion on bound trajectories, the virial theorem implies that 
$v^2/c^2$ and $GM/(rc^2)$ are of the same order, where $G$ is Newton's constant and $r$ is the separation of the two masses. Thus, corrections to Newton's potential are organized in powers of inverse speed of light, and the $n$-th order corrections contains $(v^2/c^2)^k (GM/(rc^2))^{n+1-k}$ for all $k$ between zero and $n$.

The virial theorem does not constrain unbound motion in Newton's potential. 
Requiring relativistic invariance leads to a resummation of the complete
dependence in $v^2/c^2$ at fixed order in Newton's constant, presenting the corrections as a series in $GM/(rc^2)$. 

While the origin of the velocity-dependent corrections is intuitively clear, 
the origin of those depending on Newton's constant is revealed by carefully enforcing the requirements of Bohr's correspondence principle that all 
charges are large. This implies that loop corrections can contain classical contributions and that they are given by a specific dependence on the momentum transfer~\cite{Cheung:2018wkq}, which is Fourier-conjugate to the relative distance.
For instance, $O(G^2)$ corrections include diagrams with two gravitons exchanged as well as diagrams with graviton self-interactions, 
\begin{equation}
  \label{OG2}
\begin{gathered}
\begin{tikzpicture}[scale=1.5]
\begin{feynman}
\vertex (a1) at (-1.2, -0.5) {\(m_2\)};
\vertex (a2) at (-1.2, 0.5) {\(m_1\)};
\vertex (mid1a) at (-.333,-.5);
\vertex (mid1b) at (-.333,.5);
\vertex (mid2a) at (.333,-.5);
\vertex (mid2b) at (.333,.5);
\vertex (a3) at (1, 0.5);
\vertex (a4) at (1, -0.5);
\diagram{
(a1)--[very thick] (a4),
(a2) -- [very thick] (a3) ,
(mid1a)--[thick,gluon,blue](mid1b),
(mid2a)--[thick,gluon,blue](mid2b)
};
\end{feynman}
\end{tikzpicture}
\end{gathered} +
\begin{gathered}
\begin{tikzpicture}[scale=1.5]
\begin{feynman}
\vertex (a1) at (-1, -0.5);
\vertex (a2) at (-1, 0.5);
\vertex (mid1a) at (0,0);
\vertex (mid1b) at (0,.5);
\vertex (mid2a) at (-.4,-.5);
\vertex (mid2b) at (.4,-.5);
\vertex (a3) at (1, 0.5);
\vertex (a4) at (1, -0.5);
\diagram{
(a1)--[very thick] (a4),
(a2) -- [very thick] (a3) ,
(mid1a)--[thick,gluon,blue](mid1b),
(mid2a)--[thick,gluon,blue](mid1a),
(mid2b)--[thick,gluon,blue](mid1a)
};
\end{feynman}
\end{tikzpicture}
\end{gathered} + \cdots\,,
\end{equation} 
while diagrams such as
\begin{equation}
  \label{OG3}
\begin{gathered}
\begin{tikzpicture}[scale=1]
\begin{feynman}
\vertex (a1) at (-1.4, -0.5) {\(m_2\)};
\vertex (a2) at (-1.4, 0.5) {\(m_1\)};
\vertex (mid1a) at (-.5,-.5);
\vertex (mid1b) at (-.5,.5);
\vertex (mid2a) at (.0,-.5);
\vertex (mid2b) at (.0,.5);
\vertex (mid3a) at (.5,-.5);
\vertex (mid3b) at (.5,.5);
\vertex (a3) at (1, 0.5);
\vertex (a4) at (1, -0.5);
\diagram{
(a1)--[very thick] (a4),
(a2) -- [very thick] (a3) ,
(mid1a)--[thick,gluon,blue](mid1b),
(mid2a)--[thick,gluon,blue](mid2b),
(mid3a)--[thick,gluon,blue](mid3b)
};
\end{feynman}
\end{tikzpicture}
\end{gathered} +
\begin{gathered}
\begin{tikzpicture}[scale=1]
\begin{feynman}
\vertex (a1) at (-1, -0.5);
\vertex (a2) at (-1, 0.5);
\vertex (mid1a) at (0,-.20);
\vertex (mid1b) at (0,.20);
\vertex (mid2a) at (-.5,-.5);
\vertex (mid2b) at (.5,-.5);
\vertex (mid3a) at (-.5,.5);
\vertex (mid3b) at (.5,.5);
\vertex (a3) at (1, 0.5);
\vertex (a4) at (1, -0.5);
\diagram{
(a1)--[very thick] (a4),
(a2) -- [very thick] (a3) ,
(mid1a)--[thick,gluon,blue](mid1b),
(mid2a)--[thick,gluon,blue](mid1a),
(mid2b)--[thick,gluon,blue](mid1a),
(mid3a)--[thick,gluon,blue](mid1b),
(mid3b)--[thick,gluon,blue](mid1b)
};
\end{feynman}
\end{tikzpicture}
\end{gathered} +
\begin{gathered}
\begin{tikzpicture}[scale=1]
\begin{feynman}
\vertex (a1) at (-1, -0.5);
\vertex (a2) at (-1, 0.5) ;
\vertex (mid1a) at (-.5,-.5);
\vertex (mid1b) at (-.5,.5);
\vertex (mid2a) at (-.5,0);
\vertex (mid2b) at (.5,0);
\vertex (mid3a) at (.5,-.5);
\vertex (mid3b) at (.5,.5);
\vertex (a3) at (1, 0.5);
\vertex (a4) at (1, -0.5);
\diagram{
(a1)--[very thick] (a4),
(a2) -- [very thick] (a3) ,
(mid1a)--[thick,gluon,blue](mid1b),
(mid2a)--[thick,gluon,blue](mid2b),
(mid3a)--[thick,gluon,blue](mid3b)
};
\end{feynman}
\end{tikzpicture}
\end{gathered} +
\begin{gathered}
\begin{tikzpicture}[scale=1]
\begin{feynman}
\vertex (a1) at (-1, -0.5);
\vertex (a2) at (-1, 0.5) ;
\vertex (mid1a) at (-.55,-.5);
\vertex (mid1b) at (-.55,.5);
\vertex (mid2a) at (-.55,0);
\vertex (mid2b) at (.55,0);
\vertex (mid3a) at (.55,-.5);
\vertex (mid3b) at (.55,.5);
\vertex (a3) at (1, 0.5);
\vertex (a4) at (1, -0.5);
\diagram{
(a1)--[very thick] (a4),
(a2) -- [very thick] (a3) ,
(mid1a)--[thick,gluon,blue](mid3b),
(mid3a)--[thick,gluon,blue](mid1b)
};
\end{feynman}
\end{tikzpicture}
\end{gathered} + \cdots\,,
\end{equation} 
contribute to $O(G^3)$.

Newton's potential  (\ref{newton}) can be determined from the tree-level scattering of massive particles. That same Newton potential also describes the bound state physics of, say, orbiting planets. 
This continues to hold true for corrections to this potential. While scattering amplitudes naturally yield corrections at fixed order in $G$ and all orders in velocity, they can be suitably expanded so that terms of the same order in accordance to the virial theorem are kept.
Orbiting bodies radiate gravitationally towards the eventual merger.  Close to the merger the complete velocity dependence is relevant.
For example, the complete $O(G^3)$ and $O(G^4)$ conservative Hamiltonians obtained from scattering data were analyzed by theorists of the LIGO-Virgo-KAGRA collaboration in \cite{Antonelli:2019ytb} and \cite{energetics4PM}, respectively. While the results at $O(G^4)$ are in an early stage with assumptions that require further investigation, the key messages are that they can be included in the effective one-body theory framework~\cite{Buonanno:1998gg,Buonanno:2000ef} in a way that improves comparison with numerical general relativity towards merger, and that further developments are strongly encourages.

Three-graviton Feynman rules in generic gauges can contain over 100 terms~\cite{DeWitt:1967uc}, with comparable size expressions for contact terms associated with every additional multiplicity.  The complicated expressions that appear in Feynman-graph-based calculations at higher orders in $G$ are avoided using a combination of unitarity cuts that both reduce the input for the loop integrand to products of graviton and graviton \& massive-particle tree diagrams and weed out terms with no classical contributions. The gravitational tree diagrams can be computed with little effort via  double-copy expressions such as (\ref{M4A4sq}). Calculations using the double copy were the first to push the state of the art first through $O(G^3)$ and then through  $O(G^4)$~\cite{Bern:2021dqo, Bern:2021yeh}. 

Recent progress on the application of the double copy to  gravitational wave physics is described in further detail in Section \ref{sect:precGW}. See also the dedicated Snowmass White Paper, ref.~\cite{gravWhitePaper2022}.

\vspace{3mm}

\subsection{Simplifying Calculations}

On-shell methods, such as recursion relations and unitarity, help simplify calculation of scattering amplitudes enormously. The double copy adds to this calculation power. We saw an example of this above.
Another example is furnished by loop-calculations in perturbative supergravity (SG) with the goal of understanding potential cancellations of ultraviolet (UV) divergences, for example as depicted in Table~\ref{fig:SGDivTab}. Unitarity and double-copy constructions have allowed the first direct access to multiloop order calculations in SG theories that would be otherwise inaccessible through traditional Feynman-diagram based methods. Here the double copy is  not only useful for computing   cuts in terms of much simpler tree-level super-Yang Mills (sYM) amplitudes, but manifesting color-dual $\mathcal{N}=4$ sYM integrands through four-loops~\cite{Bern:2010ue,Bern:2012uf} has allowed direct access to $\mathcal{N}\ge4$ SG integrands through the same loop order. This is discussed more in Section \ref{sect:gravUV}. 

\begin{table}[]
\begin{tabular}{@{}llll@{}}
\toprule
Loop order & $\mathcal{N}=8$ supergravity & $\mathcal{N}=5$ supergravity & $\mathcal{N}=4$ supergravity \\ \midrule
2           &  $D_c=7$ \cite{Bern:1998ug}                    &  $D_c^{(*)} > 6$ \cite{Boucher-Veronneau:2011rlc}            &  $D_c=6$ \cite{Bern:2012gh}\\
3           & $D_c=6$   \cite{Bern:2007hh}                   &    $D_c > 14/3$ \cite{Bern:2014sna}             & $D_c=14/3$ \cite{Bern:2014lha}               \\
4           & $D_c= 5 \frac{1}{2}$  \cite{Bern:2009kd}   &    $D_c > 4$ \cite{Bern:2014sna}                &  $D_c=4$ \cite{Bern:2013uka}\\
5           & $D_c= 24/5$~\cite{Bern:2018jmv}               &                    &                     \\ \bottomrule
\end{tabular}
\caption{List of direct computations that lead to bounds on the lowest spacetime dimension in which supergravity theories diverge (critical dimension) at various loop orders realized through explicit calculation using double-copy and unitarity methods.  ${}^{(*)}$ The two-loop critical dimension for $\mathcal{N}=5$ SG has not been reported in the literature, to our knowledge, but it is reasonable to expect improved behavior relative to $\mathcal{N}=4$ SG~\cite{Bern:2012gh}, much like at three and four loops.}
\label{fig:SGDivTab}
\end{table}

The double copy has impact even on simplifying calculations in gauge theories. 
This is because 
internal self-consistency of the double copy 
implies certain relations among color-ordered amplitudes.
These relations reduce the number of independent color-ordered amplitudes in the theory: 
in (super)Yang-Mills theory for example, they reduce the number of independent color-orderings at $n$ points from $(n-2)!$ to $(n-3)!$. 
This matters in particular when on-shell tree amplitudes are used as input in higher-loop calculations.  For instance, for the 4-gluon 2-loop amplitude in 
Yang-Mills theory 
there are naively 48 color-ordered cuts involving a six-point tree sewn with a four-point tree, but these are not independent. Indeed, because of self-consistency of the double copy we need only consider 6 color-ordered cuts.
These ideas, described further in the Section \ref{sect:complexity}, have not only helped with higher-loop formal studies, but hint at
future applications to amplitudes calculations of relevance to particle phenomenology.

\subsection{A Map on the Space of Field Theories}
The double copy goes beyond a relation between gravity and YM theory (\ref{GRisYM2}). In fact, there is a web of theories connected via the double copy (see Figure \ref{FigWeb} below). This includes, but is not limited to, example field theories in 
Figure \ref{fig:DCtable} that illustrates the double copy as a ``multiplication table".

\begin{figure}[t]
\begin{equation}
 \nonumber
 \begin{array}{c||c|c|c|c}
     \text{FT} \otimes \text{FT}& \text{YM} & \mathcal{N}=4~\text{sYM} & \chi\text{PT} & \text{BAS}    \\[1mm]
     \hline \hline
     \phantom{\bigg|} \text{YM}
     & ~\text{gravity$^+$} & \mathcal{N}=4~\text{SG} & ~~  \text{BI}   & \text{YM}  \\[2mm]
    \mathcal{N}=4~\text{sYM}  ~
    &~ \mathcal{N}=4~\text{SG} ~&~\mathcal{N}=8~\text{SG}  ~&~  \mathcal{N}=4~\text{sDBI} ~&~ \mathcal{N}=4~\text{sYM} \\[2mm]
     \chi\text{PT} 
     &  \text{BI}  &~~  \mathcal{N}=4~\text{sDBI} &\text{sGalileon}& \chi\text{PT}\\[2mm]
     \text{BAS} 
     & \text{YM} & \mathcal{N}=4~\text{sYM}  &~~\chi\text{PT} & \text{BAS}
 \end{array}
\end{equation}
\caption{{\footnotesize \label{fig:DCtable} Example of the web of theories connected by the double copy. As discussed in the main text, the double copy of Yang-Mills theory (YM) with itself gives gravity; the ``+" indicates that in addition to gravity one also gets the dilaton and anti-symmetric 2-form. Supersymmetry is inherited additively by the double copy; as an example $\mathcal{N}=4$ super Yang-Mills theory (sYM) with its total of 16 states --- gluons, gluinos, and scalars --- double-copies with itself to give the $16^2 = 256$ states of $\mathcal{N}=8$ supergravity (SG). Leading order 2-derivative chiral perturbation theory $\chi$PT can also act as input for the double copy and as shown in the table when it is double-copied with YM, $\mathcal{N}=4$ sYM, or itself it produces the amplitudes of a set of models relevant in various aspects of high-energy theory: Born-Infeld theory (BI) of non-linear electrodynamics, $\mathcal{N}=4$ super-Dirac Born-Infeld (sDBI), and the special Galileon (sGal) which has appeared in the contexts of modifications of gravity \cite{Nicolis:2008in,deRham:2010ik,deRham:2010kj} and as proposed subleading terms in brane actions \cite{deRham:2010eu}. Finally, the cubic bi-adjoint scalar model (BAS) plays the role of an identity element for the double copy; we discuss this further in Section \ref{sect:GenDC}.}
} 
\end{figure}

There is much variety among these models. Yang-Mills theory (YM) is a renormalizable model, gravity is not. Chiral perturbation theory  ($\chi$PT) is a low-energy effective field theory of pions whereas its double copy, the special Galileon, by itself is not a UV completable model. 
The Born-Infeld model (BI) of nonlinear electrodynamics has electromagnetic duality in 4d or, equivalently, has only helicity-conserving amplitudes, but the input theories YM and $\chi$PT do not. On top of that, $\mathcal{N}=4$ super Yang-Mills theory (sYM) is a conformal theory in 4d, while $\mathcal{N}=4$ super Dirac-Born-Infeld  (sDBI) is the low-energy effective action on a flat D3-brane in Minkowski space. Finally, with its cubic interaction the bi-adjoint scalar model (BAS) has a potential unbounded from below. It is clear that the field-theory double copy connects models with wide range of properties and applications.

\subsection{Outline}
 We begin with a review in Section~\ref{s:review}. There are a variety of different formulations of the double copy. Each representation has its own advantages and highlights different properties of the double-copy map. We present three formulations --- known in shorthand as KLT, BCJ, and CHY --- for tree amplitudes in \sect{rev:Trees} and then discuss progress on adapting them at loop-level in \sect{rev:Loop}.  We conclude our review with a brief reminder of the importance of choice of kinematic variables in \sect{rev:Kin}.

In \sect{sect:applications} we describe applications of the double copy.  Classical double copy and its applications may be the most accessible to interested readers outside of the amplitudes community so we will start there.   Besides, despite its discovery in and relevance to quantum mechanical scattering amplitudes,
one of the most compelling and remarkable aspects of the double-copy construction is that it offers fresh insight at classically relevant scales. 
\sect{sect:classicalDC} presents a discussion of classical double copy, beginning with the Kerr-Schild and Weyl double copy understanding of black holes and related solutions.  We continue with the sharpest empirical predictions of double copy in the precision gravitational wave predictions arising 
from black hole collision in \sect{sect:precGW}.  We continue the theme by considering the use of double-copy structure to lift flat space-time scattering amplitude insight to the interaction of quantum fields evolving on non-trivial classical backgrounds in \sect{sect:nontrivialBG}. 
This naturally leads into a discussion of the potential opportunities and challenges for double copy constructions in both early and late-stage cosmology in~\sect{sect:dcCosmo}.

Double-copy insight has its origins in the KLT relations originally recognized in tree-level perturbative string theory, and the dynamic interchange between QFT and string theory is strong and fruitful.  We discuss continuing prospects for the rich interplay between QFT and string theory amplitudes in \sect{sect:stringAmplitudes}.  From the local QFT perspective, string theory represents an effective UV completion in the form of a tower of  higher-derivative operators.  An important generalization of double-copy insight is to incorporate phenomenological higher-derivative operators for which string theory amplitudes represent an inspirational and clarifying goal post.  
Progress in generalizing the double copy in the KLT formalism is discussed in~\sect{sect:GenDC} and in terms of the duality between color and kinematics in  \sect{sect:higherDerivative}.  A related sharp question --- driving higher-loop progress in non-planar theories --- is whether all supergravity theories with a finite number of counterterms must require perturbative UV  completions in four-dimensions, or if they could possibly be finite.  We discuss current progress and open questions with a focus on double-copy constructions in \sect{sect:gravUV}.

Finally, one of the most attractive aspects of the double copy is how it simplifies calculations while exposing new ways to understand
what makes individual theories unique and how they can be related to other theories in sometimes surprising ways.
We explore this theme in three sections. The first discusses the reduction in complexity of independent building blocks relevant to phenomenological and formal integrands at the multiloop level in~\sect{sect:complexity}.  We continue  in \sect{sect:kinAlg} with an overview of the quest towards a sharp  understanding of the underlying off-shell kinematic algebra behind double-copy constructions as well as the simplifications we expect such a recognition to bestow.  
We close by confronting the ultimate in recycling opportunities in~\sect{sect:web}: as a small number of primary constituents can combine to build a vast web of double-copy theories, advances in these primary theories propagate through the web, and appreciating double-copy structures offers a new approach towards classifying theories by identifying shared constituents.

\section{Review: Formulations of the Double Copy}
\label{s:review}

There are various formulations of the double copy.  Here we introduce three complementary formulations of the double copy for tree-level amplitudes in theories of massless particles in the adjoint
\footnote{As we discuss later, much progress has been made on double-copy constructions involving arbitrary representations, including the fundamental, and as well as for theories with both massive and massless particles.}
of a color (or flavor) group.  These form the basis for the applications and generalizations. We then briefly discuss extensions to loop-level and choices of kinematic variables for the different types of building blocks. 

\subsection{Tree-Level: KLT, BCJ, CHY}
\label{rev:Trees}
The three tree-level formulations of the double copy are:
\begin{itemize}
\item {\bf KLT double copy}: named after Kawai, Lewellen, and Tye \cite{Kawai:1985xq}, the KLT formula is a manifestly gauge-invariant incarnation of the double copy. At $n$-point, the KLT formula for the tree amplitudes $M_n^{\text{L}\otimes \text{R}}$ in the double-copy theory $\text{L}\otimes \text{R}$ takes the form
\begin{equation}
\label{KLTdc}
M_n^{\text{L}\otimes \text{R}} = \sum_{a,b} A^\text{L}_n[a]\, S_n[a|b]\, A^\text{R}_n[b]\,,
\end{equation}
where  $A^\text{L}_n$ and $A^\text{R}_n$ are color-stripped $n$-point tree amplitudes of the ``left" and ``right" theories L and R, such as the ones in Figure \ref{fig:DCtable}. The labels $a,b$ run over two choices of $(n-3)!$ of the $n!$ possible single-trace color-orderings of $n$ particles.   Finally, $S_n[a|b]$ is the KLT kernel which is a function of the $n$-point Mandelstam variables \cite{Bern:1998sv, Bjerrum-Bohr:2010pnr}, and can be understood as the inverse of a matrix of bi-colored scalar amplitudes \cite{Cachazo:2013iea}. Connecting to the 4-point example (\ref{M4A4sq}), we see that $S_4[1234|1234] = -s_{12}s_{14}/s_{13}$.  Note that the same kernel is used in all the double-copies in Figure~\ref{fig:DCtable}. 

In string theory, $A^\text{L}_n$ and $A^\text{R}_n$ are color-stripped open-string disk amplitudes and $M_n^{\text{L}\otimes \text{R}}$ is the closed-string sphere amplitude. The string-theory version of the KLT kernel depends explicitly on the string tension via $\alpha'$ and
simplifies to the field-theory kernel $S_n[a|b]$ in the low-energy limit. Both variants of the KLT kernel can be mathematically understood in terms of intersection numbers of twisted (co-)cycles
\cite{Mizera:2017cqs, Mizera:2017rqa}.

\item {\bf BCJ double copy}: Bern, Carrasco, and Johansson \cite{Bern:2008qj} discovered a form of the double copy that relies on the proposal (proven at tree level  \cite{Bern:2010yg} and tested impressively at loop-level \cite{Bern:2010ue, Bern:2012uf}) of color-kinematics duality. The idea is to write the color-dressed amplitudes of the L and R theories as
\begin{equation} 
  \label{AnBCJ}
   A^{\rm L}_n = \sum_I \frac{n^{\rm L}_I\,c^\text{L}_I}{\prod_{j\in I} P_j^2}\,,
   \ \ \ \ \ \ \
    A^{\rm R}_n = \sum_I \frac{n^{\rm R}_I\,c^\text{R}_I}{\prod_{j\in I} P_j^2}\,,
\end{equation}
where the sum is over trivalent graphs $I$, the $c_I$'s are color factors (e.g.~$c_{ab,cd} =f^{abx} f^{cdx}$ at four points in terms of the anti-symmetric structure constants), and
the $P_j$ are the momenta on the internal lines $j$ of the trivalent diagrams summed over. 
The $n_I$ are so-called ``numerator factors" (also called ``kinematic weights") that are made from Lorentz-invariant contractions of the external momenta and polarizations / fermion wavefunctions. 

This way of writing the amplitude is highly non-unique.
BCJ proposed~\cite{Bern:2008qj} that if numerator factors can be found which obey the same Jacobi-like identities as the color-factors\footnote{It is worth pointing out that Jacobi satisfying numerators were recognized at four-point tree-level for pure Yang-Mills in the  context of motivating certain radiation zeros~\cite{Zhu:1980sz,Goebel:1980es}.  See e.g.~refs.~\cite{Harland-Lang:2015faa,Brown:2016mrh} for recent developments.}, e.g.~
\begin{equation} 
\label{colkindual}
  c_{12,34} + c_{13,42} + c_{14,23} = 0
  ~~~\iff~~~
  n_{12,34} + n_{13,42} + n_{14,23} = 0\,,
\end{equation}
where, e.g.,
\begin{equation}
n_{ab,cd}= n \left(\begin{gathered}
\begin{tikzpicture}[scale=1.5]
\begin{feynman}
\vertex (a1) at (-1, -0.5) {\(a\)};
\vertex (a2) at (-1, 0.5) {\(b\)};
\vertex (mid1) at (-0.3,0);
\vertex (mid2) at (0.3,0);
\vertex (a3) at (1, 0.5) {\(c\)};
\vertex (a4) at (1, -0.5) {\(d\)};
\diagram{
(a1)--[thick, gluon] (mid1),
(a2) -- [thick, gluon] (mid1) ,
(mid1)--[thick,gluon](mid2), 
(mid2)--[thick, gluon](a3), 
(a4) -- [thick, gluon](mid2)
};
\end{feynman}
\end{tikzpicture}
\end{gathered} \right) \,,
\end{equation}
then by replacing the color factors by these numerator factors, one finds the tree amplitude of another local field theory:
\begin{equation} 
\label{dcopyamp}
   M^{\text{L}\otimes \text{R}}_n = \sum_I \frac{n^\text{L}_I\,n^\text{R}_I}{\prod_{j\in I} P_j^2}\,.
\end{equation}
The existence of numerators $n_I$ subject to kinematic Jacobi identities as in (\ref{colkindual}) is the key prediction of the {\em color-kinematics duality}.  

Note that if instead the numerator factors $n^\text{L}_I$ had been replaced by another set of color factors $c^\text{R}_I$ in (\ref{AnBCJ}), one obtains the tree amplitudes of the cubic bi-adjoint model BAS discussed above and in Figure \ref{fig:DCtable}.

\item {\bf CHY double copy}: The scattering equation approach to scattering amplitudes by Cachazo, He, and Yuan \cite{Cachazo:2013gna, Cachazo:2013hca} gives rise to yet another formulation of the double copy. In this form, the color-stripped L and R amplitudes are formulated in terms of integrals over $n$ auxiliary variables $\sigma_i$ (punctures on the Riemann sphere) as
\begin{equation} 
\label{chycolorordered}
   A^\text{L}_n[12 \ldots n] = \int \textrm{d}\mu_n(k,\sigma)\,
   \frac{{\cal I}^\text{L}(k,\epsilon,\sigma)}{\sigma_{12}\sigma_{23} \cdots \sigma_{n1}}
\end{equation}
and similarly for $A^\text{R}$.
Here $\textrm{d}\mu_n(k,\sigma)$ is an integration measure that localizes the integral over $\sigma_i$ on the solutions to the scattering equations.\footnote{The scattering equations firstly seen in \cite{Fairlie:1972zz} read $\sum_{j \neq i} \frac{k_i \cdot k_j}{\sigma_i - \sigma_j} = 0 \ \forall \ i=1,2,\ldots,n$ in the momentum phase space of massless $n$-point amplitudes.} The function ${\cal I}^\text{L}(k,\epsilon,\sigma)$ encodes the on-shell external momenta $k_i$, the polarizations and fermion wavefunctions collectively denoted by $\epsilon_i$, and the $\sigma_i$.  
The product of $\sigma_{ij} = \sigma_i- \sigma_j$ is similar to the denominator of the Parke-Taylor expression for the $n$-gluon tree amplitude. 
(See \cite{Cachazo:2013hca, Cachazo:2014xea} for examples of the polarization-dependent part ${\cal I}^\text{L}(k,\epsilon,\sigma)$ of the integrand.)

The CHY statement of the double copy is then
\begin{equation} 
\label{chycolorordered2}
   M^{\text{L}\otimes \text{R}}_n = \int 
   \textrm{d}\mu_n(k,\sigma)\,
   {\cal I}^\text{L}(k,\epsilon,\sigma) {\cal I}^\text{R}(k,\epsilon,\sigma) \, .
\end{equation}
The soft behavior of amplitudes is more manifest in the CHY formulation. A consequence is also the now widespread  recognition\footnote{Ref.~\cite{Chen:2013fya} was the first to recognize that NLSM tree-level amplitudes satisfy the duality between color and kinematics and indeed represented a double copy. Subsequently ref.~\cite{Cachazo:2014xea} clarified the NLSM's double-copy relationship with DBI and special Galileon theories. Different approaches to explicit NLSM numerators subject to kinematic Jacobi relations can for instance be found in ref.~\cite{Du:2016tbc, Carrasco:2016ldy, Cheung:2016prv}.}  that the double copy is not limited to Yang-Mills theory and gravity, but connects a large variety of field theories such as illustrated in Figure \ref{fig:DCtable}.

\end{itemize} 
These formulations of the double copy are equivalent at tree-level. We  have presented the three versions of double-copy constructions with just enough detail to facilitate the discussion of these formulations in later sections.  An important point about all three formulations is that they either
rely upon, or predict, additional relationships between gauge
invariant color-ordered amplitudes in theories that can participate in
standard adjoint-type double-copy constructions, as we now discuss.

Consider tree-level $n$-particle gluon scattering. Factoring out color factors in a trace basis leaves $(n-1)!$ possible arrangements of the external states due to the cyclicity of color-traces.  One could, instead, keep the color-weights in adjoint $f^{abc}$ form but express them in terms of a minimal color basis due to anti-symmetry and Jacobi relations.  This is less than in the trace basis because color-traces span inclusion of the symmetric $d^{abc}$ color weights as well which are irrelevant to gluons at tree-level in pure Yang-Mills.  Looking at the coefficients of these basis color-weights yields an $(n-2)!$ element basis amongst color-ordered tree-amplitudes.  Reduction from $(n-1)!$ ordered amplitudes to $(n-2)!$ ordered amplitudes yields the Kleiss-Kuijf relations~\cite{Kleiss:1988ne,DelDuca:1999rs} which include reflection relations such as $A_n[123\ldots n] = (-1)^n A_n[n\ldots 321]$. 

The existence of the double copy suggests a further reduction. Why? Consider the KLT formulation (\ref{KLTdc}). In the sum, one gets to make a choice of $(n-3)!$ color-orderings out of the in principle $n!$ possible color-orderings. The result of the double copy cannot depend on this choice, so there have to be relations among the L and R amplitudes that ensure this ``basis-independence". These relations are, in addition to the the Kleiss-Kuijf relations, an additional set of relations known as ``BCJ relations" \cite{Bern:2008qj}. In the BCJ formulation, these relations arise from the color-kinematics duality (\ref{colkindual}). Complementary derivations 
have been based on string-theory methods \cite{Plahte:1970wy, Stieberger:2009hq, BjerrumBohr:2009rd}, on-shell recursion relations \cite{Feng:2010my, Chen:2011jxa} or BRST cohomology techniques \cite{Mafra:2015vca}.
The punchline is that the BCJ relations  reduce the number of independent single-trace amplitudes from $(n-2)!$ to $(n-3)!$. 

As a result of the combined cyclicity, Kleiss-Kuijf, and BCJ relations, rather than having 
$6!=720$ independent color-arrangements to handle for the 6-point gauge theory amplitudes, there are only $3!=6$. Similar reductions in complexity occur in other more phenomenologically relevant contexts in applications both at tree- and loop-level. 

\subsection{Loop-Level}
\label{rev:Loop}
A major 
motivation for the study of the double copy stems from loop-level applications. In each of the three formulations of the double copy in Section \ref{rev:Trees}, finding the optimal generalizations from tree-level to loop-level amplitudes is on-going research.

So far, the BCJ formulation has led to the most impressive tests and applications at the multiloop level\footnote{See refs.~\cite{Bern:2019prr,Bern:2022wqg} for recent reviews}. The basic idea is to promote the trivalent-graph expansion of tree-level amplitudes (\ref{AnBCJ}) and (\ref{dcopyamp}) to the integrands
with respect to 
the loop momenta $\ell_j$. Once the
numerators $n^{\rm L}_I,n^{\rm R}_I$ in a loop-integrand version of (\ref{AnBCJ}) obey kinematic Jacobi identities at fixed $\ell_j$, then the double-copy
formula (\ref{dcopyamp}) is claimed to yield the loop integrand of the L $\otimes$ R theory.

It follows from unitarity considerations that the BCJ double-copy construction is going to hold for loop integrands to all orders if color-dual representations can be found.   While there is evidence (at least in particular limits) that double-copy relationships hold to all orders between Yang-Mills and gravity~\cite{Oxburgh:2012zr,Saotome2012vy}, it is a central open question whether BCJ numerators exist for all loop orders in theories that satisfy the duality between color and kinematics at tree-level.   For recent attempts to prove all order relationships see refs.~\cite{Borsten:2020zgj,Borsten:2021rmh}, although there is some evidence that if there are global color-dual numerators they may have to be non-local, at least in external momenta, or may only satisfy color-dual equations cut by cut~\cite{Bern:2015ooa}.

Color-dual integrands have been found through four loops in the maximally supersymmetric theory \cite{Bern:2010ue,Bern:2012uf}, and have shaped the state-of-the-art in multiloop computations in ${\cal N}=4,5$ supergravity \cite{Bern:2012cd, Bern:2013uka, Bern:2014sna}. The  five-loop amplitude in ${\cal N}=8$ supergravity was constructed \cite{Bern:2017ucb} through a generalization of the BCJ double copy beyond kinematic Jacobi identities and trivalent graphs \cite{Bern:2017yxu}, see Section \ref{sect:gravUV} for the far-reaching implications on the UV properties. 

For the CHY double copy, the quest for loop-level generalizations is guided by the underlying ambitwistor string theories \cite{Mason:2013sva, Berkovits:2013xba, Adamo:2013tsa,Casali:2014hfa,Adamo:2015hoa}
(see \cite{Geyer:2022cey} for a review). Similar 
to the integral formulae (\ref{chycolorordered}) and (\ref{chycolorordered2}) at tree level, the ambitwistor-prescription for loop amplitudes
derives loop integrands from integrating over auxiliary variables $\sigma_i$ on a (multi-)nodal sphere \cite{Geyer:2015bja,Geyer:2015jch}. The loop-level generalizations of the polarization-dependent parts ${\cal I}^\text{L}{\cal I}^\text{R}$ in (\ref{chycolorordered2}) again take a factorized form as familiar from chiral splitting for conventional strings \cite{DHoker:1988pdl, DHoker:1989cxq}.

At tree level \cite{Cachazo:2013iea, 
Bjerrum-Bohr:2016axv, Du:2017kpo, Edison:2020ehu} and one loop \cite{He:2017spx, Edison:2020uzf}, the connection between the CHY and BCJ double copies is well understood at all multiplicities, also in the absence of supersymmetry \cite{Geyer:2017ela}. Also at two loops, the BCJ double copy has been related to the ambitwistor-framework \cite{Geyer:2016wjx, Geyer:2018xwu, Geyer:2019hnn}, and one of the
future goals is to simplify
the two-loop instances of the 
function ${\cal I}^\text{L}$ to
make the polarization dependence fully explicit at all multiplicities. At three loops, the BCJ double copy was used to propose expressions for the kinematic building blocks in both ambitwistor and conventional string theories \cite{Geyer:2021oox}, see Section \ref{sect:stringAmplitudes} for further details.

At the time of writing, the multiloop systematics of the KLT double copy is least explored. The one-loop KLT formula \cite{He:2016mzd} for field-theory amplitudes derived from ambitwistor methods \cite{He:2017spx} is tied to linearized variants of the Feynman propagators and calls for a reformulation in terms of traditional quadratic propagators.\footnote{The conversion between linearized and quadratic propagators has been discussed from a variety of perspectives \cite{Gomez:2016cqb, Gomez:2017lhy, Gomez:2017cpe,
  Ahmadiniaz:2018nvr, Agerskov:2019ryp, Farrow:2020voh}.}
Loop-level KLT formulae for closed-string amplitudes are uncharted terrain, though monodromy relations among open-string loop amplitudes and closely related methods from twisted de Rham theory \cite{Tourkine:2016bak, Hohenegger:2017kqy, Casali:2019ihm, Casali:2020knc, Stieberger:2021daa} may shed light on their existence or construction.

\subsection{Kinematic Variables}
\label{rev:Kin}
In order to advance our understanding of the double copy formulations in Section \ref{rev:Trees},
it is essential to have explicit testing grounds at different loop and leg orders. This calls for compact representations of the polarization-dependent building blocks such as the color-stripped amplitudes $A^\text{L}_n[a]$ in the KLT formula (\ref{KLTdc}),
the kinematic numerators $n_I^\text{L}$ in the BCJ double copy (\ref{dcopyamp}) and the integrands ${\cal I}^\text{L}(k,\epsilon,\sigma)$ in the CHY formula (\ref{chycolorordered2}). With a growing number of loops and legs, the compactness of explicit $A^\text{L}_n[a]$, $n_I^\text{L}$ or ${\cal I}^\text{L}(k,\epsilon,\sigma)$ crucially depends on
the kinematic variables parametrizing the external polarizations. Suitable choices of kinematic variables have repeatedly been driving forces for structural progress -- in both the amplitudes program in general and in implementations of the double copy in particular.  

As a shining example from the 80s, four-dimensional spinor-helicity variables for massless states of various spins \cite{Berends:1981rb, DeCausmaecker:1981wzb, Kleiss:1985yh, Gunion:1985vca, Xu:1986xb} paved the way towards the Parke-Taylor formula, a one-liner expression for maximally helicity violating $n$-gluon tree amplitudes \cite{Parke:1986gb}. Generalizations of spinor-helicity variables to massive external legs \cite{Perjes:1974ra,Kleiss:1986qc,Spehler:1991yw,Novaes:1991ft, Conde:2016izb,Arkani-Hamed:2017jhn} and to different spacetime dimensions \cite{Dennen:2009vk,Caron-Huot:2010nes,Boels:2012ie,Chiodaroli:2022ssi} are still under active development and continue to facilitate both the construction and compact representation of amplitude-building blocks in a vast bandwidth of theories. In planar ${\cal N}=4$ sYM, momentum twistors~\cite{Hodges:2009hk} provided an unconstrained way to specify on-shell kinematics while manifesting dual superconformal invariance (cf.\ for instance \cite{Drummond:2009fd, Mason:2009qx,Arkani-Hamed:2009nll,Mason:2010yk,Caron-Huot:2010ryg,Arkani-Hamed:2010zjl,Arkani-Hamed:2012zlh}).

Spinor helicity and twistor variables have a rich interaction with the CHY formulae. One way in which this arises is through the ambitwistor strings that underpin these formulae. Ambitwistor space is fundamentally a dimension-agnostic concept, but for fixed spacetime dimension, one can find spinorial representations of the ambitwistor geometry. Formulating ambitwistor strings in these dimension-specific representations leads to notions of `polarized' scattering equations~\cite{Witten:2004cp,Geyer:2014fka,Geyer:2018xgb,Geyer:2019ayz,Albonico:2020mge,Geyer:2020iwz}, where the usual set of scattering equations is graded by the polarization data. These polarized scattering equations can lead to amplitude formulae with dramatically fewer explicit moduli integrals (at least in certain polarization configurations, like MHV in 4-dimensions), and have links to formulae for massive scattering~\cite{Albonico:2022pmd}. 

Since the color-kinematics duality and the double copy are universally observed in any number of spacetime dimensions, it is important to construct kinematic numerators subject to Jacobi identities in terms of dimension-agnostic polarization vectors of gauge bosons. This has been a key motivation in the construction of the CHY formulae, and
recent studies of multiparticle amplitudes in (ambitwistor and conventional) string theories
at tree and loop level led to striking compactifications of their polarization dependence
in arbitrary dimensions.

In maximally supersymmetric settings, pure-spinor superspace \cite{Berkovits:2000fe, Berkovits:2006ik} led to breakthroughs in encoding the dependence of multileg amplitudes on $D \leq 10$-dimensional gauge- and supergravity-multiplet polarizations (see \cite{Mafra:2010pn} for an automated  extraction of bosonic components). Together with recursive Berends-Giele techniques \cite{Berends:1987me} and perturbiner methods \cite{Rosly:1996vr, Rosly:1997ap, Selivanov:1997aq}, pure-spinor superspace turned out to be crucial both for the simultaneous manifestations of locality and color-kinematics duality beyond four points \cite{Mafra:2011kj, Mafra:2014gja, Mafra:2015mja, Mafra:2015vca} and the construction of multileg and multiloop string amplitudes \cite{Mafra:2011nv, Mafra:2018qqe, DHoker:2020prr}.

From these developments, we expect that structural discoveries of new symmetries and relations between physical theories will keep on going hand in hand with the quest for convenient variables to manifest different simplifying features of scattering amplitudes. And at the very least, innovative choices of variables can change the load of analytic data by numerous orders of magnitude in advancing the multiloop and multileg frontiers of gauge theories, gravity or string theories.

\section{Applications}
\label{sect:applications}

\subsection{From Coulomb Charges to Schwarzschild Black Holes and Generalizations}
\label{sect:classicalDC}

The perturbative construction of solutions to  classical field equations is closely related to the construction of semi-classical tree-level scattering amplitudes~\cite{Bern:2019prr,Rosly:1996vr, Rosly:1997ap, Selivanov:1997aq, Monteiro:2011pc, Lee:2015upy, Garozzo:2018uzj, Mizera:2018jbh, Bridges:2019siz,Cheung:2021zvb}. It is perhaps unsurprising then that the double-copy structure of scattering amplitudes is  reflected in properties of classical solutions, albeit in a gauge dependent\footnote{Much like how manifesting the duality between color and kinematics requires a particular generalized choice.} manner.

Many classical metric solutions admit Kerr-Schild coordinates,
\begin{equation} 
g_{\mu\nu} = \eta_{\mu\nu} + \varphi\, k_\mu \, \tilde{k}_\nu \,,
\end{equation}  which, due to properties of $\{\varphi(x), k(x), \tilde{k}(x)\}$, have the remarkable effect of linearizing Einstein's equations. Such gravitational solutions therefore have a natural double-copy interpretation in terms of linearized solutions to gauge theory equations of motion~\cite{Monteiro:2014cda,Luna:2015paa,Luna:2016due}.
A particularly striking example of a classical double-copy construction is that the gauge field of a Coulomb charge builds the metric of the  Schwarzschild black hole~\cite{Monteiro:2014cda}. The \emph{exact} Schwarzschild metric in Kerr-Schild coordinates is
\begin{equation} 
\textrm{d}s^2 = \left(\eta_{\mu\nu} + \frac{2 GM}{r} k_\mu k_\nu\right) \, \textrm{d}x^\mu \textrm{d} x^\nu \,, 
\end{equation} 
where $k_\mu \textrm{d}x^\mu = \textrm{d}t - \textrm{d}r$. The gauge theory ``single-copy'' is
\begin{equation} 
A^a_\mu = \frac{Q}{r} k_\mu c^a\, .
\end{equation} 
This is nothing but the familiar gauge field of an abelianized point charge written in a particular  gauge~\cite{Monteiro:2014cda}, which can be sourced e.g. by massive static charges. 
Any non-abelian structure in the YM case is irrelevant in the static limit.
The exact departure from the flat space metric, encoded by the Schwarzschild solution, therefore arises from the single one-propagator diagram that generates 
the  Coulomb field, but where the abelianized color factor $c^a$ is replaced by another copy of the ``kinematic" object $k_\mu$.

Obtaining the Schwarzschild metric required the Coulomb field
to be written in a particular gauge; this can be avoided by exploiting a lesson from amplitudes and focusing on gauge-invariant observables.
Motivated by gravitational-wave physics, a formalism for computing classical observables from amplitudes has been developed~\cite{Kosower:2018adc}. 
(We will discuss this formalism in more detail in the next section.)
The field strength $F_{\mu\nu}(x)$ is an example of such an observable~\cite{Cristofoli:2021vyo}; it is a gauge-invariant quantity in electrodynamics. 
Turning to gravity, we can instead compute the linearized Riemann curvature, which is also gauge-invariant (in the linear theory).
In the static case, the field strength and the linearized curvature are~\cite{Monteiro:2020plf} on-shell Fourier integrals of three-point amplitudes
and therefore they can be written manifestly as double copies of one another.

An alternative perturbative approach to this double copy is the following. 
Suppose we take our massive sources to be very heavy and probe them by scattering very light objects off them,
gathering information in analogy to Rutherford scattering. 
These probes 
will travel on geodesics (in the gravitational case) or follow the solution of the Lorentz-force
equation of motion (in electrodynamics).
The scattering angle the probes experience can be computed from amplitudes; therefore the double copy of amplitudes
is inherited by the angles. 
To the extent that the scattering angle gives us information about the fields, we again see that there must
be a relation between charges in gauge theory and black holes in gravity~\cite{Huang:2019cja,Emond:2020lwi}. 
Remarkably, these statements generalize from Coulomb charges to magnetically-charged dyons; the double copy is a gravitational solution with both mass and NUT charge~\cite{Luna:2015paa,Huang:2019cja}. 
It is also possible to include classical spin~\cite{Monteiro:2014cda,Arkani-Hamed:2019ymq,Emond:2020lwi}, making contact with the Kerr black hole (which is the double copy of a spinning disk of charge). More general exact classical double copies are available in gravitational theories with dilaton and axion matter~\cite{Lee:2018gxc,Monteiro:2021ztt}, as well as for curved backgrounds like (A)dS using generalized Kerr-Schild metrics~\cite{BahjatAbbas2017htu,Carrillo-Gonzalez:2017iyj}.

A convenient presentation of the curvature of the Schwarzschild solution which also reveals a path to generalizing its double-copy properties to other space-times is the Weyl spinor, which is simply a spinorial formulation of the Weyl curvature tensor.
The Weyl curvature is essentially the Riemann curvature minus its traces, which are the Ricci tensor and Ricci scalar.
Both traces vanish in vacuum solutions such as Schwarzschild. 
Therefore the Weyl spinor encodes the complete information about its curvature.
One advantage of the spinorial curvature is that it is a scalar under coordinate transformations, though it nevertheless transforms under local change of the spinor basis.
The surprise is that there is a basis in which the Weyl spinor is precisely linear in the mass of the black hole.
In this basis, then, the Weyl spinor equals its linear approximation, which is nothing but an integral of the three-point amplitude.
This is suggestive of a non-perturbative double copy.

The Weyl double copy~\cite{Luna:2018dpt} is a non-perturbative relation between the Weyl spinor of a wide class of spacetimes
and the corresponding spinorial curvature, known as the Maxwell spinor, in electrodynamics.
Writing the Weyl spinor in terms of spinorial indices as $\Psi_{\alpha\beta\gamma\delta}$, 
and the Maxwell spinor as $\Phi_{\alpha\beta}$, the Weyl double copy is the relationship
\begin{equation}
\label{eqn:weylReln}
\Psi_{\alpha\beta\gamma\delta} = \frac{1}{S} \Phi_{(\alpha\beta} \Phi_{\gamma\delta)} \,.
\end{equation}
In this four-dimensional equation, the object $S$ is a scalar function, and the parentheses indicate symmetrisation over spinorial indices.  
The relationship is known to hold for large classes of Petrov type D and type N solutions~\cite{Luna:2018dpt,Godazgar:2020zbv},  as well as asymptotically in algebraically general cases~\cite{Godazgar:2021iae}.
We have also seen above that the Weyl double copy follows from the double copy of amplitudes, at least at low perturbative orders.
The conjecture is that the Weyl double copy \emph{is} the double copy of scattering amplitudes non-perturbatively in four dimensions.

A similar relationship to (\ref{eqn:weylReln}) holds~\cite{Gonzalez:2022otg,Emond:2022uaf} in three-dimensions in topologically massive theories between the Cotton spinor and the dual field strength spinor, and indeed can be derived via a dimensional reduction of the Weyl double copy under certain circumstances~\cite{Emond:2022uaf}. 
The symmetries of gravity have a double copy interpretation. In fact, symmetry generators in linearized gravitational theories have been explicitly reconstructed from gauge-theory ingredients. 
This is based on interpreting the double copy as a convolution 
of position-space gauge theory fields, both physical and ghosts~\cite{Anastasiou:2014qba,Anastasiou:2016csv,Cardoso:2016ngt,Cardoso:2016amd,Anastasiou:2017nsz,Anastasiou:2017taf,Borsten:2013bp,Anastasiou:2013hba,Anastasiou:2015vba}.Since in Kerr-Schild coordinates Einstein's equations linearize, 
the convolutional double copy should reproduce the Schwarzchild and other solution with such a description.
It can also be used to construct more general linearized solutions of Einstein's equations. For example, ref.~\cite{Luna:2020adi}
used it to construct the linearization of the Janis-Newman-Winicour solution, which apart form the metric also contains a nontrivial dilaton.

  Classical aspects of the double copy have received intense attention in the literature in recent years~\cite{White:2016jzc, Goldberger:2016iau,Luna:2016hge,DeSmet2017rve,Bahjat-Abbas:2018vgo, Gurses:2018ckx, Berman:2018hwd, Bah:2019sda,Elor:2020nqe,Luna:2020adi,Borsten:2020xbt,Easson:2020esh,Berman:2020xvs, Guevara:2020xjx,Borsten:2021zir,Gonzo:2021drq,Campiglia:2021srh,Adamo:2021dfg,Easson:2021asd}, including non-abelian classical solutions~\cite{White:2016jzc,Cheung:2020djz} and quantum double-copy predictions against a classically double-copied background~\cite{Adamo2017nia, Adamo:2018mpq}.
They offer the prospect of a more non-perturbative understanding of the double copy, and provide a glimpse of the double copy at work far from its original Minkowskian home.
For example, there is no problem formulating a classical Kerr-Schild double copy for the Coulomb charge in AdS: the result is simply the AdS-Schwarzschild solution~\cite{Luna:2015paa}.
There are rich connections to other topics, including twistor theory~\cite{White:2020sfn,Chacon:2021wbr,Farnsworth:2021wvs,Adamo:2021dfg,Guevara:2021yud,Chacon:2021hfe,Chacon:2021lox}, dualities~\cite{Cho:2019ype,Huang:2019cja,Alawadhi:2019urr,Banerjee:2019saj,Kim:2019jwm,Keeler:2020rcv}, three-dimensional physics~\cite{CarrilloGonzalez:2019gof,Alkac:2021seh,Moynihan:2021rwh,Gonzalez:2021ztm,Gonzalez:2022otg,Emond:2022uaf} and --- as we will now discuss --- classical gravitational wave physics.

\subsection{Precision Gravitational Wave Science}
\label{sect:precGW}
 
 Next generation gravitational-wave observatories will surpass the 
 incredible precision of the LIGO and Virgo detectors and demand commensurately-accurate theoretical predictions for the waveform templates 
 utilized in the detection and extraction of source parameters. 
 The required calculations, requiring contributions up to seventh order in 
 Newton's constant~\cite{Favata:2013rwa, Purrer:2019jcp}, 
 rival in complexity the exploration of high energy properties of supergravity theories~~\cite{Bern:2007hh, Bern:2012cd, Bern:2013uka,Bern:2014sna, Bern:2018jmv}. 
 In conjunction with  unitarity methods, the double copy has been 
 key to deriving new state-of-the-art results of interest\footnote{See the  dedicated Snowmass White Paper \cite{gravWhitePaper2022} for a broad discussion of QFT scattering approaches applied to gravitational wave science.} to the
 gravitational wave community.  These methods naturally complement the application of Effective Field Theory approaches to gravitational wave physics, already a mature\footnote{See, e.g., refs.~\cite{Porto:2016pyg,Levi:2018nxp} for recent reviews.} and thriving field.

The double copy and classical gravitational waves were first linked at the level of the classical double copy~\cite{Luna:2016due}. The situation involved a particularly simple gravitational Bremsstrahlung process, which could be understood using Kerr-Schild coordinates. An important breakthrough was made by Goldberger and Ridgway~\cite{Goldberger:2016iau}, who found a direct form of the double copy in the leading-order gravitational radiation generated by scattering two masses. This work was performed in a strictly classical context, involving an iterative solution of the classical equations of motion. This double copy, intuited by Goldberger and Ridgway, was connected to scattering amplitudes, specifically at five points~\cite{Luna:2017dtq}.
Advancing the Goldberger-Ridgway method to next-to-leading order required understanding how to 
disentangle classical numerator factors (which should be double-copied) from kinematic propagators (which should not). This was achieved by 
Shen through consideration of classical bi-adjoint propagation~\cite{ShenWorldLine}.  This tour-de-force result demonstrated the first appearance of the kinematic Jacobi identities  beyond field-theory scattering.
The same formalism has provided evidence of the double copy in the radiation emitted by
bound systems~\cite{Goldberger:2017vcg}.

The connection between scattering amplitudes and observables was developed into an all-orders ``KMOC'' formalism~\cite{Kosower:2018adc}. This formalism is designed to enable the extraction of observables, which are well-defined in both classical and quantum theory, from amplitudes. An initial state $|\psi\rangle$ is chosen which is classically meaningful (involving, for example, point-like particles with positions and momenta which have uncertainty negligible compared to other scales relevant to the process). Scattering amplitudes arise through time evolution to the final state $S |\psi\rangle$. Classical observables are obtained from expectation values of quantum operators in the future. Since all the dynamics is captured by amplitudes, the double copy and unitarity methods are guaranteed to be available to all orders. Spinning particles can be included in this method~\cite{Maybee:2019jus}, and the gravitational waveform itself can be extracted from the expectation value of the Riemann tensor~\cite{Cristofoli:2021vyo}. The double copy is the main source for the gravitational amplitudes required in this formalism.

Complementary recent worldline quantum field theory~\cite{Mogull:2020sak} approaches have produced state-ot-the-art results for the
Bremstrahlung waveform~\cite{Jakobsen:2021smu} including spin~\cite{Jakobsen:2021lvp} (see also~\cite{Mougiakakos:2021ckm}), as well as spinning $G^3$ observables in the scattering scenario \cite{Jakobsen:2021zvh,Jakobsen:2022fcj}. In fact, the
underlining double copy structure of the worldline approach has been investigated in~\cite{Shi:2021qsb} connecting to the classical equation of motion  solutions of ~\cite{ShenWorldLine,Luna:2017dtq}.
 
Precision observables relevant to gravitational-wave physics have been found by direct applying double copy methods in
~\cite{Bern:2019nnu, Bern:2019crd, Damgaard:2019lfh, Bautista:2019evw,Bautista:2019tdr,Herrmann:2021lqe,  Bern:2021dqo, Bautista:2021inx,Herrmann:2021tct, Brandhuber:2021eyq, Brandhuber:2021kpo, Shi:2021qsb, Bjerrum-Bohr:2021vuf, Bjerrum-Bohr:2021din}.  Indeed, the determination of the highest precision post-Minkowskian $\mathcal{O}(G^4)$ corrections to the scattering of classical non-rotating black holes involved a synthesis of effective field theory techniques, advanced multiloop integration, and the double copy applied to tree-level scattering amplitudes~\cite{Bern:2021dqo,Bern:2021yeh}. 
 This calculation determines the classically-relevant part of a three-loop four-point scattering amplitude between two massive scalars. The method exploited the double copy on the cuts, building the gravitational integrand using unitarity methods from double-copied gravitational trees.  A similar strategy 
 was used in \cite{Bern:2020buy} to obtain Compton amplitudes for arbitrary-spin particles and the ${\cal O}(G^2 S_1S_2)$ correction to the leading order amplitude for the scattering of two Kerr black holes~\cite{Vines:2017hyw}.

In massive scalar QCD, at one loop, it is possible to find a color-dual expression for a classically-relevant five-point amplitude. The relevant amplitude involves two massive particle scattering while emitting gluonic radiation~\cite{Carrasco:2020ywq}.
The existence of a color-dual expression opens a direct path to gravitational predictions~\cite{Carrasco:2021bmu} without having to build up the calculation from gravitational trees, potentially offering greatly improved scalability at higher loops.
This example is the first loop-level result involving external massive matter and manifesting the double copy at the level of the integrand. It has direct relevance to observables using, for example, the KMOC formalism.

There is by now also an increasing number of classical double-copy constructions, see e.g. refs.~\cite{Ridgway2015fdl, Goldberger:2017frp,   Carrillo-Gonzalez:2017iyj, Goldberger:2017ogt, Li2018qap, Ilderton:2018lsf, Plefka:2018dpa, Luna:2018dpt, PV:2019uuv,Huang:2019cja,Alawadhi:2019urr, Emond:2020lwi} which describe the emission of gravitational waves from various systems and in some instances can be interpreted from the perspective of scattering amplitudes.

The program to apply scattering amplitude methods to gravitational 
wave physics has produced results sought after by the gravitational wave community~\cite{Damour:2017zjx}, 
uncovered new structures and developed new tools that will enable future progress towards addressing the precision needs of future facilities, such as Einstein Telescope, the Cosmic Explorer, and LISA.
The double copy has been an integral part in this young and dynamic field,
and will be instrumental in meeting remaining challenges, such as the higher-order 
calculations for binaries of generic mass ratio.

Of particular importance are extreme mass-ratio inspirals, which can be observed over long times, expected to complete $10^4-10^5$ cycles in band for LISA~\cite{Babak:2017tow}, and may appear in different frequency windows in different detectors.   
With waveforms of sufficient precision, this much data could be used ~\cite{Berry:2019wgg}, for example: 
(1) Precision tests~\cite{Gair:2012nm} of the no-hair theorem\footnote{For an example of amplitude techniques applied to derive classical observables related to  binary systems involving helicity-0 modes see ref.~\cite{Carrillo-Gonzalez:2021mqj}.} of  general relativity.  (2) Measurement of the masses of both black holes to better than 10\% precision~\cite{Gair:2017ynp}. 
 (3) Reconstruction~\cite{Gair:2010yu} of the mass distribution of massive black holes out to redshift $z \gtrsim 4$.
(4) Precision measurement of the Hubble constant~\cite{MacLeod:2007jd}.
Current methods adapted for such systems, known as the self-force approach~\cite{Mino:1996nk,Quinn:1996am,Barack:2018yvs}, involve determining the motion of the lighter body in the exact classical field of the heavier body. 
Future observatories will be sensitive to contributions to the dynamics which are of second-order in self-force, i.e.~${\cal O}(m_\text{light}^2/m_\text{heavy}^2)$. Such contributions have not yet been completely calculated.
Approaching this problem with scattering methods motivates us to study the double copy 
in the context of scattering amplitudes in background fields. 
Further motivation comes from describing the last phase of a binary's evolution -- the ringdown -- through double-copy methods, as it may open another window on the binary's geometry prior to merger, cf.~e.g.~\cite{Hughes:2019zmt,Li:2021wgz}.

\subsection{Double Copy in Nontrivial Backgrounds}
\label{sect:nontrivialBG}

\noindent{}It is usually an unspoken assumption that the computation of scattering amplitudes is performed perturbatively around a trivial field configuration: one ordinarily considers gluon scattering in vacuum, or gravitational scattering in Minkowski spacetime. This assumption of a trivial background is sufficient for many scenarios of physical interest, but there are myriad experimental, phenomenological and theoretical contexts for which non-trivial backgrounds are better motivated. These range from the Schwinger effect in QED (see~\cite{DiPiazza:2011tq,Gonoskov:2021hwf,Fedotov:2022ely}) to QCD in strong magnetic fields (see~\cite{Kharzeev:2013jha}) to heavy ion and high-energy hadron collisions in QCD (see~\cite{Balitsky:2001gj,Iancu:2002xk,Gelis:2010nm}) to neutron star atmospheres in astrophysics~\cite{Lai:2014nma} to gravitational wave memory~\cite{Christodoulou:1991cr} and cosmology~\cite{Mukhanov:1990me} in GR.

In such scenarios, the background fields are treated as \emph{exact, non-linear, non-perturbative} solutions to the classical equations of motion, sometimes called ``strong'' backgrounds. (It is often a reasonable approximation to neglect backreaction effects on the background, or to deal with them perturbatively.) The traditional framework for this in QFT is the background field formalism (cf., \cite{Furry:1951zz,DeWitt:1967ub,tHooft:1975uxh,Abbott:1981ke}). This is a rich playground, where perturbative and non-perturbative physics meet in many interesting ways.

While the background field formalism is as old as QFT itself, performing explicit computations in even the simplest strong backgrounds is substantially more involved than in trivial backgrounds. The absence of translation invariance requires the use of a position-space formulation of Feynman rules, leading to tree-level amplitudes and loop-integrands which are no longer rational functions of kinematic data. For instance, even in highly symmetric gravitational backgrounds like (anti-)de Sitter space, rational functions are replaced by complicated transcendental functions.

This is reflected by the precision frontier for scattering in gauge theory and gravity with strong background fields. For instance, in constant-curvature symmetric spacetimes the state-of-the-art for gluon scattering\footnote{We abuse terminology by referring to such objects as ``scattering amplitudes''; in asymptotically (A)dS spacetimes, the S-matrix is replaced by boundary correlation functions or coefficients in the wavefunction of the universe, which are the analogous observables.} is only 4-points~\cite{Raju:2012zs,Rastelli:2016nze,Caron-Huot:2018kta,Alday:2020dtb,Albayrak:2020fyp,Armstrong:2020woi,Diwakar:2021juk} or 5-points~\cite{Goncalves:2019znr,Alday:2022lkk} at tree-level, while for tree-level graviton scattering it is four points in AdS$_5$~\cite{Arutyunov:1999nw,Zhiboedov:2012bm,Afkhami-Jeddi:2016ntf} and four points in AdS$_4$~\cite{Binder:2018yvd}. 
In AdS-momentum space algorithms have been developed 
for $n$-point scalar~\cite{Albayrak:2020isk}, gluon~\cite{Albayrak:2019asr} graviton~\cite{Albayrak:2019yve} tree-level correlators, and at loop level~\cite{Albayrak:2020bso}. Complete explicit expressions beyond five-points 
are yet to be found, however. 
Certain AdS gauge theory calculations rely on supersymmetry to relate gluon amplitudes to amplitudes of scalars in the same multiplet, which allows the use of powerful tools such as Mellin space.
By contrast, in Minkowski space the tree-level S-matrix of gluons or gravitons is known to \emph{arbitrary} multiplicity~\cite{Kawai:1985xq,Witten:2003nn,Roiban:2004yf,Cachazo:2012kg,Cachazo:2013hca}\footnote{We should note that there are all-multiplicity formulae for ``maximal U$(1)$-violating'' correlation functions in $\mathcal{N}=4$ sYM~\cite{Green:2020eyj,Dorigoni:2021rdo, Dorigoni:2022iem}, which can be interpreted as certain amplitudes of type IIB supergravity in AdS$_5\times S^5$.}. Likewise, recent progress for direct loop-level calculations in AdS (cf., \cite{Yuan:2018qva,Alday:2019nin,Aprile:2019rep,Drummond:2019hel,Alday:2021ajh,Herderschee:2021jbi,Gomez:2021ujt,Huang:2021xws,Heckelbacher:2022fbx,Drummond:2022dxw}) is largely restricted to 4-points\footnote{Features of four-point correlation functions to all loop orders have also been discussed, with input from 
the dual gauge theory (e.g. \cite{Bissi:2020woe,Bissi:2020wtv}).}. 
For calculations of direct experimental relevance, the situation is similar. The tree-level precision frontier is essentially at 4-points in strong-field QED (a subject that is over 75 years old) for on-going and future high-energy laser and electron/positron beam experiments, such as E320 at FACET-II, ELI, CoReLS, and LUXE (e.g., trident pair production or double Compton scattering, see \cite{DiPiazza:2011tq,Abramowicz:2019gvx,Abramowicz:2021zja,Fedotov:2022ely}).

Going beyond these low-multiplicity results is an opportunity to develop novel tools adapted to position-space problems and to not rely on special symmetry properties. It is also an opportunity to revisit the validity of perturbation theory 
in background fields and the need for resummation beyond that for trivial backgrounds. Higher-multiplicity in strong backgrounds is also of practical importance: while higher-point amplitudes come with additional factors of the ``small'' coupling constant, they are also accompanied by additional background-dressed propagators (cf., \cite{DiPiazza:2011tq,King:2015tba,Seipt:2017ckc,Fedotov:2022ely})\footnote{We would like to thank Anton Ilderton for emphasizing this point to us.}. 
These propagators contain powers of the background itself, which can compensate for the additional coupling factors. The net effect is that low-multiplicity processes do not necessarily dominate high-multiplicity ones. Furthermore, high-multiplicity scattering in curved backgrounds is related to the emission of radiation; this can be seen in terms of classical observables (e.g.~\cite{Ilderton:2013dba,Ilderton:2013tb,Adamo:2022rmp}), and for certain backgrounds is linked with resummations in high-energy scattering on a flat background (e.g.~\cite{tHooft:1987vrq,Jackiw:1991ck,Kabat:1992tb,Lodone:2009qe,Gruzinov:2014moa,Adamo:2021jxz,Adamo:2021rfq}).

With such a large knowledge gap compared to Minkowski space, it is clear that double copy can be a powerful tool for the study of gravitational scattering in curved space-times, but there are several important challenges which must be overcome. Firstly, does a notion of the double copy even exist for scattering amplitudes in background fields, and, if so, what is its general structure? This is not just a practical question: if double copy is really an intrinsic property of perturbative gravity, it should hold on any perturbative background (not just Minkowski space).

One then requires novel methods to generate suitable gauge theoretic ``data'' to seed the double copy. In background fields, it is a challenge to calculate amplitudes even at tree-level, much less arrange them into whatever the strong-field version of a color-kinematics representation might be. Clearly, novel techniques (i.e., beyond the background field expansion on space-time) are required. Finally, on both sides of any curved double copy map, new tools are needed to check and constrain results. Most modern unitarity methods break down in the presence of background fields, meaning that the usual ways of verifying whether a formula for some scattering amplitude is correct are no longer available.

While these are substantial challenges, exciting progress in recent years makes it clear that they are not insurmountable.

\paragraph{Double copy in curved spacetimes:} 
As it stands, there is no definitive prescription for what are the building blocks that lead to gravitational scattering amplitudes in curved spacetime. Do we start with gluon amplitudes (in a color-kinematics representation) in the \emph{same} spacetime? Or should we instead start with gluon amplitudes in flat space but with non-trivial \emph{gauge} background fields? If so, then which gauge background do we select to obtain graviton scattering in the desired curved spacetime\footnote{Note that this question is distinct from, but related to, the question of \emph{classical} double copy for exact solutions discussed in Section~\ref{sect:classicalDC}}?

In the first case, a substantial amount of work has focused on gravitational ``scattering'' (i.e., boundary correlation functions) in AdS, where CFT methods on the boundary can be brought to bear. Here, the proposed double copy map is constructed by ``squaring'' gluon amplitudes in the \emph{same} AdS space-time, based on observations of a double copy structure between the gluon and gravity boundary correlators in AdS momentum space~\cite{Farrow:2018yni,Li:2018wkt,Fazio:2019iit,Lipstein:2019mpu} and Mellin space~\cite{Zhou:2021gnu,Jain:2021qcl}. Additionally, there has been progress in realizing that gluon, NLSM and biadjoint scalar scattering in AdS manifests color-kinematics duality, with amplitudes obeying the BCJ relations~\cite{Albayrak:2020fyp,Armstrong:2020woi,Diwakar:2021juk,Cheung:2022pdk,Herderschee:2022ntr,Drummond:2022dxd}. These observations build on a variety of methods, ranging from Mellin representations of the AdS amplitudes~\cite{Paulos:2011ie,Fitzpatrick:2011ia} to geometric formulations of the kinematics in terms of isometry generators~\cite{Cheung:2021zvb,Cheung:2022pdk,Herderschee:2022ntr}. A key test of these methods in the coming years will be whether or not they can be used to compute gravitational scattering amplitudes in AdS beyond the state-of-the-art set with AdS/CFT techniques.

Arguments that color-kinematics duality should persist in 
any curved spacetime have been put forth
in~\cite{Cheung:2021zvb,Sivaramakrishnan:2021srm}, and it 
was shown in \cite{Cheung:2022pdk} that it does 
hold for NLSMs and the biadjoint scalar theory on symmetric spaces. 
Some explicit calculations \emph{have} been possible for plane wave spacetimes~\cite{Adamo2017nia,Adamo:2020qru}; 
these admit an S-matrix~\cite{Gibbons:1975jb}, extend to solutions of string 
theory~\cite{Amati:1988ww,Amati:1988sa,Horowitz:1989bv,Horowitz:1990sr} and can be 
viewed as local approximations to any spacetime~\cite{Penrose1976}. Here, the proposed 
double copy map does \emph{not} involve gluon scattering on the same plane wave spacetime, 
but rather gluon scattering in a gauge theoretic plane wave background \emph{in Minkowski space}. 
This gluonic plane wave scattering has been shown to obey a strong-background version of the color-kinematics 
duality~\cite{Adamo:2018mpq}, and is of interest in its own right for various topics in 
strong-field QED/QCD (see \cite{Yakimenko:2018kih,Adamo:2021jxz,Fedotov:2022ely}).

A definitive verdict on which sort of prescription is correct (indeed, it may be both -- or neither) will rely on their ability to extend calculations beyond the reach of standard background perturbation theory methods (including Witten diagrams). Recently discovered all-multiplicity formulae for gluon and graviton scattering in chiral/self-dual gauge-theoretic and gravitational plane waves~\cite{Adamo:2020syc,Adamo:2020yzi,Adamo:2022mev} should provide an important data set on which proposals can be tested in the future. 

\paragraph{Generating gauge theory data:} To improve the precision frontier of perturbative gravity in curved spacetime one must have gauge theoretic scattering data to feed into the double copy map. This means improving on the state-of-the-art for gluon scattering in strong background gauge fields as well as curved space-times. One promising toolkit for doing this is ambitwistor string theory~\cite{Mason:2013sva,Berkovits:2013xba,Adamo:2013tsa}, which underpins the CHY version of double copy~\cite{Cachazo:2013hca,Cachazo:2013iea,Cachazo:2014xea}. Ambitwistor strings can be exactly coupled to non-trivial gauge and gravitational backgrounds~\cite{Adamo:2014wea,Adamo:2018hzd,Adamo:2018ege}, and have the remarkable property that the worldsheet CFT remains solvable (i.e., local anomalies and worldsheet OPEs can be computed exactly). Ambitwistor strings can compute 3-point scattering amplitudes of gluons and gravitons in plane wave backgrounds~\cite{Adamo:2017sze}, $n$-point gluon/graviton scattering in AdS$_3$~\cite{Roehrig:2020kck}, and $n$-point scalar scattering in (A)dS spacetimes of arbitrary dimension~\cite{Eberhardt:2020ewh,Gomez:2021qfd,Gomez:2021ujt}. Developing a formalism to compute the worldsheet correlators of ambitwistor strings in generic backgrounds and target space dimensions would provide a mechanism to generate large amounts of data on both sides of any double copy correspondence for curved backgrounds.

Another promising avenue is provided by twistor theory. This trivializes the self-dual sectors of gauge theory and gravity in four-dimensions~\cite{Ward:1977ta,Penrose:1976js}, making the computation of scattering amplitudes on self-dual backgrounds tractable. For instance, twistor theory provides formulae for the complete tree-level S-matrix of Yang-Mills theory on any self-dual radiative gauge field background~\cite{Adamo:2020yzi}, and there is scope to describe gauge and gravitational scattering on \emph{any} self-dual background, including dyons, instantons and self-dual black holes. How to study non-chiral background fields with twistor theory remains an important open question. 

Finally, there has been progress in the use of worldline methods for strong background fields (cf., \cite{Edwards:2021elz,Edwards:2021uif}). An impressive example is a generating functional for $n$-point, one-loop photon scattering amplitudes in electromagnetic plane wave backgrounds~\cite{Edwards:2021vhg}. Further developments in this area (including generalizing the formalism to non-abelian gauge theories) could provide a rich source of data to seed any double-copy map for curved spacetime. 

\paragraph{Unitarity methods in background fields:} While some on-shell methods, like spinor-helicity variables in (A)dS~\cite{Maldacena:2011nz,Nagaraj:2018nxq,David:2019mos} and beyond~\cite{Adamo:2019zmk,Adamo:2020syc,Adamo:2020qru,Adamo:2022mev}, still work in curved spacetimes, the familiar toolkit of generalized unitarity fails as soon as generic background fields are introduced, since tree-amplitudes and loop-integrands are no longer rational functions. To make progress with double copy in curved spacetimes, new perspectives on constraining candidate amplitude formulae are required. In the context of (A)dS scattering, versions of BCFW recursion, ``cutting'' rules and generalized unitarity have been developed~\cite{Raju:2010by,Raju:2011mp,Raju:2012zr,Caron-Huot:2017vep,Simmons-Duffin:2017nub,Arkani-Hamed:2018kmz,Sleight:2019hfp,Sleight:2019mgd,Meltzer:2019nbs,Drummond:2020dwr,Baumann:2020dch,Meltzer:2020qbr,Melville:2021lst,Goodhew:2021oqg,Albrychiewicz:2021ndv,Baumann:2021fxj,Bonifacio:2021azc,Meltzer:2021zin,Sleight:2021plv,Baumann:2022jpr}; while promising, these have yet to deliver results for gluon or graviton scattering that surpass the state-of-the-art from standard perturbation theory or boundary CFT methods. However, it is hopefully only a matter of time before these techniques become as powerful as their flat space counterparts.

More general backgrounds are much less studied. One tantalizing avenue was recently explored in the context of plane wave backgrounds in QED, where it was shown that the simple requirement of gauge invariance imposes a factorization-like decomposition of strong field amplitudes~\cite{Ilderton:2020rgk}. While the study of such structures is still in its infancy (and yet to be generalized to non-abelian gauge theories), it shows the potential for factorization-based arguments to be lifted to background field settings. A similarly enticing idea would be to use worldsheet descriptions of curved space-time scattering (based on strings or ambitwistor strings) to study factorization properties. Here, one would simply focus on the cutting properties of the underlying worldsheet CFT~\cite{Vafa:1987ea,Polchinski:1988jq,Witten:2012bh,Adamo:2013tca}; on general grounds this should then imply a ``worldsheet factorization'' argument for the amplitudes themselves, regardless of background fields in space-time.

Even without unitarity-based tools, there are still basic consistency checks like the flat or perturbative limit (i.e., where the background is treated as a single gluon or graviton) which can be deployed. These are well-studied at tree-level and even at loops; for instance, multi-perturbative limits of loop-level amplitudes in strong field QED/QCD must match multi-collinear limits of higher-multiplicity loop amplitudes in trivial backgrounds~\cite{Adamo:2021hno}. Finally, studying flat spacetime scattering in alternatives to the usual momentum eigenstate basis could provide new insights into how to constrain curved spacetime scattering amplitudes. For 
example, 
in the conformal primary basis~\cite{Pasterski:2016qvg,Pasterski:2017kqt} there is no longer momentum conservation and tree amplitudes are not rational functions of the kinematic data. Thus, the arena of celestial holography~\cite{Pasterski:2021raf} may have important lessons to teach us about double copy in curved spacetimes~\cite{Casali:2020vuy,Casali:2020uvr,Kalyanapuram:2020epb,Pasterski:2020pdk}.  

\subsection{Double Copy and Cosmology}
\label{sect:dcCosmo}

An intriguing application of the double copy consists of providing new insights and tools for cosmology. While some initial progress has been made \cite{Bahjat-Abbas:2017htu,Carrillo-Gonzalez:2017iyj,Li:2018wkt,Farrow:2018yni,Fazio:2019iit,Lipstein:2019mpu,Albayrak:2020fyp,Armstrong:2020woi,Diwakar:2021juk,Sivaramakrishnan:2021srm,Jain:2021qcl,Herderschee:2022ntr,Cheung:2022pdk,Zhou:2021gnu,Drummond:2022dxd}, there are still many challenges to overcome. There are two primary directions that one can consider, namely early and late universe cosmology. In the following, we discuss how the double copy can contribute towards progress in both areas.

It is widely believed that the structure of the Universe that we observe today has its origin in quantum fluctuations that arose during an inflationary period (quasi-de Sitter phase). By looking at the spatial correlation functions of these fluctuations living in the future boundary of the inflationary spacetime, one can learn about the particles present in this early epoch and their dynamics, thus giving an opportunity to observe beyond the Standard Model physics. These fluctuations can be observed by looking at the statistics of the cosmic microwave background and large-scale structure. Currently, we only have measurements of the scalar 2-point function, but ongoing and future experiments will be able to provide measurements or tight constraints on tensor 2-point functions and non-gaussianities encoded in three-point functions \cite{Dore:2014cca,CMB-S4:2016ple,NASAPICO:2019thw,SimonsObservatory:2018koc,Beutler:2019ojk,Meerburg:2019qqi,Darwish:2020prn}.

While higher-point statistics will be hard to observe in the near future, one can learn valuable lessons regarding the structure of QFT in curved-spacetimes from them. The traditional computations of primordial correlation functions can become involved beyond the 2-point and tree-level cases\footnote{Note that a complementary approach dubbed the Cosmological Bootstrap approaches this problem from a different perspective using unitarity, locality, and symmetries; see the dedicated White Paper \cite{Baumann:2022jpr} for a thorough discussion on this topic.} \cite{Mukhanov:1990me,Malik:2008im,Wang:2021qez}. Therefore, a double copy relation between primordial correlators could give a new simpler method for performing calculations at loop order and for higher points in addition to providing insights on the symmetries of QFT in curved backgrounds.

As discussed also in \sect{sect:nontrivialBG}, some progress in this direction\footnote{Progress made in an AdS/CFT context \cite{Lipstein:2019mpu,Albayrak:2020fyp,Armstrong:2020woi,Diwakar:2021juk,Herderschee:2022ntr,Zhou:2021gnu,Drummond:2022dxd} and maximally symmetric spacetimes \cite{Sivaramakrishnan:2021srm,Cheung:2022pdk} can provide lessons for a dS double copy.} has been made in refs.~\cite{Farrow:2018yni,Li:2018wkt,Fazio:2019iit,Jain:2021qcl}, but a better understanding of the double copy on curved spacetimes is still required. In refs.~\cite{Farrow:2018yni,Li:2018wkt,Fazio:2019iit,Lipstein:2019mpu,Albayrak:2020fyp,Armstrong:2020woi} it was shown that the residue of the total energy pole of certain tensor (A)dS correlators is given as the square of Yang-Mills amplitudes (as expected since this residue is given by the flat space gravitational amplitude \cite{Maldacena:2011nz,Raju:2012zr}). These papers also give hints that the double copy relation can be extended beyond this special kinematic limit; thus, it would be interesting to understand how to build an explicit double copy and whether it can be solely expressed in terms of correlators. 

One should also note that the most progress has been made for conformally flat spacetimes, such as de Sitter, but inflationary spacetimes break de Sitter symmetries. Fortunately, in the slow-roll regime the symmetries are softly broken and one can compute inflationary correlators from de Sitter ones by taking soft limits \cite{Kundu:2014gxa,Arkani-Hamed:2015bza,Arkani-Hamed:2018kmz,Baumann:2019oyu}. To go beyond the slow-roll regime, one would need to be more ambitious and construct a double-copy relation in a background with broken time translations. In fact, this is an intricate challenge for constructing a double-copy relation that holds in more generic cosmological contexts.

Concerning late universe cosmology, we will focus on theories of dark energy, i.e., those that drive the accelerated expansion at late times. Whether these theories are within the web of theories related through double-copy relations is an open question. Some initial insights come from the appearance of the special galileon as a double copy. Galileon theories are higher-derivative theories with spacetime-dependent shift-symmetries that can behave as dark energy \cite{Nicolis:2008in}. Nevertheless, the special galileon corresponds to a branch that cannot drive accelerated expansion. Besides galileons, there are more general scalar and vector effective field theories that can describe dark energy and generically involve higher derivative operators, see for example refs.~\cite{Joyce:2014kja,deRham:2021efp,DeFelice:2016yws} and references therein. Explorations of effective field theories within the context of the double copy, see \sect{sect:GenDC} and \sect{sect:higherDerivative}, can shed light on whether dark energy could be obtained as a double copy. A perturbative double copy could be useful in some restricted scenarios where the theory is weakly coupled. Nevertheless, most of these theories rely on screening mechanisms that render their effects negligible within the solar system to be consistent with observations \cite{Joyce:2014kja,Brax:2021wcv}. These screening mechanisms depend on classical non-linearities becoming strong, thus only a non-perturbative double copy could shed light on these regimes.

Another intriguing class of theories that could give rise to a late accelerated expansion are those that involve massive gravitons \cite{deRham:2014zqa,Hinterbichler:2011tt,Heisenberg2021,Hinterbichler:2016try}. As an aside, we also note that massive spin-2 fields can be relevant in cosmological scenarios as possible dark matter states \cite{Blanchet:2015bia,Marzola:2017lbt,Aoki:2017cnz}. Well-behaved gravitational theories involving massive gravitons are reviewed in refs.~\cite{deRham:2014zqa,Hinterbichler:2011tt,Heisenberg2021}, a new construction with a higher strong coupling scale can be found in \cite{Gabadadze:2017jom}, and an exactly solvable two-dimensional version, which is equivalent to $T\bar{T}$ deformations, is analyzed in ref.~\cite{Tolley:2019nmm}. Calculations in these theories are highly intricate due to the complicated tensor structures appearing in their interactions. Thus, the possibility of performing computations for massive gravity theories from simpler massive gauge theories is largely appealing. It has been shown that the  standard double copy of massive gauge fields does not always correspond to a well-defined local gravitational theory \cite{Johnson:2020pny,Momeni:2020hmc}. Some constraints have been formulated so that the double copy is a healthy theory, but these constraints have several assumptions on the spectrum, interactions, or spacetime dimensions of the theory \cite{Johnson:2020pny,Momeni:2020hmc,Gonzalez:2021bes}. Most of these cases consider formulations that resemble the massless case and thus lead to theories arising from simple patterns of spontaneous symmetry breaking of massless theories. The special three-dimensional case \cite{Gonzalez:2021bes,Gonzalez:2021ztm,Gonzalez:2022otg,Moynihan:2020ejh,Moynihan:2021rwh,Burger:2021wss,Emond:2022uaf,Hang:2021oso,Ben-Shahar:2021zww} deviates from this paradigm\footnote{While the case of topologically massive theories is gauge invariant just as the massless theories, a healthy double copy can also be constructed from the square of massive Yang-Mills restricted to a special set of polarizations \cite{Gonzalez:2021bes}.} and shows that there could be different methods for obtaining healthy double-copies that do not resemble the structure of the massless one. Extending this to four-dimensional theories would be an important breakthrough in describing massive gravity theories. Beyond the cosmological implications,  theories of massive gravity could also be relevant for studying black hole physics in theories beyond general relativity \cite{Rosen:2017dvn,Rosen:2018lki,Berens:2021tzd}.

\subsection{Interplay Between QFT and String Theory}
\label{sect:stringAmplitudes}

\noindent{}String theories provide a natural framework to understand the double copy structure 
of gravitational interactions since closed-string degrees of freedom are built from tensor 
products of color-stripped open-string degrees of freedom. Gravitational states due to 
massless vibration modes of closed strings are therefore organized into double copies
of gauge multiplets from open-string excitations. As will be detailed below, this double copy 
structure of the spectrum propagates to closed-string scattering amplitudes such that
the point-particle limit\footnote{More precisely, the low-energy expansion of string amplitudes
around the point-particle limit is performed in the dimensionless combinations $\alpha’ k_i \cdot k_j$ 
formed by the inverse string tension $\alpha’$ and light-like external momenta $k_i$.} $\alpha' \rightarrow 0$ relates gravitational
interactions to bilinears in gauge-theory building blocks. Recent developments
in string perturbation theory -- see the White Paper~\cite{Berkovits:2022ivl} for an overview -- led to 
concrete multi-loop and multi-leg manifestations of the gravitational double copy 
and the closely related color-kinematics duality of gauge theories.

By the KLT relations between closed-
and open-string tree-level amplitudes \cite{Kawai:1985xq}, 
classical gravity predictions are known to reduce to gauge-theory input. 
At loop level in turn, chiral splitting \cite{DHoker:1988pdl, DHoker:1989cxq} expresses the integrand
of closed-string amplitudes at fixed loop momenta as a square of chiral halves
which individually integrate to open-string amplitudes upon 
specifying the boundary conditions for their endpoints.
Accordingly, the double-copy construction of quantum-gravity interactions 
and the color-kinematics duality of gauge theories manifest themselves 
at the level of the loop integrand \cite{Bern:2010ue}.

Manifestly gauge-invariant incarnations of the color-kinematics duality
\cite{Bern:2008qj, Boels:2011tp} descend from monodromy 
relations among color-stripped open-string amplitudes, both at tree level \cite{Plahte:1970wy, Stieberger:2009hq, BjerrumBohr:2009rd} and in loop integrands \cite{Tourkine:2016bak, Hohenegger:2017kqy, Casali:2019ihm, Casali:2020knc, Stieberger:2021daa}. The manifestly local formulation of the duality via kinematic Jacobi relations
of gauge-theory numerators was realized via point-particle limits of open strings at
tree level \cite{Mafra:2011kj, Mafra:2015vca}, one loop \cite{Mafra:2014gja, He:2015wgf, Bridges:2021ebs} and two loops \cite{Mafra:2015mja, DHoker:2020prr}. 
In field-theoretic terms, kinematic Jacobi relations within tree-level subdiagrams 
originate from a non-linear gauge transformation of perturbiner solutions to
the classical equations of motion \cite{Rosly:1996vr, Rosly:1997ap, Selivanov:1997aq}.
The construction of perturbiners in so-called BCJ gauge 
\cite{Lee:2015upy, Garozzo:2018uzj, Bridges:2019siz} is crucially inspired 
by BRST cohomology methods of the pure-spinor superstring \cite{Berkovits:2000fe}.\footnote{Also see \cite{Ahmadiniaz:2021fey, Ahmadiniaz:2021ayd} for recent worldline approaches to BCJ gauge.} Also for the NLSM, string-theory methods led to compact all-multiplicity expressions for color-kinematics dual tree-level numerators \cite{Carrasco:2016ldy, Carrasco:2016ygv}.
More general string-motivated perspectives on kinematic Jacobi identities can be obtained from
vertex operator algebra constructions
\cite{Mafra:2014oia, Fu:2018hpu}, the pure-spinor master action of ten-dimensional sYM~\cite{Ben-Shahar:2021doh} or residue theorems for moduli spaces of Riemann surfaces with marked points~\cite{Mizera:2019blq}.

The CHY formalism \cite{Cachazo:2013gna, Cachazo:2013hca} and the underlying 
ambitwistor strings \cite{Mason:2013sva, Berkovits:2013xba, Adamo:2013tsa} 
generalize the Witten-RSV  \cite{Witten:2003nn, Roiban:2004yf} and Cachazo-Skinner \cite{Cachazo:2012kg} formulae
beyond four spacetime dimensions
and share several facets of
the worldsheet description of conventional string theories. However, ambitwistor strings do not involve any analogue
of $\alpha'$ and directly compute field-theory amplitudes, see \cite{Geyer:2022cey} for a review. The tree-level realization of
the color-kinematics duality via CHY and ambitwistor methods \cite{Cachazo:2013iea, 
Bjerrum-Bohr:2016axv, Du:2017kpo, Edison:2020ehu} follows the lines of conventional-string
computations \cite{Gomez:2013wza, He:2018pol}, also see \cite{Arkani-Hamed:2017mur,Frost:2019fjn} for a closely related geometric construction in kinematic space. At loop level, the ambitwistor string yields modified Feynman propagators 
linearized in loop momentum \cite{Geyer:2015bja, Geyer:2015jch, Cachazo:2015aol} which 
introduces additional flexibility to find color-kinematics dual gauge-theory numerators
and KLT-formulae for loop integrands in supergravity \cite{He:2016mzd, He:2017spx}.

Conversely, the color-kinematics duality and double-copy structure of field theories 
provided striking insights into string amplitudes including their full-fledged $\alpha'$-dependence!
For instance, the recent proposal for three-loop four-point superstring amplitudes
\cite{Geyer:2021oox} grew out of the color-kinematics dual representation of
their sYM counterpart \cite{Bern:2010ue} and the higher-loop framework for 
ambitwistor-string amplitudes \cite{Geyer:2016wjx, Geyer:2018xwu, Geyer:2019hnn}.
The three-loop four-point expression in \cite{Geyer:2021oox} borne out of field-theory considerations 
conjecturally uplifts the low-energy limit \cite{Gomez:2013sla} 
to all orders in $\alpha'$. Similarly, the one-loop double-copy formulae due to ambitwistor strings inspired the construction of one-loop matrix elements with insertions of the tr$(D^{2k}F^n)$ and $D^{2k}R^n$ operators seen in superstring tree-level effective actions \cite{Edison:2021ebi}. These matrix elements carry key information
on the discontinuity structure in the $\alpha'$-expansion of one-loop superstring amplitudes.

String tree-level amplitudes benefit from a particularly rich interplay with field-theory structures:
disk amplitudes of open superstrings line up with the field-theory version of the KLT formula that double copies sYM with a basis of $\alpha'$-dependent moduli-space integrals \cite{Mafra:2011nv, Broedel:2013tta}. The latter can be interpreted
as (doubly-ordered) tree amplitudes of a putative effective theory of bi-colored scalars dubbed
Z-theory \cite{Carrasco:2016ldy, Mafra:2016mcc, Carrasco:2016ygv}. This generalizes the appearance of bi-adjoint $\phi^3$ \cite{Cachazo:2013iea} and NLSM amplitudes \cite{Carrasco:2016ldy} in the low-energy limit of disk integrals to infinite families of scalar higher-derivative interactions subject to the color-kinematics duality. It is a two-fold surprise
to encounter open strings as the {\it output} rather than the {\it input} of a double-copy construction
and to find all orders in $\alpha'$ resonating with a field-theory double copy!

Also, closed strings turn out to furnish field-theory double copies beyond the scope of the 
traditional string-theory KLT formula \cite{Kawai:1985xq}: tree-level amplitudes of type-II superstrings
were expressed in terms of sYM double copied with the single-valued version of 
open superstrings \cite{Schlotterer:2012ny, Stieberger:2013wea}. The underlying 
single-valued map \cite{Schnetz:2013hqa, Brown:2013gia} acts on
the multiple zeta values (MZVs) in the $\alpha'$-expansion of the disk integrals and thereby
reproduces the sphere integrals of closed-string tree amplitudes 
\cite{Stieberger:2014hba, Schlotterer:2018zce, Vanhove:2018elu, Brown:2019wna}. All the 
$\alpha'$-corrections to the open-string monodromy relations are washed out by
the single-valued map such that single-valued open superstrings obey field-theory
BCJ amplitude relations at all orders in $\alpha'$ \cite{Stieberger:2014hba}.
 
Similar field-theory double copy constructions apply to heterotic and bosonic 
strings \cite{Stieberger:2014hba, Huang:2016tag, Azevedo:2018dgo}. Open 
bosonic strings for instance arise from Z-theory 
double copied with a massive gauge theory known as $(DF)^2$+YM \cite{Azevedo:2018dgo} 
that was constructed in the context of (mass-deformed) conformal supergravity 
\cite{Johansson:2017srf}. Tree-level amplitudes of heterotic strings with any combination 
of external gauge and gravity multiplets are double copies of single-valued open superstrings with 
the $(DF)^2$+YM$+\phi^3$ theory \cite{Johansson:2017srf, Azevedo:2018dgo} which reproduces
the double-copy structure of Yang--Mills--Einstein theories \cite{Chiodaroli:2014xia} at $\alpha' \rightarrow 0$. Hence, massless tree amplitudes in a variety of perturbative string theories fit into the array of double copies in Figure \ref{overviewstring} below, where both the KLT kernel and one of the double-copy components refer to field theories. Figure \ref{overviewstring} may be viewed as a string-theory analogue of the field-theory multiplication table in Figure \ref{fig:DCtable}. 

\begin{figure}[h]
\begin{equation} \! \! \!
{\setstretch{1.75} 
\begin{array}{c||c|c|c}
{\rm string}\otimes {\rm QFT}  &\textrm{sYM} &(DF)^2 {+} \textrm{YM}&\ (DF)^2 {+} \textrm{YM}{+}\phi^3 \\\hline \hline
\textrm{Z-theory} \ & \ \textrm{open superstring} \ \, &\ \textrm{open bos.\ string} \ \, &{\setstretch{0.85} \begin{array}{c} 
\textrm{compactified open} \\
  \textrm{bosonic string}
  \end{array}}  \\
\textrm{sv}\Big(
{\setstretch{0.85} \begin{array}{c} 
\textrm{open} \\
  \textrm{superstring}
  \end{array}}
\Big)  \ & \ \textrm{closed superstring}  \ \, &\ \textrm{heterotic (gravity)} \ \, &\textrm{heterotic (gauge/gravity)}  \\
\textrm{sv}\Big(
{\setstretch{0.85} \begin{array}{c} 
\textrm{open} \\
  \textrm{bos.\ string}
  \end{array}}
\Big)  \ & \ \textrm{heterotic (gravity)} \ \, &\ \textrm{closed bos.\ string} \ \, &{\setstretch{0.85} \begin{array}{c} 
\textrm{compactified closed} \\
  \textrm{bosonic string}
  \end{array}}
\end{array}} \nonumber
\end{equation}
\caption{Double copy constructions of string amplitudes as presented in~\cite{Azevedo:2018dgo}.}
\label{overviewstring}
\end{figure}

The $\alpha'$-expansions of open-superstring \cite{Schlotterer:2012ny, Broedel:2013aza, Mafra:2016mcc} and $(DF)^2$+YM tree amplitudes pinpoint infinite families of higher-derivative operators that respect the color-kinematics duality and participate in field-theory double copies \cite{Broedel:2012rc, Stieberger:2014hba, Huang:2016tag}. The matrix elements of color-kinematics dual tr$(D^{2k}F^n)$ operators can be made fully explicit at any multiplicity by isolating the coefficients of different MZVs in open-string amplitudes: since all the $\alpha'$-corrections in the monodromy relations \cite{Plahte:1970wy, Stieberger:2009hq, BjerrumBohr:2009rd} occur in the rigid combination of $(\pi \alpha')^2$, any operator without powers of $\pi^{2}$ in its coefficient is bound to satisfy BCJ relations.\footnote{In the first place, powers of $\pi^{2}$ arise from the coefficients of even zeta values $\zeta_{2k} \in \mathbb Q \pi^{2k}$. A more precise identification of open-string amplitude contributions that obey BCJ relations relies on the $f$-alphabet description of MZVs \cite{Brown:2011ik, Schlotterer:2012ny}.} The simplest color-kinematics dual examples derived from this logic are matrix elements with single-insertions of $\alpha' {\rm tr} F^3$ from the open bosonic string and $\alpha'^3  \zeta_3  {\rm tr}( D^2 F^4+F^5)$ universal to open bosonic strings and open superstrings \cite{Broedel:2012rc}.
Higher-order examples of color-kinematics dual operators include the coefficients of all single-valued MZVs (whose double copy is realized in closed-string tree amplitudes) but also coefficients of non-single-valued MZVs such as $\alpha'^{\geq 8}  \zeta_{3,5} {\rm tr}( D^{\geq 10} F^5)$ that do not enter any closed-string amplitude. 

While the string-theory KLT and monodromy relations  apply universally to external states at arbitrary mass levels, the field-theory double copy structures in Figure \ref{overviewstring} are specific to massless string excitations. 
As a first echo for massive external states, open-superstring amplitudes with a single mass-level-one state and any number of gauge multiplets have been brought into a field-theory KLT form \cite{Guillen:2021mwp}: Z-theory amplitudes are double copied with field-theory kinematics of a colorless massive spin-two multiplet coupled to sYM and supergravity as realized by the heterotic version of chiral strings \cite{Hohm:2013jaa, Huang:2016bdd}.
It would be interesting to test this double copy for multiple external massive spin-two multiplets and to further explore their role in conformal supergravity \cite{Ferrara:2018wlb} and bimetric gravity \cite{Lust:2021jps}. Another important question (with potential input from the asymmetrically twisted variants of chiral strings \cite{Jusinskas:2021bdj}) is whether such massive double-copies generalize to higher mass levels of open-string spectra.

\noindent{\bf Future targets:} An immediate goal for the future is to investigate the loop-level systematics of (i) generating color-kinematics dual gauge-theory numerators from the $\alpha’ \rightarrow 0$ limit of string amplitudes and (ii) identifying field-theory double-copy structures that apply to the full $\alpha'$-dependence. 
\begin{itemize}
\item[(i)] The current frontier of deriving BCJ numerators from open superstrings is at the two-loop five-point level~\cite{DHoker:2020prr}. Higher-loop and -leg orders are a central research target which will require refined methods to handle the spin-structure sums of RNS superstrings or b-ghost
correlators in the pure-spinor formalism, see the dedicated White Paper~\cite{Berkovits:2022ivl}. At the same time, one-loop six-point supergravity amplitudes in general dimensions $D\leq 10$ have 
not yet been derived from sYM numerators (see e.g.~\cite{Mafra:2014gja, Bridges:2021ebs}) through the BCJ double copy without resorting to the linearized propagators of the ambitwistor-string prescription.
\item[(ii)] For open superstrings at one loop, a double-copy structure based on building blocks in the chiral-splitting procedure \cite{DHoker:1988pdl, DHoker:1989cxq} was proposed in \cite{Mafra:2017ioj, Mafra:2018qqe}, and it remains to explore their connections with loop amplitudes of Z-theory. More generally, double-copy prescriptions for loop amplitudes of open and closed strings are likely to hinge on suitable bases of moduli-space integrals that admit interpretations in terms of scalar effective field theory. A parallel goal pioneered in \cite{Zerbini:2015rss, DHoker:2015wxz, Broedel:2018izr, Zagier:2019eus} is to generalize the relations between closed strings 
and single-valued open strings beyond tree level, see \cite{Gerken:2020xfv} for a one-loop
proposal covering a conjectural basis of open-string integrals.
\end{itemize}

\noindent{\bf Big Picture:} 
In summary, double copy and color-kinematics duality were seen as surprisingly
universal phenomena interweaving string- and field-theory structures to a
growing web of theories, see Figure \ref{FigWeb} below. On the one hand,
the field-theory limit $\alpha' \rightarrow 0$ of string amplitudes keeps on sharpening our
understanding of the elegant interplay between gauge theories and gravity. On the other
hand, field theories start to return the favor and shed new light on
string perturbation theory. It will be rewarding to target string and point-particle amplitudes
in parallel in future research and to search for drastic re-formulations of their S-matrices which manifest their synergies.

\subsection{Generalizing the Double Copy and the KLT Bootstrap}
\label{sect:GenDC}
We described in the Introduction how the double copy links a variety of field theories to each other; see Figure \ref{fig:DCtable}. We can think of this as a multiplicative map on a subspace of field theories. It is interesting to understand better what it takes for such a map to work and how the double copy may be generalized to include a larger subspace of field theories. Examples:
\begin{itemize}
\item \textbf{Different representations of the color group}. Gluons are in the adjoint representation, but it is of interest too to include fundamental matter (quarks, leptons) \cite{Johansson:2014zca,Johansson:2015oia,He:2016dol, Brown:2018wss, Carrasco:2020ywq} or to study which representations and group-theory structures are compatible with the double copy, not only for the particles themselves but also for their interactions.
\item \textbf{Massive matter and mediators}.  
When masses are associated with matter particles in generic representations of the gauge group, e.g.~quarks in QCD, the generalization is relatively straightforward~\cite{Johansson:2015oia}; different masses can be bundled along with flavor and one simply associates distinct graph edges for different masses and follows group theory commutation identities to relate their weights. This color-dual perspective generalizes nicely to relations between graphs contributing to integrands at loop level~\cite{Carrasco:2020ywq}.  At tree level having different graphs with constrained algebraic relations does result in modified amplitude relations, KLT kernels, and CHY representations~\cite{Johansson:2015oia, Naculich:2014naa,Naculich:2015zha,Naculich:2015coa, Brown:2018wss}. The situation with more general massive states is more subtle with additional constraints beyond color-kinematics duality required for a physically consistent double copy~\cite{Johnson:2020pny, Momeni:2020vvr}. These constraints are automatically satisfied for spontaneous symmetry breaking in the adjoint~\cite{Chiodaroli:2015rdg,Chiodaroli:2017ehv,Momeni:2020hmc}. Consistent double-copy constraints for massive mediators is a subject of ongoing study~\cite{Gonzalez:2022mpa}, including generalizations to topological massive theories in three dimensions, see e.g.~\cite{Moynihan:2020ejh,Gonzalez:2021bes,Moynihan:2021rwh,Hang:2021oso,Li:2021yfk,Gonzalez:2021ztm,Gonzalez:2022mpa} and references therein.
\item \textbf{Higher-derivative operators}. This can be done with an eye on how to generate specific operators, as needed for example as counterterm input for loop-calculation, or systematically in the sense of what are the most general higher-derivative terms allowed \cite{Carrasco:2019yyn,Carrasco:2021ptp,Chi:2021mio}.
\end{itemize}
In this section we focus on recent progress on how to generalize the KLT form of the double copy. We consider input theories with with adjoint massless particles, such as for example Yang-Mills theory. One class of applications is the double copy in the context of effective field theory, such as 
\begin{equation}
(\text{YM + h.d.}) \otimes 
(\text{YM + h.d.})
=\text{gravity + h.d.}
\end{equation}
where h.d.~stands for  higher-derivative operators.  

It turns out that not all gauge theory operators can participate in the standard field theory double-copy construction. 
Recall from Section \ref{rev:Trees} that in order to be input for the double-copy relation, the color-ordered tree amplitudes have to satisfy the Kleiss-Kuijf (KK) and  BCJ relations. For YM theory with higher-derivative operators, $\text{tr} F^3$ satisfies the KK and BCJ relations, but 
$\text{tr} F^4$, for example, does not \cite{Broedel:2012rc}.
From a bottom-up effective field theory approach, it is interesting to understand better the selection-principle of which higher-derivative operators are allowed and which ones are not. And, to the point of this section, it is relevant to ask if the double copy can be generalized to admit a larger class of operators. 
To describe such generalizations, let us first take a closer look at the double copy to appreciate how non-trivial its existence really is. 

Consider for example the color-stripped gluon tree amplitude $A_4[1234]$. It has simple poles in the $s_{12}=(p_1+p_2)^2$ and $s_{14}=(p_1+p_4)^2$ channels, but none in the $s_{13}=(p_1+p_3)^2$ channel. If we naively square it, we get an unphysical object $A_4[1234]^2$ which has double poles at $s_{12}=0$ and $s_{14}=0$. In contrast, the 4-graviton tree amplitude $M_4(1234)$ has simple poles in all three channels, $s_{12}$, $s_{13}$, and $s_{14}$. In light of this, how can a double copy ever work? In the KLT formulation, the key ingredient is the {\em KLT kernel} $S_n$: the field theory KLT formula for the 4-point amplitude can be written as
\begin{equation}
  M_4(1234) = A_4[1234] \,S_4[1234|1234] \,A_4[1234]\,,
  ~~~\text{with}~~~S_4[1234|1234] = -\frac{s_{12}s_{14}}{s_{13}}\,.
\end{equation}
The kernel serves two crucial purposes: 
\begin{enumerate}
\item[(1)] it has zeroes that precisely cancel double poles in the product of gauge theory amplitudes, and 
\item[(2)] it provides the missing poles, in this case in the $s_{13}$-channel. 
\end{enumerate}
One thing is getting the pole structure right, another is getting the residues of those poles correct so that the resulting amplitude is indeed that of gravitons. This is all non-trivial; yet, it works. And not just at 4-point, but for any $n$ points.

The non-triviality of the double copy is important to keep in mind when we consider its possible generalizations. In the KLT formulation of the double copy, it is natural to consider modifications of the KLT kernel. The discussion above makes it  clear that changes to the KLT kernel can easily wreck properties (1) and (2) and render the construction unphysical. Moreover, spurious poles in the kernel could result in an unphysical expression for $M_4$ that is not a tree amplitude in any local theory. To address these issues, we must understand what the rules are for generalizations of the double copy kernel. 

A recent proposal \cite{Chi:2021mio} for generalizing the KLT double copy was based on the KLT algebra: the idea that the double copy has an associated identity element. Given a KLT product $\otimes$, defined by a given kernel $S_n$, the proposition is that there exists an associated identity element such that
\begin{equation}
\label{KLTalgebra0}
   \mathbf{1}  \otimes  \mathbf{1} = \mathbf{1} ~~~\text{and}~~~
   \mathbf{1}  \otimes \mathbf{R} = \mathbf{R} \,
      \,,~~~~~
    \mathbf{L} \otimes\mathbf{1}   = \mathbf{L}\,.
\end{equation}
For the field theory double copy, the identity element (sometimes called the zeroth copy) is the field theory known as cubic bi-adjoint scalar theory (BAS), as can be seen from the double multiplication table in Figure \ref{fig:DCtable}. This was first noticed in the CHY formalism  \cite{Cachazo:2013iea}, though the  BAS model also naturally arises in the BCJ language, namely when the numerator kinematic factors of the Yang-Mills amplitudes are replaced by color-factors, $n_I \to c_I$ in (\ref{AnBCJ}), one finds the tree amplitudes of the BAS model. For string theory, the amplitudes of the identity element model \cite{Mizera:2016jhj} can be understood to arise in the $\alpha'$-expansion from the BAS model with a tower of very particular higher-derivative operators with fixed coefficients. 

Written out, the meaning of $\mathbf{1}  \otimes  \mathbf{1} = \mathbf{1}$ at 4-point is that for any choices of color orderings $a, b, c, d$, the doubly color-ordered 4-point tree amplitudes $m_4$ of the identity model must satisfy
\begin{equation}
   \label{1x1is1}
   m_4[a|b] = m_4[a|c]\, S_4[c|d]\, m_4[d|b]\,.
\end{equation}
It is quite remarkable that such an identity model exists! To understand it better, it is useful to unpack the (\ref{1x1is1}). First,  note that the relation uniquely fixes all components of the kernel $S_4$ in terms of the $m_4$'s. In particular, choosing $c=b$, we see that 
$S_4[b|d] = (m_4[d|b])^{-1}$. This illustrates the unique link between the kernel and the identity model which also extends to higher points. Second,  plugging this result for the kernel back into (\ref{1x1is1}) and rearranging the equation it reads:
\begin{equation}
   \label{2by2}
    0 \,=\, 
    m_4[a|b] \, m_4[d|c] - m_4[a|c]\, \, m_4[d|b] 
    \,= \,
    \left|
    \begin{array}{cc}
    m_4[a|b] & m_4[a|c] \\
    m_4[d|c] & m_4[d|b] 
    \end{array}
    \right|
    \,.
\end{equation}
That is, any $2 \times 2$ minor of the $4! \times 4!$ matrix of all doubly color-ordered 4-point tree amplitudes of the $\mathbf{1}$-model must vanish: the matrix must have rank 1. At $n$-point the statement is that the $n! \times n!$ matrix of doubly color-ordered $n$-point tree amplitudes must have rank $(n-3)!$. This property holds for the BAS model, but adding a generic (higher-derivative) operator to BAS increases the rank above $(n-3)!$. The particular higher-derivative operators of the string-kernel in the $\alpha'$ expansion do however preserve rank $(n-3)!$. That turns out not to be the only solution.

The KLT algebra (\ref{KLTalgebra0}) gives a well-motivated pathway for generalizing the double copy: if we modify the identity element, this will --- via the equation $\mathbf{1}  \otimes  \mathbf{1} = \mathbf{1}$ ---  result in a unique new double-copy kernel, thus giving a {\em bootstrap for the KLT kernel}. Next, the requirements $\mathbf{L} \otimes\mathbf{1}   = \mathbf{L}$ and $\mathbf{1}  \otimes \mathbf{R} = \mathbf{R}$ on the amplitudes of the single-copy amplitudes become generalizations of the Kleiss-Kuijf relations and the BCJ (Bern-Carrasco-Johansson) relations (and likewise generalize the string monodromy relations). Thus the KLT double-copy bootstrap of \cite{Chi:2021mio} can be summarized as
\begin{equation}
\label{KLTalgebra}
   \underbrace{\mathbf{1}  \otimes  \mathbf{1} = \mathbf{1}}_{\text{KLT kernel bootstrap eq}} ~~~\text{and}~~~
    \underbrace{\mathbf{1}  \otimes \mathbf{R} = \mathbf{R} \,
      \,,~~~~~
    \mathbf{L} \otimes\mathbf{1}   = \mathbf{L}}_{\text{generalized KK\&BCJ / monodromy relations}} \!\!\!\!\!.
\end{equation}
These relations ensure that the result of the double copy is independent of the representation chosen for the KLT double copy. One must additionally impose locality. 

As discussed, the generalized double copy can be applied in the context of EFTs. It was shown in \cite{Chi:2021mio} how   $\mathbf{1}  \otimes  \mathbf{1} = \mathbf{1}$ uniquely fixes the KLT kernel in terms of the amplitudes of the identity model. Thus, adding all possible local  higher-derivative single-trace operators to the cubic bi-adjoint theory (BAS) and plugging it into the bootstrap equation $\mathbf{1}  \otimes  \mathbf{1} = \mathbf{1}$  selects a certain set of admissible  operators that can be added to the BAS model and it imposes certain relations among their Wilson coefficients. The result, as tested at 4- and 5-point in \cite{Chi:2021mio} and at 6-point in \cite{ACHEtoappear} indicates that the bottom-up approach to the generalized double copy gives a kernel that is more general than the strings kernel. In particular, unlike the strings kernel in which everything is fixed in terms of $\alpha'$, the generalized kernel involves a growing number of unfixed parameters at each order in the derivative expansion.

The generalized KLT double copy has been applied to to Yang-Mills theory with higher derivative operators and chiral-perturbation theory ($\chi$PT) with higher-derivative operators to get models of gravity, axion-dilaton-gravity, special Galileons, and Born-Infeld with higher-derivative corrections \cite{Chi:2021mio}. 
For example, the generalized KLT double copy allows for  $\text{tr} F^4$ with any Wilson coefficient as a higher-derivative correction to YM theory; but note that coefficient of $\text{tr} F^4$ is linked by the generalized KK and BCJ relations to the coefficients of a $\partial^2 \phi^4$ higher-derivative correction to BAS.

There are several motivations for studying generalizations of the double copy, including the need in higher-loop calculations to include counterterms in the double-copy construction. But there are also various puzzles that a better understand of the double-copy framework may help address:
\begin{itemize}
\item The Galileon scalar models have been proposed as higher-derivative corrections to the Dirac-Born-Infeld (DBI) action, however, the DBI-Galileon  does not appear to be produced by the double copy. Since $\mathcal{N}=4$ super-DBI is the result of $\chi$PT $\otimes$ ($\mathcal{N}=4$ sYM), one might have thought that simply including higher-derivative terms would produce the Galileon corrections. However, this is impossible because the 4d Galileons interactions are not compatible with $\mathcal{N}=4$ supersymmetry \cite{Elvang:2021qhq}; nor are they compatible with $\mathcal{N}=2$ supersymmetry in the presence of the DBI leading interaction \cite{Elvang:2021qhq}. There is evidence in favor of an $\mathcal{N}=1$ supersymmetric quartic Galileon \cite{Farakos:2013fne,Elvang:2017mdq}, but there is no known way to construct it using the double copy: its scalar-part, the special Galileon, simply arises from $\chi$PT $\otimes$ $\chi$PT, as can be seen from the table in Figure \ref{fig:DCtable}, but $\chi$PT is not compatible with supersymmetry. Thus, at this time, it remains a mystery how an $\mathcal{N}=1$ quartic Galileon could possibly arise from a double copy, either directly or as a higher-derivative correction to DBI. 
\item At 5-points, there is also a double-copy puzzle. The 4d quintic Galileon is a 5-point 8-derivative interaction. Yet there are no higher-derivative 5-point operators in $\chi$PT that can be double-copied to the quintic Galileon. The closest candidate is the Wess-Zumino-Witten term, but it is not compatible with the KK\&BCJ relations \cite{Elvang:2018dco}. An open question is if therefore if there exists a different version of the double copy that can produce the quintic Galileon. 
\item 4d Born-Infeld theory has electromagnetic (EM) duality. It is rather remarkable that the tree amplitudes in this theory can be produced as a double copy of $\chi$PT and Yang-Mills theory, which does not have EM duality. As symmetry of the equations of motion, but not of the Lagrangian, EM duality is only expected to be a tree-level symmetry of the amplitudes. However, it turns out that at 1-loop order there is evidence \cite{Elvang:2019twd,Elvang:2020kuj} that EM duality can be restored via finite local counterterms. As it turns out, even the simplest counterterm needed to restore EM duality in the 4-point 1-loop amplitudes cannot be produced by any known consistent form of the double copy. This is then challenging for higher-loop calculations that may use the double copy and need to take such counterterms into account. This example from Born-Infeld theory serves as a simpler toy example of a similar issue that arises in $\mathcal{N}=4$ supergravity \cite{Bern:2017rjw}. 
\end{itemize}

These bullet points outline examples of three puzzles with the standard field theory double copy. The generalizations of the double copy with higher-derivative terms added to BAS do not resolve the issues. However, the double-copy bootstrap is a more general framework and perhaps it admits a more general double copy that provide new insights and resolutions.

There is an alternative version of incorporating 
higher-derivative corrections into the double copy, namely by including kinematics into the color-factors \cite{Carrasco:2019yyn,Carrasco:2021ptp}. Recently, the connection between that approach and the  KLT bootstrap \cite{Chi:2021mio} was examined in \cite{Bonnefoy:2021qgu}. Combining the approaches may lead to new fruitful avenues of exploring the double copy. 

Finally, let us point out that the EFT generalization of the KLT kernel described here is a bottom-up approach that does not assume anything about existence of a UV model. A particularly interesting question is then to understand what constraints UV-completability brings in terms of bounds on the Wilson coefficients of the identity model associated with the KLT kernel. The modern powerful S-matrix bootstrap program  intersects the double-copy program in a very interesting way that in the future will shed light on what makes string theory so special.

\subsection{Color-Dual Effective Field Theory}
\label{sect:higherDerivative}

From an effective field theory (EFT) perspective, any new higher-energy physics will be encoded in the low-energy theory as Wilson coefficients of higher-derivative operators, motivating an understanding of what we can definitively clarify about such predictions to all orders in mass dimension.  Double-copy construction has already been used in attempts to understand potential constraints on the Wilson coefficients of higher-derivative operators in gravity theories~\cite{Bern:2017puu,Bern:2017tuc,Bern:2021ppb}.
  
Adding increasingly higher-dimension operators to encode novel UV physics traditionally means working with increasingly 
finicky and difficult operators in gauge and gravity theories.   Color-kinematics, double copy, and the success of Z-theory amplitudes (cf.\ \sect{sect:stringAmplitudes}) suggest a new approach for higher-derivative operator predictions in gauge theories: decomposing such predictions into much simpler color-dependent higher-derivative building blocks double-copied with a minimal basis of bare gauge-theory building blocks. The idea is to attempt to reduce all higher-derivative operators for gauge theories with and without massive matter to their most atomic components.  By virtue of being color-dual these higher-derivative blocks will never interfere with the gauge invariance of the complete scattering amplitudes.  If one is interested in going on to also construct  higher-derivative gravity predictions, gravitational amplitudes then follow by replacing color with known color-dual gauge-theory weights.

How should one construct (or identify) all distinct amplitudes associated with higher derivative operators of a given mass-dimension relevant to a particular $n$-field contact? A natural S-matrix approach invokes ansatze of the appropriate kinematic mass-dimension, external fields, then fixes on relevant symmetries.  Color-kinematics duality means only needing to dress a small number of cubic basis graphs at tree-level, but even this becomes expensive as mass-dimension grows.  Familiarity with the strict constraints of color-dual kinematics combined with the type of functional symmetry required to dress loop graphs suggests a more constructive route.

This perspective has recently born fruit at tree-level~\cite{Carrasco:2019yyn,Low:2019wuv,Low:2020ubn,Carrasco:2021ptp}.  There exists a simple pattern to all orders in higher-derivative  structure and an amazingly small number of primary building blocks for color-dual gauge and gravity higher-derivative  operators. A new verb has entered the color-dual story: {\em composition}.  Rather than using ansatze in Lorentz invariants and polarizations to satisfy functional color-kinematics constraints, color-dual functions can be {\em composed}~\cite{Carrasco:2019yyn} to generate new color-dual functions of arbitrarily high mass dimension.

This has profound implications for building and classifying the predictions of EFT operators as we can, at each multiplicity, find a pure-scalar color-dual function that is linear in Mandelstam invariants. This means there exists a unit step ladder to access higher-dimension operator predictions. Rather than needing to confront an ever more excruciating ansatz at higher order in mass dimension, these building blocks can be simply composed with themselves and a scalar unitary weight.  It is like building higher-derivative corrections by putting together tinker-toys. 

At four points~~\cite{Carrasco:2019yyn} it is possible to entirely climb the ladder of color-dual single-trace UV corrections to gauge theory and associated higher-derivative operators to gravity theories with a handful of building blocks.  The same four-point composition rules have been shown to also hold in the multi-trace sector~\cite{Low:2019wuv,Low:2020ubn}. The spectacular confirmation of the success of this approach involves exposing this very structure in the Z-theory description of the open super and bosonic strings at four-points. The reduction in complexity at five points~\cite{Carrasco:2021ptp} is even more profound, involving the classification of a number of relevant double-copy structures complementary to the adjoint.  A surprising primary building block at five-points -- similar to adjoint  but with central vertex antisymmetry condition relaxed, has lead to the identification of a candidate all-multiplicity linear building block: obeying adjoint relations at even multiplicity, and relaxed-adjoint relations at odd~\cite{Carrasco:2021ptp}.

The higher-derivative interactions at string tree level realize some of the color-kinematics dual operators of \cite{Carrasco:2019yyn,Carrasco:2021ptp} with particular (rational) combinations of multiple zeta values as their Wilson coefficients \cite{Schlotterer:2012ny, Broedel:2013aza, Mafra:2016mcc}.
It would be interesting to compare the entirety of loop-level effective interactions in different perturbative string theories with the operators constructed from the method in \cite{Carrasco:2019yyn,Carrasco:2021ptp}.

 Color-dual building blocks provide~\cite{Carrasco:2019yyn,Low:2019wuv,Low:2020ubn,Carrasco:2021ptp}  a powerful  tool that can be simply composed to all mass-dimension without having to resort to ansatze to construct the physical predictions of effective operators in gauge and gravity theories.   Next steps involve addressing the following sharp questions:
\begin{itemize}
\item Can these ideas be brought front and center in the description of the higher-derivative operators themselves at the heart of the action?
\item Do the tree-level composition rules change for matter with mass and in arbitrary representations of the gauge group?  Do color-dual bootstraps hold for fermionic matter and massive gauge theories?

\item What are the relevant scalar integrands at the multiloop level required to encode the predictions of higher-dimension operators of the compositional building blocks?  Four-points through four loops should be completely feasible at least at relatively low mass-dimension.  Do there exist similar composition rules at loop level that allow all-order in mass dimension building-blocks?   
\end{itemize}

\noindent{\bf Big Picture: }  Even being able to articulate these questions demonstrates the power of exploiting novel S-matrix double-copy structure to attain all-order in mass-dimension predictive control for effective gauge and gravity theories through a small number of simple building blocks, constructively generated. Leveraging double-copy structure offers a new way of thinking about higher-derivative operators in EFT---applicable across many fields of physical inquiry, one that may lead to much easier higher-loop predictions. The long view is that having these types of structures manifest in our predictions could ultimately lead to new descriptions of our physical theories, placing these atoms of prediction front-and-center.

\subsection{UV Behavior of Supergravity Theories}
\label{sect:gravUV}

Simple power-counting arguments make it clear that individual gravitational Feynman diagrams  must manifest unruly UV behavior at some loop order in four dimensions relative to gauge theory. Ultimately this is due to the dimensionality of the coupling constant, and the consequent need for additional momentum upstairs in the integrand for every gravitational interaction.  Supersymmetry is known to only provide a finite amount of protection in the UV. 
There are for example constraints from supersymmetry together with linearly and nonlinearly realized symmetries  up to 7-loop order, but not in any known way beyond, in 4d $\mathcal{N}=8$ supergravity \cite{Elvang:2010jv,Elvang:2010kc,Beisert:2010jx,Elvang:2010xn}. 
This suggests at most a delay for which loop order UV divergences must appear.  Are all pointlike (local) quantum field-theories of gravity therefore doomed to be {\em effective}, requiring perturbative completion in the UV?  Arguments based on individual Feynman diagrams only hold if there is no additional symmetry or structure enforcing cancellation between graphs.  The only known way to show that there are no missing or unrecognized  symmetries in a theory is to do the explicit calculation and find a divergence.

 From both symmetry considerations and technical accessibility, the theory to test with the best hope of perturbative finiteness in four dimensions would appear to be maximally supersymmetric supergravity ($\mathcal{N}=8$ SG).  While counterterms have been found which would be relevant starting at seven loops~\cite{Beisert:2010jx,Bossard:2011tq}, their coefficients have not yet been calculated and could indeed vanish in four dimensions.   Indeed, if one wants to celebrate optimism, there are a number of hints that suggest that four dimensions may indeed be special~\cite{Bern:2017lpv,Herrmann:2018dja,Bourjaily:2018omh,Edison:2019ovj}.   
 
 A recent milestone was the explicit calculation of the UV behavior of $\mathcal{N}=8$ SG at the five-loop order. Using the method of maximal cuts in combination with discovering~\cite{Bern:2017yxu} a crucial generalization of double-copy integrand construction, obviating the need for finding a manifest color-kinematics gauge-theory representation, the integrand of the five-loop correction to the two-to-two graviton scattering in the $\mathcal{N}=8$ SG theory  was constructed~\cite{Bern:2017ucb}. Shortly thereafter, a representation tailored to integration in the UV confirmed~\cite{Bern:2018jmv} a critical dimension of 24/5.  This is the dimension in which the  theory
 diverges due to insufficiently soft UV behavior. Notably this is the first demonstration that complete amplitudes in the maximal supergravity theory can possess a worse UV behavior than its double-copy progenitor (maximally supersymmetric gauge theory) whose formal five-loop critical dimension is 26/5. This may herald a potential higher-loop divergence in four-dimensions -- a topic of much speculation that itself awaits explicit calculation, now potentially within reach.

Perhaps more important than the UV behavior of this particular theory, performing the five-loop calculation exposed~\cite{Bern:2018jmv} entirely surprising and potentially universal consistency relations between loop-orders ({\em in different dimensions!}) for the vacuum {\em integrals} that dominate the UV.  The newly discovered consistency conditions are exactly the same in both the maximally supersymmetric gauge theory and the gravity theory. Having UV relations between loop orders suggests a path towards a  UV bootstrap program which could allow the direct probing of higher-loop behavior -- something to be explicitly tested in the near future.

Six-loops in the $\mathcal{N}=8$ theory is the obvious and critical next step to be calculated both via the new UV bootstrap approach as well as unitarity methods towards verification of this new approach.   Within the past year  the six-loop integrand for the color-dressed maximally supersymmetric gauge theory has finally been calculated~\cite{Carrasco:2021otn}. Using traditional methods this task is enormous. Completing the task required the invention of new and efficient means of extracting cut information---methods that ultimately have color-dual representations at their heart but exploit various soft-limits to relate information at a given loop level to more accessible information at lower loops, or simpler structures at higher loops. A clear next step will be to turn this gauge-theory result into a supergravity calculation and to test the range of the UV bootstrap developed at five-loops. 

Complementarily, the first UV divergence in four-dimensions ever calculated in a pure supergravity theory was at four-loops in the half-maximally supersymmetric theory~\cite{Bern:2013uka}.  This divergence has been associated with the presence of the so-called Marcus anomaly whose behavior at one-loop can be explained in terms of a double copy between maximally supersymmetric gauge theory and pure Yang-Mills~\cite{Carrasco:2013ypa}. A local counterterm has been proposed whose effects have been calculated through two loops~\cite{Bern:2017rjw,Bern:2017tuc,Bern:2019isl}.  Higher-loop calculations with this higher-derivative counterterm prove prohibitive with traditional approaches, especially if it requires higher-derivative operators for consistency. However this counterterm may be easily understood in terms of the higher-derivative building blocks described in \sect{sect:higherDerivative} -- involving a double copy seeded by one vector weight  proportional to ${F^3}$.  Ultimately we are interested in whether the anomaly tempering counterterm provides for finiteness to all orders, or whether the theory requires~\cite{Carrasco:2022lbm} an infinite number of counterterms.

\noindent{\bf Big Picture:}  The question is to understand definitively whether or not QFT in four-dimensions can admit a perturbatively finite QFT of gravity without the need for infinitely many counterterms --- or extended structure as they are understood in the context of string theory.  These are ``big-theory"-type projects: multi-year calculations that push current technology to its limit, forcing innovation and recognition of previously unappreciated structure in order to arrive at an answer.  
 More loops is different, and considering supersymmetric theories offers a simpler playground to build tools and identify potentially helpful structure.  Sometimes this structure is generic, and sometimes it is very special.   The ability to only consider graphs without one-loop triangle and bubble subdiagrams is special to gravity theories whose single-copy involve maximally supersymmetric gauge theory.  On the other hand, the color-kinematics duality and double copy appear to be 
properties 
of field theories not only beyond supersymmetry, but indeed beyond gauge and gravity theories. It is an open question as to how general the UV bootstrap of ~\cite{Bern:2018jmv} might be.  It is worth noting that pushing the envelope of finite-field related methods~(cf.~for example \cite{DeLaurentis:2022otd} and references therein) applied to simplify the integral-level book-keeping will be a nontrivial concrete outcome of this line of investigation. 

\subsection{Simplifying Calculations in Gauge Theories  with Matter}
\label{sect:complexity}
One research thrust has been to consider applying the duality between color and kinematics towards simplifying Standard-Model type integrands. The following is a summary of current progress:
\begin{itemize}
\item  QCD with massive matter in arbitrary representations coupled to massless gauge theories admits~\cite{Johansson:2015oia, He:2016dol, Brown:2018wss,Johansson:2019dnu} color-dual representations at tree-level, with associated reduction in number of independent ordered (stripped) tree-amplitudes.
\item Massive scalar QCD with matter in arbitrary representations is known to be color-dual through one-loop five-points~\cite{Johansson:2015oia,Plefka:2019wyg,Carrasco:2020ywq}. 
\item Supersymmetric generalizations, (S)QCD, with external glue and matter in the fundamental are color-dual through two-loop four-points~\cite{Johansson:2017bfl, Kalin:2018thp, Kalin:2019vjc, Duhr:2019ywc}.
\item While one must be careful with the duality between color and kinematics for generic massive gauge theories~\cite{Momeni:2020vvr,Johnson:2020pny}, spontaneous symmetry breaking (``Higgsing") in the adjoint is color-dual at tree-level, and indeed massive gauge theories are understandable at tree-level as a consistent dimensional reduction~\cite{Chiodaroli:2015rdg,Chiodaroli:2017ehv}.
\end{itemize}
It is absolutely fair to ask what this program could buy us in terms of precision QCD and Standard Model calculations more generally. 

The relevant quantitative predictions for hard processes in QCD are obtained~\cite{Gehrmann:2021qex} using perturbation theory, after accounting for the non-perturbative bound state dynamics of partons and hadron fragmentation. The most immediate and omnipresent challenge when performing multiloop calculations towards IR safe observables is integration~\cite{Caola:2022ayt}, especially over virtual loop momenta involving multiple internal mass scales.  While there has been a tremendous amount of progress, new ideas and approaches to integration remains an active area of pressing interest~\cite{Weinzierl:2022eaz, Caola:2022ayt,Bourjaily:2022bwx, Abreu:2022mfk, Blumlein:2022zkr}. When applicable, the duality between color and kinematics appears to hold at the integrand level and so does not confront, at least in any direct sense, the primary challenge of loop integration.  Nevertheless, perturbative calculation relies on organization into local Feynman integrals, and therefore even at the integrand level, as loop level and multiplicity increase, we have a problem: the number of contributing diagrams grows factorially.  At the integrand construction level, the duality between color and kinematics has the potential to simplify the situation as we will describe.  

Unitarity, or on-shell, methods~\cite{Bern:1994zx,Bern:1994cg,Bern:1995db,Bern:1997sc,Britto:2004nc,Bern:2007ct} have become standard in the current era of precision  calculation. The idea is to avoid calculating with unwieldy arbitrary-gauge off-shell Feynman rules that carry around unphysical information that must cancel in final gauge invariant observables.  Instead we employ compact on-shell gauge-invariant quantities like tree-level scattering amplitudes.  It is sufficient to verify an integrand if it satisfies all unitarity cuts to tree-level results.  Since cuts satisfy a spanning relationship, only a relatively small number of cuts need to actually be performed to verify an integrand.  Furthermore, because gauge theory tree-amplitudes can be expressed in terms of a color-basis, we only need to consider simpler color-ordered (or color-stripped) cuts.  Having an easy approach to verification leads to natural construction techniques, whereby one builds the integrand so as to simply satisfy all cuts.  Furthermore when a minimal basis of integrals is known, such as one-loop, it is possible to directly target the coefficients of integrals themselves via cut techniques.  It is important to be able to capture $D$-dimensional information when using variants of dimensional regularization, so often cuts are performed in $D$-dimensions.  Furthermore there are subtleties~\cite{Bern:1995db} around massive tadpoles that are in principle only directly extractable via on-shell methods via forward-limit cuts, but could in principle receive cut information propagated from more accessible graphs by color-dual kinematic relations.

Consider a four-point two-loop gluon QCD calculation with quarks in the fundamental.  At higher multiplicity or loop level the gain can be even more impressive, but NNLO processes in many ways still represent the state of the art.  One of the most complicated cuts is between two 5-point trees. The most complicated part of this calculation will involve gluons crossing the cut.  Prior to the recognition that there are in fact only $(m-3)!$ distinct color-ordered amplitudes at $m$-points under the BCJ relations
as described in Section \ref{rev:Trees}, 36 different color-ordered cuts would in principle need to be performed as per the Kleiss-Kuijf basis.  Now only four distinct cuts need be considered.  
A similar reduction holds for the cut between a six-point tree and a four-point tree.  For example, if both the external and the cut lines are gluons, before the duality between color and kinematics we would have had to worry about 48 color-order cuts, while now we only need to consider six.\footnote{
For a cut between a six-point tree and a four-point tree of a four-fermion amplitude with one cut fermion and two cut gluon lines, the duality reduces the number of cuts from 24 to 6, cf. Tables 2 and 4 of~\cite{Johansson:2015oia}.
}  
These gains arise simply from the theory being color-dual at tree-level.

\begin{figure}[t]
    \includegraphics[width=1\textwidth]{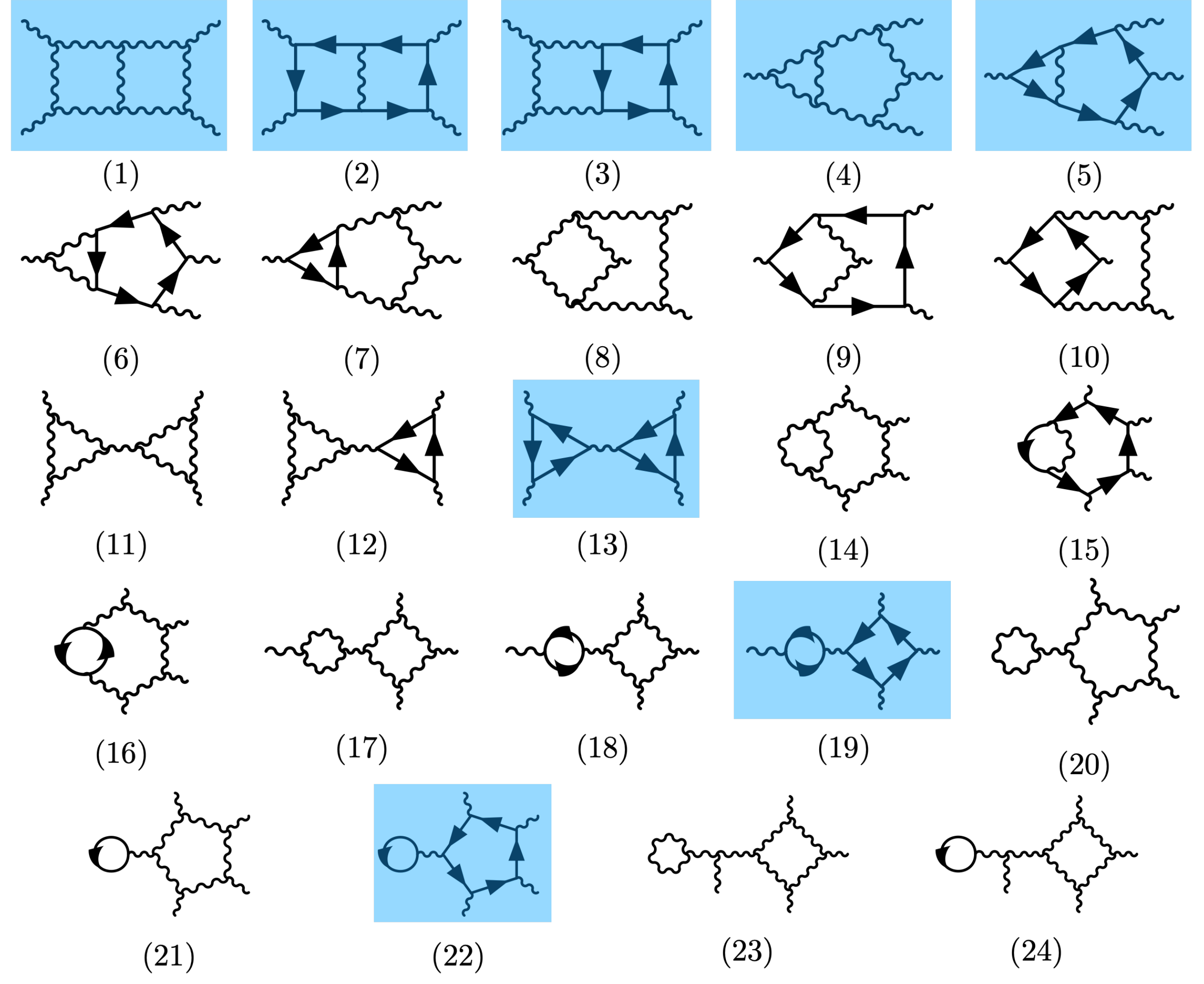}
  \caption{\small A complete list of non-vanishing graphs contributing to four-point 2-loop $\mathcal{N}=2$ SQCD taken from ref.~\cite{Johansson:2017bfl}. The eight basis graphs whose kinematic weights combine to dress all graphs are~(1)--(5), (13), (19) and (22).     
\label{fig:sqcd2loop}
}
\end{figure}

What if we can make the duality between color-and-kinematics manifest directly at the level of the integrand? 
The scattering amplitude for supersymmetric QCD with massive matter in the fundamental for the two-loop correction has been constructed~\cite{Johansson:2017bfl}  and integrated~\cite{Duhr:2019ywc}, including the maximally helicity violating case  (MHV).  MHV, at four-points, refers to the situation when two of the external gluons are positive helicity and two external gluons are negative helicity.  As depicted in Figure~\ref{fig:sqcd2loop}, there are 24 distinct graph topologies that could contribute to the amplitude, appearing with all distinct permutations of labels. In order to build the amplitude, each graph must be mapped to its relevant color-weights, kinematic weights, and propagators for all relevant labelings.  This procedure of mapping a graph to its specific weights is called dressing the graph, and the weights are often called dressings.  An advantage of graph-based unitarity approaches is that one only needs to constrain dressings for individual topologies, allowing automorphism invariance to account for all distinct labeling permutations. The goal of cut construction is to find the appropriate dressings for each topology such that all cuts are satisfied. 

For the two-loop MHV $\mathcal{N}=2$ SQCD amplitude it is possible to write a spanning ansatz for kinematic dressings of each graph in terms of a minimal basis of Lorentz invariants involving momenta and external polarizations.  Any parameter not constrained by physical cuts will inevitably turn out to be a generalized gauge choice that will never contribute upon integration. There are about 4.5k parameters per graph.  If it had been  impossible to make the duality between color and kinematics manifest at the level of the integrand this would mean a total ansatz size for the amplitude on the order of 105k parameters.  The duality between color and kinematics however relates the kinematic dressings of graphs to each other.  For example, the double-triangle kinematic weights are given as differences between the kinematic weights of double boxes,
\begin{align}
  \begin{aligned}
    n_{11}(1234;\ell_1,\ell_2)&=
    n_1(1234;\ell_1,\ell_2)-n_1(1243;\ell_1,p_3+p_4-\ell_2),\\
    n\bigg(\includegraph{9}{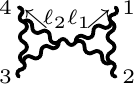}\bigg)&=
    n\bigg(\includegraph{9}{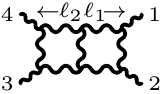}\bigg)-n\bigg(\includegraph{10}{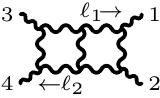}\bigg).
  \end{aligned}
\end{align}
The authors of ref.~\cite{Johansson:2017bfl} used such linear relations to reduce all kinematic weights to those of just eight basis graphs.   This reduces the entire size of the kinematic ansatz for the integrand to 35k parameters, of which many are drastically constrained due to symmetries of the graphs before even admitting cut data.  The resulting integrand was subsequently integrated in ref.~\cite{Duhr:2019ywc}. 

Even more dramatic reductions occur when all particles transform in the same representation of the gauge group.  The $\mathcal{N}=4$ super Yang-Mills amplitude requires only $\sim80$ non-vanishing topologies at the  four-loop correction to four-point scattering.  The reason the number of topologies is so small is because maximal supersymmetry removes a need to consider any graphs with internal one-loop bubble or triangle subdiagrams.  All of the kinematic dressings can be expressed in terms of linear functions of a single~\cite{Bern:2012uf} nonplanar basis graph by exploiting color-dual Jacobi-like relations. 

There is often residual generalized gauge freedom upon finding color-dual numerators for a given theory at the integrand level.  This means that certain parameters cancel for any physical cut or upon integration.  This freedom can be exploited to make manifest UV behavior graph by graph  as per three and four loops in $\mathcal{N}=4$ super Yang-Mills ~\cite{Bern:2010ue,Bern:2012uf}. Intriguingly this freedom can be used to expose IR behavior as well, as in the example of MHV two-loop SQCD, cf.~the original integrands of ref.~\cite{Johansson:2017bfl} vs the relatively IR safe integrand of ref.~\cite{Kalin:2019vjc}.  

It is worth noting that the act of discovering simplicity is often driven by the needs of encountering tremendous complexity. We are well past the days where the boundary of perturbative calculation proceeded without automation.  Pushing the boundary of perturbative calculation often means the development and application of massively parallel computational techniques, analytic as well as numeric.  This has been critical for a number of projects from event generation ~\cite{Childers:2015tyv}, to supergravity calculations~\cite{Bern:2018jmv}, to precision gravitational waves~\cite{Bern:2021dqo,Bern:2021yeh}, to cosmological large scale structure~\cite{Carrasco:2013mua}.   The amount of analytic data generated in integrand construction can be enormous.  The size of the six-loop integrand of the friendliest gauge theory, the $\mathcal{N}=4$ super Yang-Mills theory, is about a gigabyte~\cite{Carrasco:2021otn}.  Constructing it meant evaluating and considering orders of magnitude more data in the form of unitarity cuts.  It will not be long before analytic calculations approach the terabyte scale.  

Success in these calculations yields not only new understanding of the language of perturbative field theory but also the development of tools to handle large-scale analytic data, aiding researchers in identifying meaningful patterns. This expertise pays dividends. The ultimate goal, of course, is to extract from this data, when possible, the correct reformulation minimizing the need for such large intermediary stages. Indeed historically, this is how the duality between color and kinematics was first identified~\cite{Bern:2008qj} in the midst of performing a four-loop calculation in the $\mathcal{N}=8$ maximal supergravity theory~\cite{Bern:2009kd}.

\subsection{Towards Understanding the Kinematic Algebra}
\label{sect:kinAlg} 
The celebrated double copy slogan, eqn.~(\ref{GRisYM2}), could be expressed as: {\em ``understanding gravity is no more complicated than understanding gauge theory.''}  A corollary, emerging from attempts to understand the duality between color and kinematics at the level of equations of motion~\cite{Monteiro:2011pc,Bjerrum-Bohr:2012kaa,Lee:2015upy, Garozzo:2018uzj, Bridges:2019siz,Chen:2019ywi,Cheung:2020djz,Chen:2021chy,Cheung:2021zvb} and the action~\cite{Cheung:2016prv, Ben-Shahar:2021doh,Ben-Shahar:2021zww}, could be phrased {\em ``gauge theory 
need not 
be more complicated than a scalar theory.''} 
To fully realize the idea, it is necessary to write the non-trivial kinematic details of gauge theory as a Lie-algebraic structure that is dual to the color Lie algebra. This is know as the {\em kinematic algebra}.         

A detailed understanding of the kinematic algebra would be transformative from many possible perspectives. Given a diagram with any number of loops or external legs, we could in principle write down its numerator by simply contracting appropriate kinematic ``structure constants''. The diagrams themselves would be convenient devices for specifying invariant tensors associated with the kinematic algebra.
Gravity would be realized as a peculiar Yang-Mills theory: one in which the gauge algebra is equal to the kinematic algebra. Loop integrations could potentially be reinterpreted as kinematic traces. The study of the space of functions emerging from loop integration may take on a wholly new algebraic character. Finally, the mathematical structure of the kinematic algebra could reveal new hidden symmetries or clarify details about the presumed emergence of spacetime.

In special situations it has been possible to identify a simple kinematic algebra. For example, it is useful to restrict attention to the ``self-dual'' sector of pure gauge theory and gravity~\cite{Monteiro:2011pc,Boels:2013bi,Krasnov:2021cva}.
This restriction can be interpreted as choosing to scatter only gluons or gravitons of a 
particular helicity, say positive helicity.
At tree level this implies the vanishing of all amplitudes beyond three points. However, a closely-related quantity is non-vanishing: the solution of the classical equations of motion sourced by an ensemble of plane waves.
In general, this solution is a generating function of tree amplitudes, 
and can be thought of as a sum over tree graphs with exactly one external off-shell leg.

The expansion of the self-dual classical solution in powers of couplings can be organized in terms of trivalent diagrams. In the Yang-Mills case, each diagram corresponds to a particular kinematic and color weights. Remarkably, it is possible to choose a gauge in which the kinematic weights, computed from the equations of motion,
manifestly satisfy the same algebraic relations as the Jacobi relations enjoyed by the corresponding color factors.
A closely related gauge choice is available in gravity, revealing that the gravitational
solution is indeed a double copy of the gauge solution to all perturbative orders in the self-dual theories.

It is possible to understand the details of the self-dual kinematic algebra: it is an
algebra of area-preserving diffeomorphisms of a two-dimensional plane embedded in four-dimensional spacetime. The kinematic structure constants appear as two copies in the three-point vertex of gravity, and thus in the gravitational equations of motion. 
While it is no great mystery that that a diffeomorphism algebra plays a central role in self-dual gravity,
it is more surprising that the same algebra controls self-dual Yang-Mills theory. There are some indications that the origin of this fact may be traced back to the Sugawara construction of a related CFT in two dimensions~\cite{Cheung:2016iub}.  The self-dual sector has recently been generalized to other theories via Moyal deformations and relaxing the symmetry~\cite{Chacon:2020fmr}.
An interpretation of the kinematic algebra based on the Drinfeld double of the Lie algebra of vector fields was discussed in reference~\cite{Fu:2016plh}.

In more recent work, it has been shown~\cite{Ben-Shahar:2021zww} that the kinematic algebra includes volume-preserving diffeomorphisms in the special case of three-dimensional Chern-Simons theory. In contrast to the above self-dual case, which does not constitute a consistent quantum theory, the complete action of pure Chern-Simons theory gives rise to off-shell Feynman rules that manifest the duality between color and kinematics. This holds both at tree level and any loop order. However, similar to the self-dual case, the Feynman diagrams in pure Chern-Simons theory non-trivially conspire to give vanishing amplitudes when taken on shell. The off-shell correlation functions are non-vanishing in Lorenz gauge and provides the first fully off-shell realization of the duality.    

In the self-dual case, since we choose to study only one helicity, it follows that we can describe the theory
in terms of a scalar field (at the expense of manifest Lorentz invariance).
The self-dual kinematic algebra is therefore relevant to a kind of scalar theory.
Closely related constructions are available in other scalar theories, notably
the non-linear sigma model~\cite{Cheung:2016prv,Cheung:2017yef, Mizera:2018jbh}.

Another lesson from the self-dual theory is that it can be useful to think about color-kinematics duality, and the double copy, at the level of the classical equations of motion.   Indeed, using traditional field theory methods in a first-order formalism,  ref.~\cite{Cheung:2021zvb} not only identified the relevant algebraic relations for Yang-Mills covariantly, but also produced associated explicit realizations of all multiplicity color-dual numerators at tree-level. Explicit color-dual solutions have found an off-shell realization through classical perturbiner solutions \cite{Rosly:1996vr, Rosly:1997ap, Selivanov:1997aq} in so-called BCJ gauge \cite{Lee:2015upy, Garozzo:2018uzj, Bridges:2019siz}.
Just this year there has additionally been new insight into double-copying off shell currents~\cite{Cho:2021nim} and the kinematic algebra~\cite{Bonezzi:2022yuh} from the perspective of double-field theory. 

If we zoom into any differentiable manifold associated with a continuous group we will see its defining local generators obeying a Lie algebra. Indeed, scattering amplitudes in the NLSM can be shown to encode such a field-space geometry~\cite{Alonso:2015fsp,Alonso:2016oah}. Could we from this perspective understand the origin of the kinematic numerators describing various theories by relating them to scalar interactions in a theory-specific field-space geometry? This idea was recently explored~\cite{Cheung:2022vnd} by writing massless bosonic tree-amplitudes for general theories in terms of NLSM amplitudes and replacing the field-space geometry with a notion of kinematic-space geometry. 

Despite significant recent progress, the kinematic algebra of Yang-Mills theory remains an enigmatic problem. A central obstacle to overcome is the vast generalized gauge freedom that is associated to kinematic numerators, and additional principles are needed for narrowing down possible constructions. Stratifying the kinematic numerators according to the MHV sectors of Yang-Mills theory was shown to be a powerful organization principle in refs.~\cite{Chen:2019ywi,Chen:2021chy}, which permitted a formulation of the kinematic algebra up to the next-to-MHV level. This formulation was motivated by current algebras and produced all-multiplicity numerators using a handful of ``fusion product'' rules for the generators of the algebra. The approach later expanded in a combinatorial direction in refs.~\cite{Brandhuber:2021eyq,Brandhuber:2021bsf}, where a well-known quasi-shuffle Hopf algebra was shown to be isomorphic to the fusion product of generators that give gauge-invariant tree-level kinematic numerators, to all multiplicity and all MHV sectors. These kinematic numerators correspond to a heavy-mass effective theory coupled to Yang-Mills theory, and include poles of the heavy-mass particles in the numerators. This mild non-locality is what makes it possible to give a manifestly gauge-invariant formulation, and echoes the construction of gauge-invariant numerators in ref.~\cite{Cheung:2021zvb}. Curiously, the number of terms in the numerators (as well as number of generators) are given by the Fubini numbers. 

{\bf Big picture}. While still in its infancy, the exploration of the kinematic algebra of Yang-Mills theory, and of other gauge and scalar theories, constitute a cornucopia of rich mathematical structures that deserve further attention. As briefly reviewed, there are many options for formulating the kinematic algebra: through differential operators, kinematic structure constants, geometric maps, field equations, and at the Lagrangian level.  We hope  these  interlocking perspectives   will eventually crack open the problem of finding an elegant, mathematically deep, and practical formulation of the kinematic algebra underlying color-kinematics duality. Once fully revealed, the algebra invites deep physical interpretations and possible consequences for new physics.

\subsection{Unifying the Web of Theories}
\label{sect:web}

\begin{figure}[t]
\begin{center}
  \includegraphics[width=1.01\textwidth]{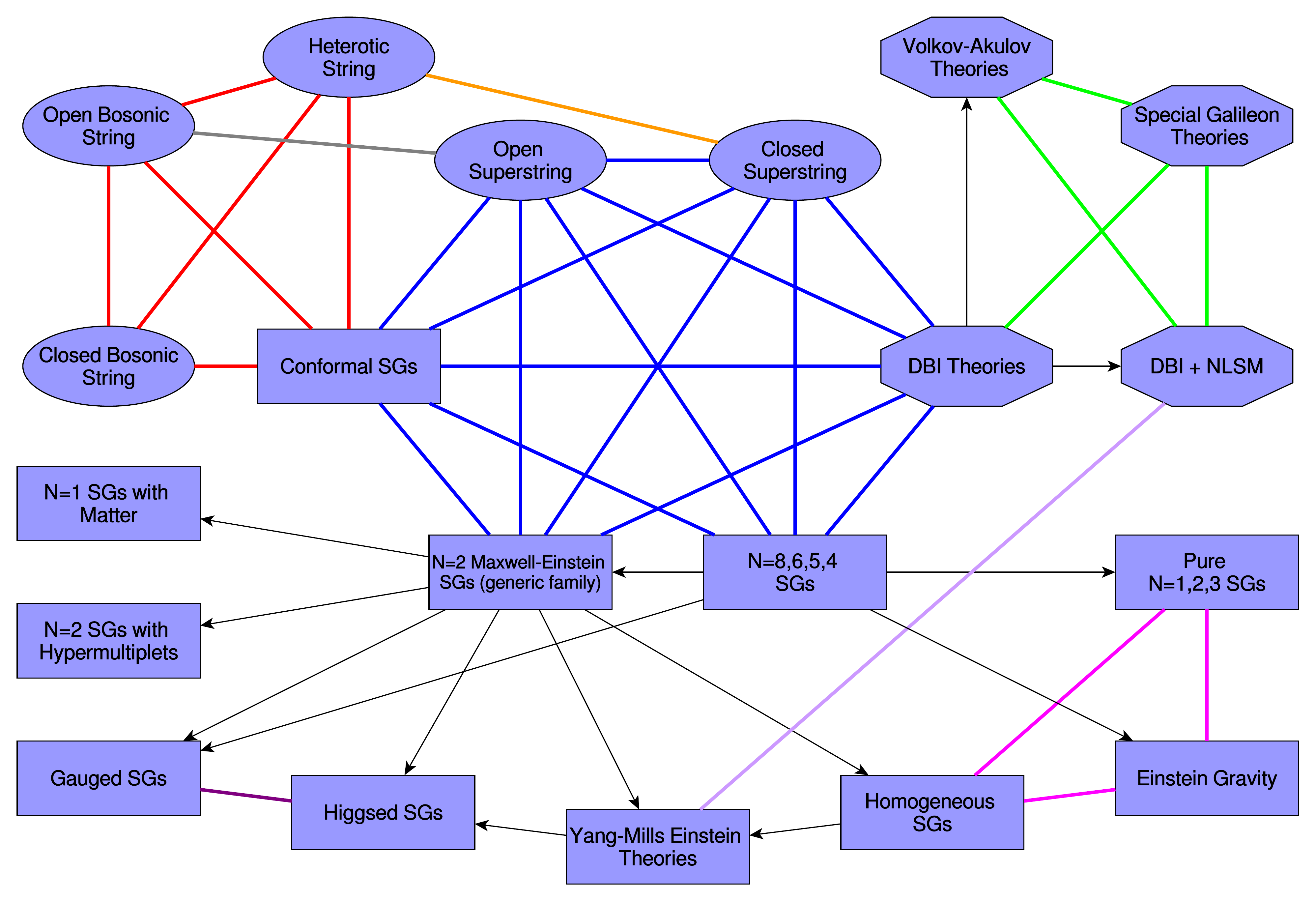} \caption{Web of theories as presented in ref.~\cite{Bern:2019prr}. Nodes represent the main
 double-copy-constructible theories, including gravitational theories (rectangular nodes), string theories
 (oval nodes) and non-gravitational theories (octagonal
 nodes). Undirected links connect theories with  a
 common gauge-theory factor. Directed links
 are drawn between theories constructed by modifying both gauge-theory
 factors. \label{FigWeb}}
\end{center}
\end{figure}

From its original formulation, it has been clear that the double-copy structure does not rely on the presence of  supersymmetry. Much early work focused on using the double copy to facilitate calculations in maximal and half-maximal supergravity, but that was mostly a consequence of the increased simplicity of these theories, as well as of the interest in examining their UV properties. Indeed, from the very beginning, examples of non-supersymmetric theories admitting double-copy constructions have been analyzed. These include the double copy of pure Yang-Mills theory, which yields Einstein gravity with some additional states (in four dimensions, a complex scalar field, which can be seen as the double copy of gluons of opposite polarizations). Over the years, more complicated examples of the construction have emerged, leading to the realization that the double-copy structure might be considerably more general than originally envisioned, as depicted in Figure~\ref{FigWeb}.
However, an important question remains open on exactly how general the double-copy structure is.

The space of all possible gravitational theories is extremely difficult to chart and, in order to delve deeper into the above question, we need to use some classifying principle. Once again, supersymmetry comes to the rescue: the supergravity community has long taken on the task of charting the space of all possible supergravity theories, uncovering and exploiting some beautiful geometrical structures in the process \cite{freedman2012supergravity}. This classification begins from theories with a large number of supersymmetries, including maximal ${\cal N}=8$ supergravity\footnote{Here we specify the amount of supersymmetry in four spacetime dimensions. The reader should be aware that most of these theories can also be specified in higher dimensions.}~\cite{Cremmer:1979up} and half-maximal ${\cal N}=4$ supergravity ~\cite{Cremmer:1977tt,Das:1977uy}. Because of the high amount of symmetry, the former theory is unique. The latter can be specified completely by a single parameter --- the number of vector fields in the theory. Once the field content is known, the symmetries are so constraining that all interactions are fixed. 

However, when supersymmetry is reduced to ${\cal N}=2$, supergravity theories are much less constrained \cite{Gunaydin1983bi,Gunaydin1984ak,Gunaydin1984nt,Gunaydin1986fg}. In particular, the spectrum alone no longer qualifies a theory completely, and additional information on the interactions is needed. 
The simplest theories with this property constitute a fundamental test for the application of the double copy to generic theories of gravity. These are the so-called homogeneous ${\cal N}=2$ Maxwell-Einstein supergravities. After earlier progress in describing particular cases \cite{Carrasco:2012ca,Chiodaroli2013upa,Cachazo:2014xea}, a general double-copy construction for these theories was formulated in 2015 in ref.\ \cite{Chiodaroli:2015wal}. These theories were previously explicitly classified in the supergravity literature \cite{deWit1991nm}.  
Remarkably, the construction of ref.\ \cite{Chiodaroli:2015wal} succeeds  in realizing all theories in the supergravity classification \cite{deWit1991nm} as double copies of suitably-chosen gauge-theory factors, demonstrating that the double copy is a property of very general classes of theories. In particular, the key ingredient in the construction for homogeneous theories is the fact that the gauge theories involved can admit extra non-adjoint representations in which some of the matter fields transform while still obeying the duality between color and kinematics. This idea has also been used to present constructions giving pure theories \cite{Johansson:2014zca} and ${\cal N}=2$ theories with hypermultiplets \cite{Chiodaroli:2015wal,Anastasiou:2017nsz}. In turn, the double copy has been used to conduct highly-nontrivial loop calculations in these theories along the lines of Refs. \cite{Johansson:2017bfl,Ben-Shahar:2018uie,Kalin:2018thp}. 

In the past  years, more and more gravitational theories have been recognized to secretly admit a double-copy description. A particularly important class is given by Yang-Mills-Einstein theories, 
some simple supergravities which include non-abelian gauge interactions. Aside from the double-copy method \cite{Bern1999bx,Chiodaroli:2014xia,Chiodaroli:2015rdg,Chiodaroli:2016jqw,Fu:2017uzt,Chiodaroli:2017ngp,Wu:2021exa,He:2021lro,Dong:2021qai}, a variety of approaches have been employed in studying the amplitudes in these theories, facilitating contact between closely-related subfields \cite{Cachazo2014nsa,Cachazo:2014xea,Casali2015vta,Adamo:2015gia,Stieberger2016lng,Nandan2016pya,SchlottererEYMHeterotic,TengFengBCJNumerators,CheungUnifyingRelations,Roehrig:2017wvh,Du2017gnh,Mazloumi:2022lga,Porkert:2022efy}. 

Another very important class of constructions arises in case of the so-called gauged supergravities. These are theories obtained from the more conventional (ungauged) supergravity theories by promoting part of the R-symmetry to a local symmetry under which some of the supergravity fields are charged. This  operation results in an array of interesting physical features, including the possibility of non-trivial scalar potentials, spontaneously-broken supersymmetry (in case the theory admits Minkowski vacua) and the possibility of anti-de Sitter vacua \cite{Samtleben2008pe}. Most importantly, gauged supergravities are a subject of current investigation from the supergravity community, with the recent discovery of various families of novel theories ~\cite{DallAgata:2011aa,DallAgata:2012mfj,Catino:2013ppa,Dallagata:2021lsc,Bobev:2020ttg,Krishnan:2020sfg}. 
While the study of gauged supergravities from the vantage point of the double copy is still in the early stages \cite{Chiodaroli:2017ehv,Chiodaroli:2018dbu}, this is one of the instances in which amplitude methods in general and the double-copy technique in particular can provide new insights to solve challenging problems that have long remained open, as well as potential for cross fertilization between different fields in theoretical physics. 
Another notable set of constructions is the one for conformal supergravities \cite{Johansson:2017srf,Johansson:2018ues}.

The net result of a significant body of work over the last few years is a wide web of theories connected by a double-copy construction (see Refs.\ \cite{Bern:2019prr,Bern:2022wqg} for recent reviews). This web illustrates the power of the double copy as a classifying principle. In the coming years, the double copy will be applied to theories with minimal or no supersymmetry. Much more information is required to specify these theories in comparison with the instances of  the double copy known so far. The open challenge is to map this abundance of supergravity information into gauge-theory data, as it was done in the case of homogeneous supergravities.

{\bf Big picture}. While the results so far have raised the prospect of the double copy being a  general feature of gravitational interactions, extension of the web of double-copy-constructible theories to include generic theories with minimal or no supersymmetry will be the fundamental test to determine whether this is the case. At a conceptual level, this is a question with deep implications in terms of gravitational interactions admitting simpler descriptions using gauge-theory building blocks. 
Additionally, the near future will reveal whether the double copy can be used as an organizing principle to chart the space of all possible theories. As previously emphasized, this organizing principle is independent from supersymmetry --- we started from analyzing supersymmetric theories simply because they are usually simpler and more under control than their non-supersymmetric relatives.  More work also needs to be done in understanding the physical significance of the connections between theories revealed by the double copy. The double copy organization of the web of theories connects theories sharing a common gauge-theory factor, but the physics behind this connection often remains elusive and necessitates further study. Finally, the double copy remains above all a formidable computational tool. As more and more instances of this construction are identified, the doors open to the possibility of conducting highly non-trivial precision calculations in these theories.

\section*{Acknowledgments}
\label{sect:ack}

TA is supported by a Royal Society University Research Fellowship and by the Leverhulme Trust (RPG-2020-386). 
JJMC is supported in part by the DOE under contract DE-SC0021485 and by the Alfred P. Sloan Foundation.
MCG is supported by the European Union’s Horizon 2020 Research Council grant 724659 MassiveCosmo ERC–2016–COG and the STFC grants ST/P000762/1 and ST/T000791/1. 
MC is supported by the Swedish Research Council under grant 2019-05283.
HE is supported in part by Department of Energy grant DE-SC0007859.  
HJ is supported in part by the Knut and Alice Wallenberg Foundation under grants KAW 2018.0116 (From Scattering Amplitudes to Gravitational Waves) and KAW 2018.0162, and the Ragnar S\"oderberg Foundation (Swedish Foundations’ Starting Grant).
DOC is supported by the STFC grant ``Particle Physics at the Higgs Centre''.
RR is supported by the US Department of Energy under Grant No. DE-SC00019066.
OS is supported by the European Research Council under ERC-STG-804286 UNISCAMP. 
This work was supported in part by the National Science Foundation under Grant No. NSF PHY-1748958.

\section*{Solicited Feedback}
\label{sect:feedback}
Snowmass is a community planning exercise, and this document aspires to represent the excitement and interests of a growing and thermal community. 
We gratefully acknowledge the following for valuable feedback in the construction of this white paper: Daniel Baumann, Justin Berman, Zvi Bern, Ji-Yian Du, Alex Edison, Michael Graesser, Daniel Green, Aidan Herderschee, Song He, Anton Ilderton, Callum Jones, Arthur Lipstein,  James Mangan, Sebastian Mizera, Gustav Mogull, Ricardo Monteiro, Alexander Ochirov, Julio Parra-Martinez, Riccardo Penco, Frank Petriello, Jan Plefka, Fei Teng, Andrew Tolley, Mark Trodden, and Chris White.

\bibliographystyle{JHEP}

\bibliography{dblcpy_WP}

\providecommand{\href}[2]{#2}\begingroup\raggedright\begin{thebibliography}{100}

\bibitem{Travaglini:2022uwo}
G.~Travaglini et~al., \emph{{The SAGEX Review on Scattering Amplitudes}},
  \href{https://arxiv.org/abs/2203.13011}{{\ttfamily 2203.13011}}.

\bibitem{Kawai:1985xq}
H.~Kawai, D.~Lewellen and S.~Tye, \emph{{A Relation Between Tree Amplitudes of
  Closed and Open Strings}},
  \href{https://doi.org/10.1016/0550-3213(86)90362-7}{\emph{Nucl. Phys. B}
  {\bfseries 269} (1986) 1}.

\bibitem{Bern:2008qj}
Z.~Bern, J.~J.~M. Carrasco and H.~Johansson, \emph{{New Relations for
  Gauge-Theory Amplitudes}},
  \href{https://doi.org/10.1103/PhysRevD.78.085011}{\emph{Phys. Rev.}
  {\bfseries D78} (2008) 085011}
  [\href{https://arxiv.org/abs/0805.3993}{{\ttfamily 0805.3993}}].

\bibitem{Bern:2010ue}
Z.~Bern, J.~J.~M. Carrasco and H.~Johansson, \emph{{Perturbative Quantum
  Gravity as a Double Copy of Gauge Theory}},
  \href{https://doi.org/10.1103/PhysRevLett.105.061602}{\emph{Phys. Rev. Lett.}
  {\bfseries 105} (2010) 061602}
  [\href{https://arxiv.org/abs/1004.0476}{{\ttfamily 1004.0476}}].

\bibitem{Cachazo:2013gna}
F.~Cachazo, S.~He and E.~Y. Yuan, \emph{{Scattering equations and
  Kawai-Lewellen-Tye orthogonality}},
  \href{https://doi.org/10.1103/PhysRevD.90.065001}{\emph{Phys. Rev.}
  {\bfseries D90} (2014) 065001}
  [\href{https://arxiv.org/abs/1306.6575}{{\ttfamily 1306.6575}}].

\bibitem{Cachazo:2013hca}
F.~Cachazo, S.~He and E.~Y. Yuan, \emph{{Scattering of Massless Particles in
  Arbitrary Dimensions}},
  \href{https://doi.org/10.1103/PhysRevLett.113.171601}{\emph{Phys. Rev. Lett.}
  {\bfseries 113} (2014) 171601}
  [\href{https://arxiv.org/abs/1307.2199}{{\ttfamily 1307.2199}}].

\bibitem{Cheung:2018wkq}
C.~Cheung, I.~Z. Rothstein and M.~P. Solon, \emph{{From Scattering Amplitudes
  to Classical Potentials in the Post-Minkowskian Expansion}},
  \href{https://doi.org/10.1103/PhysRevLett.121.251101}{\emph{Phys. Rev. Lett.}
  {\bfseries 121} (2018) 251101}
  [\href{https://arxiv.org/abs/1808.02489}{{\ttfamily 1808.02489}}].

\bibitem{Antonelli:2019ytb}
A.~Antonelli, A.~Buonanno, J.~Steinhoff, M.~van~de Meent and J.~Vines,
  \emph{{Energetics of two-body Hamiltonians in post-Minkowskian gravity}},
  \href{https://doi.org/10.1103/PhysRevD.99.104004}{\emph{Phys. Rev. D}
  {\bfseries 99} (2019) 104004}
  [\href{https://arxiv.org/abs/1901.07102}{{\ttfamily 1901.07102}}].

\bibitem{energetics4PM}
M.~Khalil, A.~Buonanno, J.~Steinhoff and J.~Vines, \emph{{Energetics and
  scattering of gravitational two-body systems at fourth post-Minkowskian
  order}},  \href{https://arxiv.org/abs/2204.05047}{{\ttfamily 2204.05047}}.

\bibitem{Buonanno:1998gg}
A.~Buonanno and T.~Damour, \emph{{Effective one-body approach to general
  relativistic two-body dynamics}},
  \href{https://doi.org/10.1103/PhysRevD.59.084006}{\emph{Phys. Rev. D}
  {\bfseries 59} (1999) 084006}
  [\href{https://arxiv.org/abs/gr-qc/9811091}{{\ttfamily gr-qc/9811091}}].

\bibitem{Buonanno:2000ef}
A.~Buonanno and T.~Damour, \emph{{Transition from inspiral to plunge in binary
  black hole coalescences}},
  \href{https://doi.org/10.1103/PhysRevD.62.064015}{\emph{Phys. Rev. D}
  {\bfseries 62} (2000) 064015}
  [\href{https://arxiv.org/abs/gr-qc/0001013}{{\ttfamily gr-qc/0001013}}].

\bibitem{DeWitt:1967uc}
B.~S. DeWitt, \emph{{Quantum Theory of Gravity. 3. Applications of the
  Covariant Theory}},
  \href{https://doi.org/10.1103/PhysRev.162.1239}{\emph{Phys. Rev.} {\bfseries
  162} (1967) 1239}.

\bibitem{Bern:2021dqo}
Z.~Bern, J.~Parra-Martinez, R.~Roiban, M.~S. Ruf, C.-H. Shen, M.~P. Solon
  et~al., \emph{{Scattering Amplitudes and Conservative Binary Dynamics at
  ${\cal O}(G^4)$}},
  \href{https://doi.org/10.1103/PhysRevLett.126.171601}{\emph{Phys. Rev. Lett.}
  {\bfseries 126} (2021) 171601}
  [\href{https://arxiv.org/abs/2101.07254}{{\ttfamily 2101.07254}}].

\bibitem{Bern:2021yeh}
Z.~Bern, J.~Parra-Martinez, R.~Roiban, M.~S. Ruf, C.-H. Shen, M.~P. Solon
  et~al., \emph{{Scattering Amplitudes, the Tail Effect, and Conservative
  Binary Dynamics at $O(G^4)$}},
  \href{https://arxiv.org/abs/2112.10750}{{\ttfamily 2112.10750}}.

\bibitem{gravWhitePaper2022}
A.~Buonanno, M.~Khalil, D.~O'Connell, R.~Roiban, M.~Solon, M.~Zeng et~al.,
  \emph{{Snowmass White Paper: Gravitational Waves and Scattering Amplitudes}},
   \href{https://arxiv.org/abs/2204.05194}{{\ttfamily 2204.05194}}.

\bibitem{Bern:2012uf}
Z.~Bern, J.~J.~M. Carrasco, L.~J. Dixon, H.~Johansson and R.~Roiban,
  \emph{{Simplifying Multiloop Integrands and Ultraviolet Divergences of Gauge
  Theory and Gravity Amplitudes}},
  \href{https://doi.org/10.1103/PhysRevD.85.105014}{\emph{Phys. Rev.}
  {\bfseries D85} (2012) 105014}
  [\href{https://arxiv.org/abs/1201.5366}{{\ttfamily 1201.5366}}].

\bibitem{Bern:1998ug}
Z.~Bern, L.~J. Dixon, D.~C. Dunbar, M.~Perelstein and J.~S. Rozowsky, \emph{{On
  the relationship between Yang-Mills theory and gravity and its implication
  for ultraviolet divergences}},
  \href{https://doi.org/10.1016/S0550-3213(98)00420-9}{\emph{Nucl. Phys. B}
  {\bfseries 530} (1998) 401}
  [\href{https://arxiv.org/abs/hep-th/9802162}{{\ttfamily hep-th/9802162}}].

\bibitem{Boucher-Veronneau:2011rlc}
C.~Boucher-Veronneau and L.~J. Dixon, \emph{{$\mathcal{N} \ge 4$ Supergravity
  Amplitudes from Gauge Theory at Two Loops}},
  \href{https://doi.org/10.1007/JHEP12(2011)046}{\emph{JHEP} {\bfseries 12}
  (2011) 046} [\href{https://arxiv.org/abs/1110.1132}{{\ttfamily 1110.1132}}].

\bibitem{Bern:2012gh}
Z.~Bern, S.~Davies, T.~Dennen and Y.-t. Huang, \emph{{Ultraviolet cancellations
  in half-maximal supergravity as a consequence of the double-copy structure}},
  \href{https://doi.org/10.1103/PhysRevD.86.105014}{\emph{Phys. Rev.}
  {\bfseries D86} (2012) 105014}
  [\href{https://arxiv.org/abs/1209.2472}{{\ttfamily 1209.2472}}].

\bibitem{Bern:2007hh}
Z.~Bern, J.~J. Carrasco, L.~J. Dixon, H.~Johansson, D.~A. Kosower and
  R.~Roiban, \emph{{Three-Loop Superfiniteness of N=8 Supergravity}},
  \href{https://doi.org/10.1103/PhysRevLett.98.161303}{\emph{Phys. Rev. Lett.}
  {\bfseries 98} (2007) 161303}
  [\href{https://arxiv.org/abs/hep-th/0702112}{{\ttfamily hep-th/0702112}}].

\bibitem{Bern:2014sna}
Z.~Bern, S.~Davies and T.~Dennen, \emph{{Enhanced ultraviolet cancellations in
  $\mathcal N=5$ supergravity at four loops}},
  \href{https://doi.org/10.1103/PhysRevD.90.105011}{\emph{Phys. Rev. D}
  {\bfseries 90} (2014) 105011}
  [\href{https://arxiv.org/abs/1409.3089}{{\ttfamily 1409.3089}}].

\bibitem{Bern:2014lha}
Z.~Bern, S.~Davies and T.~Dennen, \emph{{The Ultraviolet Critical Dimension of
  Half-Maximal Supergravity at Three Loops}},
  \href{https://arxiv.org/abs/1412.2441}{{\ttfamily 1412.2441}}.

\bibitem{Bern:2009kd}
Z.~Bern, J.~J. Carrasco, L.~J. Dixon, H.~Johansson and R.~Roiban, \emph{{The
  Ultraviolet Behavior of N=8 Supergravity at Four Loops}},
  \href{https://doi.org/10.1103/PhysRevLett.103.081301}{\emph{Phys. Rev. Lett.}
  {\bfseries 103} (2009) 081301}
  [\href{https://arxiv.org/abs/0905.2326}{{\ttfamily 0905.2326}}].

\bibitem{Bern:2013uka}
Z.~Bern, S.~Davies, T.~Dennen, A.~V. Smirnov and V.~A. Smirnov,
  \emph{{Ultraviolet Properties of N=4 Supergravity at Four Loops}},
  \href{https://doi.org/10.1103/PhysRevLett.111.231302}{\emph{Phys. Rev. Lett.}
  {\bfseries 111} (2013) 231302}
  [\href{https://arxiv.org/abs/1309.2498}{{\ttfamily 1309.2498}}].

\bibitem{Bern:2018jmv}
Z.~Bern, J.~J. Carrasco, W.-M. Chen, A.~Edison, H.~Johansson, J.~Parra-Martinez
  et~al., \emph{{Ultraviolet Properties of $\mathcal N = 8$ Supergravity at
  Five Loops}}, \href{https://doi.org/10.1103/PhysRevD.98.086021}{\emph{Phys.
  Rev.} {\bfseries D98} (2018) 086021}
  [\href{https://arxiv.org/abs/1804.09311}{{\ttfamily 1804.09311}}].

\bibitem{Nicolis:2008in}
A.~Nicolis, R.~Rattazzi and E.~Trincherini, \emph{{The Galileon as a local
  modification of gravity}},
  \href{https://doi.org/10.1103/PhysRevD.79.064036}{\emph{Phys. Rev. D}
  {\bfseries 79} (2009) 064036}
  [\href{https://arxiv.org/abs/0811.2197}{{\ttfamily 0811.2197}}].

\bibitem{deRham:2010ik}
C.~de~Rham and G.~Gabadadze, \emph{{Generalization of the Fierz-Pauli Action}},
  \href{https://doi.org/10.1103/PhysRevD.82.044020}{\emph{Phys. Rev. D}
  {\bfseries 82} (2010) 044020}
  [\href{https://arxiv.org/abs/1007.0443}{{\ttfamily 1007.0443}}].

\bibitem{deRham:2010kj}
C.~de~Rham, G.~Gabadadze and A.~J. Tolley, \emph{{Resummation of Massive
  Gravity}}, \href{https://doi.org/10.1103/PhysRevLett.106.231101}{\emph{Phys.
  Rev. Lett.} {\bfseries 106} (2011) 231101}
  [\href{https://arxiv.org/abs/1011.1232}{{\ttfamily 1011.1232}}].

\bibitem{deRham:2010eu}
C.~de~Rham and A.~J. Tolley, \emph{{DBI and the Galileon reunited}},
  \href{https://doi.org/10.1088/1475-7516/2010/05/015}{\emph{JCAP} {\bfseries
  05} (2010) 015} [\href{https://arxiv.org/abs/1003.5917}{{\ttfamily
  1003.5917}}].

\bibitem{Bern:1998sv}
Z.~Bern, L.~J. Dixon, M.~Perelstein and J.~Rozowsky, \emph{{Multileg one loop
  gravity amplitudes from gauge theory}},
  \href{https://doi.org/10.1016/S0550-3213(99)00029-2}{\emph{Nucl. Phys. B}
  {\bfseries 546} (1999) 423}
  [\href{https://arxiv.org/abs/hep-th/9811140}{{\ttfamily hep-th/9811140}}].

\bibitem{Bjerrum-Bohr:2010pnr}
N.~E.~J. Bjerrum-Bohr, P.~H. Damgaard, T.~Sondergaard and P.~Vanhove,
  \emph{{The Momentum Kernel of Gauge and Gravity Theories}},
  \href{https://doi.org/10.1007/JHEP01(2011)001}{\emph{JHEP} {\bfseries 01}
  (2011) 001} [\href{https://arxiv.org/abs/1010.3933}{{\ttfamily 1010.3933}}].

\bibitem{Cachazo:2013iea}
F.~Cachazo, S.~He and E.~Y. Yuan, \emph{{Scattering of Massless Particles:
  Scalars, Gluons and Gravitons}},
  \href{https://doi.org/10.1007/JHEP07(2014)033}{\emph{JHEP} {\bfseries 07}
  (2014) 033} [\href{https://arxiv.org/abs/1309.0885}{{\ttfamily 1309.0885}}].

\bibitem{Mizera:2017cqs}
S.~Mizera, \emph{{Combinatorics and Topology of Kawai-Lewellen-Tye Relations}},
  \href{https://doi.org/10.1007/JHEP08(2017)097}{\emph{JHEP} {\bfseries 08}
  (2017) 097} [\href{https://arxiv.org/abs/1706.08527}{{\ttfamily
  1706.08527}}].

\bibitem{Mizera:2017rqa}
S.~Mizera, \emph{{Scattering Amplitudes from Intersection Theory}},
  \href{https://doi.org/10.1103/PhysRevLett.120.141602}{\emph{Phys. Rev. Lett.}
  {\bfseries 120} (2018) 141602}
  [\href{https://arxiv.org/abs/1711.00469}{{\ttfamily 1711.00469}}].

\bibitem{Bern:2010yg}
Z.~Bern, T.~Dennen, Y.-t. Huang and M.~Kiermaier, \emph{{Gravity as the Square
  of Gauge Theory}},
  \href{https://doi.org/10.1103/PhysRevD.82.065003}{\emph{Phys. Rev. D}
  {\bfseries 82} (2010) 065003}
  [\href{https://arxiv.org/abs/1004.0693}{{\ttfamily 1004.0693}}].

\bibitem{Zhu:1980sz}
D.-p. Zhu, \emph{{Zeros in Scattering Amplitudes and the Structure of
  Nonabelian Gauge Theories}},
  \href{https://doi.org/10.1103/PhysRevD.22.2266}{\emph{Phys. Rev. D}
  {\bfseries 22} (1980) 2266}.

\bibitem{Goebel:1980es}
C.~J. Goebel, F.~Halzen and J.~P. Leveille, \emph{{Angular zeros of Brown,
  Mikaelian, Sahdev, and Samuel and the factorization of tree amplitudes in
  gauge theories}}, \href{https://doi.org/10.1103/PhysRevD.23.2682}{\emph{Phys.
  Rev. D} {\bfseries 23} (1981) 2682}.

\bibitem{Harland-Lang:2015faa}
L.~A. Harland-Lang, \emph{{Planar radiation zeros in five-parton QCD
  amplitudes}}, \href{https://doi.org/10.1007/JHEP05(2015)146}{\emph{JHEP}
  {\bfseries 05} (2015) 146}
  [\href{https://arxiv.org/abs/1503.06798}{{\ttfamily 1503.06798}}].

\bibitem{Brown:2016mrh}
R.~W. Brown and S.~G. Naculich, \emph{{BCJ relations from a new symmetry of
  gauge-theory amplitudes}},
  \href{https://doi.org/10.1007/JHEP10(2016)130}{\emph{JHEP} {\bfseries 10}
  (2016) 130} [\href{https://arxiv.org/abs/1608.04387}{{\ttfamily
  1608.04387}}].

\bibitem{Fairlie:1972zz}
D.~B. Fairlie and D.~E. Roberts, \emph{{Dual Models without Tachyons - a new
  approach}},  1972.

\bibitem{Cachazo:2014xea}
F.~Cachazo, S.~He and E.~Y. Yuan, \emph{{Scattering Equations and Matrices:
  From Einstein To Yang-Mills, DBI and NLSM}},
  \href{https://doi.org/10.1007/JHEP07(2015)149}{\emph{JHEP} {\bfseries 07}
  (2015) 149} [\href{https://arxiv.org/abs/1412.3479}{{\ttfamily 1412.3479}}].

\bibitem{Chen:2013fya}
G.~Chen and Y.-J. Du, \emph{{Amplitude Relations in Non-linear Sigma Model}},
  \href{https://doi.org/10.1007/JHEP01(2014)061}{\emph{JHEP} {\bfseries 01}
  (2014) 061} [\href{https://arxiv.org/abs/1311.1133}{{\ttfamily 1311.1133}}].

\bibitem{Du:2016tbc}
Y.-J. Du and C.-H. Fu, \emph{{Explicit BCJ numerators of nonlinear simga
  model}}, \href{https://doi.org/10.1007/JHEP09(2016)174}{\emph{JHEP}
  {\bfseries 09} (2016) 174}
  [\href{https://arxiv.org/abs/1606.05846}{{\ttfamily 1606.05846}}].

\bibitem{Carrasco:2016ldy}
J.~J.~M. Carrasco, C.~R. Mafra and O.~Schlotterer, \emph{{Abelian Z-theory:
  NLSM amplitudes and $\alpha$'-corrections from the open string}},
  \href{https://doi.org/10.1007/JHEP06(2017)093}{\emph{JHEP} {\bfseries 06}
  (2017) 093} [\href{https://arxiv.org/abs/1608.02569}{{\ttfamily
  1608.02569}}].

\bibitem{Cheung:2016prv}
C.~Cheung and C.-H. Shen, \emph{{Symmetry for Flavor-Kinematics Duality from an
  Action}}, \href{https://doi.org/10.1103/PhysRevLett.118.121601}{\emph{Phys.
  Rev. Lett.} {\bfseries 118} (2017) 121601}
  [\href{https://arxiv.org/abs/1612.00868}{{\ttfamily 1612.00868}}].

\bibitem{Kleiss:1988ne}
R.~Kleiss and H.~Kuijf, \emph{{Multi - Gluon Cross-sections and Five Jet
  Production at Hadron Colliders}},
  \href{https://doi.org/10.1016/0550-3213(89)90574-9}{\emph{Nucl. Phys. B}
  {\bfseries 312} (1989) 616}.

\bibitem{DelDuca:1999rs}
V.~Del~Duca, L.~J. Dixon and F.~Maltoni, \emph{{New color decompositions for
  gauge amplitudes at tree and loop level}},
  \href{https://doi.org/10.1016/S0550-3213(99)00809-3}{\emph{Nucl. Phys. B}
  {\bfseries 571} (2000) 51}
  [\href{https://arxiv.org/abs/hep-ph/9910563}{{\ttfamily hep-ph/9910563}}].

\bibitem{Plahte:1970wy}
E.~Plahte, \emph{{Symmetry properties of dual tree-graph n-point amplitudes}},
  \href{https://doi.org/10.1007/BF02824716}{\emph{Nuovo Cim. A} {\bfseries 66}
  (1970) 713}.

\bibitem{Stieberger:2009hq}
S.~Stieberger, \emph{{Open \& Closed vs. Pure Open String Disk Amplitudes}},
  \href{https://arxiv.org/abs/0907.2211}{{\ttfamily 0907.2211}}.

\bibitem{BjerrumBohr:2009rd}
N.~E.~J. Bjerrum-Bohr, P.~H. Damgaard and P.~Vanhove, \emph{{Minimal Basis for
  Gauge Theory Amplitudes}},
  \href{https://doi.org/10.1103/PhysRevLett.103.161602}{\emph{Phys. Rev. Lett.}
  {\bfseries 103} (2009) 161602}
  [\href{https://arxiv.org/abs/0907.1425}{{\ttfamily 0907.1425}}].

\bibitem{Feng:2010my}
B.~Feng, R.~Huang and Y.~Jia, \emph{{Gauge Amplitude Identities by On-shell
  Recursion Relation in S-matrix Program}},
  \href{https://doi.org/10.1016/j.physletb.2010.11.011}{\emph{Phys. Lett. B}
  {\bfseries 695} (2011) 350}
  [\href{https://arxiv.org/abs/1004.3417}{{\ttfamily 1004.3417}}].

\bibitem{Chen:2011jxa}
Y.-X. Chen, Y.-J. Du and B.~Feng, \emph{{A Proof of the Explicit Minimal-basis
  Expansion of Tree Amplitudes in Gauge Field Theory}},
  \href{https://doi.org/10.1007/JHEP02(2011)112}{\emph{JHEP} {\bfseries 02}
  (2011) 112} [\href{https://arxiv.org/abs/1101.0009}{{\ttfamily 1101.0009}}].

\bibitem{Mafra:2015vca}
C.~R. Mafra and O.~Schlotterer, \emph{{Berends-Giele recursions and the BCJ
  duality in superspace and components}},
  \href{https://doi.org/10.1007/JHEP03(2016)097}{\emph{JHEP} {\bfseries 03}
  (2016) 097} [\href{https://arxiv.org/abs/1510.08846}{{\ttfamily
  1510.08846}}].

\bibitem{Bern:2019prr}
Z.~Bern, J.~J. Carrasco, M.~Chiodaroli, H.~Johansson and R.~Roiban, \emph{{The
  Duality Between Color and Kinematics and its Applications}},
  \href{https://arxiv.org/abs/1909.01358}{{\ttfamily 1909.01358}}.

\bibitem{Bern:2022wqg}
Z.~Bern, J.~J. Carrasco, M.~Chiodaroli, H.~Johansson and R.~Roiban, \emph{{The
  SAGEX Review on Scattering Amplitudes, Chapter 2: An Invitation to
  Color-Kinematics Duality and the Double Copy}},
  \href{https://arxiv.org/abs/2203.13013}{{\ttfamily 2203.13013}}.

\bibitem{Oxburgh:2012zr}
S.~Oxburgh and C.~D. White, \emph{{BCJ duality and the double copy in the soft
  limit}}, \href{https://doi.org/10.1007/JHEP02(2013)127}{\emph{JHEP}
  {\bfseries 02} (2013) 127} [\href{https://arxiv.org/abs/1210.1110}{{\ttfamily
  1210.1110}}].

\bibitem{Saotome2012vy}
R.~Saotome and R.~Akhoury, \emph{{Relationship Between Gravity and Gauge
  Scattering in the High Energy Limit}},
  \href{https://doi.org/10.1007/JHEP01(2013)123}{\emph{JHEP} {\bfseries 01}
  (2013) 123} [\href{https://arxiv.org/abs/1210.8111}{{\ttfamily 1210.8111}}].

\bibitem{Borsten:2020zgj}
L.~Borsten, B.~Jurco, H.~Kim, T.~Macrelli, C.~Saemann and M.~Wolf,
  \emph{{Becchi-Rouet-Stora-Tyutin-Lagrangian Double Copy of Yang-Mills
  Theory}}, \href{https://doi.org/10.1103/PhysRevLett.126.191601}{\emph{Phys.
  Rev. Lett.} {\bfseries 126} (2021) 191601}
  [\href{https://arxiv.org/abs/2007.13803}{{\ttfamily 2007.13803}}].

\bibitem{Borsten:2021rmh}
L.~Borsten, B.~Jurco, H.~Kim, T.~Macrelli, C.~Saemann and M.~Wolf,
  \emph{{Tree-Level Color-Kinematics Duality Implies Loop-Level
  Color-Kinematics Duality}},
  \href{https://arxiv.org/abs/2108.03030}{{\ttfamily 2108.03030}}.

\bibitem{Bern:2015ooa}
Z.~Bern, S.~Davies and J.~Nohle, \emph{{Double-Copy Constructions and Unitarity
  Cuts}}, \href{https://doi.org/10.1103/PhysRevD.93.105015}{\emph{Phys. Rev. D}
  {\bfseries 93} (2016) 105015}
  [\href{https://arxiv.org/abs/1510.03448}{{\ttfamily 1510.03448}}].

\bibitem{Bern:2012cd}
Z.~Bern, S.~Davies, T.~Dennen and Y.-t. Huang, \emph{{Absence of Three-Loop
  Four-Point Divergences in N=4 Supergravity}},
  \href{https://doi.org/10.1103/PhysRevLett.108.201301}{\emph{Phys. Rev. Lett.}
  {\bfseries 108} (2012) 201301}
  [\href{https://arxiv.org/abs/1202.3423}{{\ttfamily 1202.3423}}].

\bibitem{Bern:2017ucb}
Z.~Bern, J.~J.~M. Carrasco, W.-M. Chen, H.~Johansson, R.~Roiban and M.~Zeng,
  \emph{{Five-loop four-point integrand of $N=8$ supergravity as a generalized
  double copy}}, \href{https://doi.org/10.1103/PhysRevD.96.126012}{\emph{Phys.
  Rev.} {\bfseries D96} (2017) 126012}
  [\href{https://arxiv.org/abs/1708.06807}{{\ttfamily 1708.06807}}].

\bibitem{Bern:2017yxu}
Z.~Bern, J.~J. Carrasco, W.-M. Chen, H.~Johansson and R.~Roiban, \emph{{Gravity
  Amplitudes as Generalized Double Copies of Gauge-Theory Amplitudes}},
  \href{https://doi.org/10.1103/PhysRevLett.118.181602}{\emph{Phys. Rev. Lett.}
  {\bfseries 118} (2017) 181602}
  [\href{https://arxiv.org/abs/1701.02519}{{\ttfamily 1701.02519}}].

\bibitem{Mason:2013sva}
L.~Mason and D.~Skinner, \emph{{Ambitwistor strings and the scattering
  equations}}, \href{https://doi.org/10.1007/JHEP07(2014)048}{\emph{JHEP}
  {\bfseries 07} (2014) 048} [\href{https://arxiv.org/abs/1311.2564}{{\ttfamily
  1311.2564}}].

\bibitem{Berkovits:2013xba}
N.~Berkovits, \emph{{Infinite Tension Limit of the Pure Spinor Superstring}},
  \href{https://doi.org/10.1007/JHEP03(2014)017}{\emph{JHEP} {\bfseries 03}
  (2014) 017} [\href{https://arxiv.org/abs/1311.4156}{{\ttfamily 1311.4156}}].

\bibitem{Adamo:2013tsa}
T.~Adamo, E.~Casali and D.~Skinner, \emph{{Ambitwistor strings and the
  scattering equations at one loop}},
  \href{https://doi.org/10.1007/JHEP04(2014)104}{\emph{JHEP} {\bfseries 04}
  (2014) 104} [\href{https://arxiv.org/abs/1312.3828}{{\ttfamily 1312.3828}}].

\bibitem{Casali:2014hfa}
E.~Casali and P.~Tourkine, \emph{{Infrared behaviour of the one-loop scattering
  equations and supergravity integrands}},
  \href{https://doi.org/10.1007/JHEP04(2015)013}{\emph{JHEP} {\bfseries 04}
  (2015) 013} [\href{https://arxiv.org/abs/1412.3787}{{\ttfamily 1412.3787}}].

\bibitem{Adamo:2015hoa}
T.~Adamo and E.~Casali, \emph{{Scattering equations, supergravity integrands,
  and pure spinors}},
  \href{https://doi.org/10.1007/JHEP05(2015)120}{\emph{JHEP} {\bfseries 05}
  (2015) 120} [\href{https://arxiv.org/abs/1502.06826}{{\ttfamily
  1502.06826}}].

\bibitem{Geyer:2022cey}
Y.~Geyer and L.~Mason, \emph{{The SAGEX Review on Scattering Amplitudes,
  Chapter 6: Ambitwistor Strings and Amplitudes from the Worldsheet}},
  \href{https://arxiv.org/abs/2203.13017}{{\ttfamily 2203.13017}}.

\bibitem{Geyer:2015bja}
Y.~Geyer, L.~Mason, R.~Monteiro and P.~Tourkine, \emph{{Loop Integrands for
  Scattering Amplitudes from the Riemann Sphere}},
  \href{https://doi.org/10.1103/PhysRevLett.115.121603}{\emph{Phys. Rev. Lett.}
  {\bfseries 115} (2015) 121603}
  [\href{https://arxiv.org/abs/1507.00321}{{\ttfamily 1507.00321}}].

\bibitem{Geyer:2015jch}
Y.~Geyer, L.~Mason, R.~Monteiro and P.~Tourkine, \emph{{One-loop amplitudes on
  the Riemann sphere}},
  \href{https://doi.org/10.1007/JHEP03(2016)114}{\emph{JHEP} {\bfseries 03}
  (2016) 114} [\href{https://arxiv.org/abs/1511.06315}{{\ttfamily
  1511.06315}}].

\bibitem{DHoker:1988pdl}
E.~D'Hoker and D.~H. Phong, \emph{{The Geometry of String Perturbation
  Theory}}, \href{https://doi.org/10.1103/RevModPhys.60.917}{\emph{Rev. Mod.
  Phys.} {\bfseries 60} (1988) 917}.

\bibitem{DHoker:1989cxq}
E.~D'Hoker and D.~Phong, \emph{{Conformal Scalar Fields and Chiral Splitting on
  Superriemann Surfaces}},
  \href{https://doi.org/10.1007/BF01218413}{\emph{Commun. Math. Phys.}
  {\bfseries 125} (1989) 469}.

\bibitem{Bjerrum-Bohr:2016axv}
N.~E.~J. Bjerrum-Bohr, J.~L. Bourjaily, P.~H. Damgaard and B.~Feng,
  \emph{{Manifesting Color-Kinematics Duality in the Scattering Equation
  Formalism}}, \href{https://doi.org/10.1007/JHEP09(2016)094}{\emph{JHEP}
  {\bfseries 09} (2016) 094}
  [\href{https://arxiv.org/abs/1608.00006}{{\ttfamily 1608.00006}}].

\bibitem{Du:2017kpo}
Y.-J. Du and F.~Teng, \emph{{BCJ numerators from reduced Pfaffian}},
  \href{https://doi.org/10.1007/JHEP04(2017)033}{\emph{JHEP} {\bfseries 04}
  (2017) 033} [\href{https://arxiv.org/abs/1703.05717}{{\ttfamily
  1703.05717}}].

\bibitem{Edison:2020ehu}
A.~Edison and F.~Teng, \emph{{Efficient Calculation of Crossing Symmetric BCJ
  Tree Numerators}}, \href{https://doi.org/10.1007/JHEP12(2020)138}{\emph{JHEP}
  {\bfseries 12} (2020) 138}
  [\href{https://arxiv.org/abs/2005.03638}{{\ttfamily 2005.03638}}].

\bibitem{He:2017spx}
S.~He, O.~Schlotterer and Y.~Zhang, \emph{{New BCJ representations for one-loop
  amplitudes in gauge theories and gravity}},
  \href{https://doi.org/10.1016/j.nuclphysb.2018.03.003}{\emph{Nucl. Phys.}
  {\bfseries B930} (2018) 328}
  [\href{https://arxiv.org/abs/1706.00640}{{\ttfamily 1706.00640}}].

\bibitem{Edison:2020uzf}
A.~Edison, S.~He, O.~Schlotterer and F.~Teng, \emph{{One-loop Correlators and
  BCJ Numerators from Forward Limits}},
  \href{https://doi.org/10.1007/JHEP09(2020)079}{\emph{JHEP} {\bfseries 09}
  (2020) 079} [\href{https://arxiv.org/abs/2005.03639}{{\ttfamily
  2005.03639}}].

\bibitem{Geyer:2017ela}
Y.~Geyer and R.~Monteiro, \emph{{Gluons and gravitons at one loop from
  ambitwistor strings}},
  \href{https://doi.org/10.1007/JHEP03(2018)068}{\emph{JHEP} {\bfseries 03}
  (2018) 068} [\href{https://arxiv.org/abs/1711.09923}{{\ttfamily
  1711.09923}}].

\bibitem{Geyer:2016wjx}
Y.~Geyer, L.~Mason, R.~Monteiro and P.~Tourkine, \emph{{Two-Loop Scattering
  Amplitudes from the Riemann Sphere}},
  \href{https://doi.org/10.1103/PhysRevD.94.125029}{\emph{Phys. Rev.}
  {\bfseries D94} (2016) 125029}
  [\href{https://arxiv.org/abs/1607.08887}{{\ttfamily 1607.08887}}].

\bibitem{Geyer:2018xwu}
Y.~Geyer and R.~Monteiro, \emph{{Two-Loop Scattering Amplitudes from
  Ambitwistor Strings: from Genus Two to the Nodal Riemann Sphere}},
  \href{https://doi.org/10.1007/JHEP11(2018)008}{\emph{JHEP} {\bfseries 11}
  (2018) 008} [\href{https://arxiv.org/abs/1805.05344}{{\ttfamily
  1805.05344}}].

\bibitem{Geyer:2019hnn}
Y.~Geyer, R.~Monteiro and R.~Stark-Much\~ao, \emph{{Two-Loop Scattering
  Amplitudes: Double-Forward Limit and Colour-Kinematics Duality}},
  \href{https://doi.org/10.1007/JHEP12(2019)049}{\emph{JHEP} {\bfseries 12}
  (2019) 049} [\href{https://arxiv.org/abs/1908.05221}{{\ttfamily
  1908.05221}}].

\bibitem{Geyer:2021oox}
Y.~Geyer, R.~Monteiro and R.~Stark-Much\~ao, \emph{{Superstring Loop Amplitudes
  from the Field Theory Limit}},
  \href{https://doi.org/10.1103/PhysRevLett.127.211603}{\emph{Phys. Rev. Lett.}
  {\bfseries 127} (2021) 211603}
  [\href{https://arxiv.org/abs/2106.03968}{{\ttfamily 2106.03968}}].

\bibitem{He:2016mzd}
S.~He and O.~Schlotterer, \emph{{New Relations for Gauge-Theory and Gravity
  Amplitudes at Loop Level}},
  \href{https://doi.org/10.1103/PhysRevLett.118.161601}{\emph{Phys. Rev. Lett.}
  {\bfseries 118} (2017) 161601}
  [\href{https://arxiv.org/abs/1612.00417}{{\ttfamily 1612.00417}}].

\bibitem{Gomez:2016cqb}
H.~Gomez, S.~Mizera and G.~Zhang, \emph{{CHY Loop Integrands from Holomorphic
  Forms}}, \href{https://doi.org/10.1007/JHEP03(2017)092}{\emph{JHEP}
  {\bfseries 03} (2017) 092}
  [\href{https://arxiv.org/abs/1612.06854}{{\ttfamily 1612.06854}}].

\bibitem{Gomez:2017lhy}
H.~Gomez, \emph{{Quadratic Feynman Loop Integrands From Massless Scattering
  Equations}}, \href{https://doi.org/10.1103/PhysRevD.95.106006}{\emph{Phys.
  Rev. D} {\bfseries 95} (2017) 106006}
  [\href{https://arxiv.org/abs/1703.04714}{{\ttfamily 1703.04714}}].

\bibitem{Gomez:2017cpe}
H.~Gomez, C.~Lopez-Arcos and P.~Talavera, \emph{{One-loop Parke-Taylor factors
  for quadratic propagators from massless scattering equations}},
  \href{https://doi.org/10.1007/JHEP10(2017)175}{\emph{JHEP} {\bfseries 10}
  (2017) 175} [\href{https://arxiv.org/abs/1707.08584}{{\ttfamily
  1707.08584}}].

\bibitem{Ahmadiniaz:2018nvr}
N.~Ahmadiniaz, H.~Gomez and C.~Lopez-Arcos, \emph{{Non-planar one-loop
  Parke-Taylor factors in the CHY approach for quadratic propagators}},
  \href{https://doi.org/10.1007/JHEP05(2018)055}{\emph{JHEP} {\bfseries 05}
  (2018) 055} [\href{https://arxiv.org/abs/1802.00015}{{\ttfamily
  1802.00015}}].

\bibitem{Agerskov:2019ryp}
J.~Agerskov, N.~E.~J. Bjerrum-Bohr, H.~Gomez and C.~Lopez-Arcos,
  \emph{{One-Loop Yang-Mills Integrands from Scattering Equations}},
  \href{https://doi.org/10.1103/PhysRevD.102.045023}{\emph{Phys. Rev. D}
  {\bfseries 102} (2020) 045023}
  [\href{https://arxiv.org/abs/1910.03602}{{\ttfamily 1910.03602}}].

\bibitem{Farrow:2020voh}
J.~A. Farrow, Y.~Geyer, A.~E. Lipstein, R.~Monteiro and R.~Stark-Much\~ao,
  \emph{{Propagators, BCFW recursion and new scattering equations at one
  loop}}, \href{https://doi.org/10.1007/JHEP10(2020)074}{\emph{JHEP} {\bfseries
  10} (2020) 074} [\href{https://arxiv.org/abs/2007.00623}{{\ttfamily
  2007.00623}}].

\bibitem{Tourkine:2016bak}
P.~Tourkine and P.~Vanhove, \emph{{Higher-loop amplitude monodromy relations in
  string and gauge theory}},
  \href{https://doi.org/10.1103/PhysRevLett.117.211601}{\emph{Phys. Rev. Lett.}
  {\bfseries 117} (2016) 211601}
  [\href{https://arxiv.org/abs/1608.01665}{{\ttfamily 1608.01665}}].

\bibitem{Hohenegger:2017kqy}
S.~Hohenegger and S.~Stieberger, \emph{{Monodromy Relations in Higher-Loop
  String Amplitudes}},
  \href{https://doi.org/10.1016/j.nuclphysb.2017.09.020}{\emph{Nucl. Phys.}
  {\bfseries B925} (2017) 63}
  [\href{https://arxiv.org/abs/1702.04963}{{\ttfamily 1702.04963}}].

\bibitem{Casali:2019ihm}
E.~Casali, S.~Mizera and P.~Tourkine, \emph{{Monodromy relations from twisted
  homology}}, \href{https://doi.org/10.1007/JHEP12(2019)087}{\emph{JHEP}
  {\bfseries 12} (2019) 087}
  [\href{https://arxiv.org/abs/1910.08514}{{\ttfamily 1910.08514}}].

\bibitem{Casali:2020knc}
E.~Casali, S.~Mizera and P.~Tourkine, \emph{{Loop amplitudes monodromy
  relations and color-kinematics duality}},
  \href{https://doi.org/10.1007/JHEP03(2021)048}{\emph{JHEP} {\bfseries 03}
  (2021) 048} [\href{https://arxiv.org/abs/2005.05329}{{\ttfamily
  2005.05329}}].

\bibitem{Stieberger:2021daa}
S.~Stieberger, \emph{{Open \& Closed vs. Pure Open String One-Loop
  Amplitudes}},  \href{https://arxiv.org/abs/2105.06888}{{\ttfamily
  2105.06888}}.

\bibitem{Berends:1981rb}
F.~A. Berends, R.~Kleiss, P.~De~Causmaecker, R.~Gastmans and T.~T. Wu,
  \emph{{Single Bremsstrahlung Processes in Gauge Theories}},
  \href{https://doi.org/10.1016/0370-2693(81)90685-7}{\emph{Phys. Lett. B}
  {\bfseries 103} (1981) 124}.

\bibitem{DeCausmaecker:1981wzb}
P.~De~Causmaecker, R.~Gastmans, W.~Troost and T.~T. Wu, \emph{{Helicity
  Amplitudes for Massless QED}},
  \href{https://doi.org/10.1016/0370-2693(81)91025-X}{\emph{Phys. Lett. B}
  {\bfseries 105} (1981) 215}.

\bibitem{Kleiss:1985yh}
R.~Kleiss and W.~J. Stirling, \emph{{Spinor Techniques for Calculating p anti-p
  ---\ensuremath{>} W+- / Z0 + Jets}},
  \href{https://doi.org/10.1016/0550-3213(85)90285-8}{\emph{Nucl. Phys. B}
  {\bfseries 262} (1985) 235}.

\bibitem{Gunion:1985vca}
J.~F. Gunion and Z.~Kunszt, \emph{{Improved Analytic Techniques for Tree Graph
  Calculations and the G g q anti-q Lepton anti-Lepton Subprocess}},
  \href{https://doi.org/10.1016/0370-2693(85)90774-9}{\emph{Phys. Lett. B}
  {\bfseries 161} (1985) 333}.

\bibitem{Xu:1986xb}
Z.~Xu, D.-H. Zhang and L.~Chang, \emph{{Helicity Amplitudes for Multiple
  Bremsstrahlung in Massless Nonabelian Gauge Theories}},
  \href{https://doi.org/10.1016/0550-3213(87)90479-2}{\emph{Nucl. Phys. B}
  {\bfseries 291} (1987) 392}.

\bibitem{Parke:1986gb}
S.~J. Parke and T.~R. Taylor, \emph{{An Amplitude for $n$ Gluon Scattering}},
  \href{https://doi.org/10.1103/PhysRevLett.56.2459}{\emph{Phys. Rev. Lett.}
  {\bfseries 56} (1986) 2459}.

\bibitem{Perjes:1974ra}
Z.~Perjes, \emph{{Twistor Variables of Relativistic Mechanics}},
  \href{https://doi.org/10.1103/PhysRevD.11.2031}{\emph{Phys. Rev. D}
  {\bfseries 11} (1975) 2031}.

\bibitem{Kleiss:1986qc}
R.~Kleiss and W.~J. Stirling, \emph{{Cross-sections for the Production of an
  Arbitrary Number of Photons in Electron - Positron Annihilation}},
  \href{https://doi.org/10.1016/0370-2693(86)90454-5}{\emph{Phys. Lett. B}
  {\bfseries 179} (1986) 159}.

\bibitem{Spehler:1991yw}
D.~Spehler and S.~F. Novaes, \emph{{Helicity wave functions for massless and
  massive spin-2 particles}},
  \href{https://doi.org/10.1103/PhysRevD.44.3990}{\emph{Phys. Rev. D}
  {\bfseries 44} (1991) 3990}.

\bibitem{Novaes:1991ft}
S.~F. Novaes and D.~Spehler, \emph{{Weyl-Van Der Waerden spinor technic for
  spin 3/2 fermions}},
  \href{https://doi.org/10.1016/0550-3213(92)90689-9}{\emph{Nucl. Phys. B}
  {\bfseries 371} (1992) 618}.

\bibitem{Conde:2016izb}
E.~Conde, E.~Joung and K.~Mkrtchyan, \emph{{Spinor-Helicity Three-Point
  Amplitudes from Local Cubic Interactions}},
  \href{https://doi.org/10.1007/JHEP08(2016)040}{\emph{JHEP} {\bfseries 08}
  (2016) 040} [\href{https://arxiv.org/abs/1605.07402}{{\ttfamily
  1605.07402}}].

\bibitem{Arkani-Hamed:2017jhn}
N.~Arkani-Hamed, T.-C. Huang and Y.-t. Huang, \emph{{Scattering amplitudes for
  all masses and spins}},
  \href{https://doi.org/10.1007/JHEP11(2021)070}{\emph{JHEP} {\bfseries 11}
  (2021) 070} [\href{https://arxiv.org/abs/1709.04891}{{\ttfamily
  1709.04891}}].

\bibitem{Dennen:2009vk}
T.~Dennen, Y.-t. Huang and W.~Siegel, \emph{{Supertwistor space for 6D maximal
  super Yang-Mills}},
  \href{https://doi.org/10.1007/JHEP04(2010)127}{\emph{JHEP} {\bfseries 04}
  (2010) 127} [\href{https://arxiv.org/abs/0910.2688}{{\ttfamily 0910.2688}}].

\bibitem{Caron-Huot:2010nes}
S.~Caron-Huot and D.~O'Connell, \emph{{Spinor Helicity and Dual Conformal
  Symmetry in Ten Dimensions}},
  \href{https://doi.org/10.1007/JHEP08(2011)014}{\emph{JHEP} {\bfseries 08}
  (2011) 014} [\href{https://arxiv.org/abs/1010.5487}{{\ttfamily 1010.5487}}].

\bibitem{Boels:2012ie}
R.~H. Boels and D.~O'Connell, \emph{{Simple superamplitudes in higher
  dimensions}}, \href{https://doi.org/10.1007/JHEP06(2012)163}{\emph{JHEP}
  {\bfseries 06} (2012) 163} [\href{https://arxiv.org/abs/1201.2653}{{\ttfamily
  1201.2653}}].

\bibitem{Chiodaroli:2022ssi}
M.~Chiodaroli, M.~Gunaydin, H.~Johansson and R.~Roiban, \emph{{Spinor-helicity
  formalism for massive and massless amplitudes in five dimensions}},
  \href{https://arxiv.org/abs/2202.08257}{{\ttfamily 2202.08257}}.

\bibitem{Hodges:2009hk}
A.~Hodges, \emph{{Eliminating spurious poles from gauge-theoretic amplitudes}},
  \href{https://doi.org/10.1007/JHEP05(2013)135}{\emph{JHEP} {\bfseries 05}
  (2013) 135} [\href{https://arxiv.org/abs/0905.1473}{{\ttfamily 0905.1473}}].

\bibitem{Drummond:2009fd}
J.~M. Drummond, J.~M. Henn and J.~Plefka, \emph{{Yangian symmetry of scattering
  amplitudes in N=4 super Yang-Mills theory}},
  \href{https://doi.org/10.1088/1126-6708/2009/05/046}{\emph{JHEP} {\bfseries
  05} (2009) 046} [\href{https://arxiv.org/abs/0902.2987}{{\ttfamily
  0902.2987}}].

\bibitem{Mason:2009qx}
L.~J. Mason and D.~Skinner, \emph{{Dual Superconformal Invariance, Momentum
  Twistors and Grassmannians}},
  \href{https://doi.org/10.1088/1126-6708/2009/11/045}{\emph{JHEP} {\bfseries
  11} (2009) 045} [\href{https://arxiv.org/abs/0909.0250}{{\ttfamily
  0909.0250}}].

\bibitem{Arkani-Hamed:2009nll}
N.~Arkani-Hamed, F.~Cachazo and C.~Cheung, \emph{{The Grassmannian Origin Of
  Dual Superconformal Invariance}},
  \href{https://doi.org/10.1007/JHEP03(2010)036}{\emph{JHEP} {\bfseries 03}
  (2010) 036} [\href{https://arxiv.org/abs/0909.0483}{{\ttfamily 0909.0483}}].

\bibitem{Mason:2010yk}
L.~J. Mason and D.~Skinner, \emph{{The Complete Planar S-matrix of N=4 SYM as a
  Wilson Loop in Twistor Space}},
  \href{https://doi.org/10.1007/JHEP12(2010)018}{\emph{JHEP} {\bfseries 12}
  (2010) 018} [\href{https://arxiv.org/abs/1009.2225}{{\ttfamily 1009.2225}}].

\bibitem{Caron-Huot:2010ryg}
S.~Caron-Huot, \emph{{Notes on the scattering amplitude / Wilson loop
  duality}}, \href{https://doi.org/10.1007/JHEP07(2011)058}{\emph{JHEP}
  {\bfseries 07} (2011) 058} [\href{https://arxiv.org/abs/1010.1167}{{\ttfamily
  1010.1167}}].

\bibitem{Arkani-Hamed:2010zjl}
N.~Arkani-Hamed, J.~L. Bourjaily, F.~Cachazo, S.~Caron-Huot and J.~Trnka,
  \emph{{The All-Loop Integrand For Scattering Amplitudes in Planar N=4 SYM}},
  \href{https://doi.org/10.1007/JHEP01(2011)041}{\emph{JHEP} {\bfseries 01}
  (2011) 041} [\href{https://arxiv.org/abs/1008.2958}{{\ttfamily 1008.2958}}].

\bibitem{Arkani-Hamed:2012zlh}
N.~Arkani-Hamed, J.~L. Bourjaily, F.~Cachazo, A.~B. Goncharov, A.~Postnikov and
  J.~Trnka, \emph{{Grassmannian Geometry of Scattering Amplitudes}}. Cambridge
  University Press, 4, 2016,
  \href{https://doi.org/10.1017/CBO9781316091548}{10.1017/CBO9781316091548},
  [\href{https://arxiv.org/abs/1212.5605}{{\ttfamily 1212.5605}}].

\bibitem{Witten:2004cp}
E.~Witten, \emph{{Parity invariance for strings in twistor space}},
  \href{https://doi.org/10.4310/ATMP.2004.v8.n5.a1}{\emph{Adv. Theor. Math.
  Phys.} {\bfseries 8} (2004) 779}
  [\href{https://arxiv.org/abs/hep-th/0403199}{{\ttfamily hep-th/0403199}}].

\bibitem{Geyer:2014fka}
Y.~Geyer, A.~E. Lipstein and L.~J. Mason, \emph{{Ambitwistor Strings in Four
  Dimensions}},
  \href{https://doi.org/10.1103/PhysRevLett.113.081602}{\emph{Phys. Rev. Lett.}
  {\bfseries 113} (2014) 081602}
  [\href{https://arxiv.org/abs/1404.6219}{{\ttfamily 1404.6219}}].

\bibitem{Geyer:2018xgb}
Y.~Geyer and L.~Mason, \emph{{Polarized Scattering Equations for 6D
  Superamplitudes}},
  \href{https://doi.org/10.1103/PhysRevLett.122.101601}{\emph{Phys. Rev. Lett.}
  {\bfseries 122} (2019) 101601}
  [\href{https://arxiv.org/abs/1812.05548}{{\ttfamily 1812.05548}}].

\bibitem{Geyer:2019ayz}
Y.~Geyer and L.~Mason, \emph{{Supersymmetric S-matrices from the worldsheet in
  10 \textbackslash{}\& 11d}},
  \href{https://doi.org/10.1016/j.physletb.2020.135361}{\emph{Phys. Lett. B}
  {\bfseries 804} (2020) 135361}
  [\href{https://arxiv.org/abs/1901.00134}{{\ttfamily 1901.00134}}].

\bibitem{Albonico:2020mge}
G.~Albonico, Y.~Geyer and L.~Mason, \emph{{Recursion and worldsheet formulae
  for 6d superamplitudes}},
  \href{https://doi.org/10.1007/JHEP08(2020)066}{\emph{JHEP} {\bfseries 08}
  (2020) 066} [\href{https://arxiv.org/abs/2001.05928}{{\ttfamily
  2001.05928}}].

\bibitem{Geyer:2020iwz}
Y.~Geyer, L.~Mason and D.~Skinner, \emph{{Ambitwistor strings in six and five
  dimensions}}, \href{https://doi.org/10.1007/JHEP08(2021)153}{\emph{JHEP}
  {\bfseries 08} (2021) 153}
  [\href{https://arxiv.org/abs/2012.15172}{{\ttfamily 2012.15172}}].

\bibitem{Albonico:2022pmd}
G.~Albonico, Y.~Geyer and L.~Mason, \emph{{From twistor-particle models to
  massive amplitudes}},  \href{https://arxiv.org/abs/2203.08087}{{\ttfamily
  2203.08087}}.

\bibitem{Berkovits:2000fe}
N.~Berkovits, \emph{{Super Poincare covariant quantization of the
  superstring}},
  \href{https://doi.org/10.1088/1126-6708/2000/04/018}{\emph{JHEP} {\bfseries
  04} (2000) 018} [\href{https://arxiv.org/abs/hep-th/0001035}{{\ttfamily
  hep-th/0001035}}].

\bibitem{Berkovits:2006ik}
N.~Berkovits, \emph{{Explaining Pure Spinor Superspace}},
  \href{https://arxiv.org/abs/hep-th/0612021}{{\ttfamily hep-th/0612021}}.

\bibitem{Mafra:2010pn}
C.~R. Mafra, \emph{{PSS: A FORM Program to Evaluate Pure Spinor Superspace
  Expressions}},  \href{https://arxiv.org/abs/1007.4999}{{\ttfamily
  1007.4999}}.

\bibitem{Berends:1987me}
F.~A. Berends and W.~T. Giele, \emph{{Recursive Calculations for Processes with
  n Gluons}}, \href{https://doi.org/10.1016/0550-3213(88)90442-7}{\emph{Nucl.
  Phys. B} {\bfseries 306} (1988) 759}.

\bibitem{Rosly:1996vr}
A.~A. Rosly and K.~G. Selivanov, \emph{{On amplitudes in selfdual sector of
  Yang-Mills theory}},
  \href{https://doi.org/10.1016/S0370-2693(97)00268-2}{\emph{Phys. Lett. B}
  {\bfseries 399} (1997) 135}
  [\href{https://arxiv.org/abs/hep-th/9611101}{{\ttfamily hep-th/9611101}}].

\bibitem{Rosly:1997ap}
A.~A. Rosly and K.~G. Selivanov, \emph{{Gravitational SD perturbiner}},
  \href{https://arxiv.org/abs/hep-th/9710196}{{\ttfamily hep-th/9710196}}.

\bibitem{Selivanov:1997aq}
K.~G. Selivanov, \emph{{SD perturbiner in Yang-Mills + gravity}},
  \href{https://doi.org/10.1016/S0370-2693(97)01514-1}{\emph{Phys. Lett. B}
  {\bfseries 420} (1998) 274}
  [\href{https://arxiv.org/abs/hep-th/9710197}{{\ttfamily hep-th/9710197}}].

\bibitem{Mafra:2011kj}
C.~Mafra, O.~Schlotterer and S.~Stieberger, \emph{{Explicit BCJ Numerators from
  Pure Spinors}}, \href{https://doi.org/10.1007/JHEP07(2011)092}{\emph{JHEP}
  {\bfseries 07} (2011) 092} [\href{https://arxiv.org/abs/1104.5224}{{\ttfamily
  1104.5224}}].

\bibitem{Mafra:2014gja}
C.~Mafra and O.~Schlotterer, \emph{{Towards one-loop SYM amplitudes from the
  pure spinor BRST cohomology}},
  \href{https://doi.org/10.1002/prop.201400076}{\emph{Fortsch. Phys.}
  {\bfseries 63} (2015) 105} [\href{https://arxiv.org/abs/1410.0668}{{\ttfamily
  1410.0668}}].

\bibitem{Mafra:2015mja}
C.~R. Mafra and O.~Schlotterer, \emph{{Two-loop five-point amplitudes of super
  Yang-Mills and supergravity in pure spinor superspace}},
  \href{https://doi.org/10.1007/JHEP10(2015)124}{\emph{JHEP} {\bfseries 10}
  (2015) 124} [\href{https://arxiv.org/abs/1505.02746}{{\ttfamily
  1505.02746}}].

\bibitem{Mafra:2011nv}
C.~Mafra, O.~Schlotterer and S.~Stieberger, \emph{{Complete N-Point Superstring
  Disk Amplitude I. Pure Spinor Computation}},
  \href{https://doi.org/10.1016/j.nuclphysb.2013.04.023}{\emph{Nucl. Phys.}
  {\bfseries B873} (2013) 419}
  [\href{https://arxiv.org/abs/1106.2645}{{\ttfamily 1106.2645}}].

\bibitem{Mafra:2018qqe}
C.~R. Mafra and O.~Schlotterer, \emph{{Towards the n-point one-loop superstring
  amplitude. Part III. One-loop correlators and their double-copy structure}},
  \href{https://doi.org/10.1007/JHEP08(2019)092}{\emph{JHEP} {\bfseries 08}
  (2019) 092} [\href{https://arxiv.org/abs/1812.10971}{{\ttfamily
  1812.10971}}].

\bibitem{DHoker:2020prr}
E.~D'Hoker, C.~Mafra, B.~Pioline and O.~Schlotterer, \emph{{Two-loop
  superstring five-point amplitudes. Part I. Construction via chiral splitting
  and pure spinors}},
  \href{https://doi.org/10.1007/JHEP08(2020)135}{\emph{JHEP} {\bfseries 08}
  (2020) 135} [\href{https://arxiv.org/abs/2006.05270}{{\ttfamily
  2006.05270}}].

\bibitem{Monteiro:2011pc}
R.~Monteiro and D.~O'Connell, \emph{{The Kinematic Algebra From the Self-Dual
  Sector}}, \href{https://doi.org/10.1007/JHEP07(2011)007}{\emph{JHEP}
  {\bfseries 07} (2011) 007} [\href{https://arxiv.org/abs/1105.2565}{{\ttfamily
  1105.2565}}].

\bibitem{Lee:2015upy}
S.~Lee, C.~R. Mafra and O.~Schlotterer, \emph{{Non-linear gauge transformations
  in $D=10$ SYM theory and the BCJ duality}},
  \href{https://doi.org/10.1007/JHEP03(2016)090}{\emph{JHEP} {\bfseries 03}
  (2016) 090} [\href{https://arxiv.org/abs/1510.08843}{{\ttfamily
  1510.08843}}].

\bibitem{Garozzo:2018uzj}
L.~M. Garozzo, L.~Queimada and O.~Schlotterer, \emph{{Berends-Giele currents in
  Bern-Carrasco-Johansson gauge for $F^3$- and $F^4$-deformed Yang-Mills
  amplitudes}}, \href{https://doi.org/10.1007/JHEP02(2019)078}{\emph{JHEP}
  {\bfseries 02} (2019) 078}
  [\href{https://arxiv.org/abs/1809.08103}{{\ttfamily 1809.08103}}].

\bibitem{Mizera:2018jbh}
S.~Mizera and B.~Skrzypek, \emph{{Perturbiner Methods for Effective Field
  Theories and the Double Copy}},
  \href{https://doi.org/10.1007/JHEP10(2018)018}{\emph{JHEP} {\bfseries 10}
  (2018) 018} [\href{https://arxiv.org/abs/1809.02096}{{\ttfamily
  1809.02096}}].

\bibitem{Bridges:2019siz}
E.~Bridges and C.~R. Mafra, \emph{{Algorithmic construction of SYM
  multiparticle superfields in the BCJ gauge}},
  \href{https://doi.org/10.1007/JHEP10(2019)022}{\emph{JHEP} {\bfseries 10}
  (2019) 022} [\href{https://arxiv.org/abs/1906.12252}{{\ttfamily
  1906.12252}}].

\bibitem{Cheung:2021zvb}
C.~Cheung and J.~Mangan, \emph{{Covariant color-kinematics duality}},
  \href{https://doi.org/10.1007/JHEP11(2021)069}{\emph{JHEP} {\bfseries 11}
  (2021) 069} [\href{https://arxiv.org/abs/2108.02276}{{\ttfamily
  2108.02276}}].

\bibitem{Monteiro:2014cda}
R.~Monteiro, D.~O'Connell and C.~D. White, \emph{{Black holes and the double
  copy}}, \href{https://doi.org/10.1007/JHEP12(2014)056}{\emph{JHEP} {\bfseries
  12} (2014) 056} [\href{https://arxiv.org/abs/1410.0239}{{\ttfamily
  1410.0239}}].

\bibitem{Luna:2015paa}
A.~Luna, R.~Monteiro, D.~O'Connell and C.~D. White, \emph{{The classical double
  copy for TaubNUT spacetime}},
  \href{https://doi.org/10.1016/j.physletb.2015.09.021}{\emph{Phys. Lett.}
  {\bfseries B750} (2015) 272}
  [\href{https://arxiv.org/abs/1507.01869}{{\ttfamily 1507.01869}}].

\bibitem{Luna:2016due}
A.~Luna, R.~Monteiro, I.~Nicholson, D.~O'Connell and C.~D. White, \emph{{The
  double copy: Bremsstrahlung and accelerating black holes}},
  \href{https://doi.org/10.1007/JHEP06(2016)023}{\emph{JHEP} {\bfseries 06}
  (2016) 023} [\href{https://arxiv.org/abs/1603.05737}{{\ttfamily
  1603.05737}}].

\bibitem{Kosower:2018adc}
D.~A. Kosower, B.~Maybee and D.~O'Connell, \emph{{Amplitudes, Observables, and
  Classical Scattering}},
  \href{https://doi.org/10.1007/JHEP02(2019)137}{\emph{JHEP} {\bfseries 02}
  (2019) 137} [\href{https://arxiv.org/abs/1811.10950}{{\ttfamily
  1811.10950}}].

\bibitem{Cristofoli:2021vyo}
A.~Cristofoli, R.~Gonzo, D.~A. Kosower and D.~O'Connell, \emph{{Waveforms from
  Amplitudes}},  \href{https://arxiv.org/abs/2107.10193}{{\ttfamily
  2107.10193}}.

\bibitem{Monteiro:2020plf}
R.~Monteiro, D.~O'Connell, D.~Peinador~Veiga and M.~Sergola, \emph{{Classical
  solutions and their double copy in split signature}},
  \href{https://doi.org/10.1007/JHEP05(2021)268}{\emph{JHEP} {\bfseries 05}
  (2021) 268} [\href{https://arxiv.org/abs/2012.11190}{{\ttfamily
  2012.11190}}].

\bibitem{Huang:2019cja}
Y.-T. Huang, U.~Kol and D.~O'Connell, \emph{{Double copy of electric-magnetic
  duality}}, \href{https://doi.org/10.1103/PhysRevD.102.046005}{\emph{Phys.
  Rev. D} {\bfseries 102} (2020) 046005}
  [\href{https://arxiv.org/abs/1911.06318}{{\ttfamily 1911.06318}}].

\bibitem{Emond:2020lwi}
W.~T. Emond, Y.-T. Huang, U.~Kol, N.~Moynihan and D.~O'Connell,
  \emph{{Amplitudes from Coulomb to Kerr-Taub-NUT}},
  \href{https://arxiv.org/abs/2010.07861}{{\ttfamily 2010.07861}}.

\bibitem{Arkani-Hamed:2019ymq}
N.~Arkani-Hamed, Y.-t. Huang and D.~O'Connell, \emph{{Kerr black holes as
  elementary particles}},
  \href{https://doi.org/10.1007/JHEP01(2020)046}{\emph{JHEP} {\bfseries 01}
  (2020) 046} [\href{https://arxiv.org/abs/1906.10100}{{\ttfamily
  1906.10100}}].

\bibitem{Lee:2018gxc}
K.~Lee, \emph{{Kerr-Schild Double Field Theory and Classical Double Copy}},
  \href{https://doi.org/10.1007/JHEP10(2018)027}{\emph{JHEP} {\bfseries 10}
  (2018) 027} [\href{https://arxiv.org/abs/1807.08443}{{\ttfamily
  1807.08443}}].

\bibitem{Monteiro:2021ztt}
R.~Monteiro, S.~Nagy, D.~O'Connell, D.~Peinador~Veiga and M.~Sergola,
  \emph{{NS-NS Spacetimes from Amplitudes}},
  \href{https://arxiv.org/abs/2112.08336}{{\ttfamily 2112.08336}}.

\bibitem{BahjatAbbas2017htu}
N.~Bahjat-Abbas, A.~Luna and C.~D. White, \emph{{The Kerr-Schild double copy in
  curved spacetime}},
  \href{https://doi.org/10.1007/JHEP12(2017)004}{\emph{JHEP} {\bfseries 12}
  (2017) 004} [\href{https://arxiv.org/abs/1710.01953}{{\ttfamily
  1710.01953}}].

\bibitem{Carrillo-Gonzalez:2017iyj}
M.~Carrillo-Gonz\'alez, R.~Penco and M.~Trodden, \emph{{The classical double
  copy in maximally symmetric spacetimes}},
  \href{https://doi.org/10.1007/JHEP04(2018)028}{\emph{JHEP} {\bfseries 04}
  (2018) 028} [\href{https://arxiv.org/abs/1711.01296}{{\ttfamily
  1711.01296}}].

\bibitem{Luna:2018dpt}
A.~Luna, R.~Monteiro, I.~Nicholson and D.~O'Connell, \emph{{Type D Spacetimes
  and the Weyl Double Copy}},
  \href{https://doi.org/10.1088/1361-6382/ab03e6}{\emph{Class. Quant. Grav.}
  {\bfseries 36} (2019) 065003}
  [\href{https://arxiv.org/abs/1810.08183}{{\ttfamily 1810.08183}}].

\bibitem{Godazgar:2020zbv}
H.~Godazgar, M.~Godazgar, R.~Monteiro, D.~Peinador~Veiga and C.~N. Pope,
  \emph{{Weyl Double Copy for Gravitational Waves}},
  \href{https://doi.org/10.1103/PhysRevLett.126.101103}{\emph{Phys. Rev. Lett.}
  {\bfseries 126} (2021) 101103}
  [\href{https://arxiv.org/abs/2010.02925}{{\ttfamily 2010.02925}}].

\bibitem{Godazgar:2021iae}
H.~Godazgar, M.~Godazgar, R.~Monteiro, D.~Peinador~Veiga and C.~N. Pope,
  \emph{{Asymptotic Weyl double copy}},
  \href{https://doi.org/10.1007/JHEP11(2021)126}{\emph{JHEP} {\bfseries 11}
  (2021) 126} [\href{https://arxiv.org/abs/2109.07866}{{\ttfamily
  2109.07866}}].

\bibitem{Gonzalez:2022otg}
M.~C. Gonz\'alez, A.~Momeni and J.~Rumbutis, \emph{{Cotton Double Copy for
  Gravitational Waves}},  \href{https://arxiv.org/abs/2202.10476}{{\ttfamily
  2202.10476}}.

\bibitem{Emond:2022uaf}
W.~T. Emond and N.~Moynihan, \emph{{Scattering Amplitudes and The Cotton Double
  Copy}},  \href{https://arxiv.org/abs/2202.10499}{{\ttfamily 2202.10499}}.

\bibitem{Anastasiou:2014qba}
A.~Anastasiou, L.~Borsten, M.~J. Duff, L.~J. Hughes and S.~Nagy,
  \emph{{Yang-Mills origin of gravitational symmetries}},
  \href{https://doi.org/10.1103/PhysRevLett.113.231606}{\emph{Phys. Rev. Lett.}
  {\bfseries 113} (2014) 231606}
  [\href{https://arxiv.org/abs/1408.4434}{{\ttfamily 1408.4434}}].

\bibitem{Anastasiou:2016csv}
A.~Anastasiou, L.~Borsten, M.~J. Duff, M.~J. Hughes, A.~Marrani, S.~Nagy
  et~al., \emph{{Twin supergravities from Yang-Mills theory squared}},
  \href{https://doi.org/10.1103/PhysRevD.96.026013}{\emph{Phys. Rev. D}
  {\bfseries 96} (2017) 026013}
  [\href{https://arxiv.org/abs/1610.07192}{{\ttfamily 1610.07192}}].

\bibitem{Cardoso:2016ngt}
G.~L. Cardoso, S.~Nagy and S.~Nampuri, \emph{{A double copy for $ \mathcal{N}=2
  $ supergravity: a linearised tale told on-shell}},
  \href{https://doi.org/10.1007/JHEP10(2016)127}{\emph{JHEP} {\bfseries 10}
  (2016) 127} [\href{https://arxiv.org/abs/1609.05022}{{\ttfamily
  1609.05022}}].

\bibitem{Cardoso:2016amd}
G.~Cardoso, S.~Nagy and S.~Nampuri, \emph{{Multi-centered $ \mathcal{N}=2 $ BPS
  black holes: a double copy description}},
  \href{https://doi.org/10.1007/JHEP04(2017)037}{\emph{JHEP} {\bfseries 04}
  (2017) 037} [\href{https://arxiv.org/abs/1611.04409}{{\ttfamily
  1611.04409}}].

\bibitem{Anastasiou:2017nsz}
A.~Anastasiou, L.~Borsten, M.~J. Duff, A.~Marrani, S.~Nagy and M.~Zoccali,
  \emph{{Are all supergravity theories Yang\textendash{}Mills squared?}},
  \href{https://doi.org/10.1016/j.nuclphysb.2018.07.023}{\emph{Nucl. Phys. B}
  {\bfseries 934} (2018) 606}
  [\href{https://arxiv.org/abs/1707.03234}{{\ttfamily 1707.03234}}].

\bibitem{Anastasiou:2017taf}
A.~Anastasiou, L.~Borsten, M.~J. Duff, A.~Marrani, S.~Nagy and M.~Zoccali,
  \emph{{The Mile High Magic Pyramid}}, {\emph{Contemp. Math.} {\bfseries 721}
  (2019) 1} [\href{https://arxiv.org/abs/1711.08476}{{\ttfamily 1711.08476}}].

\bibitem{Borsten:2013bp}
L.~Borsten, M.~J. Duff, L.~J. Hughes and S.~Nagy, \emph{{Magic Square from
  Yang-Mills Squared}},
  \href{https://doi.org/10.1103/PhysRevLett.112.131601}{\emph{Phys. Rev. Lett.}
  {\bfseries 112} (2014) 131601}
  [\href{https://arxiv.org/abs/1301.4176}{{\ttfamily 1301.4176}}].

\bibitem{Anastasiou:2013hba}
A.~Anastasiou, L.~Borsten, M.~J. Duff, L.~J. Hughes and S.~Nagy, \emph{{A magic
  pyramid of supergravities}},
  \href{https://doi.org/10.1007/JHEP04(2014)178}{\emph{JHEP} {\bfseries 04}
  (2014) 178} [\href{https://arxiv.org/abs/1312.6523}{{\ttfamily 1312.6523}}].

\bibitem{Anastasiou:2015vba}
A.~Anastasiou, L.~Borsten, M.~J. Hughes and S.~Nagy, \emph{{Global symmetries
  of Yang-Mills squared in various dimensions}},
  \href{https://doi.org/10.1007/JHEP01(2016)148}{\emph{JHEP} {\bfseries 01}
  (2016) 148} [\href{https://arxiv.org/abs/1502.05359}{{\ttfamily
  1502.05359}}].

\bibitem{Luna:2020adi}
A.~Luna, S.~Nagy and C.~White, \emph{{The convolutional double copy: a case
  study with a point}},
  \href{https://doi.org/10.1007/JHEP09(2020)062}{\emph{JHEP} {\bfseries 09}
  (2020) 062} [\href{https://arxiv.org/abs/2004.11254}{{\ttfamily
  2004.11254}}].

\bibitem{White:2016jzc}
C.~D. White, \emph{{Exact solutions for the biadjoint scalar field}},
  \href{https://doi.org/10.1016/j.physletb.2016.10.052}{\emph{Phys. Lett. B}
  {\bfseries 763} (2016) 365}
  [\href{https://arxiv.org/abs/1606.04724}{{\ttfamily 1606.04724}}].

\bibitem{Goldberger:2016iau}
W.~D. Goldberger and A.~K. Ridgway, \emph{{Radiation and the classical double
  copy for color charges}},
  \href{https://doi.org/10.1103/PhysRevD.95.125010}{\emph{Phys. Rev.}
  {\bfseries D95} (2017) 125010}
  [\href{https://arxiv.org/abs/1611.03493}{{\ttfamily 1611.03493}}].

\bibitem{Luna:2016hge}
A.~Luna, R.~Monteiro, I.~Nicholson, A.~Ochirov, D.~O'Connell, N.~Westerberg
  et~al., \emph{{Perturbative spacetimes from Yang-Mills theory}},
  \href{https://doi.org/10.1007/JHEP04(2017)069}{\emph{JHEP} {\bfseries 04}
  (2017) 069} [\href{https://arxiv.org/abs/1611.07508}{{\ttfamily
  1611.07508}}].

\bibitem{DeSmet2017rve}
P.-J. De~Smet and C.~D. White, \emph{{Extended solutions for the biadjoint
  scalar field}},
  \href{https://doi.org/10.1016/j.physletb.2017.11.007}{\emph{Phys. Lett. B}
  {\bfseries 775} (2017) 163}
  [\href{https://arxiv.org/abs/1708.01103}{{\ttfamily 1708.01103}}].

\bibitem{Bahjat-Abbas:2018vgo}
N.~Bahjat-Abbas, R.~Stark-Much\~ao and C.~D. White, \emph{{Biadjoint wires}},
  \href{https://doi.org/10.1016/j.physletb.2018.11.026}{\emph{Phys. Lett. B}
  {\bfseries 788} (2019) 274}
  [\href{https://arxiv.org/abs/1810.08118}{{\ttfamily 1810.08118}}].

\bibitem{Gurses:2018ckx}
M.~Gurses and B.~Tekin, \emph{{Classical Double Copy: Kerr-Schild-Kundt metrics
  from Yang-Mills Theory}},
  \href{https://doi.org/10.1103/PhysRevD.98.126017}{\emph{Phys. Rev. D}
  {\bfseries 98} (2018) 126017}
  [\href{https://arxiv.org/abs/1810.03411}{{\ttfamily 1810.03411}}].

\bibitem{Berman:2018hwd}
D.~S. Berman, E.~Chac\'on, A.~Luna and C.~D. White, \emph{{The self-dual
  classical double copy, and the Eguchi-Hanson instanton}},
  \href{https://doi.org/10.1007/JHEP01(2019)107}{\emph{JHEP} {\bfseries 01}
  (2019) 107} [\href{https://arxiv.org/abs/1809.04063}{{\ttfamily
  1809.04063}}].

\bibitem{Bah:2019sda}
I.~Bah, R.~Dempsey and P.~Weck, \emph{{Kerr-Schild Double Copy and Complex
  Worldlines}}, \href{https://doi.org/10.1007/JHEP02(2020)180}{\emph{JHEP}
  {\bfseries 02} (2020) 180}
  [\href{https://arxiv.org/abs/1910.04197}{{\ttfamily 1910.04197}}].

\bibitem{Elor:2020nqe}
G.~Elor, K.~Farnsworth, M.~L. Graesser and G.~Herczeg, \emph{{The
  Newman-Penrose Map and the Classical Double Copy}},
  \href{https://doi.org/10.1007/JHEP12(2020)121}{\emph{JHEP} {\bfseries 12}
  (2020) 121} [\href{https://arxiv.org/abs/2006.08630}{{\ttfamily
  2006.08630}}].

\bibitem{Borsten:2020xbt}
L.~Borsten and S.~Nagy, \emph{{The pure BRST Einstein-Hilbert Lagrangian from
  the double-copy to cubic order}},
  \href{https://doi.org/10.1007/JHEP07(2020)093}{\emph{JHEP} {\bfseries 07}
  (2020) 093} [\href{https://arxiv.org/abs/2004.14945}{{\ttfamily
  2004.14945}}].

\bibitem{Easson:2020esh}
D.~A. Easson, C.~Keeler and T.~Manton, \emph{{Classical double copy of
  nonsingular black holes}},
  \href{https://doi.org/10.1103/PhysRevD.102.086015}{\emph{Phys. Rev. D}
  {\bfseries 102} (2020) 086015}
  [\href{https://arxiv.org/abs/2007.16186}{{\ttfamily 2007.16186}}].

\bibitem{Berman:2020xvs}
D.~S. Berman, K.~Kim and K.~Lee, \emph{{The classical double copy for M-theory
  from a Kerr-Schild ansatz for exceptional field theory}},
  \href{https://doi.org/10.1007/JHEP04(2021)071}{\emph{JHEP} {\bfseries 04}
  (2021) 071} [\href{https://arxiv.org/abs/2010.08255}{{\ttfamily
  2010.08255}}].

\bibitem{Guevara:2020xjx}
A.~Guevara, B.~Maybee, A.~Ochirov, D.~O'connell and J.~Vines, \emph{{A
  worldsheet for Kerr}},
  \href{https://doi.org/10.1007/JHEP03(2021)201}{\emph{JHEP} {\bfseries 03}
  (2021) 201} [\href{https://arxiv.org/abs/2012.11570}{{\ttfamily
  2012.11570}}].

\bibitem{Borsten:2021zir}
L.~Borsten, I.~Jubb, V.~Makwana and S.~Nagy, \emph{{Gauge \texttimes{} gauge =
  gravity on homogeneous spaces using tensor convolutions}},
  \href{https://doi.org/10.1007/JHEP06(2021)117}{\emph{JHEP} {\bfseries 06}
  (2021) 117} [\href{https://arxiv.org/abs/2104.01135}{{\ttfamily
  2104.01135}}].

\bibitem{Gonzo:2021drq}
R.~Gonzo and C.~Shi, \emph{{Geodesics from classical double copy}},
  \href{https://doi.org/10.1103/PhysRevD.104.105012}{\emph{Phys. Rev. D}
  {\bfseries 104} (2021) 105012}
  [\href{https://arxiv.org/abs/2109.01072}{{\ttfamily 2109.01072}}].

\bibitem{Campiglia:2021srh}
M.~Campiglia and S.~Nagy, \emph{{A double copy for asymptotic symmetries in the
  self-dual sector}},
  \href{https://doi.org/10.1007/JHEP03(2021)262}{\emph{JHEP} {\bfseries 03}
  (2021) 262} [\href{https://arxiv.org/abs/2102.01680}{{\ttfamily
  2102.01680}}].

\bibitem{Adamo:2021dfg}
T.~Adamo and U.~Kol, \emph{{Classical double copy at null infinity}},
  \href{https://arxiv.org/abs/2109.07832}{{\ttfamily 2109.07832}}.

\bibitem{Easson:2021asd}
D.~A. Easson, T.~Manton and A.~Svesko, \emph{{Sources in the Weyl Double
  Copy}}, \href{https://doi.org/10.1103/PhysRevLett.127.271101}{\emph{Phys.
  Rev. Lett.} {\bfseries 127} (2021) 271101}
  [\href{https://arxiv.org/abs/2110.02293}{{\ttfamily 2110.02293}}].

\bibitem{Cheung:2020djz}
C.~Cheung and J.~Mangan, \emph{{Scattering Amplitudes and the Navier-Stokes
  Equation}},  \href{https://arxiv.org/abs/2010.15970}{{\ttfamily 2010.15970}}.

\bibitem{Adamo2017nia}
T.~Adamo, E.~Casali, L.~Mason and S.~Nekovar, \emph{{Scattering on plane waves
  and the double copy}},
  \href{https://doi.org/10.1088/1361-6382/aa9961}{\emph{Class. Quant. Grav.}
  {\bfseries 35} (2018) 015004}
  [\href{https://arxiv.org/abs/1706.08925}{{\ttfamily 1706.08925}}].

\bibitem{Adamo:2018mpq}
T.~Adamo, E.~Casali, L.~Mason and S.~Nekovar, \emph{{Plane wave backgrounds and
  colour-kinematics duality}},
  \href{https://doi.org/10.1007/JHEP02(2019)198}{\emph{JHEP} {\bfseries 02}
  (2019) 198} [\href{https://arxiv.org/abs/1810.05115}{{\ttfamily
  1810.05115}}].

\bibitem{White:2020sfn}
C.~D. White, \emph{{Twistorial Foundation for the Classical Double Copy}},
  \href{https://doi.org/10.1103/PhysRevLett.126.061602}{\emph{Phys. Rev. Lett.}
  {\bfseries 126} (2021) 061602}
  [\href{https://arxiv.org/abs/2012.02479}{{\ttfamily 2012.02479}}].

\bibitem{Chacon:2021wbr}
E.~Chac\'on, S.~Nagy and C.~D. White, \emph{{The Weyl double copy from twistor
  space}}, \href{https://doi.org/10.1007/JHEP05(2021)239}{\emph{JHEP}
  {\bfseries 05} (2021) 2239}
  [\href{https://arxiv.org/abs/2103.16441}{{\ttfamily 2103.16441}}].

\bibitem{Farnsworth:2021wvs}
K.~Farnsworth, M.~L. Graesser and G.~Herczeg, \emph{{Twistor Space Origins of
  the Newman-Penrose Map}},  \href{https://arxiv.org/abs/2104.09525}{{\ttfamily
  2104.09525}}.

\bibitem{Guevara:2021yud}
A.~Guevara, \emph{{Reconstructing Classical Spacetimes from the S-Matrix in
  Twistor Space}},  \href{https://arxiv.org/abs/2112.05111}{{\ttfamily
  2112.05111}}.

\bibitem{Chacon:2021hfe}
E.~Chac\'on, A.~Luna and C.~D. White, \emph{{The double copy of the multipole
  expansion}},  \href{https://arxiv.org/abs/2108.07702}{{\ttfamily
  2108.07702}}.

\bibitem{Chacon:2021lox}
E.~Chac\'on, S.~Nagy and C.~D. White, \emph{{Alternative formulations of the
  twistor double copy}},
  \href{https://doi.org/10.1007/JHEP03(2022)180}{\emph{JHEP} {\bfseries 03}
  (2022) 180} [\href{https://arxiv.org/abs/2112.06764}{{\ttfamily
  2112.06764}}].

\bibitem{Cho:2019ype}
W.~Cho and K.~Lee, \emph{{Heterotic Kerr-Schild Double Field Theory and
  Classical Double Copy}},
  \href{https://doi.org/10.1007/JHEP07(2019)030}{\emph{JHEP} {\bfseries 07}
  (2019) 030} [\href{https://arxiv.org/abs/1904.11650}{{\ttfamily
  1904.11650}}].

\bibitem{Alawadhi:2019urr}
R.~Alawadhi, D.~S. Berman, B.~Spence and D.~Peinador~Veiga, \emph{{S-duality
  and the double copy}},
  \href{https://doi.org/10.1007/JHEP03(2020)059}{\emph{JHEP} {\bfseries 03}
  (2020) 059} [\href{https://arxiv.org/abs/1911.06797}{{\ttfamily
  1911.06797}}].

\bibitem{Banerjee:2019saj}
A.~Banerjee, E.~O. Colg\'ain, J.~A. Rosabal and H.~Yavartanoo, \emph{{Ehlers as
  EM duality in the double copy}},
  \href{https://doi.org/10.1103/PhysRevD.102.126017}{\emph{Phys. Rev. D}
  {\bfseries 102} (2020) 126017}
  [\href{https://arxiv.org/abs/1912.02597}{{\ttfamily 1912.02597}}].

\bibitem{Kim:2019jwm}
K.~Kim, K.~Lee, R.~Monteiro, I.~Nicholson and D.~Peinador~Veiga, \emph{{The
  Classical Double Copy of a Point Charge}},
  \href{https://doi.org/10.1007/JHEP02(2020)046}{\emph{JHEP} {\bfseries 02}
  (2020) 046} [\href{https://arxiv.org/abs/1912.02177}{{\ttfamily
  1912.02177}}].

\bibitem{Keeler:2020rcv}
C.~Keeler, T.~Manton and N.~Monga, \emph{{From Navier-Stokes to Maxwell via
  Einstein}}, \href{https://doi.org/10.1007/JHEP08(2020)147}{\emph{JHEP}
  {\bfseries 08} (2020) 147}
  [\href{https://arxiv.org/abs/2005.04242}{{\ttfamily 2005.04242}}].

\bibitem{CarrilloGonzalez:2019gof}
M.~Carrillo~Gonz\'alez, B.~Melcher, K.~Ratliff, S.~Watson and C.~D. White,
  \emph{{The classical double copy in three spacetime dimensions}},
  \href{https://doi.org/10.1007/JHEP07(2019)167}{\emph{JHEP} {\bfseries 07}
  (2019) 167} [\href{https://arxiv.org/abs/1904.11001}{{\ttfamily
  1904.11001}}].

\bibitem{Alkac:2021seh}
G.~Alkac, M.~K. Gumus and M.~A. Olpak, \emph{{Kerr-Schild double copy of the
  Coulomb solution in three dimensions}},
  \href{https://doi.org/10.1103/PhysRevD.104.044034}{\emph{Phys. Rev. D}
  {\bfseries 104} (2021) 044034}
  [\href{https://arxiv.org/abs/2105.11550}{{\ttfamily 2105.11550}}].

\bibitem{Moynihan:2021rwh}
N.~Moynihan, \emph{{Massive Covariant Colour-Kinematics in 3D}},
  \href{https://arxiv.org/abs/2110.02209}{{\ttfamily 2110.02209}}.

\bibitem{Gonzalez:2021ztm}
M.~C. Gonz\'alez, A.~Momeni and J.~Rumbutis, \emph{{Massive Double Copy in the
  High-Energy Limit}},  \href{https://arxiv.org/abs/2112.08401}{{\ttfamily
  2112.08401}}.

\bibitem{Favata:2013rwa}
M.~Favata, \emph{{Systematic parameter errors in inspiraling neutron star
  binaries}}, \href{https://doi.org/10.1103/PhysRevLett.112.101101}{\emph{Phys.
  Rev. Lett.} {\bfseries 112} (2014) 101101}
  [\href{https://arxiv.org/abs/1310.8288}{{\ttfamily 1310.8288}}].

\bibitem{Purrer:2019jcp}
M.~P\"urrer and C.-J. Haster, \emph{{Gravitational waveform accuracy
  requirements for future ground-based detectors}},
  \href{https://doi.org/10.1103/PhysRevResearch.2.023151}{\emph{Phys. Rev.
  Res.} {\bfseries 2} (2020) 023151}
  [\href{https://arxiv.org/abs/1912.10055}{{\ttfamily 1912.10055}}].

\bibitem{Porto:2016pyg}
R.~A. Porto, \emph{{The effective field theorist\textquoteright{}s approach to
  gravitational dynamics}},
  \href{https://doi.org/10.1016/j.physrep.2016.04.003}{\emph{Phys. Rept.}
  {\bfseries 633} (2016) 1} [\href{https://arxiv.org/abs/1601.04914}{{\ttfamily
  1601.04914}}].

\bibitem{Levi:2018nxp}
M.~Levi, \emph{{Effective Field Theories of Post-Newtonian Gravity: A
  comprehensive review}},
  \href{https://doi.org/10.1088/1361-6633/ab12bc}{\emph{Rept. Prog. Phys.}
  {\bfseries 83} (2020) 075901}
  [\href{https://arxiv.org/abs/1807.01699}{{\ttfamily 1807.01699}}].

\bibitem{Luna:2017dtq}
A.~Luna, I.~Nicholson, D.~O'Connell and C.~D. White, \emph{{Inelastic Black
  Hole Scattering from Charged Scalar Amplitudes}},
  \href{https://doi.org/10.1007/JHEP03(2018)044}{\emph{JHEP} {\bfseries 03}
  (2018) 044} [\href{https://arxiv.org/abs/1711.03901}{{\ttfamily
  1711.03901}}].

\bibitem{ShenWorldLine}
C.-H. Shen, \emph{{Gravitational Radiation from Color-Kinematics Duality}},
  \href{https://doi.org/10.1007/JHEP11(2018)162}{\emph{JHEP} {\bfseries 11}
  (2018) 162} [\href{https://arxiv.org/abs/1806.07388}{{\ttfamily
  1806.07388}}].

\bibitem{Goldberger:2017vcg}
W.~D. Goldberger and A.~K. Ridgway, \emph{{Bound states and the classical
  double copy}}, \href{https://doi.org/10.1103/PhysRevD.97.085019}{\emph{Phys.
  Rev.} {\bfseries D97} (2018) 085019}
  [\href{https://arxiv.org/abs/1711.09493}{{\ttfamily 1711.09493}}].

\bibitem{Maybee:2019jus}
B.~Maybee, D.~O'Connell and J.~Vines, \emph{{Observables and amplitudes for
  spinning particles and black holes}},
  \href{https://doi.org/10.1007/JHEP12(2019)156}{\emph{JHEP} {\bfseries 12}
  (2019) 156} [\href{https://arxiv.org/abs/1906.09260}{{\ttfamily
  1906.09260}}].

\bibitem{Mogull:2020sak}
G.~Mogull, J.~Plefka and J.~Steinhoff, \emph{{Classical black hole scattering
  from a worldline quantum field theory}},
  \href{https://doi.org/10.1007/JHEP02(2021)048}{\emph{JHEP} {\bfseries 02}
  (2021) 048} [\href{https://arxiv.org/abs/2010.02865}{{\ttfamily
  2010.02865}}].

\bibitem{Jakobsen:2021smu}
G.~U. Jakobsen, G.~Mogull, J.~Plefka and J.~Steinhoff, \emph{{Classical
  Gravitational Bremsstrahlung from a Worldline Quantum Field Theory}},
  \href{https://doi.org/10.1103/PhysRevLett.126.201103}{\emph{Phys. Rev. Lett.}
  {\bfseries 126} (2021) 201103}
  [\href{https://arxiv.org/abs/2101.12688}{{\ttfamily 2101.12688}}].

\bibitem{Jakobsen:2021lvp}
G.~U. Jakobsen, G.~Mogull, J.~Plefka and J.~Steinhoff, \emph{{Gravitational
  Bremsstrahlung and Hidden Supersymmetry of Spinning Bodies}},
  \href{https://doi.org/10.1103/PhysRevLett.128.011101}{\emph{Phys. Rev. Lett.}
  {\bfseries 128} (2022) 011101}
  [\href{https://arxiv.org/abs/2106.10256}{{\ttfamily 2106.10256}}].

\bibitem{Mougiakakos:2021ckm}
S.~Mougiakakos, M.~M. Riva and F.~Vernizzi, \emph{{Gravitational Bremsstrahlung
  in the post-Minkowskian effective field theory}},
  \href{https://doi.org/10.1103/PhysRevD.104.024041}{\emph{Phys. Rev. D}
  {\bfseries 104} (2021) 024041}
  [\href{https://arxiv.org/abs/2102.08339}{{\ttfamily 2102.08339}}].

\bibitem{Jakobsen:2021zvh}
G.~U. Jakobsen, G.~Mogull, J.~Plefka and J.~Steinhoff, \emph{{SUSY in the sky
  with gravitons}}, \href{https://doi.org/10.1007/JHEP01(2022)027}{\emph{JHEP}
  {\bfseries 01} (2022) 027}
  [\href{https://arxiv.org/abs/2109.04465}{{\ttfamily 2109.04465}}].

\bibitem{Jakobsen:2022fcj}
G.~U. Jakobsen and G.~Mogull, \emph{{Conservative and radiative dynamics of
  spinning bodies at third post-Minkowskian order using worldline quantum field
  theory}},  \href{https://arxiv.org/abs/2201.07778}{{\ttfamily 2201.07778}}.

\bibitem{Shi:2021qsb}
C.~Shi and J.~Plefka, \emph{{Classical double copy of worldline quantum field
  theory}}, \href{https://doi.org/10.1103/PhysRevD.105.026007}{\emph{Phys. Rev.
  D} {\bfseries 105} (2022) 026007}
  [\href{https://arxiv.org/abs/2109.10345}{{\ttfamily 2109.10345}}].

\bibitem{Bern:2019nnu}
Z.~Bern, C.~Cheung, R.~Roiban, C.-H. Shen, M.~P. Solon and M.~Zeng,
  \emph{{Scattering Amplitudes and the Conservative Hamiltonian for Binary
  Systems at Third Post-Minkowskian Order}},
  \href{https://doi.org/10.1103/PhysRevLett.122.201603}{\emph{Phys. Rev. Lett.}
  {\bfseries 122} (2019) 201603}
  [\href{https://arxiv.org/abs/1901.04424}{{\ttfamily 1901.04424}}].

\bibitem{Bern:2019crd}
Z.~Bern, C.~Cheung, R.~Roiban, C.-H. Shen, M.~P. Solon and M.~Zeng,
  \emph{{Black Hole Binary Dynamics from the Double Copy and Effective
  Theory}}, \href{https://doi.org/10.1007/JHEP10(2019)206}{\emph{JHEP}
  {\bfseries 10} (2019) 206}
  [\href{https://arxiv.org/abs/1908.01493}{{\ttfamily 1908.01493}}].

\bibitem{Damgaard:2019lfh}
P.~H. Damgaard, K.~Haddad and A.~Helset, \emph{{Heavy Black Hole Effective
  Theory}}, \href{https://doi.org/10.1007/JHEP11(2019)070}{\emph{JHEP}
  {\bfseries 11} (2019) 070}
  [\href{https://arxiv.org/abs/1908.10308}{{\ttfamily 1908.10308}}].

\bibitem{Bautista:2019evw}
Y.~F. Bautista and A.~Guevara, \emph{{On the double copy for spinning matter}},
  \href{https://doi.org/10.1007/JHEP11(2021)184}{\emph{JHEP} {\bfseries 11}
  (2021) 184} [\href{https://arxiv.org/abs/1908.11349}{{\ttfamily
  1908.11349}}].

\bibitem{Bautista:2019tdr}
Y.~F. Bautista and A.~Guevara, \emph{{From Scattering Amplitudes to Classical
  Physics: Universality, Double Copy and Soft Theorems}},
  \href{https://arxiv.org/abs/1903.12419}{{\ttfamily 1903.12419}}.

\bibitem{Herrmann:2021lqe}
E.~Herrmann, J.~Parra-Martinez, M.~S. Ruf and M.~Zeng, \emph{{Gravitational
  Bremsstrahlung from Reverse Unitarity}},
  \href{https://doi.org/10.1103/PhysRevLett.126.201602}{\emph{Phys. Rev. Lett.}
  {\bfseries 126} (2021) 201602}
  [\href{https://arxiv.org/abs/2101.07255}{{\ttfamily 2101.07255}}].

\bibitem{Bautista:2021inx}
Y.~F. Bautista and N.~Siemonsen, \emph{{Post-Newtonian waveforms from spinning
  scattering amplitudes}},
  \href{https://doi.org/10.1007/JHEP01(2022)006}{\emph{JHEP} {\bfseries 01}
  (2022) 006} [\href{https://arxiv.org/abs/2110.12537}{{\ttfamily
  2110.12537}}].

\bibitem{Herrmann:2021tct}
E.~Herrmann, J.~Parra-Martinez, M.~S. Ruf and M.~Zeng, \emph{{Radiative
  classical gravitational observables at $ \mathcal{O} $(G$^{3}$) from
  scattering amplitudes}},
  \href{https://doi.org/10.1007/JHEP10(2021)148}{\emph{JHEP} {\bfseries 10}
  (2021) 148} [\href{https://arxiv.org/abs/2104.03957}{{\ttfamily
  2104.03957}}].

\bibitem{Brandhuber:2021eyq}
A.~Brandhuber, G.~Chen, G.~Travaglini and C.~Wen, \emph{{Classical
  gravitational scattering from a gauge-invariant double copy}},
  \href{https://doi.org/10.1007/JHEP10(2021)118}{\emph{JHEP} {\bfseries 10}
  (2021) 118} [\href{https://arxiv.org/abs/2108.04216}{{\ttfamily
  2108.04216}}].

\bibitem{Brandhuber:2021kpo}
A.~Brandhuber, G.~Chen, G.~Travaglini and C.~Wen, \emph{{A new gauge-invariant
  double copy for heavy-mass effective theory}},
  \href{https://doi.org/10.1007/JHEP07(2021)047}{\emph{JHEP} {\bfseries 07}
  (2021) 047} [\href{https://arxiv.org/abs/2104.11206}{{\ttfamily
  2104.11206}}].

\bibitem{Bjerrum-Bohr:2021vuf}
N.~E.~J. Bjerrum-Bohr, P.~H. Damgaard, L.~Plant\'e and P.~Vanhove,
  \emph{{Classical gravity from loop amplitudes}},
  \href{https://doi.org/10.1103/PhysRevD.104.026009}{\emph{Phys. Rev. D}
  {\bfseries 104} (2021) 026009}
  [\href{https://arxiv.org/abs/2104.04510}{{\ttfamily 2104.04510}}].

\bibitem{Bjerrum-Bohr:2021din}
N.~E.~J. Bjerrum-Bohr, P.~H. Damgaard, L.~Plant\'e and P.~Vanhove, \emph{{The
  amplitude for classical gravitational scattering at third Post-Minkowskian
  order}}, \href{https://doi.org/10.1007/JHEP08(2021)172}{\emph{JHEP}
  {\bfseries 08} (2021) 172}
  [\href{https://arxiv.org/abs/2105.05218}{{\ttfamily 2105.05218}}].

\bibitem{Bern:2020buy}
Z.~Bern, A.~Luna, R.~Roiban, C.-H. Shen and M.~Zeng, \emph{{Spinning black hole
  binary dynamics, scattering amplitudes, and effective field theory}},
  \href{https://doi.org/10.1103/PhysRevD.104.065014}{\emph{Phys. Rev. D}
  {\bfseries 104} (2021) 065014}
  [\href{https://arxiv.org/abs/2005.03071}{{\ttfamily 2005.03071}}].

\bibitem{Vines:2017hyw}
J.~Vines, \emph{{Scattering of two spinning black holes in post-Minkowskian
  gravity, to all orders in spin, and effective-one-body mappings}},
  \href{https://doi.org/10.1088/1361-6382/aaa3a8}{\emph{Class. Quant. Grav.}
  {\bfseries 35} (2018) 084002}
  [\href{https://arxiv.org/abs/1709.06016}{{\ttfamily 1709.06016}}].

\bibitem{Carrasco:2020ywq}
J.~J.~M. Carrasco and I.~A. Vazquez-Holm, \emph{{Loop-Level Double-Copy for
  Massive Quantum Particles}},
  \href{https://doi.org/10.1103/PhysRevD.103.045002}{\emph{Phys. Rev. D}
  {\bfseries 103} (2021) 045002}
  [\href{https://arxiv.org/abs/2010.13435}{{\ttfamily 2010.13435}}].

\bibitem{Carrasco:2021bmu}
J.~J.~M. Carrasco and I.~A. Vazquez-Holm, \emph{{Extracting Einstein from the
  loop-level double-copy}},
  \href{https://doi.org/10.1007/JHEP11(2021)088}{\emph{JHEP} {\bfseries 11}
  (2021) 088} [\href{https://arxiv.org/abs/2108.06798}{{\ttfamily
  2108.06798}}].

\bibitem{Ridgway2015fdl}
A.~K. Ridgway and M.~B. Wise, \emph{{Static Spherically Symmetric Kerr-Schild
  Metrics and Implications for the Classical Double Copy}},
  \href{https://doi.org/10.1103/PhysRevD.94.044023}{\emph{Phys. Rev. D}
  {\bfseries 94} (2016) 044023}
  [\href{https://arxiv.org/abs/1512.02243}{{\ttfamily 1512.02243}}].

\bibitem{Goldberger:2017frp}
W.~D. Goldberger, S.~G. Prabhu and J.~O. Thompson, \emph{{Classical gluon and
  graviton radiation from the bi-adjoint scalar double copy}},
  \href{https://doi.org/10.1103/PhysRevD.96.065009}{\emph{Phys. Rev.}
  {\bfseries D96} (2017) 065009}
  [\href{https://arxiv.org/abs/1705.09263}{{\ttfamily 1705.09263}}].

\bibitem{Goldberger:2017ogt}
W.~D. Goldberger, J.~Li and S.~G. Prabhu, \emph{{Spinning particles, axion
  radiation, and the classical double copy}},
  \href{https://doi.org/10.1103/PhysRevD.97.105018}{\emph{Phys. Rev.}
  {\bfseries D97} (2018) 105018}
  [\href{https://arxiv.org/abs/1712.09250}{{\ttfamily 1712.09250}}].

\bibitem{Li2018qap}
J.~Li and S.~G. Prabhu, \emph{{Gravitational radiation from the classical
  spinning double copy}},
  \href{https://doi.org/10.1103/PhysRevD.97.105019}{\emph{Phys. Rev. D}
  {\bfseries 97} (2018) 105019}
  [\href{https://arxiv.org/abs/1803.02405}{{\ttfamily 1803.02405}}].

\bibitem{Ilderton:2018lsf}
A.~Ilderton, \emph{{Screw-symmetric gravitational waves: a double copy of the
  vortex}}, \href{https://doi.org/10.1016/j.physletb.2018.04.069}{\emph{Phys.
  Lett. B} {\bfseries 782} (2018) 22}
  [\href{https://arxiv.org/abs/1804.07290}{{\ttfamily 1804.07290}}].

\bibitem{Plefka:2018dpa}
J.~Plefka, J.~Steinhoff and W.~Wormsbecher, \emph{{Effective action of dilaton
  gravity as the classical double copy of Yang-Mills theory}},
  \href{https://doi.org/10.1103/PhysRevD.99.024021}{\emph{Phys. Rev. D}
  {\bfseries 99} (2019) 024021}
  [\href{https://arxiv.org/abs/1807.09859}{{\ttfamily 1807.09859}}].

\bibitem{PV:2019uuv}
A.~P.~V. and A.~Manu, \emph{{Classical double copy from Color Kinematics
  duality: A proof in the soft limit}},
  \href{https://doi.org/10.1103/PhysRevD.101.046014}{\emph{Phys. Rev. D}
  {\bfseries 101} (2020) 046014}
  [\href{https://arxiv.org/abs/1907.10021}{{\ttfamily 1907.10021}}].

\bibitem{Damour:2017zjx}
T.~Damour, \emph{{High-energy gravitational scattering and the general
  relativistic two-body problem}},
  \href{https://doi.org/10.1103/PhysRevD.97.044038}{\emph{Phys. Rev. D}
  {\bfseries 97} (2018) 044038}
  [\href{https://arxiv.org/abs/1710.10599}{{\ttfamily 1710.10599}}].

\bibitem{Babak:2017tow}
S.~Babak, J.~Gair, A.~Sesana, E.~Barausse, C.~F. Sopuerta, C.~P.~L. Berry
  et~al., \emph{{Science with the space-based interferometer LISA. V: Extreme
  mass-ratio inspirals}},
  \href{https://doi.org/10.1103/PhysRevD.95.103012}{\emph{Phys. Rev. D}
  {\bfseries 95} (2017) 103012}
  [\href{https://arxiv.org/abs/1703.09722}{{\ttfamily 1703.09722}}].

\bibitem{Berry:2019wgg}
C.~P.~L. Berry, S.~A. Hughes, C.~F. Sopuerta, A.~J.~K. Chua, A.~Heffernan,
  K.~Holley-Bockelmann et~al., \emph{{The unique potential of extreme
  mass-ratio inspirals for gravitational-wave astronomy}},
  \href{https://arxiv.org/abs/1903.03686}{{\ttfamily 1903.03686}}.

\bibitem{Gair:2012nm}
J.~R. Gair, M.~Vallisneri, S.~L. Larson and J.~G. Baker, \emph{{Testing General
  Relativity with Low-Frequency, Space-Based Gravitational-Wave Detectors}},
  \href{https://doi.org/10.12942/lrr-2013-7}{\emph{Living Rev. Rel.} {\bfseries
  16} (2013) 7} [\href{https://arxiv.org/abs/1212.5575}{{\ttfamily
  1212.5575}}].

\bibitem{Carrillo-Gonzalez:2021mqj}
M.~Carrillo-Gonz\'alez, C.~de~Rham and A.~J. Tolley, \emph{{Scattering
  amplitudes for binary systems beyond GR}},
  \href{https://doi.org/10.1007/JHEP11(2021)087}{\emph{JHEP} {\bfseries 11}
  (2021) 087} [\href{https://arxiv.org/abs/2107.11384}{{\ttfamily
  2107.11384}}].

\bibitem{Gair:2017ynp}
J.~R. Gair, S.~Babak, A.~Sesana, P.~Amaro-Seoane, E.~Barausse, C.~P.~L. Berry
  et~al., \emph{{Prospects for observing extreme-mass-ratio inspirals with
  LISA}}, \href{https://doi.org/10.1088/1742-6596/840/1/012021}{\emph{J. Phys.
  Conf. Ser.} {\bfseries 840} (2017) 012021}
  [\href{https://arxiv.org/abs/1704.00009}{{\ttfamily 1704.00009}}].

\bibitem{Gair:2010yu}
J.~R. Gair, C.~Tang and M.~Volonteri, \emph{{LISA extreme-mass-ratio inspiral
  events as probes of the black hole mass function}},
  \href{https://doi.org/10.1103/PhysRevD.81.104014}{\emph{Phys. Rev. D}
  {\bfseries 81} (2010) 104014}
  [\href{https://arxiv.org/abs/1004.1921}{{\ttfamily 1004.1921}}].

\bibitem{MacLeod:2007jd}
C.~L. MacLeod and C.~J. Hogan, \emph{{Precision of Hubble constant derived
  using black hole binary absolute distances and statistical redshift
  information}}, \href{https://doi.org/10.1103/PhysRevD.77.043512}{\emph{Phys.
  Rev. D} {\bfseries 77} (2008) 043512}
  [\href{https://arxiv.org/abs/0712.0618}{{\ttfamily 0712.0618}}].

\bibitem{Mino:1996nk}
Y.~Mino, M.~Sasaki and T.~Tanaka, \emph{{Gravitational radiation reaction to a
  particle motion}},
  \href{https://doi.org/10.1103/PhysRevD.55.3457}{\emph{Phys. Rev. D}
  {\bfseries 55} (1997) 3457}
  [\href{https://arxiv.org/abs/gr-qc/9606018}{{\ttfamily gr-qc/9606018}}].

\bibitem{Quinn:1996am}
T.~C. Quinn and R.~M. Wald, \emph{{An Axiomatic approach to electromagnetic and
  gravitational radiation reaction of particles in curved space-time}},
  \href{https://doi.org/10.1103/PhysRevD.56.3381}{\emph{Phys. Rev. D}
  {\bfseries 56} (1997) 3381}
  [\href{https://arxiv.org/abs/gr-qc/9610053}{{\ttfamily gr-qc/9610053}}].

\bibitem{Barack:2018yvs}
L.~Barack and A.~Pound, \emph{{Self-force and radiation reaction in general
  relativity}}, \href{https://doi.org/10.1088/1361-6633/aae552}{\emph{Rept.
  Prog. Phys.} {\bfseries 82} (2019) 016904}
  [\href{https://arxiv.org/abs/1805.10385}{{\ttfamily 1805.10385}}].

\bibitem{Hughes:2019zmt}
S.~A. Hughes, A.~Apte, G.~Khanna and H.~Lim, \emph{{Learning about black hole
  binaries from their ringdown spectra}},
  \href{https://doi.org/10.1103/PhysRevLett.123.161101}{\emph{Phys. Rev. Lett.}
  {\bfseries 123} (2019) 161101}
  [\href{https://arxiv.org/abs/1901.05900}{{\ttfamily 1901.05900}}].

\bibitem{Li:2021wgz}
X.~Li, L.~Sun, R.~K.~L. Lo, E.~Payne and Y.~Chen, \emph{{Angular emission
  patterns of remnant black holes}},
  \href{https://doi.org/10.1103/PhysRevD.105.024016}{\emph{Phys. Rev. D}
  {\bfseries 105} (2022) 024016}
  [\href{https://arxiv.org/abs/2110.03116}{{\ttfamily 2110.03116}}].

\bibitem{DiPiazza:2011tq}
A.~Di~Piazza, C.~Muller, K.~Z. Hatsagortsyan and C.~H. Keitel, \emph{{Extremely
  high-intensity laser interactions with fundamental quantum systems}},
  \href{https://doi.org/10.1103/RevModPhys.84.1177}{\emph{Rev. Mod. Phys.}
  {\bfseries 84} (2012) 1177}
  [\href{https://arxiv.org/abs/1111.3886}{{\ttfamily 1111.3886}}].

\bibitem{Gonoskov:2021hwf}
A.~Gonoskov, T.~G. Blackburn, M.~Marklund and S.~S. Bulanov, \emph{{Charged
  particle motion and radiation in strong electromagnetic fields}},
  \href{https://arxiv.org/abs/2107.02161}{{\ttfamily 2107.02161}}.

\bibitem{Fedotov:2022ely}
A.~Fedotov, A.~Ilderton, F.~Karbstein, B.~King, D.~Seipt, H.~Taya et~al.,
  \emph{{Advances in QED with intense background fields}},
  \href{https://arxiv.org/abs/2203.00019}{{\ttfamily 2203.00019}}.

\bibitem{Kharzeev:2013jha}
D.~Kharzeev, K.~Landsteiner, A.~Schmitt and H.-U. Yee, eds., \emph{{Strongly
  Interacting Matter in Magnetic Fields}}, vol.~871. 2013,
  \href{https://doi.org/10.1007/978-3-642-37305-3}{10.1007/978-3-642-37305-3}.

\bibitem{Balitsky:2001gj}
I.~Balitsky, \emph{{High-energy QCD and Wilson lines}},
  \href{https://arxiv.org/abs/hep-ph/0101042}{{\ttfamily hep-ph/0101042}}.

\bibitem{Iancu:2002xk}
E.~Iancu, A.~Leonidov and L.~McLerran, \emph{{The Color glass condensate: An
  Introduction}},  in \emph{{Cargese Summer School on QCD Perspectives on Hot
  and Dense Matter}}, pp.~73--145, 2, 2002,
  \href{https://arxiv.org/abs/hep-ph/0202270}{{\ttfamily hep-ph/0202270}}.

\bibitem{Gelis:2010nm}
F.~Gelis, E.~Iancu, J.~Jalilian-Marian and R.~Venugopalan, \emph{{The Color
  Glass Condensate}},
  \href{https://doi.org/10.1146/annurev.nucl.010909.083629}{\emph{Ann. Rev.
  Nucl. Part. Sci.} {\bfseries 60} (2010) 463}
  [\href{https://arxiv.org/abs/1002.0333}{{\ttfamily 1002.0333}}].

\bibitem{Lai:2014nma}
D.~Lai, \emph{{Physics in Very Strong Magnetic Fields: Introduction and
  Overview}}, \href{https://doi.org/10.1007/s11214-015-0137-z}{\emph{Space Sci.
  Rev.} {\bfseries 191} (2015) 13}
  [\href{https://arxiv.org/abs/1411.7995}{{\ttfamily 1411.7995}}].

\bibitem{Christodoulou:1991cr}
D.~Christodoulou, \emph{{Nonlinear nature of gravitation and gravitational wave
  experiments}}, \href{https://doi.org/10.1103/PhysRevLett.67.1486}{\emph{Phys.
  Rev. Lett.} {\bfseries 67} (1991) 1486}.

\bibitem{Mukhanov:1990me}
V.~F. Mukhanov, H.~A. Feldman and R.~H. Brandenberger, \emph{{Theory of
  cosmological perturbations. Part 1. Classical perturbations. Part 2. Quantum
  theory of perturbations. Part 3. Extensions}},
  \href{https://doi.org/10.1016/0370-1573(92)90044-Z}{\emph{Phys. Rept.}
  {\bfseries 215} (1992) 203}.

\bibitem{Furry:1951zz}
W.~H. Furry, \emph{{On Bound States and Scattering in Positron Theory}},
  \href{https://doi.org/10.1103/PhysRev.81.915}{\emph{Phys. Rev.} {\bfseries
  81} (1951) 115}.

\bibitem{DeWitt:1967ub}
B.~S. DeWitt, \emph{{Quantum Theory of Gravity. 2. The Manifestly Covariant
  Theory}}, \href{https://doi.org/10.1103/PhysRev.162.1195}{\emph{Phys. Rev.}
  {\bfseries 162} (1967) 1195}.

\bibitem{tHooft:1975uxh}
G.~'t~Hooft, \emph{{The Background Field Method in Gauge Field Theories}},  in
  \emph{{12th Annual Winter School of Theoretical Physics}}, 1975.

\bibitem{Abbott:1981ke}
L.~F. Abbott, \emph{{Introduction to the Background Field Method}}, {\emph{Acta
  Phys. Polon. B} {\bfseries 13} (1982) 33}.

\bibitem{Raju:2012zs}
S.~Raju, \emph{{Four Point Functions of the Stress Tensor and Conserved
  Currents in AdS$_4$/CFT$_3$}},
  \href{https://doi.org/10.1103/PhysRevD.85.126008}{\emph{Phys. Rev. D}
  {\bfseries 85} (2012) 126008}
  [\href{https://arxiv.org/abs/1201.6452}{{\ttfamily 1201.6452}}].

\bibitem{Rastelli:2016nze}
L.~Rastelli and X.~Zhou, \emph{{Mellin amplitudes for $AdS_5\times S^5$}},
  \href{https://doi.org/10.1103/PhysRevLett.118.091602}{\emph{Phys. Rev. Lett.}
  {\bfseries 118} (2017) 091602}
  [\href{https://arxiv.org/abs/1608.06624}{{\ttfamily 1608.06624}}].

\bibitem{Caron-Huot:2018kta}
S.~Caron-Huot and A.-K. Trinh, \emph{{All tree-level correlators in
  AdS$_{5}$\texttimes{}S$_{5}$ supergravity: hidden ten-dimensional conformal
  symmetry}}, \href{https://doi.org/10.1007/JHEP01(2019)196}{\emph{JHEP}
  {\bfseries 01} (2019) 196}
  [\href{https://arxiv.org/abs/1809.09173}{{\ttfamily 1809.09173}}].

\bibitem{Alday:2020dtb}
L.~F. Alday and X.~Zhou, \emph{{All Holographic Four-Point Functions in All
  Maximally Supersymmetric CFTs}},
  \href{https://doi.org/10.1103/PhysRevX.11.011056}{\emph{Phys. Rev. X}
  {\bfseries 11} (2021) 011056}
  [\href{https://arxiv.org/abs/2006.12505}{{\ttfamily 2006.12505}}].

\bibitem{Albayrak:2020fyp}
S.~Albayrak, S.~Kharel and D.~Meltzer, \emph{{On duality of color and
  kinematics in (A)dS momentum space}},
  \href{https://doi.org/10.1007/JHEP03(2021)249}{\emph{JHEP} {\bfseries 03}
  (2021) 249} [\href{https://arxiv.org/abs/2012.10460}{{\ttfamily
  2012.10460}}].

\bibitem{Armstrong:2020woi}
C.~Armstrong, A.~E. Lipstein and J.~Mei, \emph{{Color/kinematics duality in
  AdS$_{4}$}}, \href{https://doi.org/10.1007/JHEP02(2021)194}{\emph{JHEP}
  {\bfseries 02} (2021) 194}
  [\href{https://arxiv.org/abs/2012.02059}{{\ttfamily 2012.02059}}].

\bibitem{Diwakar:2021juk}
P.~Diwakar, A.~Herderschee, R.~Roiban and F.~Teng, \emph{{BCJ amplitude
  relations for Anti-de Sitter boundary correlators in embedding space}},
  \href{https://doi.org/10.1007/JHEP10(2021)141}{\emph{JHEP} {\bfseries 10}
  (2021) 141} [\href{https://arxiv.org/abs/2106.10822}{{\ttfamily
  2106.10822}}].

\bibitem{Goncalves:2019znr}
V.~Gon\c{c}alves, R.~Pereira and X.~Zhou, \emph{{$20'$ Five-Point Function from
  $AdS_5\times S^5$ Supergravity}},
  \href{https://doi.org/10.1007/JHEP10(2019)247}{\emph{JHEP} {\bfseries 10}
  (2019) 247} [\href{https://arxiv.org/abs/1906.05305}{{\ttfamily
  1906.05305}}].

\bibitem{Alday:2022lkk}
L.~F. Alday, V.~Gon\c{c}alves and X.~Zhou, \emph{{Super Gluon Five-Point
  Amplitudes in AdS Space}},
  \href{https://arxiv.org/abs/2201.04422}{{\ttfamily 2201.04422}}.

\bibitem{Arutyunov:1999nw}
G.~Arutyunov and S.~Frolov, \emph{{Three point Green function of the stress
  energy tensor in the AdS / CFT correspondence}},
  \href{https://doi.org/10.1103/PhysRevD.60.026004}{\emph{Phys. Rev. D}
  {\bfseries 60} (1999) 026004}
  [\href{https://arxiv.org/abs/hep-th/9901121}{{\ttfamily hep-th/9901121}}].

\bibitem{Zhiboedov:2012bm}
A.~Zhiboedov, \emph{{A note on three-point functions of conserved currents}},
  \href{https://arxiv.org/abs/1206.6370}{{\ttfamily 1206.6370}}.

\bibitem{Afkhami-Jeddi:2016ntf}
N.~Afkhami-Jeddi, T.~Hartman, S.~Kundu and A.~Tajdini, \emph{{Einstein gravity
  3-point functions from conformal field theory}},
  \href{https://doi.org/10.1007/JHEP12(2017)049}{\emph{JHEP} {\bfseries 12}
  (2017) 049} [\href{https://arxiv.org/abs/1610.09378}{{\ttfamily
  1610.09378}}].

\bibitem{Binder:2018yvd}
D.~J. Binder, S.~M. Chester and S.~S. Pufu, \emph{{Absence of $D^4 R^4$ in
  M-Theory From ABJM}},
  \href{https://doi.org/10.1007/JHEP04(2020)052}{\emph{JHEP} {\bfseries 04}
  (2020) 052} [\href{https://arxiv.org/abs/1808.10554}{{\ttfamily
  1808.10554}}].

\bibitem{Albayrak:2020isk}
S.~Albayrak, C.~Chowdhury and S.~Kharel, \emph{{Study of momentum space scalar
  amplitudes in AdS spacetime}},
  \href{https://doi.org/10.1103/PhysRevD.101.124043}{\emph{Phys. Rev. D}
  {\bfseries 101} (2020) 124043}
  [\href{https://arxiv.org/abs/2001.06777}{{\ttfamily 2001.06777}}].

\bibitem{Albayrak:2019asr}
S.~Albayrak, C.~Chowdhury and S.~Kharel, \emph{{New relation for Witten
  diagrams}}, \href{https://doi.org/10.1007/JHEP10(2019)274}{\emph{JHEP}
  {\bfseries 10} (2019) 274}
  [\href{https://arxiv.org/abs/1904.10043}{{\ttfamily 1904.10043}}].

\bibitem{Albayrak:2019yve}
S.~Albayrak and S.~Kharel, \emph{{Towards the higher point holographic momentum
  space amplitudes. Part II. Gravitons}},
  \href{https://doi.org/10.1007/JHEP12(2019)135}{\emph{JHEP} {\bfseries 12}
  (2019) 135} [\href{https://arxiv.org/abs/1908.01835}{{\ttfamily
  1908.01835}}].

\bibitem{Albayrak:2020bso}
S.~Albayrak and S.~Kharel, \emph{{Spinning loop amplitudes in
  anti\textendash{}de Sitter space}},
  \href{https://doi.org/10.1103/PhysRevD.103.026004}{\emph{Phys. Rev. D}
  {\bfseries 103} (2021) 026004}
  [\href{https://arxiv.org/abs/2006.12540}{{\ttfamily 2006.12540}}].

\bibitem{Witten:2003nn}
E.~Witten, \emph{{Perturbative gauge theory as a string theory in twistor
  space}}, \href{https://doi.org/10.1007/s00220-004-1187-3}{\emph{Commun. Math.
  Phys.} {\bfseries 252} (2004) 189}
  [\href{https://arxiv.org/abs/hep-th/0312171}{{\ttfamily hep-th/0312171}}].

\bibitem{Roiban:2004yf}
R.~Roiban, M.~Spradlin and A.~Volovich, \emph{{On the tree level S matrix of
  Yang-Mills theory}},
  \href{https://doi.org/10.1103/PhysRevD.70.026009}{\emph{Phys. Rev. D}
  {\bfseries 70} (2004) 026009}
  [\href{https://arxiv.org/abs/hep-th/0403190}{{\ttfamily hep-th/0403190}}].

\bibitem{Cachazo:2012kg}
F.~Cachazo and D.~Skinner, \emph{{Gravity from Rational Curves in Twistor
  Space}}, \href{https://doi.org/10.1103/PhysRevLett.110.161301}{\emph{Phys.
  Rev. Lett.} {\bfseries 110} (2013) 161301}
  [\href{https://arxiv.org/abs/1207.0741}{{\ttfamily 1207.0741}}].

\bibitem{Green:2020eyj}
M.~B. Green and C.~Wen, \emph{{Maximal U(1)$_{Y}$-violating n-point correlators
  in $ \mathcal{N} $ = 4 super-Yang-Mills theory}},
  \href{https://doi.org/10.1007/JHEP02(2021)042}{\emph{JHEP} {\bfseries 02}
  (2021) 042} [\href{https://arxiv.org/abs/2009.01211}{{\ttfamily
  2009.01211}}].

\bibitem{Dorigoni:2021rdo}
D.~Dorigoni, M.~B. Green and C.~Wen, \emph{{Exact expressions for $n$-point
  maximal $U(1)_Y$-violating integrated correlators in $SU(N)$ $\mathcal{N}=4$
  SYM}}, \href{https://doi.org/10.1007/JHEP11(2021)132}{\emph{JHEP} {\bfseries
  11} (2021) 132} [\href{https://arxiv.org/abs/2109.08086}{{\ttfamily
  2109.08086}}].

\bibitem{Dorigoni:2022iem}
D.~Dorigoni, M.~B. Green and C.~Wen, \emph{{The SAGEX Review on Scattering
  Amplitudes, Chapter 10: Modular covariance of type IIB string amplitudes and
  their $\mathcal{N}=4$ supersymmetric Yang-Mills duals}},
  \href{https://arxiv.org/abs/2203.13021}{{\ttfamily 2203.13021}}.

\bibitem{Yuan:2018qva}
E.~Y. Yuan, \emph{{Simplicity in AdS Perturbative Dynamics}},
  \href{https://arxiv.org/abs/1801.07283}{{\ttfamily 1801.07283}}.

\bibitem{Alday:2019nin}
L.~F. Alday and X.~Zhou, \emph{{Simplicity of AdS Supergravity at One Loop}},
  \href{https://doi.org/10.1007/JHEP09(2020)008}{\emph{JHEP} {\bfseries 09}
  (2020) 008} [\href{https://arxiv.org/abs/1912.02663}{{\ttfamily
  1912.02663}}].

\bibitem{Aprile:2019rep}
F.~Aprile, J.~Drummond, P.~Heslop and H.~Paul, \emph{{One-loop amplitudes in
  AdS$_{5}$\texttimes{}S$^{5}$ supergravity from $ \mathcal{N} $ = 4 SYM at
  strong coupling}}, \href{https://doi.org/10.1007/JHEP03(2020)190}{\emph{JHEP}
  {\bfseries 03} (2020) 190}
  [\href{https://arxiv.org/abs/1912.01047}{{\ttfamily 1912.01047}}].

\bibitem{Drummond:2019hel}
J.~M. Drummond and H.~Paul, \emph{{One-loop string corrections to AdS
  amplitudes from CFT}},
  \href{https://doi.org/10.1007/JHEP03(2021)038}{\emph{JHEP} {\bfseries 03}
  (2021) 038} [\href{https://arxiv.org/abs/1912.07632}{{\ttfamily
  1912.07632}}].

\bibitem{Alday:2021ajh}
L.~F. Alday, A.~Bissi and X.~Zhou, \emph{{One-loop gluon amplitudes in AdS}},
  \href{https://doi.org/10.1007/JHEP02(2022)105}{\emph{JHEP} {\bfseries 02}
  (2022) 105} [\href{https://arxiv.org/abs/2110.09861}{{\ttfamily
  2110.09861}}].

\bibitem{Herderschee:2021jbi}
A.~Herderschee, \emph{{A New Framework for Higher Loop Witten Diagrams}},
  \href{https://arxiv.org/abs/2112.08226}{{\ttfamily 2112.08226}}.

\bibitem{Gomez:2021ujt}
H.~Gomez, R.~L. Jusinskas and A.~Lipstein, \emph{{Cosmological Scattering
  Equations at Tree-level and One-loop}},
  \href{https://arxiv.org/abs/2112.12695}{{\ttfamily 2112.12695}}.

\bibitem{Huang:2021xws}
Z.~Huang and E.~Y. Yuan, \emph{{Graviton Scattering in
  $\mathrm{AdS}_5\times\mathrm{S}^5$ at Two Loops}},
  \href{https://arxiv.org/abs/2112.15174}{{\ttfamily 2112.15174}}.

\bibitem{Heckelbacher:2022fbx}
T.~Heckelbacher, I.~Sachs, E.~Skvortsov and P.~Vanhove, \emph{{Analytical
  evaluation of AdS${}_4$ Witten diagrams as flat space multi-loop Feynman
  integrals}},  \href{https://arxiv.org/abs/2201.09626}{{\ttfamily
  2201.09626}}.

\bibitem{Drummond:2022dxw}
J.~M. Drummond and H.~Paul, \emph{{Two-loop supergravity on AdS$_5\times$S$^5$
  from CFT}},  \href{https://arxiv.org/abs/2204.01829}{{\ttfamily 2204.01829}}.

\bibitem{Bissi:2020woe}
A.~Bissi, G.~Fardelli and A.~Georgoudis, \emph{{All loop structures in
  supergravity amplitudes on AdS5 \texttimes{} S5 from CFT}},
  \href{https://doi.org/10.1088/1751-8121/ac0ebf}{\emph{J. Phys. A} {\bfseries
  54} (2021) 324002} [\href{https://arxiv.org/abs/2010.12557}{{\ttfamily
  2010.12557}}].

\bibitem{Bissi:2020wtv}
A.~Bissi, G.~Fardelli and A.~Georgoudis, \emph{{Towards all loop supergravity
  amplitudes on AdS5\texttimes{}S5}},
  \href{https://doi.org/10.1103/PhysRevD.104.L041901}{\emph{Phys. Rev. D}
  {\bfseries 104} (2021) L041901}
  [\href{https://arxiv.org/abs/2002.04604}{{\ttfamily 2002.04604}}].

\bibitem{Abramowicz:2019gvx}
H.~Abramowicz et~al., \emph{{Letter of Intent for the LUXE Experiment}},
  \href{https://arxiv.org/abs/1909.00860}{{\ttfamily 1909.00860}}.

\bibitem{Abramowicz:2021zja}
H.~Abramowicz et~al., \emph{{Conceptual design report for the LUXE
  experiment}},
  \href{https://doi.org/10.1140/epjs/s11734-021-00249-z}{\emph{Eur. Phys. J.
  ST} {\bfseries 230} (2021) 2445}
  [\href{https://arxiv.org/abs/2102.02032}{{\ttfamily 2102.02032}}].

\bibitem{King:2015tba}
B.~King and T.~Heinzl, \emph{{Measuring Vacuum Polarisation with High Power
  Lasers}}, \href{https://doi.org/10.1017/hpl.2016.1}{\emph{High Power Laser
  Sci. Eng.} {\bfseries 4} (2016) }
  [\href{https://arxiv.org/abs/1510.08456}{{\ttfamily 1510.08456}}].

\bibitem{Seipt:2017ckc}
D.~Seipt, \emph{{Volkov States and Non-linear Compton Scattering in Short and
  Intense Laser Pulses}},  in \emph{{Quantum Field Theory at the Limits}: {from
  Strong Fields to Heavy Quarks}}, pp.~24--43, 2017,
  \href{https://arxiv.org/abs/1701.03692}{{\ttfamily 1701.03692}},
  \href{https://doi.org/10.3204/DESY-PROC-2016-04/Seipt}{DOI}.

\bibitem{Ilderton:2013dba}
A.~Ilderton and G.~Torgrimsson, \emph{{Radiation reaction from QED: lightfront
  perturbation theory in a plane wave background}},
  \href{https://doi.org/10.1103/PhysRevD.88.025021}{\emph{Phys. Rev. D}
  {\bfseries 88} (2013) 025021}
  [\href{https://arxiv.org/abs/1304.6842}{{\ttfamily 1304.6842}}].

\bibitem{Ilderton:2013tb}
A.~Ilderton and G.~Torgrimsson, \emph{{Radiation reaction in strong field
  QED}}, \href{https://doi.org/10.1016/j.physletb.2013.07.045}{\emph{Phys.
  Lett. B} {\bfseries 725} (2013) 481}
  [\href{https://arxiv.org/abs/1301.6499}{{\ttfamily 1301.6499}}].

\bibitem{Adamo:2022rmp}
T.~Adamo, A.~Cristofoli and A.~Ilderton, \emph{{Classical physics from
  amplitudes on curved backgrounds}},
  \href{https://arxiv.org/abs/2203.13785}{{\ttfamily 2203.13785}}.

\bibitem{tHooft:1987vrq}
G.~'t~Hooft, \emph{{Graviton Dominance in Ultrahigh-Energy Scattering}},
  \href{https://doi.org/10.1016/0370-2693(87)90159-6}{\emph{Phys. Lett. B}
  {\bfseries 198} (1987) 61}.

\bibitem{Jackiw:1991ck}
R.~Jackiw, D.~N. Kabat and M.~Ortiz, \emph{{Electromagnetic fields of a
  massless particle and the eikonal}},
  \href{https://doi.org/10.1016/0370-2693(92)90971-6}{\emph{Phys. Lett. B}
  {\bfseries 277} (1992) 148}
  [\href{https://arxiv.org/abs/hep-th/9112020}{{\ttfamily hep-th/9112020}}].

\bibitem{Kabat:1992tb}
D.~N. Kabat and M.~Ortiz, \emph{{Eikonal quantum gravity and Planckian
  scattering}}, \href{https://doi.org/10.1016/0550-3213(92)90627-N}{\emph{Nucl.
  Phys. B} {\bfseries 388} (1992) 570}
  [\href{https://arxiv.org/abs/hep-th/9203082}{{\ttfamily hep-th/9203082}}].

\bibitem{Lodone:2009qe}
P.~Lodone and V.~S. Rychkov, \emph{{Radiation Problem in Transplanckian
  Scattering}},
  \href{https://doi.org/10.1088/1126-6708/2009/12/036}{\emph{JHEP} {\bfseries
  12} (2009) 036} [\href{https://arxiv.org/abs/0909.3519}{{\ttfamily
  0909.3519}}].

\bibitem{Gruzinov:2014moa}
A.~Gruzinov and G.~Veneziano, \emph{{Gravitational Radiation from Massless
  Particle Collisions}},
  \href{https://doi.org/10.1088/0264-9381/33/12/125012}{\emph{Class. Quant.
  Grav.} {\bfseries 33} (2016) 125012}
  [\href{https://arxiv.org/abs/1409.4555}{{\ttfamily 1409.4555}}].

\bibitem{Adamo:2021jxz}
T.~Adamo, A.~Ilderton and A.~J. MacLeod, \emph{{Particle-beam scattering from
  strong-field QED}},
  \href{https://doi.org/10.1103/PhysRevD.104.116013}{\emph{Phys. Rev. D}
  {\bfseries 104} (2021) 116013}
  [\href{https://arxiv.org/abs/2110.02567}{{\ttfamily 2110.02567}}].

\bibitem{Adamo:2021rfq}
T.~Adamo, A.~Cristofoli and P.~Tourkine, \emph{{Eikonal amplitudes from curved
  backgrounds}},  \href{https://arxiv.org/abs/2112.09113}{{\ttfamily
  2112.09113}}.

\bibitem{Farrow:2018yni}
J.~A. Farrow, A.~E. Lipstein and P.~McFadden, \emph{{Double copy structure of
  CFT correlators}}, \href{https://doi.org/10.1007/JHEP02(2019)130}{\emph{JHEP}
  {\bfseries 02} (2019) 130}
  [\href{https://arxiv.org/abs/1812.11129}{{\ttfamily 1812.11129}}].

\bibitem{Li:2018wkt}
S.~Y. Li, Y.~Wang and S.~Zhou, \emph{{KLT-Like Behaviour of Inflationary
  Graviton Correlators}},
  \href{https://doi.org/10.1088/1475-7516/2018/12/023}{\emph{JCAP} {\bfseries
  12} (2018) 023} [\href{https://arxiv.org/abs/1806.06242}{{\ttfamily
  1806.06242}}].

\bibitem{Fazio:2019iit}
A.~R. Fazio, \emph{{Cosmological correlators, In\textendash{}In formalism and
  double copy}}, \href{https://doi.org/10.1142/S0217732320500765}{\emph{Mod.
  Phys. Lett. A} {\bfseries 35} (2020) 2050076}
  [\href{https://arxiv.org/abs/1909.07343}{{\ttfamily 1909.07343}}].

\bibitem{Lipstein:2019mpu}
A.~E. Lipstein and P.~McFadden, \emph{{Double copy structure and the flat space
  limit of conformal correlators in even dimensions}},
  \href{https://doi.org/10.1103/PhysRevD.101.125006}{\emph{Phys. Rev. D}
  {\bfseries 101} (2020) 125006}
  [\href{https://arxiv.org/abs/1912.10046}{{\ttfamily 1912.10046}}].

\bibitem{Zhou:2021gnu}
X.~Zhou, \emph{{Double Copy Relation in AdS Space}},
  \href{https://doi.org/10.1103/PhysRevLett.127.141601}{\emph{Phys. Rev. Lett.}
  {\bfseries 127} (2021) 141601}
  [\href{https://arxiv.org/abs/2106.07651}{{\ttfamily 2106.07651}}].

\bibitem{Jain:2021qcl}
S.~Jain, R.~R. John, A.~Mehta, A.~A. Nizami and A.~Suresh, \emph{{Double copy
  structure of parity-violating CFT correlators}},
  \href{https://doi.org/10.1007/JHEP07(2021)033}{\emph{JHEP} {\bfseries 07}
  (2021) 033} [\href{https://arxiv.org/abs/2104.12803}{{\ttfamily
  2104.12803}}].

\bibitem{Cheung:2022pdk}
C.~Cheung, J.~Parra-Martinez and A.~Sivaramakrishnan, \emph{{On-shell
  Correlators and Color-Kinematics Duality in Curved Symmetric Spacetimes}},
  \href{https://arxiv.org/abs/2201.05147}{{\ttfamily 2201.05147}}.

\bibitem{Herderschee:2022ntr}
A.~Herderschee, R.~Roiban and F.~Teng, \emph{{On the Differential
  Representation and Color-Kinematics Duality of AdS Boundary Correlators}},
  \href{https://arxiv.org/abs/2201.05067}{{\ttfamily 2201.05067}}.

\bibitem{Drummond:2022dxd}
J.~M. Drummond, R.~Glew and M.~Santagata, \emph{{BCJ relations in ${AdS}_5
  \times S^3$ and the double-trace spectrum of super gluons}},
  \href{https://arxiv.org/abs/2202.09837}{{\ttfamily 2202.09837}}.

\bibitem{Paulos:2011ie}
M.~F. Paulos, \emph{{Towards Feynman rules for Mellin amplitudes}},
  \href{https://doi.org/10.1007/JHEP10(2011)074}{\emph{JHEP} {\bfseries 10}
  (2011) 074} [\href{https://arxiv.org/abs/1107.1504}{{\ttfamily 1107.1504}}].

\bibitem{Fitzpatrick:2011ia}
A.~L. Fitzpatrick, J.~Kaplan, J.~Penedones, S.~Raju and B.~C. van Rees,
  \emph{{A Natural Language for AdS/CFT Correlators}},
  \href{https://doi.org/10.1007/JHEP11(2011)095}{\emph{JHEP} {\bfseries 11}
  (2011) 095} [\href{https://arxiv.org/abs/1107.1499}{{\ttfamily 1107.1499}}].

\bibitem{Sivaramakrishnan:2021srm}
A.~Sivaramakrishnan, \emph{{Towards Color-Kinematics Duality in Generic
  Spacetimes}},  \href{https://arxiv.org/abs/2110.15356}{{\ttfamily
  2110.15356}}.

\bibitem{Adamo:2020qru}
T.~Adamo and A.~Ilderton, \emph{{Classical and quantum double copy of
  back-reaction}}, \href{https://doi.org/10.1007/JHEP09(2020)200}{\emph{JHEP}
  {\bfseries 09} (2020) 200}
  [\href{https://arxiv.org/abs/2005.05807}{{\ttfamily 2005.05807}}].

\bibitem{Gibbons:1975jb}
G.~W. Gibbons, \emph{{Quantized Fields Propagating in Plane Wave Space-Times}},
  \href{https://doi.org/10.1007/BF01629249}{\emph{Commun. Math. Phys.}
  {\bfseries 45} (1975) 191}.

\bibitem{Amati:1988ww}
D.~Amati and C.~Klimcik, \emph{{Strings in a Shock Wave Background and
  Generation of Curved Geometry from Flat Space String Theory}},
  \href{https://doi.org/10.1016/0370-2693(88)90355-3}{\emph{Phys. Lett. B}
  {\bfseries 210} (1988) 92}.

\bibitem{Amati:1988sa}
D.~Amati and C.~Klimcik, \emph{{Nonperturbative Computation of the Weyl Anomaly
  for a Class of Nontrivial Backgrounds}},
  \href{https://doi.org/10.1016/0370-2693(89)91092-7}{\emph{Phys. Lett. B}
  {\bfseries 219} (1989) 443}.

\bibitem{Horowitz:1989bv}
G.~T. Horowitz and A.~R. Steif, \emph{{Space-Time Singularities in String
  Theory}}, \href{https://doi.org/10.1103/PhysRevLett.64.260}{\emph{Phys. Rev.
  Lett.} {\bfseries 64} (1990) 260}.

\bibitem{Horowitz:1990sr}
G.~T. Horowitz and A.~R. Steif, \emph{{Strings in Strong Gravitational
  Fields}}, \href{https://doi.org/10.1103/PhysRevD.42.1950}{\emph{Phys. Rev. D}
  {\bfseries 42} (1990) 1950}.

\bibitem{Penrose1976}
R.~Penrose, \emph{Any Space-Time has a Plane Wave as a Limit}, pp.~271--275.
\newblock Springer Netherlands, 1976.

\bibitem{Yakimenko:2018kih}
V.~Yakimenko et~al., \emph{{Prospect of Studying Nonperturbative QED with
  Beam-Beam Collisions}},
  \href{https://doi.org/10.1103/PhysRevLett.122.190404}{\emph{Phys. Rev. Lett.}
  {\bfseries 122} (2019) 190404}
  [\href{https://arxiv.org/abs/1807.09271}{{\ttfamily 1807.09271}}].

\bibitem{Adamo:2020syc}
T.~Adamo, L.~Mason and A.~Sharma, \emph{{MHV scattering of gluons and gravitons
  in chiral strong fields}},
  \href{https://doi.org/10.1103/PhysRevLett.125.041602}{\emph{Phys. Rev. Lett.}
  {\bfseries 125} (2020) 041602}
  [\href{https://arxiv.org/abs/2003.13501}{{\ttfamily 2003.13501}}].

\bibitem{Adamo:2020yzi}
T.~Adamo, L.~Mason and A.~Sharma, \emph{{Gluon scattering on self-dual
  radiative gauge fields}},  \href{https://arxiv.org/abs/2010.14996}{{\ttfamily
  2010.14996}}.

\bibitem{Adamo:2022mev}
T.~Adamo, L.~Mason and A.~Sharma, \emph{{Graviton scattering in self-dual
  radiative space-times}},  \href{https://arxiv.org/abs/2203.02238}{{\ttfamily
  2203.02238}}.

\bibitem{Adamo:2014wea}
T.~Adamo, E.~Casali and D.~Skinner, \emph{{A Worldsheet Theory for
  Supergravity}}, \href{https://doi.org/10.1007/JHEP02(2015)116}{\emph{JHEP}
  {\bfseries 02} (2015) 116} [\href{https://arxiv.org/abs/1409.5656}{{\ttfamily
  1409.5656}}].

\bibitem{Adamo:2018hzd}
T.~Adamo, E.~Casali and S.~Nekovar, \emph{{Yang-Mills theory from the
  worldsheet}}, \href{https://doi.org/10.1103/PhysRevD.98.086022}{\emph{Phys.
  Rev. D} {\bfseries 98} (2018) 086022}
  [\href{https://arxiv.org/abs/1807.09171}{{\ttfamily 1807.09171}}].

\bibitem{Adamo:2018ege}
T.~Adamo, E.~Casali and S.~Nekovar, \emph{{Ambitwistor string vertex operators
  on curved backgrounds}},
  \href{https://doi.org/10.1007/JHEP01(2019)213}{\emph{JHEP} {\bfseries 01}
  (2019) 213} [\href{https://arxiv.org/abs/1809.04489}{{\ttfamily
  1809.04489}}].

\bibitem{Adamo:2017sze}
T.~Adamo, E.~Casali, L.~Mason and S.~Nekovar, \emph{{Amplitudes on plane waves
  from ambitwistor strings}},
  \href{https://doi.org/10.1007/JHEP11(2017)160}{\emph{JHEP} {\bfseries 11}
  (2017) 160} [\href{https://arxiv.org/abs/1708.09249}{{\ttfamily
  1708.09249}}].

\bibitem{Roehrig:2020kck}
K.~Roehrig and D.~Skinner, \emph{{Ambitwistor strings and the scattering
  equations on AdS$_{3}$\texttimes{}S$^{3}$}},
  \href{https://doi.org/10.1007/JHEP02(2022)073}{\emph{JHEP} {\bfseries 02}
  (2022) 073} [\href{https://arxiv.org/abs/2007.07234}{{\ttfamily
  2007.07234}}].

\bibitem{Eberhardt:2020ewh}
L.~Eberhardt, S.~Komatsu and S.~Mizera, \emph{{Scattering equations in AdS:
  scalar correlators in arbitrary dimensions}},
  \href{https://doi.org/10.1007/JHEP11(2020)158}{\emph{JHEP} {\bfseries 11}
  (2020) 158} [\href{https://arxiv.org/abs/2007.06574}{{\ttfamily
  2007.06574}}].

\bibitem{Gomez:2021qfd}
H.~Gomez, R.~L. Jusinskas and A.~Lipstein, \emph{{Cosmological Scattering
  Equations}},
  \href{https://doi.org/10.1103/PhysRevLett.127.251604}{\emph{Phys. Rev. Lett.}
  {\bfseries 127} (2021) 251604}
  [\href{https://arxiv.org/abs/2106.11903}{{\ttfamily 2106.11903}}].

\bibitem{Ward:1977ta}
R.~S. Ward, \emph{{On Selfdual gauge fields}},
  \href{https://doi.org/10.1016/0375-9601(77)90842-8}{\emph{Phys. Lett. A}
  {\bfseries 61} (1977) 81}.

\bibitem{Penrose:1976js}
R.~Penrose, \emph{{Nonlinear Gravitons and Curved Twistor Theory}},
  \href{https://doi.org/10.1007/BF00762011}{\emph{Gen. Rel. Grav.} {\bfseries
  7} (1976) 31}.

\bibitem{Edwards:2021elz}
J.~P. Edwards, C.~M. Mata, U.~M\"uller and C.~Schubert, \emph{{New Techniques
  for Worldline Integration}},
  \href{https://doi.org/10.3842/SIGMA.2021.065}{\emph{SIGMA} {\bfseries 17}
  (2021) 065} [\href{https://arxiv.org/abs/2106.12071}{{\ttfamily
  2106.12071}}].

\bibitem{Edwards:2021uif}
J.~P. Edwards and C.~Schubert, \emph{{Plane Wave Backgrounds in the Worldline
  Formalism}},  in \emph{{29th International Laser Physics Workshop}}, 12,
  2021, \href{https://arxiv.org/abs/2112.13944}{{\ttfamily 2112.13944}}.

\bibitem{Edwards:2021vhg}
J.~P. Edwards and C.~Schubert, \emph{{N-photon amplitudes in a plane-wave
  background}},
  \href{https://doi.org/10.1016/j.physletb.2021.136696}{\emph{Phys. Lett. B}
  {\bfseries 822} (2021) 136696}
  [\href{https://arxiv.org/abs/2105.08173}{{\ttfamily 2105.08173}}].

\bibitem{Maldacena:2011nz}
J.~M. Maldacena and G.~L. Pimentel, \emph{{On graviton non-Gaussianities during
  inflation}}, \href{https://doi.org/10.1007/JHEP09(2011)045}{\emph{JHEP}
  {\bfseries 09} (2011) 045} [\href{https://arxiv.org/abs/1104.2846}{{\ttfamily
  1104.2846}}].

\bibitem{Nagaraj:2018nxq}
B.~Nagaraj and D.~Ponomarev, \emph{{Spinor-Helicity Formalism for Massless
  Fields in AdS$_4$}},
  \href{https://doi.org/10.1103/PhysRevLett.122.101602}{\emph{Phys. Rev. Lett.}
  {\bfseries 122} (2019) 101602}
  [\href{https://arxiv.org/abs/1811.08438}{{\ttfamily 1811.08438}}].

\bibitem{David:2019mos}
A.~David, N.~Fischer and Y.~Neiman, \emph{{Spinor-helicity variables for
  cosmological horizons in de Sitter space}},
  \href{https://doi.org/10.1103/PhysRevD.100.045005}{\emph{Phys. Rev. D}
  {\bfseries 100} (2019) 045005}
  [\href{https://arxiv.org/abs/1906.01058}{{\ttfamily 1906.01058}}].

\bibitem{Adamo:2019zmk}
T.~Adamo and A.~Ilderton, \emph{{Gluon helicity flip in a plane wave
  background}}, \href{https://doi.org/10.1007/JHEP06(2019)015}{\emph{JHEP}
  {\bfseries 06} (2019) 015}
  [\href{https://arxiv.org/abs/1903.01491}{{\ttfamily 1903.01491}}].

\bibitem{Raju:2010by}
S.~Raju, \emph{{BCFW for Witten Diagrams}},
  \href{https://doi.org/10.1103/PhysRevLett.106.091601}{\emph{Phys. Rev. Lett.}
  {\bfseries 106} (2011) 091601}
  [\href{https://arxiv.org/abs/1011.0780}{{\ttfamily 1011.0780}}].

\bibitem{Raju:2011mp}
S.~Raju, \emph{{Recursion Relations for AdS/CFT Correlators}},
  \href{https://doi.org/10.1103/PhysRevD.83.126002}{\emph{Phys. Rev. D}
  {\bfseries 83} (2011) 126002}
  [\href{https://arxiv.org/abs/1102.4724}{{\ttfamily 1102.4724}}].

\bibitem{Raju:2012zr}
S.~Raju, \emph{{New Recursion Relations and a Flat Space Limit for AdS/CFT
  Correlators}}, \href{https://doi.org/10.1103/PhysRevD.85.126009}{\emph{Phys.
  Rev. D} {\bfseries 85} (2012) 126009}
  [\href{https://arxiv.org/abs/1201.6449}{{\ttfamily 1201.6449}}].

\bibitem{Caron-Huot:2017vep}
S.~Caron-Huot, \emph{{Analyticity in Spin in Conformal Theories}},
  \href{https://doi.org/10.1007/JHEP09(2017)078}{\emph{JHEP} {\bfseries 09}
  (2017) 078} [\href{https://arxiv.org/abs/1703.00278}{{\ttfamily
  1703.00278}}].

\bibitem{Simmons-Duffin:2017nub}
D.~Simmons-Duffin, D.~Stanford and E.~Witten, \emph{{A spacetime derivation of
  the Lorentzian OPE inversion formula}},
  \href{https://doi.org/10.1007/JHEP07(2018)085}{\emph{JHEP} {\bfseries 07}
  (2018) 085} [\href{https://arxiv.org/abs/1711.03816}{{\ttfamily
  1711.03816}}].

\bibitem{Arkani-Hamed:2018kmz}
N.~Arkani-Hamed, D.~Baumann, H.~Lee and G.~L. Pimentel, \emph{{The Cosmological
  Bootstrap: Inflationary Correlators from Symmetries and Singularities}},
  \href{https://doi.org/10.1007/JHEP04(2020)105}{\emph{JHEP} {\bfseries 04}
  (2020) 105} [\href{https://arxiv.org/abs/1811.00024}{{\ttfamily
  1811.00024}}].

\bibitem{Sleight:2019hfp}
C.~Sleight and M.~Taronna, \emph{{Bootstrapping Inflationary Correlators in
  Mellin Space}}, \href{https://doi.org/10.1007/JHEP02(2020)098}{\emph{JHEP}
  {\bfseries 02} (2020) 098}
  [\href{https://arxiv.org/abs/1907.01143}{{\ttfamily 1907.01143}}].

\bibitem{Sleight:2019mgd}
C.~Sleight, \emph{{A Mellin Space Approach to Cosmological Correlators}},
  \href{https://doi.org/10.1007/JHEP01(2020)090}{\emph{JHEP} {\bfseries 01}
  (2020) 090} [\href{https://arxiv.org/abs/1906.12302}{{\ttfamily
  1906.12302}}].

\bibitem{Meltzer:2019nbs}
D.~Meltzer, E.~Perlmutter and A.~Sivaramakrishnan, \emph{{Unitarity Methods in
  AdS/CFT}}, \href{https://doi.org/10.1007/JHEP03(2020)061}{\emph{JHEP}
  {\bfseries 03} (2020) 061}
  [\href{https://arxiv.org/abs/1912.09521}{{\ttfamily 1912.09521}}].

\bibitem{Drummond:2020dwr}
J.~M. Drummond, H.~Paul and M.~Santagata, \emph{{Bootstrapping string theory on
  AdS$_5 \times S^5$}},  \href{https://arxiv.org/abs/2004.07282}{{\ttfamily
  2004.07282}}.

\bibitem{Baumann:2020dch}
D.~Baumann, C.~Duaso~Pueyo, A.~Joyce, H.~Lee and G.~L. Pimentel, \emph{{The
  Cosmological Bootstrap: Spinning Correlators from Symmetries and
  Factorization}},
  \href{https://doi.org/10.21468/SciPostPhys.11.3.071}{\emph{SciPost Phys.}
  {\bfseries 11} (2021) 071}
  [\href{https://arxiv.org/abs/2005.04234}{{\ttfamily 2005.04234}}].

\bibitem{Meltzer:2020qbr}
D.~Meltzer and A.~Sivaramakrishnan, \emph{{CFT unitarity and the AdS Cutkosky
  rules}}, \href{https://doi.org/10.1007/JHEP11(2020)073}{\emph{JHEP}
  {\bfseries 11} (2020) 073}
  [\href{https://arxiv.org/abs/2008.11730}{{\ttfamily 2008.11730}}].

\bibitem{Melville:2021lst}
S.~Melville and E.~Pajer, \emph{{Cosmological Cutting Rules}},
  \href{https://doi.org/10.1007/JHEP05(2021)249}{\emph{JHEP} {\bfseries 05}
  (2021) 249} [\href{https://arxiv.org/abs/2103.09832}{{\ttfamily
  2103.09832}}].

\bibitem{Goodhew:2021oqg}
H.~Goodhew, S.~Jazayeri, M.~H. Gordon~Lee and E.~Pajer, \emph{{Cutting
  cosmological correlators}},
  \href{https://doi.org/10.1088/1475-7516/2021/08/003}{\emph{JCAP} {\bfseries
  08} (2021) 003} [\href{https://arxiv.org/abs/2104.06587}{{\ttfamily
  2104.06587}}].

\bibitem{Albrychiewicz:2021ndv}
E.~Albrychiewicz, Y.~Neiman and M.~Tsulaia, \emph{{MHV amplitudes and BCFW
  recursion for Yang-Mills theory in the de Sitter static patch}},
  \href{https://doi.org/10.1007/JHEP09(2021)176}{\emph{JHEP} {\bfseries 09}
  (2021) 176} [\href{https://arxiv.org/abs/2105.07572}{{\ttfamily
  2105.07572}}].

\bibitem{Baumann:2021fxj}
D.~Baumann, W.-M. Chen, C.~Duaso~Pueyo, A.~Joyce, H.~Lee and G.~L. Pimentel,
  \emph{{Linking the Singularities of Cosmological Correlators}},
  \href{https://arxiv.org/abs/2106.05294}{{\ttfamily 2106.05294}}.

\bibitem{Bonifacio:2021azc}
J.~Bonifacio, E.~Pajer and D.-G. Wang, \emph{{From amplitudes to contact
  cosmological correlators}},
  \href{https://doi.org/10.1007/JHEP10(2021)001}{\emph{JHEP} {\bfseries 10}
  (2021) 001} [\href{https://arxiv.org/abs/2106.15468}{{\ttfamily
  2106.15468}}].

\bibitem{Meltzer:2021zin}
D.~Meltzer, \emph{{The inflationary wavefunction from analyticity and
  factorization}},
  \href{https://doi.org/10.1088/1475-7516/2021/12/018}{\emph{JCAP} {\bfseries
  12} (2021) 018} [\href{https://arxiv.org/abs/2107.10266}{{\ttfamily
  2107.10266}}].

\bibitem{Sleight:2021plv}
C.~Sleight and M.~Taronna, \emph{{From dS to AdS and back}},
  \href{https://doi.org/10.1007/JHEP12(2021)074}{\emph{JHEP} {\bfseries 12}
  (2021) 074} [\href{https://arxiv.org/abs/2109.02725}{{\ttfamily
  2109.02725}}].

\bibitem{Baumann:2022jpr}
D.~Baumann, D.~Green, A.~Joyce, E.~Pajer, G.~L. Pimentel, C.~Sleight et~al.,
  \emph{{Snowmass White Paper: The Cosmological Bootstrap}},  in \emph{{2022
  Snowmass Summer Study}}, 3, 2022,
  \href{https://arxiv.org/abs/2203.08121}{{\ttfamily 2203.08121}}.

\bibitem{Ilderton:2020rgk}
A.~Ilderton and A.~J. MacLeod, \emph{{The analytic structure of amplitudes on
  backgrounds from gauge invariance and the infra-red}},
  \href{https://doi.org/10.1007/JHEP04(2020)078}{\emph{JHEP} {\bfseries 04}
  (2020) 078} [\href{https://arxiv.org/abs/2001.10553}{{\ttfamily
  2001.10553}}].

\bibitem{Vafa:1987ea}
C.~Vafa, \emph{{Conformal Theories and Punctured Surfaces}},
  \href{https://doi.org/10.1016/0370-2693(87)91358-X}{\emph{Phys. Lett. B}
  {\bfseries 199} (1987) 195}.

\bibitem{Polchinski:1988jq}
J.~Polchinski, \emph{{Factorization of Bosonic String Amplitudes}},
  \href{https://doi.org/10.1016/0550-3213(88)90522-6}{\emph{Nucl. Phys. B}
  {\bfseries 307} (1988) 61}.

\bibitem{Witten:2012bh}
E.~Witten, \emph{{Superstring Perturbation Theory Revisited}},
  \href{https://arxiv.org/abs/1209.5461}{{\ttfamily 1209.5461}}.

\bibitem{Adamo:2013tca}
T.~Adamo, \emph{{Worldsheet factorization for twistor-strings}},
  \href{https://doi.org/10.1007/JHEP04(2014)080}{\emph{JHEP} {\bfseries 04}
  (2014) 080} [\href{https://arxiv.org/abs/1310.8602}{{\ttfamily 1310.8602}}].

\bibitem{Adamo:2021hno}
T.~Adamo, A.~Ilderton and A.~J. MacLeod, \emph{{One-loop multicollinear limits
  from 2-point amplitudes on self-dual backgrounds}},
  \href{https://doi.org/10.1007/JHEP12(2021)207}{\emph{JHEP} {\bfseries 12}
  (2021) 207} [\href{https://arxiv.org/abs/2103.12850}{{\ttfamily
  2103.12850}}].

\bibitem{Pasterski:2016qvg}
S.~Pasterski, S.-H. Shao and A.~Strominger, \emph{{Flat Space Amplitudes and
  Conformal Symmetry of the Celestial Sphere}},
  \href{https://doi.org/10.1103/PhysRevD.96.065026}{\emph{Phys. Rev. D}
  {\bfseries 96} (2017) 065026}
  [\href{https://arxiv.org/abs/1701.00049}{{\ttfamily 1701.00049}}].

\bibitem{Pasterski:2017kqt}
S.~Pasterski and S.-H. Shao, \emph{{Conformal basis for flat space
  amplitudes}}, \href{https://doi.org/10.1103/PhysRevD.96.065022}{\emph{Phys.
  Rev. D} {\bfseries 96} (2017) 065022}
  [\href{https://arxiv.org/abs/1705.01027}{{\ttfamily 1705.01027}}].

\bibitem{Pasterski:2021raf}
S.~Pasterski, M.~Pate and A.-M. Raclariu, \emph{{Celestial Holography}},  in
  \emph{{2022 Snowmass Summer Study}}, 11, 2021,
  \href{https://arxiv.org/abs/2111.11392}{{\ttfamily 2111.11392}}.

\bibitem{Casali:2020vuy}
E.~Casali and A.~Puhm, \emph{{Double Copy for Celestial Amplitudes}},
  \href{https://doi.org/10.1103/PhysRevLett.126.101602}{\emph{Phys. Rev. Lett.}
  {\bfseries 126} (2021) 101602}
  [\href{https://arxiv.org/abs/2007.15027}{{\ttfamily 2007.15027}}].

\bibitem{Casali:2020uvr}
E.~Casali and A.~Sharma, \emph{{Celestial double copy from the worldsheet}},
  \href{https://doi.org/10.1007/JHEP05(2021)157}{\emph{JHEP} {\bfseries 05}
  (2021) 157} [\href{https://arxiv.org/abs/2011.10052}{{\ttfamily
  2011.10052}}].

\bibitem{Kalyanapuram:2020epb}
N.~Kalyanapuram, \emph{{Soft Gravity by Squaring Soft QED on the Celestial
  Sphere}}, \href{https://doi.org/10.1103/PhysRevD.103.085016}{\emph{Phys. Rev.
  D} {\bfseries 103} (2021) 085016}
  [\href{https://arxiv.org/abs/2011.11412}{{\ttfamily 2011.11412}}].

\bibitem{Pasterski:2020pdk}
S.~Pasterski and A.~Puhm, \emph{{Shifting spin on the celestial sphere}},
  \href{https://doi.org/10.1103/PhysRevD.104.086020}{\emph{Phys. Rev. D}
  {\bfseries 104} (2021) 086020}
  [\href{https://arxiv.org/abs/2012.15694}{{\ttfamily 2012.15694}}].

\bibitem{Bahjat-Abbas:2017htu}
N.~Bahjat-Abbas, A.~Luna and C.~D. White, \emph{{The Kerr-Schild double copy in
  curved spacetime}},
  \href{https://doi.org/10.1007/JHEP12(2017)004}{\emph{JHEP} {\bfseries 12}
  (2017) 004} [\href{https://arxiv.org/abs/1710.01953}{{\ttfamily
  1710.01953}}].

\bibitem{Dore:2014cca}
O.~Dor\'e et~al., \emph{{Cosmology with the SPHEREX All-Sky Spectral Survey}},
  \href{https://arxiv.org/abs/1412.4872}{{\ttfamily 1412.4872}}.

\bibitem{CMB-S4:2016ple}
{\scshape CMB-S4} collaboration, \emph{{CMB-S4 Science Book, First Edition}},
  \href{https://arxiv.org/abs/1610.02743}{{\ttfamily 1610.02743}}.

\bibitem{NASAPICO:2019thw}
{\scshape NASA PICO} collaboration, \emph{{PICO: Probe of Inflation and Cosmic
  Origins}},  \href{https://arxiv.org/abs/1902.10541}{{\ttfamily 1902.10541}}.

\bibitem{SimonsObservatory:2018koc}
{\scshape Simons Observatory} collaboration, \emph{{The Simons Observatory:
  Science goals and forecasts}},
  \href{https://doi.org/10.1088/1475-7516/2019/02/056}{\emph{JCAP} {\bfseries
  02} (2019) 056} [\href{https://arxiv.org/abs/1808.07445}{{\ttfamily
  1808.07445}}].

\bibitem{Beutler:2019ojk}
F.~Beutler, M.~Biagetti, D.~Green, A.~Slosar and B.~Wallisch, \emph{{Primordial
  Features from Linear to Nonlinear Scales}},
  \href{https://doi.org/10.1103/PhysRevResearch.1.033209}{\emph{Phys. Rev.
  Res.} {\bfseries 1} (2019) 033209}
  [\href{https://arxiv.org/abs/1906.08758}{{\ttfamily 1906.08758}}].

\bibitem{Meerburg:2019qqi}
P.~D. Meerburg et~al., \emph{{Primordial Non-Gaussianity}},
  \href{https://arxiv.org/abs/1903.04409}{{\ttfamily 1903.04409}}.

\bibitem{Darwish:2020prn}
O.~Darwish, S.~Foreman, M.~M. Abidi, T.~Baldauf, B.~D. Sherwin and P.~D.
  Meerburg, \emph{{Density reconstruction from biased tracers and its
  application to primordial non-Gaussianity}},
  \href{https://doi.org/10.1103/PhysRevD.104.123520}{\emph{Phys. Rev. D}
  {\bfseries 104} (2021) 123520}
  [\href{https://arxiv.org/abs/2007.08472}{{\ttfamily 2007.08472}}].

\bibitem{Malik:2008im}
K.~A. Malik and D.~Wands, \emph{{Cosmological perturbations}},
  \href{https://doi.org/10.1016/j.physrep.2009.03.001}{\emph{Phys. Rept.}
  {\bfseries 475} (2009) 1} [\href{https://arxiv.org/abs/0809.4944}{{\ttfamily
  0809.4944}}].

\bibitem{Wang:2021qez}
L.-T. Wang, Z.-Z. Xianyu and Y.-M. Zhong, \emph{{Precision calculation of
  inflation correlators at one loop}},
  \href{https://doi.org/10.1007/JHEP02(2022)085}{\emph{JHEP} {\bfseries 02}
  (2022) 085} [\href{https://arxiv.org/abs/2109.14635}{{\ttfamily
  2109.14635}}].

\bibitem{Kundu:2014gxa}
N.~Kundu, A.~Shukla and S.~P. Trivedi, \emph{{Constraints from Conformal
  Symmetry on the Three Point Scalar Correlator in Inflation}},
  \href{https://doi.org/10.1007/JHEP04(2015)061}{\emph{JHEP} {\bfseries 04}
  (2015) 061} [\href{https://arxiv.org/abs/1410.2606}{{\ttfamily 1410.2606}}].

\bibitem{Arkani-Hamed:2015bza}
N.~Arkani-Hamed and J.~Maldacena, \emph{{Cosmological Collider Physics}},
  \href{https://arxiv.org/abs/1503.08043}{{\ttfamily 1503.08043}}.

\bibitem{Baumann:2019oyu}
D.~Baumann, C.~Duaso~Pueyo, A.~Joyce, H.~Lee and G.~L. Pimentel, \emph{{The
  cosmological bootstrap: weight-shifting operators and scalar seeds}},
  \href{https://doi.org/10.1007/JHEP12(2020)204}{\emph{JHEP} {\bfseries 12}
  (2020) 204} [\href{https://arxiv.org/abs/1910.14051}{{\ttfamily
  1910.14051}}].

\bibitem{Joyce:2014kja}
A.~Joyce, B.~Jain, J.~Khoury and M.~Trodden, \emph{{Beyond the Cosmological
  Standard Model}},
  \href{https://doi.org/10.1016/j.physrep.2014.12.002}{\emph{Phys. Rept.}
  {\bfseries 568} (2015) 1} [\href{https://arxiv.org/abs/1407.0059}{{\ttfamily
  1407.0059}}].

\bibitem{deRham:2021efp}
C.~de~Rham, S.~Garcia-Saenz, L.~Heisenberg and V.~Pozsgay, \emph{{Cosmology of
  Extended Proca-Nuevo}},  \href{https://arxiv.org/abs/2110.14327}{{\ttfamily
  2110.14327}}.

\bibitem{DeFelice:2016yws}
A.~De~Felice, L.~Heisenberg, R.~Kase, S.~Mukohyama, S.~Tsujikawa and Y.-l.
  Zhang, \emph{{Cosmology in generalized Proca theories}},
  \href{https://doi.org/10.1088/1475-7516/2016/06/048}{\emph{JCAP} {\bfseries
  06} (2016) 048} [\href{https://arxiv.org/abs/1603.05806}{{\ttfamily
  1603.05806}}].

\bibitem{Brax:2021wcv}
P.~Brax, S.~Casas, H.~Desmond and B.~Elder, \emph{{Testing Screened Modified
  Gravity}}, \href{https://doi.org/10.3390/universe8010011}{\emph{Universe}
  {\bfseries 8} (2021) 11} [\href{https://arxiv.org/abs/2201.10817}{{\ttfamily
  2201.10817}}].

\bibitem{deRham:2014zqa}
C.~de~Rham, \emph{{Massive Gravity}},
  \href{https://doi.org/10.12942/lrr-2014-7}{\emph{Living Rev. Rel.} {\bfseries
  17} (2014) 7} [\href{https://arxiv.org/abs/1401.4173}{{\ttfamily
  1401.4173}}].

\bibitem{Hinterbichler:2011tt}
K.~Hinterbichler, \emph{{Theoretical Aspects of Massive Gravity}},
  \href{https://doi.org/10.1103/RevModPhys.84.671}{\emph{Rev. Mod. Phys.}
  {\bfseries 84} (2012) 671} [\href{https://arxiv.org/abs/1105.3735}{{\ttfamily
  1105.3735}}].

\bibitem{Heisenberg2021}
L.~Heisenberg, \emph{Massive Gravity and Bigravity}, pp.~89--96.
\newblock Springer International Publishing, Cham, 2021.
\newblock 10.1007/978-3-030-83715-0\textunderscore7.

\bibitem{Hinterbichler:2016try}
K.~Hinterbichler, \emph{{Cosmology of Massive Gravity and its Extensions}},  in
  \emph{{51st Rencontres de Moriond on Cosmology}}, pp.~223--232, 2016,
  \href{https://arxiv.org/abs/1701.02873}{{\ttfamily 1701.02873}}.

\bibitem{Blanchet:2015bia}
L.~Blanchet and L.~Heisenberg, \emph{{Dipolar Dark Matter with Massive
  Bigravity}}, \href{https://doi.org/10.1088/1475-7516/2015/12/026}{\emph{JCAP}
  {\bfseries 12} (2015) 026}
  [\href{https://arxiv.org/abs/1505.05146}{{\ttfamily 1505.05146}}].

\bibitem{Marzola:2017lbt}
L.~Marzola, M.~Raidal and F.~R. Urban, \emph{{Oscillating Spin-2 Dark Matter}},
  \href{https://doi.org/10.1103/PhysRevD.97.024010}{\emph{Phys. Rev. D}
  {\bfseries 97} (2018) 024010}
  [\href{https://arxiv.org/abs/1708.04253}{{\ttfamily 1708.04253}}].

\bibitem{Aoki:2017cnz}
K.~Aoki and K.-i. Maeda, \emph{{Condensate of Massive Graviton and Dark
  Matter}}, \href{https://doi.org/10.1103/PhysRevD.97.044002}{\emph{Phys. Rev.
  D} {\bfseries 97} (2018) 044002}
  [\href{https://arxiv.org/abs/1707.05003}{{\ttfamily 1707.05003}}].

\bibitem{Gabadadze:2017jom}
G.~Gabadadze, \emph{{Scale-up of $\Lambda_3$: Massive gravity with a higher
  strong interaction scale}},
  \href{https://doi.org/10.1103/PhysRevD.96.084018}{\emph{Phys. Rev. D}
  {\bfseries 96} (2017) 084018}
  [\href{https://arxiv.org/abs/1707.01739}{{\ttfamily 1707.01739}}].

\bibitem{Tolley:2019nmm}
A.~J. Tolley, \emph{{$ T\overline{T} $ deformations, massive gravity and
  non-critical strings}},
  \href{https://doi.org/10.1007/JHEP06(2020)050}{\emph{JHEP} {\bfseries 06}
  (2020) 050} [\href{https://arxiv.org/abs/1911.06142}{{\ttfamily
  1911.06142}}].

\bibitem{Johnson:2020pny}
L.~A. Johnson, C.~R.~T. Jones and S.~Paranjape, \emph{{Constraints on a Massive
  Double-Copy and Applications to Massive Gravity}},
  \href{https://doi.org/10.1007/JHEP02(2021)148}{\emph{JHEP} {\bfseries 02}
  (2021) 148} [\href{https://arxiv.org/abs/2004.12948}{{\ttfamily
  2004.12948}}].

\bibitem{Momeni:2020hmc}
A.~Momeni, J.~Rumbutis and A.~J. Tolley, \emph{{Kaluza-Klein from
  colour-kinematics duality for massive fields}},
  \href{https://doi.org/10.1007/JHEP08(2021)081}{\emph{JHEP} {\bfseries 08}
  (2021) 081} [\href{https://arxiv.org/abs/2012.09711}{{\ttfamily
  2012.09711}}].

\bibitem{Gonzalez:2021bes}
M.~C. Gonz\'alez, A.~Momeni and J.~Rumbutis, \emph{{Massive double copy in
  three spacetime dimensions}},
  \href{https://doi.org/10.1007/JHEP08(2021)116}{\emph{JHEP} {\bfseries 08}
  (2021) 116} [\href{https://arxiv.org/abs/2107.00611}{{\ttfamily
  2107.00611}}].

\bibitem{Moynihan:2020ejh}
N.~Moynihan, \emph{{Scattering Amplitudes and the Double Copy in Topologically
  Massive Theories}},
  \href{https://doi.org/10.1007/JHEP12(2020)163}{\emph{JHEP} {\bfseries 12}
  (2020) 163} [\href{https://arxiv.org/abs/2006.15957}{{\ttfamily
  2006.15957}}].

\bibitem{Burger:2021wss}
D.~J. Burger, W.~T. Emond and N.~Moynihan, \emph{{Anyons and the double copy}},
  \href{https://doi.org/10.1007/JHEP01(2022)017}{\emph{JHEP} {\bfseries 01}
  (2022) 017} [\href{https://arxiv.org/abs/2103.10416}{{\ttfamily
  2103.10416}}].

\bibitem{Hang:2021oso}
Y.-F. Hang, H.-J. He and C.~Shen, \emph{{Structure of Chern-Simons scattering
  amplitudes from topological equivalence theorem and double-copy}},
  \href{https://doi.org/10.1007/JHEP01(2022)153}{\emph{JHEP} {\bfseries 01}
  (2022) 153} [\href{https://arxiv.org/abs/2110.05399}{{\ttfamily
  2110.05399}}].

\bibitem{Ben-Shahar:2021zww}
M.~Ben-Shahar and H.~Johansson, \emph{{Off-Shell Color-Kinematics Duality for
  Chern-Simons}},  \href{https://arxiv.org/abs/2112.11452}{{\ttfamily
  2112.11452}}.

\bibitem{Rosen:2017dvn}
R.~A. Rosen, \emph{{Non-Singular Black Holes in Massive Gravity: Time-Dependent
  Solutions}}, \href{https://doi.org/10.1007/JHEP10(2017)206}{\emph{JHEP}
  {\bfseries 10} (2017) 206}
  [\href{https://arxiv.org/abs/1702.06543}{{\ttfamily 1702.06543}}].

\bibitem{Rosen:2018lki}
R.~A. Rosen, \emph{{Black Hole Mechanics for Massive Gravitons}},
  \href{https://doi.org/10.1103/PhysRevD.98.104008}{\emph{Phys. Rev. D}
  {\bfseries 98} (2018) 104008}
  [\href{https://arxiv.org/abs/1805.12135}{{\ttfamily 1805.12135}}].

\bibitem{Berens:2021tzd}
R.~Berens, L.~Krauth and R.~A. Rosen, \emph{{Gravitational collapse in massive
  gravity in de Sitter spacetime}},
  \href{https://doi.org/10.1103/PhysRevD.105.064057}{\emph{Phys. Rev. D}
  {\bfseries 105} (2022) 064057}
  [\href{https://arxiv.org/abs/2109.10411}{{\ttfamily 2109.10411}}].

\bibitem{Berkovits:2022ivl}
N.~Berkovits, E.~D'Hoker, M.~B. Green, H.~Johansson and O.~Schlotterer,
  \emph{{Snowmass White Paper: String Perturbation Theory}},  in \emph{{2022
  Snowmass Summer Study}}, 3, 2022,
  \href{https://arxiv.org/abs/2203.09099}{{\ttfamily 2203.09099}}.

\bibitem{Boels:2011tp}
R.~H. Boels and R.~S. Isermann, \emph{{New relations for scattering amplitudes
  in Yang-Mills theory at loop level}},
  \href{https://doi.org/10.1103/PhysRevD.85.021701}{\emph{Phys. Rev. D}
  {\bfseries 85} (2012) 021701}
  [\href{https://arxiv.org/abs/1109.5888}{{\ttfamily 1109.5888}}].

\bibitem{He:2015wgf}
S.~He, R.~Monteiro and O.~Schlotterer, \emph{{String-inspired BCJ numerators
  for one-loop MHV amplitudes}},
  \href{https://doi.org/10.1007/JHEP01(2016)171}{\emph{JHEP} {\bfseries 01}
  (2016) 171} [\href{https://arxiv.org/abs/1507.06288}{{\ttfamily
  1507.06288}}].

\bibitem{Bridges:2021ebs}
E.~Bridges and C.~R. Mafra, \emph{{Local BCJ numerators for ten-dimensional SYM
  at one loop}}, \href{https://doi.org/10.1007/JHEP07(2021)031}{\emph{JHEP}
  {\bfseries 07} (2021) 031}
  [\href{https://arxiv.org/abs/2102.12943}{{\ttfamily 2102.12943}}].

\bibitem{Ahmadiniaz:2021fey}
N.~Ahmadiniaz, F.~M. Balli, C.~Lopez-Arcos, A.~Q. Velez and C.~Schubert,
  \emph{{Color-kinematics duality from the Bern-Kosower formalism}},
  \href{https://doi.org/10.1103/PhysRevD.104.L041702}{\emph{Phys. Rev. D}
  {\bfseries 104} (2021) L041702}
  [\href{https://arxiv.org/abs/2105.06745}{{\ttfamily 2105.06745}}].

\bibitem{Ahmadiniaz:2021ayd}
N.~Ahmadiniaz, F.~M. Balli, O.~Corradini, C.~Lopez-Arcos, A.~Q. Velez and
  C.~Schubert, \emph{{Manifest colour-kinematics duality and double-copy in the
  string-based formalism}},
  \href{https://doi.org/10.1016/j.nuclphysb.2022.115690}{\emph{Nucl. Phys. B}
  {\bfseries 975} (2022) 115690}
  [\href{https://arxiv.org/abs/2110.04853}{{\ttfamily 2110.04853}}].

\bibitem{Carrasco:2016ygv}
J.~J.~M. Carrasco, C.~R. Mafra and O.~Schlotterer, \emph{{Semi-abelian
  Z-theory: NLSM$+\phi^{3}$ from the open string}},
  \href{https://doi.org/10.1007/JHEP08(2017)135}{\emph{JHEP} {\bfseries 08}
  (2017) 135} [\href{https://arxiv.org/abs/1612.06446}{{\ttfamily
  1612.06446}}].

\bibitem{Mafra:2014oia}
C.~R. Mafra and O.~Schlotterer, \emph{{Multiparticle SYM equations of motion
  and pure spinor BRST blocks}},
  \href{https://doi.org/10.1007/JHEP07(2014)153}{\emph{JHEP} {\bfseries 07}
  (2014) 153} [\href{https://arxiv.org/abs/1404.4986}{{\ttfamily 1404.4986}}].

\bibitem{Fu:2018hpu}
C.-H. Fu, P.~Vanhove and Y.~Wang, \emph{{A Vertex Operator Algebra Construction
  of the Colour-Kinematics Dual numerator}},
  \href{https://doi.org/10.1007/JHEP09(2018)141}{\emph{JHEP} {\bfseries 09}
  (2018) 141} [\href{https://arxiv.org/abs/1806.09584}{{\ttfamily
  1806.09584}}].

\bibitem{Ben-Shahar:2021doh}
M.~Ben-Shahar and M.~Guillen, \emph{{10D super-Yang-Mills scattering amplitudes
  from its pure spinor action}},
  \href{https://doi.org/10.1007/JHEP12(2021)014}{\emph{JHEP} {\bfseries 12}
  (2021) 014} [\href{https://arxiv.org/abs/2108.11708}{{\ttfamily
  2108.11708}}].

\bibitem{Mizera:2019blq}
S.~Mizera, \emph{{Kinematic Jacobi Identity is a Residue Theorem: Geometry of
  Color-Kinematics Duality for Gauge and Gravity Amplitudes}},
  \href{https://doi.org/10.1103/PhysRevLett.124.141601}{\emph{Phys. Rev. Lett.}
  {\bfseries 124} (2020) 141601}
  [\href{https://arxiv.org/abs/1912.03397}{{\ttfamily 1912.03397}}].

\bibitem{Gomez:2013wza}
H.~Gomez and E.~Y. Yuan, \emph{{N-point tree-level scattering amplitude in the
  new Berkovits` string}},
  \href{https://doi.org/10.1007/JHEP04(2014)046}{\emph{JHEP} {\bfseries 04}
  (2014) 046} [\href{https://arxiv.org/abs/1312.5485}{{\ttfamily 1312.5485}}].

\bibitem{He:2018pol}
S.~He, F.~Teng and Y.~Zhang, \emph{{String amplitudes from field-theory
  amplitudes and vice versa}},
  \href{https://doi.org/10.1103/PhysRevLett.122.211603}{\emph{Phys. Rev. Lett.}
  {\bfseries 122} (2019) 211603}
  [\href{https://arxiv.org/abs/1812.03369}{{\ttfamily 1812.03369}}].

\bibitem{Arkani-Hamed:2017mur}
N.~Arkani-Hamed, Y.~Bai, S.~He and G.~Yan, \emph{{Scattering forms and the
  positive geometry of kinematics, color and the worldsheet}},
  \href{https://doi.org/10.1007/JHEP05(2018)096}{\emph{JHEP} {\bfseries 05}
  (2018) 096} [\href{https://arxiv.org/abs/1711.09102}{{\ttfamily
  1711.09102}}].

\bibitem{Frost:2019fjn}
H.~Frost and L.~Mason, \emph{{Lie Polynomials and a Twistorial Correspondence
  for Amplitudes}},  \href{https://arxiv.org/abs/1912.04198}{{\ttfamily
  1912.04198}}.

\bibitem{Cachazo:2015aol}
F.~Cachazo, S.~He and E.~Y. Yuan, \emph{{One-Loop Corrections from Higher
  Dimensional Tree Amplitudes}},
  \href{https://doi.org/10.1007/JHEP08(2016)008}{\emph{JHEP} {\bfseries 08}
  (2016) 008} [\href{https://arxiv.org/abs/1512.05001}{{\ttfamily
  1512.05001}}].

\bibitem{Gomez:2013sla}
H.~Gomez and C.~R. Mafra, \emph{{The closed-string 3-loop amplitude and
  S-duality}}, \href{https://doi.org/10.1007/JHEP10(2013)217}{\emph{JHEP}
  {\bfseries 10} (2013) 217} [\href{https://arxiv.org/abs/1308.6567}{{\ttfamily
  1308.6567}}].

\bibitem{Edison:2021ebi}
A.~Edison, M.~Guillen, H.~Johansson, O.~Schlotterer and F.~Teng,
  \emph{{One-loop matrix elements of effective superstring interactions:
  \ensuremath{\alpha}'-expanding loop integrands}},
  \href{https://doi.org/10.1007/JHEP12(2021)007}{\emph{JHEP} {\bfseries 12}
  (2021) 007} [\href{https://arxiv.org/abs/2107.08009}{{\ttfamily
  2107.08009}}].

\bibitem{Broedel:2013tta}
J.~Broedel, O.~Schlotterer and S.~Stieberger, \emph{{Polylogarithms, Multiple
  Zeta Values and Superstring Amplitudes}},
  \href{https://doi.org/10.1002/prop.201300019}{\emph{Fortsch. Phys.}
  {\bfseries 61} (2013) 812} [\href{https://arxiv.org/abs/1304.7267}{{\ttfamily
  1304.7267}}].

\bibitem{Mafra:2016mcc}
C.~R. Mafra and O.~Schlotterer, \emph{{Non-abelian $Z$-theory: Berends-Giele
  recursion for the $\alpha'$-expansion of disk integrals}},
  \href{https://doi.org/10.1007/JHEP01(2017)031}{\emph{JHEP} {\bfseries 01}
  (2017) 031} [\href{https://arxiv.org/abs/1609.07078}{{\ttfamily
  1609.07078}}].

\bibitem{Schlotterer:2012ny}
O.~Schlotterer and S.~Stieberger, \emph{{Motivic Multiple Zeta Values and
  Superstring Amplitudes}},
  \href{https://doi.org/10.1088/1751-8113/46/47/475401}{\emph{J. Phys. A}
  {\bfseries 46} (2013) 475401}
  [\href{https://arxiv.org/abs/1205.1516}{{\ttfamily 1205.1516}}].

\bibitem{Stieberger:2013wea}
S.~Stieberger, \emph{{Closed superstring amplitudes, single-valued multiple
  zeta values and the Deligne associator}},
  \href{https://doi.org/10.1088/1751-8113/47/15/155401}{\emph{J. Phys.}
  {\bfseries A47} (2014) 155401}
  [\href{https://arxiv.org/abs/1310.3259}{{\ttfamily 1310.3259}}].

\bibitem{Schnetz:2013hqa}
O.~Schnetz, \emph{{Graphical functions and single-valued multiple
  polylogarithms}},
  \href{https://doi.org/10.4310/CNTP.2014.v8.n4.a1}{\emph{Commun. Num. Theor.
  Phys.} {\bfseries 08} (2014) 589}
  [\href{https://arxiv.org/abs/1302.6445}{{\ttfamily 1302.6445}}].

\bibitem{Brown:2013gia}
F.~Brown, \emph{{Single-valued Motivic Periods and Multiple Zeta Values}},
  \href{https://doi.org/10.1017/fms.2014.18}{\emph{SIGMA} {\bfseries 2} (2014)
  e25} [\href{https://arxiv.org/abs/1309.5309}{{\ttfamily 1309.5309}}].

\bibitem{Stieberger:2014hba}
S.~Stieberger and T.~Taylor, \emph{{Closed String Amplitudes as Single-Valued
  Open String Amplitudes}},
  \href{https://doi.org/10.1016/j.nuclphysb.2014.02.005}{\emph{Nucl. Phys.}
  {\bfseries B881} (2014) 269}
  [\href{https://arxiv.org/abs/1401.1218}{{\ttfamily 1401.1218}}].

\bibitem{Schlotterer:2018zce}
O.~Schlotterer and O.~Schnetz, \emph{{Closed strings as single-valued open
  strings: A genus-zero derivation}},
  \href{https://doi.org/10.1088/1751-8121/aaea14}{\emph{J. Phys. A} {\bfseries
  52} (2019) 045401} [\href{https://arxiv.org/abs/1808.00713}{{\ttfamily
  1808.00713}}].

\bibitem{Vanhove:2018elu}
P.~Vanhove and F.~Zerbini, \emph{{Single-valued hyperlogarithms, correlation
  functions and closed string amplitudes}},
  \href{https://arxiv.org/abs/1812.03018}{{\ttfamily 1812.03018}}.

\bibitem{Brown:2019wna}
F.~Brown and C.~Dupont, \emph{{Single-valued integration and superstring
  amplitudes in genus zero}},
  \href{https://doi.org/10.1007/s00220-021-03969-4}{\emph{Commun. Math. Phys.}
  {\bfseries 382} (2021) 815}
  [\href{https://arxiv.org/abs/1910.01107}{{\ttfamily 1910.01107}}].

\bibitem{Huang:2016tag}
Y.-t. Huang, O.~Schlotterer and C.~Wen, \emph{{Universality in string
  interactions}}, \href{https://doi.org/10.1007/JHEP09(2016)155}{\emph{JHEP}
  {\bfseries 09} (2016) 155}
  [\href{https://arxiv.org/abs/1602.01674}{{\ttfamily 1602.01674}}].

\bibitem{Azevedo:2018dgo}
T.~Azevedo, M.~Chiodaroli, H.~Johansson and O.~Schlotterer, \emph{{Heterotic
  and bosonic string amplitudes via field theory}},
  \href{https://doi.org/10.1007/JHEP10(2018)012}{\emph{JHEP} {\bfseries 10}
  (2018) 012} [\href{https://arxiv.org/abs/1803.05452}{{\ttfamily
  1803.05452}}].

\bibitem{Johansson:2017srf}
H.~Johansson and J.~Nohle, \emph{{Conformal Gravity from Gauge Theory}},
  \href{https://arxiv.org/abs/1707.02965}{{\ttfamily 1707.02965}}.

\bibitem{Chiodaroli:2014xia}
M.~Chiodaroli, M.~G\"unaydin, H.~Johansson and R.~Roiban, \emph{{Scattering
  amplitudes in $ \mathcal{N}=2 $ Maxwell-Einstein and Yang-Mills/Einstein
  supergravity}}, \href{https://doi.org/10.1007/JHEP01(2015)081}{\emph{JHEP}
  {\bfseries 01} (2015) 081} [\href{https://arxiv.org/abs/1408.0764}{{\ttfamily
  1408.0764}}].

\bibitem{Broedel:2013aza}
J.~Broedel, O.~Schlotterer, S.~Stieberger and T.~Terasoma, \emph{{All order
  $\alpha^{\prime}$-expansion of superstring trees from the Drinfeld
  associator}}, \href{https://doi.org/10.1103/PhysRevD.89.066014}{\emph{Phys.
  Rev. D} {\bfseries 89} (2014) 066014}
  [\href{https://arxiv.org/abs/1304.7304}{{\ttfamily 1304.7304}}].

\bibitem{Broedel:2012rc}
J.~Broedel and L.~J. Dixon, \emph{{Color-kinematics duality and double-copy
  construction for amplitudes from higher-dimension operators}},
  \href{https://doi.org/10.1007/JHEP10(2012)091}{\emph{JHEP} {\bfseries 10}
  (2012) 091} [\href{https://arxiv.org/abs/1208.0876}{{\ttfamily 1208.0876}}].

\bibitem{Brown:2011ik}
F.~C.~S. Brown, \emph{On the decomposition of motivic multiple zeta values},
  in \emph{Galois-{T}eichm\"uller theory and arithmetic geometry}, vol.~63 of
  \emph{Adv. Stud. Pure Math.}, pp.~31--58, Math. Soc. Japan, Tokyo, (2012),
  \href{https://arxiv.org/abs/1102.1310}{{\ttfamily 1102.1310}}.

\bibitem{Guillen:2021mwp}
M.~Guillen, H.~Johansson, R.~L. Jusinskas and O.~Schlotterer, \emph{{Scattering
  Massive String Resonances through Field-Theory Methods}},
  \href{https://doi.org/10.1103/PhysRevLett.127.051601}{\emph{Phys. Rev. Lett.}
  {\bfseries 127} (2021) 051601}
  [\href{https://arxiv.org/abs/2104.03314}{{\ttfamily 2104.03314}}].

\bibitem{Hohm:2013jaa}
O.~Hohm, W.~Siegel and B.~Zwiebach, \emph{{Doubled $\alpha'$-geometry}},
  \href{https://doi.org/10.1007/JHEP02(2014)065}{\emph{JHEP} {\bfseries 02}
  (2014) 065} [\href{https://arxiv.org/abs/1306.2970}{{\ttfamily 1306.2970}}].

\bibitem{Huang:2016bdd}
Y.-t. Huang, W.~Siegel and E.~Y. Yuan, \emph{{Factorization of Chiral String
  Amplitudes}}, \href{https://doi.org/10.1007/JHEP09(2016)101}{\emph{JHEP}
  {\bfseries 09} (2016) 101}
  [\href{https://arxiv.org/abs/1603.02588}{{\ttfamily 1603.02588}}].

\bibitem{Ferrara:2018wlb}
S.~Ferrara, A.~Kehagias and D.~L\"ust, \emph{{Bimetric, Conformal Supergravity
  and its Superstring Embedding}},
  \href{https://doi.org/10.1007/JHEP05(2019)100}{\emph{JHEP} {\bfseries 05}
  (2019) 100} [\href{https://arxiv.org/abs/1810.08147}{{\ttfamily
  1810.08147}}].

\bibitem{Lust:2021jps}
D.~Lust, C.~Markou, P.~Mazloumi and S.~Stieberger, \emph{{Extracting bigravity
  from string theory}},
  \href{https://doi.org/10.1007/JHEP12(2021)220}{\emph{JHEP} {\bfseries 12}
  (2021) 220} [\href{https://arxiv.org/abs/2106.04614}{{\ttfamily
  2106.04614}}].

\bibitem{Jusinskas:2021bdj}
R.~L. Jusinskas, \emph{{Asymmetrically twisted strings}},
  \href{https://arxiv.org/abs/2108.13426}{{\ttfamily 2108.13426}}.

\bibitem{Mafra:2017ioj}
C.~Mafra and O.~Schlotterer, \emph{{Double-Copy Structure of One-Loop
  Open-String Amplitudes}},
  \href{https://doi.org/10.1103/PhysRevLett.121.011601}{\emph{Phys. Rev. Lett.}
  {\bfseries 121} (2018) 011601}
  [\href{https://arxiv.org/abs/1711.09104}{{\ttfamily 1711.09104}}].

\bibitem{Zerbini:2015rss}
F.~Zerbini, \emph{{Single-valued multiple zeta values in genus 1 superstring
  amplitudes}},
  \href{https://doi.org/10.4310/CNTP.2016.v10.n4.a2}{\emph{Commun. Num. Theor.
  Phys.} {\bfseries 10} (2016) 703}
  [\href{https://arxiv.org/abs/1512.05689}{{\ttfamily 1512.05689}}].

\bibitem{DHoker:2015wxz}
E.~D'Hoker, M.~B. Green, O.~G\"urdogan and P.~Vanhove, \emph{{Modular Graph
  Functions}}, \href{https://doi.org/10.4310/CNTP.2017.v11.n1.a4}{\emph{Commun.
  Num. Theor. Phys.} {\bfseries 11} (2017) 165}
  [\href{https://arxiv.org/abs/1512.06779}{{\ttfamily 1512.06779}}].

\bibitem{Broedel:2018izr}
J.~Broedel, O.~Schlotterer and F.~Zerbini, \emph{{From elliptic multiple zeta
  values to modular graph functions: open and closed strings at one loop}},
  \href{https://doi.org/10.1007/JHEP01(2019)155}{\emph{JHEP} {\bfseries 01}
  (2019) 155} [\href{https://arxiv.org/abs/1803.00527}{{\ttfamily
  1803.00527}}].

\bibitem{Zagier:2019eus}
D.~Zagier and F.~Zerbini, \emph{{Genus-zero and genus-one string amplitudes and
  special multiple zeta values}},
  \href{https://doi.org/10.4310/CNTP.2020.v14.n2.a4}{\emph{Commun. Num. Theor.
  Phys.} {\bfseries 14} (2020) 413}
  [\href{https://arxiv.org/abs/1906.12339}{{\ttfamily 1906.12339}}].

\bibitem{Gerken:2020xfv}
J.~E. Gerken, A.~Kleinschmidt, C.~R. Mafra, O.~Schlotterer and B.~Verbeek,
  \emph{{Towards closed strings as single-valued open strings at genus one}},
  \href{https://doi.org/10.1088/1751-8121/abe58b}{\emph{J. Phys. A} {\bfseries
  55} (2022) 025401} [\href{https://arxiv.org/abs/2010.10558}{{\ttfamily
  2010.10558}}].

\bibitem{Johansson:2014zca}
H.~Johansson and A.~Ochirov, \emph{{Pure Gravities via Color-Kinematics Duality
  for Fundamental Matter}},
  \href{https://doi.org/10.1007/JHEP11(2015)046}{\emph{JHEP} {\bfseries 11}
  (2015) 046} [\href{https://arxiv.org/abs/1407.4772}{{\ttfamily 1407.4772}}].

\bibitem{Johansson:2015oia}
H.~Johansson and A.~Ochirov, \emph{{Color-Kinematics Duality for QCD
  Amplitudes}}, \href{https://doi.org/10.1007/JHEP01(2016)170}{\emph{JHEP}
  {\bfseries 01} (2016) 170}
  [\href{https://arxiv.org/abs/1507.00332}{{\ttfamily 1507.00332}}].

\bibitem{He:2016dol}
S.~He and Y.~Zhang, \emph{{Connected formulas for amplitudes in standard
  model}}, \href{https://doi.org/10.1007/JHEP03(2017)093}{\emph{JHEP}
  {\bfseries 03} (2017) 093}
  [\href{https://arxiv.org/abs/1607.02843}{{\ttfamily 1607.02843}}].

\bibitem{Brown:2018wss}
R.~W. Brown and S.~G. Naculich, \emph{{KLT-type relations for QCD and bicolor
  amplitudes from color-factor symmetry}},
  \href{https://doi.org/10.1007/JHEP03(2018)057}{\emph{JHEP} {\bfseries 03}
  (2018) 057} [\href{https://arxiv.org/abs/1802.01620}{{\ttfamily
  1802.01620}}].

\bibitem{Naculich:2014naa}
S.~G. Naculich, \emph{{Scattering equations and BCJ relations for gauge and
  gravitational amplitudes with massive scalar particles}},
  \href{https://doi.org/10.1007/JHEP09(2014)029}{\emph{JHEP} {\bfseries 09}
  (2014) 029} [\href{https://arxiv.org/abs/1407.7836}{{\ttfamily 1407.7836}}].

\bibitem{Naculich:2015zha}
S.~G. Naculich, \emph{{CHY representations for gauge theory and gravity
  amplitudes with up to three massive particles}},
  \href{https://doi.org/10.1007/JHEP05(2015)050}{\emph{JHEP} {\bfseries 05}
  (2015) 050} [\href{https://arxiv.org/abs/1501.03500}{{\ttfamily
  1501.03500}}].

\bibitem{Naculich:2015coa}
S.~G. Naculich, \emph{{Amplitudes for massive vector and scalar bosons in
  spontaneously-broken gauge theory from the CHY representation}},
  \href{https://doi.org/10.1007/JHEP09(2015)122}{\emph{JHEP} {\bfseries 09}
  (2015) 122} [\href{https://arxiv.org/abs/1506.06134}{{\ttfamily
  1506.06134}}].

\bibitem{Momeni:2020vvr}
A.~Momeni, J.~Rumbutis and A.~J. Tolley, \emph{{Massive Gravity from Double
  Copy}}, \href{https://doi.org/10.1007/JHEP12(2020)030}{\emph{JHEP} {\bfseries
  12} (2020) 030} [\href{https://arxiv.org/abs/2004.07853}{{\ttfamily
  2004.07853}}].

\bibitem{Chiodaroli:2015rdg}
M.~Chiodaroli, M.~Gunaydin, H.~Johansson and R.~Roiban, \emph{{Spontaneously
  Broken Yang-Mills-Einstein Supergravities as Double Copies}},
  \href{https://doi.org/10.1007/JHEP06(2017)064}{\emph{JHEP} {\bfseries 06}
  (2017) 064} [\href{https://arxiv.org/abs/1511.01740}{{\ttfamily
  1511.01740}}].

\bibitem{Chiodaroli:2017ehv}
M.~Chiodaroli, M.~Gunaydin, H.~Johansson and R.~Roiban, \emph{{Gauged
  Supergravities and Spontaneous Supersymmetry Breaking from the Double Copy
  Construction}},
  \href{https://doi.org/10.1103/PhysRevLett.120.171601}{\emph{Phys. Rev. Lett.}
  {\bfseries 120} (2018) 171601}
  [\href{https://arxiv.org/abs/1710.08796}{{\ttfamily 1710.08796}}].

\bibitem{Gonzalez:2022mpa}
M.~C. Gonz\'alez, Q.~Liang and M.~Trodden, \emph{{Double Copy for Massive
  Scalar Field Theories}},  \href{https://arxiv.org/abs/2202.00620}{{\ttfamily
  2202.00620}}.

\bibitem{Li:2021yfk}
Y.~Li, Y.-F. Hang, H.-J. He and S.~He, \emph{{Scattering amplitudes of
  Kaluza-Klein strings and extended massive double-copy}},
  \href{https://doi.org/10.1007/JHEP02(2022)120}{\emph{JHEP} {\bfseries 02}
  (2022) 120} [\href{https://arxiv.org/abs/2111.12042}{{\ttfamily
  2111.12042}}].

\bibitem{Carrasco:2019yyn}
J.~J.~M. Carrasco, L.~Rodina, Z.~Yin and S.~Zekioglu, \emph{{Simple encoding of
  higher derivative gauge and gravity counterterms}},
  \href{https://doi.org/10.1103/PhysRevLett.125.251602}{\emph{Phys. Rev. Lett.}
  {\bfseries 125} (2020) 251602}
  [\href{https://arxiv.org/abs/1910.12850}{{\ttfamily 1910.12850}}].

\bibitem{Carrasco:2021ptp}
J.~J.~M. Carrasco, L.~Rodina and S.~Zekioglu, \emph{{Composing effective
  prediction at five points}},
  \href{https://doi.org/10.1007/JHEP06(2021)169}{\emph{JHEP} {\bfseries 06}
  (2021) 169} [\href{https://arxiv.org/abs/2104.08370}{{\ttfamily
  2104.08370}}].

\bibitem{Chi:2021mio}
H.-H. Chi, H.~Elvang, A.~Herderschee, C.~R.~T. Jones and S.~Paranjape,
  \emph{{Generalizations of the double-copy: the KLT bootstrap}},
  \href{https://doi.org/10.1007/JHEP03(2022)077}{\emph{JHEP} {\bfseries 03}
  (2022) 077} [\href{https://arxiv.org/abs/2106.12600}{{\ttfamily
  2106.12600}}].

\bibitem{Mizera:2016jhj}
S.~Mizera, \emph{{Inverse of the String Theory KLT Kernel}},
  \href{https://doi.org/10.1007/JHEP06(2017)084}{\emph{JHEP} {\bfseries 06}
  (2017) 084} [\href{https://arxiv.org/abs/1610.04230}{{\ttfamily
  1610.04230}}].

\bibitem{ACHEtoappear}
A.~S.-K. Chen and H.~Elvang{\emph{, to appear$\! \!$} }.

\bibitem{Elvang:2021qhq}
H.~Elvang and M.~D. Mitchell, \emph{{On Extended Supersymmetry of 4d Galileons
  and 3-Brane Effective Actions}},
  \href{https://arxiv.org/abs/2111.12686}{{\ttfamily 2111.12686}}.

\bibitem{Farakos:2013fne}
F.~Farakos, C.~Germani and A.~Kehagias, \emph{{On ghost-free supersymmetric
  galileons}}, \href{https://doi.org/10.1007/JHEP11(2013)045}{\emph{JHEP}
  {\bfseries 11} (2013) 045} [\href{https://arxiv.org/abs/1306.2961}{{\ttfamily
  1306.2961}}].

\bibitem{Elvang:2017mdq}
H.~Elvang, M.~Hadjiantonis, C.~R.~T. Jones and S.~Paranjape, \emph{{On the
  Supersymmetrization of Galileon Theories in Four Dimensions}},
  \href{https://doi.org/10.1016/j.physletb.2018.04.032}{\emph{Phys. Lett. B}
  {\bfseries 781} (2018) 656}
  [\href{https://arxiv.org/abs/1712.09937}{{\ttfamily 1712.09937}}].

\bibitem{Elvang:2018dco}
H.~Elvang, M.~Hadjiantonis, C.~R.~T. Jones and S.~Paranjape, \emph{{Soft
  Bootstrap and Supersymmetry}},
  \href{https://doi.org/10.1007/JHEP01(2019)195}{\emph{JHEP} {\bfseries 01}
  (2019) 195} [\href{https://arxiv.org/abs/1806.06079}{{\ttfamily
  1806.06079}}].

\bibitem{Elvang:2019twd}
H.~Elvang, M.~Hadjiantonis, C.~R.~T. Jones and S.~Paranjape,
  \emph{{All-Multiplicity One-Loop Amplitudes in Born-Infeld Electrodynamics
  from Generalized Unitarity}},
  \href{https://doi.org/10.1007/JHEP03(2020)009}{\emph{JHEP} {\bfseries 03}
  (2020) 009} [\href{https://arxiv.org/abs/1906.05321}{{\ttfamily
  1906.05321}}].

\bibitem{Elvang:2020kuj}
H.~Elvang, M.~Hadjiantonis, C.~R.~T. Jones and S.~Paranjape,
  \emph{{Electromagnetic Duality and D3-Brane Scattering Amplitudes Beyond
  Leading Order}}, \href{https://doi.org/10.1007/JHEP04(2021)173}{\emph{JHEP}
  {\bfseries 04} (2021) 173}
  [\href{https://arxiv.org/abs/2006.08928}{{\ttfamily 2006.08928}}].

\bibitem{Bern:2017rjw}
Z.~Bern, J.~Parra-Martinez and R.~Roiban, \emph{{Canceling the U(1) Anomaly in
  the $S$ Matrix of $N$=4 Supergravity}},
  \href{https://doi.org/10.1103/PhysRevLett.121.101604}{\emph{Phys. Rev. Lett.}
  {\bfseries 121} (2018) 101604}
  [\href{https://arxiv.org/abs/1712.03928}{{\ttfamily 1712.03928}}].

\bibitem{Bonnefoy:2021qgu}
Q.~Bonnefoy, G.~Durieux, C.~Grojean, C.~S. Machado and J.~Roosmale~Nepveu,
  \emph{{The seeds of EFT double copy}},
  \href{https://arxiv.org/abs/2112.11453}{{\ttfamily 2112.11453}}.

\bibitem{Bern:2017puu}
Z.~Bern, H.-H. Chi, L.~Dixon and A.~Edison, \emph{{Two-Loop Renormalization of
  Quantum Gravity Simplified}},
  \href{https://doi.org/10.1103/PhysRevD.95.046013}{\emph{Phys. Rev. D}
  {\bfseries 95} (2017) 046013}
  [\href{https://arxiv.org/abs/1701.02422}{{\ttfamily 1701.02422}}].

\bibitem{Bern:2017tuc}
Z.~Bern, A.~Edison, D.~Kosower and J.~Parra-Martinez, \emph{{Curvature-squared
  multiplets, evanescent effects, and the U(1) anomaly in $N=4$ supergravity}},
  \href{https://doi.org/10.1103/PhysRevD.96.066004}{\emph{Phys. Rev.}
  {\bfseries D96} (2017) 066004}
  [\href{https://arxiv.org/abs/1706.01486}{{\ttfamily 1706.01486}}].

\bibitem{Bern:2021ppb}
Z.~Bern, D.~Kosmopoulos and A.~Zhiboedov, \emph{{Gravitational effective field
  theory islands, low-spin dominance, and the four-graviton amplitude}},
  \href{https://doi.org/10.1088/1751-8121/ac0e51}{\emph{J. Phys. A} {\bfseries
  54} (2021) 344002} [\href{https://arxiv.org/abs/2103.12728}{{\ttfamily
  2103.12728}}].

\bibitem{Low:2019wuv}
I.~Low and Z.~Yin, \emph{{New Flavor-Kinematics Dualities and Extensions of
  Nonlinear Sigma Models}},
  \href{https://doi.org/10.1016/j.physletb.2020.135544}{\emph{Phys. Lett. B}
  {\bfseries 807} (2020) 135544}
  [\href{https://arxiv.org/abs/1911.08490}{{\ttfamily 1911.08490}}].

\bibitem{Low:2020ubn}
I.~Low, L.~Rodina and Z.~Yin, \emph{{Double Copy in Higher Derivative Operators
  of Nambu-Goldstone Bosons}},
  \href{https://doi.org/10.1103/PhysRevD.103.025004}{\emph{Phys. Rev. D}
  {\bfseries 103} (2021) 025004}
  [\href{https://arxiv.org/abs/2009.00008}{{\ttfamily 2009.00008}}].

\bibitem{Elvang:2010jv}
H.~Elvang, D.~Z. Freedman and M.~Kiermaier, \emph{{A simple approach to
  counterterms in N=8 supergravity}},
  \href{https://doi.org/10.1007/JHEP11(2010)016}{\emph{JHEP} {\bfseries 11}
  (2010) 016} [\href{https://arxiv.org/abs/1003.5018}{{\ttfamily 1003.5018}}].

\bibitem{Elvang:2010kc}
H.~Elvang and M.~Kiermaier, \emph{{Stringy KLT relations, global symmetries,
  and $E_{7(7)}$ violation}},
  \href{https://doi.org/10.1007/JHEP10(2010)108}{\emph{JHEP} {\bfseries 10}
  (2010) 108} [\href{https://arxiv.org/abs/1007.4813}{{\ttfamily 1007.4813}}].

\bibitem{Beisert:2010jx}
N.~Beisert, H.~Elvang, D.~Z. Freedman, M.~Kiermaier, A.~Morales and
  S.~Stieberger, \emph{{E7(7) constraints on counterterms in N=8
  supergravity}},
  \href{https://doi.org/10.1016/j.physletb.2010.09.069}{\emph{Phys. Lett. B}
  {\bfseries 694} (2011) 265}
  [\href{https://arxiv.org/abs/1009.1643}{{\ttfamily 1009.1643}}].

\bibitem{Elvang:2010xn}
H.~Elvang, D.~Z. Freedman and M.~Kiermaier, \emph{{SUSY Ward identities,
  Superamplitudes, and Counterterms}},
  \href{https://doi.org/10.1088/1751-8113/44/45/454009}{\emph{J. Phys. A}
  {\bfseries 44} (2011) 454009}
  [\href{https://arxiv.org/abs/1012.3401}{{\ttfamily 1012.3401}}].

\bibitem{Bossard:2011tq}
G.~Bossard, P.~S. Howe, K.~S. Stelle and P.~Vanhove, \emph{{The vanishing
  volume of D=4 superspace}},
  \href{https://doi.org/10.1088/0264-9381/28/21/215005}{\emph{Class. Quant.
  Grav.} {\bfseries 28} (2011) 215005}
  [\href{https://arxiv.org/abs/1105.6087}{{\ttfamily 1105.6087}}].

\bibitem{Bern:2017lpv}
Z.~Bern, M.~Enciso, J.~Parra-Martinez and M.~Zeng, \emph{{Manifesting enhanced
  cancellations in supergravity: integrands versus integrals}},
  \href{https://doi.org/10.1007/JHEP05(2017)137}{\emph{JHEP} {\bfseries 05}
  (2017) 137} [\href{https://arxiv.org/abs/1703.08927}{{\ttfamily
  1703.08927}}].

\bibitem{Herrmann:2018dja}
E.~Herrmann and J.~Trnka, \emph{{UV cancellations in gravity loop integrands}},
  \href{https://doi.org/10.1007/JHEP02(2019)084}{\emph{JHEP} {\bfseries 02}
  (2019) 084} [\href{https://arxiv.org/abs/1808.10446}{{\ttfamily
  1808.10446}}].

\bibitem{Bourjaily:2018omh}
J.~L. Bourjaily, E.~Herrmann and J.~Trnka, \emph{{Maximally supersymmetric
  amplitudes at infinite loop momentum}},
  \href{https://doi.org/10.1103/PhysRevD.99.066006}{\emph{Phys. Rev. D}
  {\bfseries 99} (2019) 066006}
  [\href{https://arxiv.org/abs/1812.11185}{{\ttfamily 1812.11185}}].

\bibitem{Edison:2019ovj}
A.~Edison, E.~Herrmann, J.~Parra-Martinez and J.~Trnka, \emph{{Gravity loop
  integrands from the ultraviolet}},
  \href{https://doi.org/10.21468/SciPostPhys.10.1.016}{\emph{SciPost Phys.}
  {\bfseries 10} (2021) 016}
  [\href{https://arxiv.org/abs/1909.02003}{{\ttfamily 1909.02003}}].

\bibitem{Carrasco:2021otn}
J.~J.~M. Carrasco, A.~Edison and H.~Johansson, \emph{{Maximal Super-Yang-Mills
  at Six Loops via Novel Integrand Bootstrap}},
  \href{https://arxiv.org/abs/2112.05178}{{\ttfamily 2112.05178}}.

\bibitem{Carrasco:2013ypa}
J.~J.~M. Carrasco, R.~Kallosh, R.~Roiban and A.~A. Tseytlin, \emph{{On the U(1)
  duality anomaly and the S-matrix of N=4 supergravity}},
  \href{https://doi.org/10.1007/JHEP07(2013)029}{\emph{JHEP} {\bfseries 07}
  (2013) 029} [\href{https://arxiv.org/abs/1303.6219}{{\ttfamily 1303.6219}}].

\bibitem{Bern:2019isl}
Z.~Bern, D.~Kosower and J.~Parra-Martinez, \emph{{Two-loop n-point anomalous
  amplitudes in N=4 supergravity}},
  \href{https://doi.org/10.1098/rspa.2019.0722}{\emph{Proc. Roy. Soc. Lond. A}
  {\bfseries 476} (2020) 20190722}
  [\href{https://arxiv.org/abs/1905.05151}{{\ttfamily 1905.05151}}].

\bibitem{Carrasco:2022lbm}
J.~J.~M. Carrasco, M.~Lewandowski and N.~H. Pavao, \emph{{The color-dual fate
  of N=4 supergravity}},  \href{https://arxiv.org/abs/2203.03592}{{\ttfamily
  2203.03592}}.

\bibitem{DeLaurentis:2022otd}
G.~De~Laurentis and B.~Page, \emph{{Ans\"atze for Scattering Amplitudes from
  $p$-adic Numbers and Algebraic Geometry}},
  \href{https://arxiv.org/abs/2203.04269}{{\ttfamily 2203.04269}}.

\bibitem{Johansson:2019dnu}
H.~Johansson and A.~Ochirov, \emph{{Double copy for massive quantum particles
  with spin}}, \href{https://doi.org/10.1007/JHEP09(2019)040}{\emph{JHEP}
  {\bfseries 09} (2019) 040}
  [\href{https://arxiv.org/abs/1906.12292}{{\ttfamily 1906.12292}}].

\bibitem{Plefka:2019wyg}
J.~Plefka, C.~Shi and T.~Wang, \emph{{Double copy of massive scalar QCD}},
  \href{https://doi.org/10.1103/PhysRevD.101.066004}{\emph{Phys. Rev. D}
  {\bfseries 101} (2020) 066004}
  [\href{https://arxiv.org/abs/1911.06785}{{\ttfamily 1911.06785}}].

\bibitem{Johansson:2017bfl}
H.~Johansson, G.~K\"alin and G.~Mogull, \emph{{Two-loop supersymmetric QCD and
  half-maximal supergravity amplitudes}},
  \href{https://doi.org/10.1007/JHEP09(2017)019}{\emph{JHEP} {\bfseries 09}
  (2017) 019} [\href{https://arxiv.org/abs/1706.09381}{{\ttfamily
  1706.09381}}].

\bibitem{Kalin:2018thp}
G.~K\"alin, G.~Mogull and A.~Ochirov, \emph{{Two-loop $ \mathcal{N} $ = 2 SQCD
  amplitudes with external matter from iterated cuts}},
  \href{https://doi.org/10.1007/JHEP07(2019)120}{\emph{JHEP} {\bfseries 07}
  (2019) 120} [\href{https://arxiv.org/abs/1811.09604}{{\ttfamily
  1811.09604}}].

\bibitem{Kalin:2019vjc}
G.~K\"alin, G.~Mogull, A.~Ochirov and B.~Verbeek, \emph{{Infrared and
  transcendental structure of two-loop supersymmetric QCD amplitudes}},
  \href{https://doi.org/10.1007/JHEP01(2020)068}{\emph{JHEP} {\bfseries 01}
  (2020) 068} [\href{https://arxiv.org/abs/1911.10218}{{\ttfamily
  1911.10218}}].

\bibitem{Duhr:2019ywc}
C.~Duhr, H.~Johansson, G.~K\"alin, G.~Mogull and B.~Verbeek, \emph{{Full-Color
  Two-Loop Four-Gluon Amplitude in $\mathcal{N}$=2 Supersymmetric QCD}},
  \href{https://doi.org/10.1103/PhysRevLett.123.241601}{\emph{Phys. Rev. Lett.}
  {\bfseries 123} (2019) 241601}
  [\href{https://arxiv.org/abs/1904.05299}{{\ttfamily 1904.05299}}].

\bibitem{Gehrmann:2021qex}
T.~Gehrmann and B.~Malaescu, \emph{{Precision QCD Physics at the LHC}},
  \href{https://arxiv.org/abs/2111.02319}{{\ttfamily 2111.02319}}.

\bibitem{Caola:2022ayt}
F.~Caola, W.~Chen, C.~Duhr, X.~Liu, B.~Mistlberger, F.~Petriello et~al.,
  \emph{{The Path forward to N$^3$LO}},  in \emph{{2022 Snowmass Summer
  Study}}, 3, 2022, \href{https://arxiv.org/abs/2203.06730}{{\ttfamily
  2203.06730}}.

\bibitem{Weinzierl:2022eaz}
S.~Weinzierl, \emph{{Feynman Integrals}},
  \href{https://arxiv.org/abs/2201.03593}{{\ttfamily 2201.03593}}.

\bibitem{Bourjaily:2022bwx}
J.~L. Bourjaily et~al., \emph{{Functions Beyond Multiple Polylogarithms for
  Precision Collider Physics}},  in \emph{{2022 Snowmass Summer Study}}, 3,
  2022, \href{https://arxiv.org/abs/2203.07088}{{\ttfamily 2203.07088}}.

\bibitem{Abreu:2022mfk}
S.~Abreu, R.~Britto and C.~Duhr, \emph{{The SAGEX Review on Scattering
  Amplitudes, Chapter 3: Mathematical structures in Feynman integrals}},
  \href{https://arxiv.org/abs/2203.13014}{{\ttfamily 2203.13014}}.

\bibitem{Blumlein:2022zkr}
J.~Bl\"umlein and C.~Schneider, \emph{{The SAGEX Review on Scattering
  Amplitudes, Chapter 4: Multi-loop Feynman Integrals}},
  \href{https://arxiv.org/abs/2203.13015}{{\ttfamily 2203.13015}}.

\bibitem{Bern:1994zx}
Z.~Bern, L.~J. Dixon, D.~C. Dunbar and D.~A. Kosower, \emph{{One loop n point
  gauge theory amplitudes, unitarity and collinear limits}},
  \href{https://doi.org/10.1016/0550-3213(94)90179-1}{\emph{Nucl. Phys. B}
  {\bfseries 425} (1994) 217}
  [\href{https://arxiv.org/abs/hep-ph/9403226}{{\ttfamily hep-ph/9403226}}].

\bibitem{Bern:1994cg}
Z.~Bern, L.~J. Dixon, D.~C. Dunbar and D.~A. Kosower, \emph{{Fusing gauge
  theory tree amplitudes into loop amplitudes}},
  \href{https://doi.org/10.1016/0550-3213(94)00488-Z}{\emph{Nucl. Phys. B}
  {\bfseries 435} (1995) 59}
  [\href{https://arxiv.org/abs/hep-ph/9409265}{{\ttfamily hep-ph/9409265}}].

\bibitem{Bern:1995db}
Z.~Bern and A.~G. Morgan, \emph{{Massive loop amplitudes from unitarity}},
  \href{https://doi.org/10.1016/0550-3213(96)00078-8}{\emph{Nucl. Phys. B}
  {\bfseries 467} (1996) 479}
  [\href{https://arxiv.org/abs/hep-ph/9511336}{{\ttfamily hep-ph/9511336}}].

\bibitem{Bern:1997sc}
Z.~Bern, L.~J. Dixon and D.~A. Kosower, \emph{{One loop amplitudes for e+ e- to
  four partons}},
  \href{https://doi.org/10.1016/S0550-3213(97)00703-7}{\emph{Nucl. Phys. B}
  {\bfseries 513} (1998) 3}
  [\href{https://arxiv.org/abs/hep-ph/9708239}{{\ttfamily hep-ph/9708239}}].

\bibitem{Britto:2004nc}
R.~Britto, F.~Cachazo and B.~Feng, \emph{{Generalized unitarity and one-loop
  amplitudes in N=4 super-Yang-Mills}},
  \href{https://doi.org/10.1016/j.nuclphysb.2005.07.014}{\emph{Nucl. Phys. B}
  {\bfseries 725} (2005) 275}
  [\href{https://arxiv.org/abs/hep-th/0412103}{{\ttfamily hep-th/0412103}}].

\bibitem{Bern:2007ct}
Z.~Bern, J.~J.~M. Carrasco, H.~Johansson and D.~A. Kosower, \emph{{Maximally
  supersymmetric planar Yang-Mills amplitudes at five loops}},
  \href{https://doi.org/10.1103/PhysRevD.76.125020}{\emph{Phys. Rev. D}
  {\bfseries 76} (2007) 125020}
  [\href{https://arxiv.org/abs/0705.1864}{{\ttfamily 0705.1864}}].

\bibitem{Childers:2015tyv}
J.~T. Childers, T.~D. Uram, T.~J. LeCompte, M.~E. Papka and D.~P. Benjamin,
  \emph{{Adapting the serial Alpgen parton-interaction generator to simulate
  LHC collisions on millions of parallel threads}},
  \href{https://doi.org/10.1016/j.cpc.2016.09.013}{\emph{Comput. Phys. Commun.}
  {\bfseries 210} (2017) 54}
  [\href{https://arxiv.org/abs/1511.07312}{{\ttfamily 1511.07312}}].

\bibitem{Carrasco:2013mua}
J.~J.~M. Carrasco, S.~Foreman, D.~Green and L.~Senatore, \emph{{The Effective
  Field Theory of Large Scale Structures at Two Loops}},
  \href{https://doi.org/10.1088/1475-7516/2014/07/057}{\emph{JCAP} {\bfseries
  1407} (2014) 057} [\href{https://arxiv.org/abs/1310.0464}{{\ttfamily
  1310.0464}}].

\bibitem{Bjerrum-Bohr:2012kaa}
N.~E.~J. Bjerrum-Bohr, P.~H. Damgaard, R.~Monteiro and D.~O'Connell,
  \emph{{Algebras for Amplitudes}},
  \href{https://doi.org/10.1007/JHEP06(2012)061}{\emph{JHEP} {\bfseries 06}
  (2012) 061} [\href{https://arxiv.org/abs/1203.0944}{{\ttfamily 1203.0944}}].

\bibitem{Chen:2019ywi}
G.~Chen, H.~Johansson, F.~Teng and T.~Wang, \emph{{On the kinematic algebra for
  BCJ numerators beyond the MHV sector}},
  \href{https://doi.org/10.1007/JHEP11(2019)055}{\emph{JHEP} {\bfseries 11}
  (2019) 055} [\href{https://arxiv.org/abs/1906.10683}{{\ttfamily
  1906.10683}}].

\bibitem{Chen:2021chy}
G.~Chen, H.~Johansson, F.~Teng and T.~Wang, \emph{{Next-to-MHV Yang-Mills
  kinematic algebra}},
  \href{https://doi.org/10.1007/JHEP10(2021)042}{\emph{JHEP} {\bfseries 10}
  (2021) 042} [\href{https://arxiv.org/abs/2104.12726}{{\ttfamily
  2104.12726}}].

\bibitem{Boels:2013bi}
R.~H. Boels, R.~S. Isermann, R.~Monteiro and D.~O'Connell,
  \emph{{Colour-Kinematics Duality for One-Loop Rational Amplitudes}},
  \href{https://doi.org/10.1007/JHEP04(2013)107}{\emph{JHEP} {\bfseries 04}
  (2013) 107} [\href{https://arxiv.org/abs/1301.4165}{{\ttfamily 1301.4165}}].

\bibitem{Krasnov:2021cva}
K.~Krasnov and E.~Skvortsov, \emph{{Flat self-dual gravity}},
  \href{https://doi.org/10.1007/JHEP08(2021)082}{\emph{JHEP} {\bfseries 08}
  (2021) 082} [\href{https://arxiv.org/abs/2106.01397}{{\ttfamily
  2106.01397}}].

\bibitem{Cheung:2016iub}
C.~Cheung, A.~de~la Fuente and R.~Sundrum, \emph{{4D scattering amplitudes and
  asymptotic symmetries from 2D CFT}},
  \href{https://doi.org/10.1007/JHEP01(2017)112}{\emph{JHEP} {\bfseries 01}
  (2017) 112} [\href{https://arxiv.org/abs/1609.00732}{{\ttfamily
  1609.00732}}].

\bibitem{Chacon:2020fmr}
E.~Chac\'on, H.~Garc\'\i{}a-Compe\'an, A.~Luna, R.~Monteiro and C.~D. White,
  \emph{{New heavenly double copies}},
  \href{https://doi.org/10.1007/JHEP03(2021)247}{\emph{JHEP} {\bfseries 03}
  (2021) 247} [\href{https://arxiv.org/abs/2008.09603}{{\ttfamily
  2008.09603}}].

\bibitem{Fu:2016plh}
C.-H. Fu and K.~Krasnov, \emph{{Colour-Kinematics duality and the Drinfeld
  double of the Lie algebra of diffeomorphisms}},
  \href{https://doi.org/10.1007/JHEP01(2017)075}{\emph{JHEP} {\bfseries 01}
  (2017) 075} [\href{https://arxiv.org/abs/1603.02033}{{\ttfamily
  1603.02033}}].

\bibitem{Cheung:2017yef}
C.~Cheung, G.~N. Remmen, C.-H. Shen and C.~Wen, \emph{{Pions as Gluons in
  Higher Dimensions}},
  \href{https://doi.org/10.1007/JHEP04(2018)129}{\emph{JHEP} {\bfseries 04}
  (2018) 129} [\href{https://arxiv.org/abs/1709.04932}{{\ttfamily
  1709.04932}}].

\bibitem{Cho:2021nim}
K.~Cho, K.~Kim and K.~Lee, \emph{{The off-shell recursion for gravity and the
  classical double copy for currents}},
  \href{https://doi.org/10.1007/JHEP01(2022)186}{\emph{JHEP} {\bfseries 01}
  (2022) 186} [\href{https://arxiv.org/abs/2109.06392}{{\ttfamily
  2109.06392}}].

\bibitem{Bonezzi:2022yuh}
R.~Bonezzi, F.~Diaz-Jaramillo and O.~Hohm, \emph{{The Gauge Structure of Double
  Field Theory follows from Yang-Mills Theory}},
  \href{https://arxiv.org/abs/2203.07397}{{\ttfamily 2203.07397}}.

\bibitem{Alonso:2015fsp}
R.~Alonso, E.~E. Jenkins and A.~V. Manohar, \emph{{A Geometric Formulation of
  Higgs Effective Field Theory: Measuring the Curvature of Scalar Field
  Space}}, \href{https://doi.org/10.1016/j.physletb.2016.01.041}{\emph{Phys.
  Lett. B} {\bfseries 754} (2016) 335}
  [\href{https://arxiv.org/abs/1511.00724}{{\ttfamily 1511.00724}}].

\bibitem{Alonso:2016oah}
R.~Alonso, E.~E. Jenkins and A.~V. Manohar, \emph{{Geometry of the Scalar
  Sector}}, \href{https://doi.org/10.1007/JHEP08(2016)101}{\emph{JHEP}
  {\bfseries 08} (2016) 101}
  [\href{https://arxiv.org/abs/1605.03602}{{\ttfamily 1605.03602}}].

\bibitem{Cheung:2022vnd}
C.~Cheung, A.~Helset and J.~Parra-Martinez, \emph{{Geometry-Kinematics
  Duality}},  \href{https://arxiv.org/abs/2202.06972}{{\ttfamily 2202.06972}}.

\bibitem{Brandhuber:2021bsf}
A.~Brandhuber, G.~Chen, H.~Johansson, G.~Travaglini and C.~Wen,
  \emph{{Kinematic Hopf Algebra for Bern-Carrasco-Johansson Numerators in
  Heavy-Mass Effective Field Theory and Yang-Mills Theory}},
  \href{https://doi.org/10.1103/PhysRevLett.128.121601}{\emph{Phys. Rev. Lett.}
  {\bfseries 128} (2022) 121601}
  [\href{https://arxiv.org/abs/2111.15649}{{\ttfamily 2111.15649}}].

\bibitem{freedman2012supergravity}
D.~Freedman and A.~Van~Proeyen, \emph{Supergravity}. Cambridge University
  Press, 2012.

\bibitem{Cremmer:1979up}
E.~Cremmer and B.~Julia, \emph{{The SO(8) Supergravity}},
  \href{https://doi.org/10.1016/0550-3213(79)90331-6}{\emph{Nucl. Phys.}
  {\bfseries B159} (1979) 141}.

\bibitem{Cremmer:1977tt}
E.~Cremmer, J.~Scherk and S.~Ferrara, \emph{{SU(4) invariant supergravity
  theory}}, \href{https://doi.org/10.1016/0370-2693(78)90060-6}{\emph{Phys.
  Lett.} {\bfseries 74B} (1978) 61}.

\bibitem{Das:1977uy}
A.~K. Das, \emph{{SO(4) invariant extended supergravity}},
  \href{https://doi.org/10.1103/PhysRevD.15.2805}{\emph{Phys. Rev.} {\bfseries
  D15} (1977) 2805}.

\bibitem{Gunaydin1983bi}
M.~G\"unaydin, G.~Sierra and P.~K. Townsend, \emph{{The Geometry of ${\cal
  N}=2$ Maxwell-Einstein supergravity and Jordan algebras}},
  \href{https://doi.org/10.1016/0550-3213(84)90142-1}{\emph{Nucl. Phys.}
  {\bfseries B242} (1984) 244}.

\bibitem{Gunaydin1984ak}
M.~G\"unaydin, G.~Sierra and P.~K. Townsend, \emph{{Gauging the $d = 5$
  Maxwell-Einstein supergravity theories: More on Jordan algebras}},
  \href{https://doi.org/10.1016/0550-3213(85)90547-4}{\emph{Nucl. Phys.}
  {\bfseries B253} (1985) 573}.

\bibitem{Gunaydin1984nt}
M.~G\"unaydin, G.~Sierra and P.~K. Townsend, \emph{{Quantization of the gauge
  coupling constant in a five-dimensional {Yang-Mills} / Einstein supergravity
  theory}}, \href{https://doi.org/10.1103/PhysRevLett.53.322}{\emph{Phys. Rev.
  Lett.} {\bfseries 53} (1984) 322}.

\bibitem{Gunaydin1986fg}
M.~G\"unaydin, G.~Sierra and P.~K. Townsend, \emph{{More on $d=5$
  Maxwell-Einstein supergravity: Symmetric spaces and kinks}},
  \href{https://doi.org/10.1088/0264-9381/3/5/007}{\emph{Class. Quant. Grav.}
  {\bfseries 3} (1986) 763}.

\bibitem{Carrasco:2012ca}
J.~J.~M. Carrasco, M.~Chiodaroli, M.~G{\"u}naydin and R.~Roiban,
  \emph{{One-loop four-point amplitudes in pure and matter-coupled $N \le 4$
  supergravity}}, \href{https://doi.org/10.1007/JHEP03(2013)056}{\emph{JHEP}
  {\bfseries 03} (2013) 056} [\href{https://arxiv.org/abs/1212.1146}{{\ttfamily
  1212.1146}}].

\bibitem{Chiodaroli2013upa}
M.~Chiodaroli, Q.~Jin and R.~Roiban, \emph{{Color/kinematics duality for
  general abelian orbifolds of ${\cal N}=4$ super-Yang-Mills theory}},
  \href{https://doi.org/10.1007/JHEP01(2014)152}{\emph{JHEP} {\bfseries 01}
  (2014) 152} [\href{https://arxiv.org/abs/1311.3600}{{\ttfamily 1311.3600}}].

\bibitem{Chiodaroli:2015wal}
M.~Chiodaroli, M.~Gunaydin, H.~Johansson and R.~Roiban, \emph{{Complete
  construction of magical, symmetric and homogeneous N=2 supergravities as
  double copies of gauge theories}},
  \href{https://doi.org/10.1103/PhysRevLett.117.011603}{\emph{Phys. Rev. Lett.}
  {\bfseries 117} (2016) 011603}
  [\href{https://arxiv.org/abs/1512.09130}{{\ttfamily 1512.09130}}].

\bibitem{deWit1991nm}
B.~de~Wit and A.~Van~Proeyen, \emph{{Special geometry, cubic polynomials and
  homogeneous quaternionic spaces}},
  \href{https://doi.org/10.1007/BF02097627}{\emph{Commun. Math. Phys.}
  {\bfseries 149} (1992) 307}
  [\href{https://arxiv.org/abs/hep-th/9112027}{{\ttfamily hep-th/9112027}}].

\bibitem{Ben-Shahar:2018uie}
M.~Ben-Shahar and M.~Chiodaroli, \emph{{One-loop amplitudes for $ \mathcal{N} $
  = 2 homogeneous supergravities}},
  \href{https://doi.org/10.1007/JHEP03(2019)153}{\emph{JHEP} {\bfseries 03}
  (2019) 153} [\href{https://arxiv.org/abs/1812.00402}{{\ttfamily
  1812.00402}}].

\bibitem{Bern1999bx}
Z.~Bern, A.~De~Freitas and H.~L. Wong, \emph{{On the coupling of gravitons to
  matter}}, \href{https://doi.org/10.1103/PhysRevLett.84.3531}{\emph{Phys. Rev.
  Lett.} {\bfseries 84} (2000) 3531}
  [\href{https://arxiv.org/abs/hep-th/9912033}{{\ttfamily hep-th/9912033}}].

\bibitem{Chiodaroli:2016jqw}
M.~Chiodaroli, \emph{{Simplifying amplitudes in Maxwell-Einstein and
  Yang-Mills-Einstein supergravities}},  7, 2016,
  \href{https://arxiv.org/abs/1607.04129}{{\ttfamily 1607.04129}},
  \href{https://doi.org/10.1515/9783110452150-011}{DOI}.

\bibitem{Fu:2017uzt}
C.-H. Fu, Y.-J. Du, R.~Huang and B.~Feng, \emph{{Expansion of
  Einstein-Yang-Mills Amplitude}},
  \href{https://doi.org/10.1007/JHEP09(2017)021}{\emph{JHEP} {\bfseries 09}
  (2017) 021} [\href{https://arxiv.org/abs/1702.08158}{{\ttfamily
  1702.08158}}].

\bibitem{Chiodaroli:2017ngp}
M.~Chiodaroli, M.~Gunaydin, H.~Johansson and R.~Roiban, \emph{{Explicit
  Formulae for Yang-Mills-Einstein Amplitudes from the Double Copy}},
  \href{https://doi.org/10.1007/JHEP07(2017)002}{\emph{JHEP} {\bfseries 07}
  (2017) 002} [\href{https://arxiv.org/abs/1703.00421}{{\ttfamily
  1703.00421}}].

\bibitem{Wu:2021exa}
K.~Wu and Y.-J. Du, \emph{{Off-shell extended graphic rule and the expansion of
  Berends-Giele currents in Yang-Mills theory}},
  \href{https://doi.org/10.1007/JHEP01(2022)162}{\emph{JHEP} {\bfseries 01}
  (2022) 162} [\href{https://arxiv.org/abs/2109.14462}{{\ttfamily
  2109.14462}}].

\bibitem{He:2021lro}
S.~He, L.~Hou, J.~Tian and Y.~Zhang, \emph{{Kinematic numerators from the
  worldsheet: cubic trees from labelled trees}},
  \href{https://doi.org/10.1007/JHEP08(2021)118}{\emph{JHEP} {\bfseries 08}
  (2021) 118} [\href{https://arxiv.org/abs/2103.15810}{{\ttfamily
  2103.15810}}].

\bibitem{Dong:2021qai}
J.~Dong, S.~He and L.~Hou, \emph{{Universal expansions of scattering amplitudes
  for gravitons, gluons and Goldstone particles}},
  \href{https://arxiv.org/abs/2111.10525}{{\ttfamily 2111.10525}}.

\bibitem{Cachazo2014nsa}
F.~Cachazo, S.~He and E.~Y. Yuan, \emph{{Einstein-Yang-Mills scattering
  amplitudes from scattering equations}},
  \href{https://doi.org/10.1007/JHEP01(2015)121}{\emph{JHEP} {\bfseries 01}
  (2015) 121} [\href{https://arxiv.org/abs/1409.8256}{{\ttfamily 1409.8256}}].

\bibitem{Casali2015vta}
E.~Casali, Y.~Geyer, L.~Mason, R.~Monteiro and K.~A. Roehrig, \emph{{New
  Ambitwistor string theories}},
  \href{https://doi.org/10.1007/JHEP11(2015)038}{\emph{JHEP} {\bfseries 11}
  (2015) 038} [\href{https://arxiv.org/abs/1506.08771}{{\ttfamily
  1506.08771}}].

\bibitem{Adamo:2015gia}
T.~Adamo, E.~Casali, K.~A. Roehrig and D.~Skinner, \emph{{On tree amplitudes of
  supersymmetric Einstein-Yang-Mills theory}},
  \href{https://doi.org/10.1007/JHEP12(2015)177}{\emph{JHEP} {\bfseries 12}
  (2015) 177} [\href{https://arxiv.org/abs/1507.02207}{{\ttfamily
  1507.02207}}].

\bibitem{Stieberger2016lng}
S.~Stieberger and T.~R. Taylor, \emph{{New relations for Einstein--Yang--Mills
  amplitudes}},
  \href{https://doi.org/10.1016/j.nuclphysb.2016.09.014}{\emph{Nucl. Phys.}
  {\bfseries B913} (2016) 151}
  [\href{https://arxiv.org/abs/1606.09616}{{\ttfamily 1606.09616}}].

\bibitem{Nandan2016pya}
D.~Nandan, J.~Plefka, O.~Schlotterer and C.~Wen, \emph{{Einstein-Yang-Mills
  from pure Yang-Mills amplitudes}},
  \href{https://doi.org/10.1007/JHEP10(2016)070}{\emph{JHEP} {\bfseries 10}
  (2016) 070} [\href{https://arxiv.org/abs/1607.05701}{{\ttfamily
  1607.05701}}].

\bibitem{SchlottererEYMHeterotic}
O.~Schlotterer, \emph{{Amplitude relations in heterotic string theory and
  Einstein-Yang-Mills}},
  \href{https://doi.org/10.1007/JHEP11(2016)074}{\emph{JHEP} {\bfseries 11}
  (2016) 074} [\href{https://arxiv.org/abs/1608.00130}{{\ttfamily
  1608.00130}}].

\bibitem{TengFengBCJNumerators}
F.~Teng and B.~Feng, \emph{{Expanding Einstein-Yang-Mills by Yang-Mills in CHY
  frame}}, \href{https://doi.org/10.1007/JHEP05(2017)075}{\emph{JHEP}
  {\bfseries 05} (2017) 075}
  [\href{https://arxiv.org/abs/1703.01269}{{\ttfamily 1703.01269}}].

\bibitem{CheungUnifyingRelations}
C.~Cheung, C.-H. Shen and C.~Wen, \emph{{Unifying relations for scattering
  amplitudes}}, \href{https://doi.org/10.1007/JHEP02(2018)095}{\emph{JHEP}
  {\bfseries 02} (2018) 095}
  [\href{https://arxiv.org/abs/1705.03025}{{\ttfamily 1705.03025}}].

\bibitem{Roehrig:2017wvh}
K.~A. Roehrig, \emph{{Chiral splitting and $ \mathcal{N}=4 $
  Einstein-Yang-Mills tree amplitudes in 4d}},
  \href{https://doi.org/10.1007/JHEP08(2017)033}{\emph{JHEP} {\bfseries 08}
  (2017) 033} [\href{https://arxiv.org/abs/1705.09315}{{\ttfamily
  1705.09315}}].

\bibitem{Du2017gnh}
Y.-J. Du, B.~Feng and F.~Teng, \emph{{Expansion of all multitrace tree level
  EYM amplitudes}}, \href{https://doi.org/10.1007/JHEP12(2017)038}{\emph{JHEP}
  {\bfseries 12} (2017) 038}
  [\href{https://arxiv.org/abs/1708.04514}{{\ttfamily 1708.04514}}].

\bibitem{Mazloumi:2022lga}
P.~Mazloumi and S.~Stieberger, \emph{{Einstein Yang-Mills Amplitudes from
  Intersections of Twisted Forms}},
  \href{https://arxiv.org/abs/2201.00837}{{\ttfamily 2201.00837}}.

\bibitem{Porkert:2022efy}
F.~Porkert and O.~Schlotterer, \emph{{One-loop amplitudes in
  Einstein-Yang-Mills from forward limits}},
  \href{https://arxiv.org/abs/2201.12072}{{\ttfamily 2201.12072}}.

\bibitem{Samtleben2008pe}
H.~Samtleben, \emph{{Lectures on gauged supergravity and flux
  compactifications}},
  \href{https://doi.org/10.1088/0264-9381/25/21/214002}{\emph{Class. Quant.
  Grav.} {\bfseries 25} (2008) 214002}
  [\href{https://arxiv.org/abs/0808.4076}{{\ttfamily 0808.4076}}].

\bibitem{DallAgata:2011aa}
G.~Dall'Agata and G.~Inverso, \emph{{On the Vacua of N = 8 Gauged Supergravity
  in 4 Dimensions}},
  \href{https://doi.org/10.1016/j.nuclphysb.2012.01.023}{\emph{Nucl. Phys. B}
  {\bfseries 859} (2012) 70} [\href{https://arxiv.org/abs/1112.3345}{{\ttfamily
  1112.3345}}].

\bibitem{DallAgata:2012mfj}
G.~Dall'Agata, G.~Inverso and M.~Trigiante, \emph{{Evidence for a family of
  SO(8) gauged supergravity theories}},
  \href{https://doi.org/10.1103/PhysRevLett.109.201301}{\emph{Phys. Rev. Lett.}
  {\bfseries 109} (2012) 201301}
  [\href{https://arxiv.org/abs/1209.0760}{{\ttfamily 1209.0760}}].

\bibitem{Catino:2013ppa}
F.~Catino, G.~Dall'Agata, G.~Inverso and F.~Zwirner, \emph{{On the moduli space
  of spontaneously broken $N = 8$ supergravity}},
  \href{https://doi.org/10.1007/JHEP09(2013)040}{\emph{JHEP} {\bfseries 09}
  (2013) 040} [\href{https://arxiv.org/abs/1307.4389}{{\ttfamily 1307.4389}}].

\bibitem{Dallagata:2021lsc}
G.~Dall'agata, G.~Inverso and D.~Partipilo, \emph{{Old and new vacua of 5D
  maximal supergravity}},
  \href{https://doi.org/10.1007/JHEP04(2021)039}{\emph{JHEP} {\bfseries 04}
  (2021) 039} [\href{https://arxiv.org/abs/2101.04149}{{\ttfamily
  2101.04149}}].

\bibitem{Bobev:2020ttg}
N.~Bobev, T.~Fischbacher, F.~F. Gautason and K.~Pilch, \emph{{A cornucopia of
  AdS$_{5}$ vacua}}, \href{https://doi.org/10.1007/JHEP07(2020)240}{\emph{JHEP}
  {\bfseries 07} (2020) 240}
  [\href{https://arxiv.org/abs/2003.03979}{{\ttfamily 2003.03979}}].

\bibitem{Krishnan:2020sfg}
C.~Krishnan, V.~Mohan and S.~Ray, \emph{{Machine Learning ${\cal N}=8, D=5$
  Gauged Supergravity}},
  \href{https://doi.org/10.1002/prop.202000027}{\emph{Fortsch. Phys.}
  {\bfseries 68} (2020) 2000027}
  [\href{https://arxiv.org/abs/2002.12927}{{\ttfamily 2002.12927}}].

\bibitem{Chiodaroli:2018dbu}
M.~Chiodaroli, M.~G\"unaydin, H.~Johansson and R.~Roiban, \emph{{Non-Abelian
  gauged supergravities as double copies}},
  \href{https://doi.org/10.1007/JHEP06(2019)099}{\emph{JHEP} {\bfseries 06}
  (2019) 099} [\href{https://arxiv.org/abs/1812.10434}{{\ttfamily
  1812.10434}}].

\bibitem{Johansson:2018ues}
H.~Johansson, G.~Mogull and F.~Teng, \emph{{Unraveling conformal gravity
  amplitudes}}, \href{https://doi.org/10.1007/JHEP09(2018)080}{\emph{JHEP}
  {\bfseries 09} (2018) 080}
  [\href{https://arxiv.org/abs/1806.05124}{{\ttfamily 1806.05124}}].

\end{thebibliography}\endgroup

\end{document}